\documentclass[12pt]{article}
\makeatletter \@addtoreset{equation}{section} \makeatother
\renewcommand{\theequation}{\thesection.\arabic{equation}}
\newcommand{\ba}{\begin{array}}
\newcommand{\ea}{\end{array}}
\newcommand{\beq}{\begin{equation}}
\newcommand{\eeq}{\end{equation}}
\newcommand{\bea}{\begin{eqnarray}}
\newcommand{\eea}{\end{eqnarray}}

\def\bce{\begin{center}}
\def\ece{\end{center}}

\def\nonu{\nonumber}

\def\pa{\partial}

\def\be{\beta}

\def\ep{\epsilon}

\def\la{\lambda}

\def\si{\sigma}

\def\eps6{{\displaystyle \mathop{\epsilon}^{6}}{}}
\def\g6{{\displaystyle \mathop{g}^{6}}{}}

\def\nab6{{\displaystyle \mathop{\nabla}^{6}}{}}

\def\0{{\sst{(0)}}}
\def\1{{\sst{(1)}}}
\def\2{{\sst{(2)}}}
\def\3{{\sst{(3)}}}
\def\4{{\sst{(4)}}}
\def\5{{\sst{(5)}}}
\def\6{{\sst{(6)}}}
\def\7{{\sst{(7)}}}
\def\8{{\sst{(8)}}}

\def\ba{\begin{array}}
\def\ea{\end{array}}
\def\beq{\begin{equation}}
\def\eeq{\end{equation}}
\def\be{\begin{equation}}
\def\ee{\end{equation}}

\def\la{\lambda}
\def\eps{\epsilon}

\def\ba{\begin{array}}
\def\ea{\end{array}}
\def\beq{\begin{equation}}
\def\eeq{\end{equation}}
\def\be{\begin{equation}}
\def\ee{\end{equation}}

\def\la{\lambda}
\def\eps{\epsilon}

\def\eps6{{\displaystyle \mathop{\epsilon}^{6}}{}}

\def\nab6{{\displaystyle \mathop{\nabla}^{6}}{}}

\newcommand{\bean}{\begin{eqnarray*}}
\newcommand{\eean}{\end{eqnarray*}}

\begin{document}
\thispagestyle{empty} \addtocounter{page}{-1}
   \begin{flushright}
\end{flushright}

\vspace*{1.3cm}
  
\centerline{ \Large \bf The Next
  $16$ Higher Spin Currents and
}
\vspace*{0.3cm}
\centerline{ \Large \bf  Three-Point Functions
  in the Large ${\cal N}=4$ Holography} 
\vspace*{1.5cm}
\centerline{{\bf  Changhyun Ahn}, {\bf Dong-gyu Kim} and {\bf Man Hea Kim}
} 
\vspace*{1.0cm} 
\centerline{\it 
Department of Physics, Kyungpook National University, Taegu
41566, Korea} 
\vspace*{0.8cm} 
\centerline{\tt ahn, ehdrb430, manhea@knu.ac.kr
} 
\vskip2cm

\centerline{\bf Abstract}
\vspace*{0.5cm}

By using the known operator product expansions (OPEs)
between the lowest $16$  higher spin currents
of spins $(1, \frac{3}{2}, \frac{3}{2}, \frac{3}{2}, \frac{3}{2},
2,2,2,2,2,2, \frac{5}{2}, \frac{5}{2}, \frac{5}{2}, \frac{5}{2}, 3)$
in an extension of the large ${\cal N}=4$ linear superconformal
algebra, one determines the OPEs  
between the lowest $16$  higher spin currents
in an extension of the large ${\cal N}=4$ nonlinear superconformal
algebra for generic $N$ and $k$. The Wolf space coset contains the
group $G =SU(N+2)$ and the affine Kac-Moody spin $1$ current
has the level $k$. The next $16$ higher spin currents
of spins $(2,\frac{5}{2}, \frac{5}{2}, \frac{5}{2}, \frac{5}{2},
3,3,3,3,3,3, \frac{7}{2}, \frac{7}{2}, \frac{7}{2}, \frac{7}{2},4)$
arise in the
above OPEs.
The most general lowest higher spin $2$ current
in this multiplet
can be determined in terms of affine Kac-Moody spin $\frac{1}{2}, 1$
currents.
By careful analysis of the zero mode (higher spin)
eigenvalue equations,
the three-point functions of bosonic higher spin $2, 3, 4$
currents with two scalars
are obtained for finite $N$ and $k$.
Furthermore, we also analyze the three-point functions
of bosonic higher spin $2, 3, 4$ currents in the extension of
 the large ${\cal N}=4$ linear superconformal
algebra.
It turns out that the three-point functions of higher spin $2,3$
currents in the two cases are equal to each other at finite $N$
and $k$.
Under the large $(N,k)$ 't Hooft limit, the two descriptions for the
three-point functions of higher spin $4$ current
coincide with each other.
The higher spin extension of $SO(4)$ Knizhnik Bershadsky
algebra is described.

\baselineskip=18pt
\newpage
\renewcommand{\theequation}
{\arabic{section}\mbox{.}\arabic{equation}}

\tableofcontents


\section{ Introduction}


The operator product expansions (OPEs)
between the lowest $16$  higher spin currents
of spins $(1,\frac{3}{2}, \frac{3}{2}, \frac{3}{2}, \frac{3}{2},
2, 2, 2, 2, 2, 2, \frac{5}{2}, \frac{5}{2}, \frac{5}{2}, \frac{5}{2},3)$
in an extension of the large ${\cal N}=4$ linear superconformal algebra
\cite{STVplb,npb1988,Schoutens88,ST,Saulina}
have been found in \cite{AK1509} \footnote{The terminology `large'
  here (we do not take any limit)
  is nothing to do with the one in the
  large $N$ 't Hooft like limit later.}.
The right hand sides of these OPEs can be written in terms of 
the above lowest $16$ higher spin currents (up to quadratic),
the $16$ currents which 
generate the large ${\cal N}=4$ linear superconformal algebra
and the next $16$ higher spin currents
of spins
\bea
(2,\frac{5}{2}, \frac{5}{2}, \frac{5}{2}, \frac{5}{2},
3, 3, 3, 3, 3, 3, \frac{7}{2}, \frac{7}{2}, \frac{7}{2}, \frac{7}{2},4).
\label{second}
\eea
For fixed $N$ with
the group $G=SU(N+2)$ in the ${\cal N}=4$ coset model
\cite{ST},
the explicit results for the lowest (and next) $16$ higher spin currents
in terms of WZW currents are given in \cite{Ahn1504}.
According to the construction of \cite{GS},
one can decouple the spin $1$ current and the four
spin $\frac{1}{2}$ currents (among $16$ currents of
the large ${\cal N}=4$ linear superconformal algebra) and then
one obtains $11$ currents which
generate the large ${\cal N}=4$ nonlinear superconformal algebra
\cite{GS,VP,GPTV,GK}.
One can apply the mechanism of \cite{GS} to the above
lowest $16$  higher spin currents
in an extension of the large ${\cal N}=4$ linear superconformal algebra
and one obtains the 
lowest $16$  higher spin currents \cite{BCG}
implicitly (which are regular in the OPEs
with the above spin $1$ current and the four
spin $\frac{1}{2}$ currents)
in an extension of the large ${\cal N}=4$ nonlinear superconformal
algebra
\footnote{The higher spin currents (with boldface notation)
  in the `linear' version are defined as
  the ones in the extension of the large ${\cal N}=4$ `linear' superconformal
  algebra while the higher spin currents in the `nonlinear' version
  are defined as  the ones in the extension of the large
  ${\cal N}=4$ `nonlinear' superconformal
  algebra. The OPEs between the higher spin currents in the linear
  (and nonlinear) version are nonlinear.}.
For fixed $N$,
the explicit results for the lowest (and next) $16$ higher spin currents
in terms of WZW currents
were given in \cite{Ahn1311,Ahn1408} while for general $N$,
the lowest $16$ higher spin currents were given in \cite{AK1411}
in terms of WZW currents implicitly.

One way to obtain
the OPEs between the lowest $16$ higher spin currents
in the nonlinear version
is to use the known OPEs (in previous paragraph)
between the lowest $16$ higher spin currents
in the linear version by using the explicit relations \cite{BCG} between
these two
kinds of $16$ lowest higher spin currents
\footnote{One can also try to obtain these OPEs from the results of
  \cite{Ahn1408} by introducing the arbitrary coefficients in the right
  hand sides of the OPEs and using the Jacobi identities. }.
See also the equation (\ref{Phinonandlin}).
The next step is to reexpress the right hand sides of the OPEs
(written in terms of $16$ currents, the lowest $16$ higher spin currents
and other terms coming from the next $16$ higher spin
currents in the linear version)
in terms of $11$ currents and the lowest (and the next)
$16$ higher spin currents in the nonlinear
version with the help of the explicit relations between them \cite{GS,BCG}.
Of course, the right hand sides of the OPEs
do not contain the above
 spin $1$ current and the four
 spin $\frac{1}{2}$ currents and therefore they have simple form
 compared to the ones in the linear version \cite{AK1509}.
 Furthermore, there are also
 extra terms (the product of the above $11$ currents and
 the lowest $16$ higher spin currents which did not appear in the
 linear version)
 as well as the terms obtained by simply replacing the terms
 appearing in the linear version with them in the nonlinear version.

 In the context of the large ${\cal N}=4$ holography \cite{GG1305},
 one of the consistency checks
 for this duality  is to check the matching of the correlation functions
 both in the matrix extended higher spin theories on $AdS_3$
 and in the large ${\cal N}=4$ coset theories \cite{ST} in two dimensions
 \footnote{There is also other consistency check which can be described
   as the matching of BPS spectrum in both sides. Recently, in \cite{EGGL},
   by analyzing the BPS spectrum of string theory and supergravity
   theory on $AdS_3 \times {\bf S}^3 \times {\bf S}^3 \times
   {\bf S}^1$, it has been found that
   the BPS spectra
   of both descriptions agree. It would be interesting to obtain the
   BPS spectrum in the ${\cal N}=4$ coset theories
   and see whether this matches with the findings
   in \cite{EGGL}. See also \cite{Gaberdiel17}.}.
 See also the relevant works in \cite{GG1406,GG1501,GG1512}.
 The three-point functions of the (bosonic) higher spin $1,2,3$ currents
 (which are member of the above  lowest $16$ higher spin currents)
  both in the linear and nonlinear versions
 with two scalars have been found in \cite{AK1506}.
 One of the main features of this construction is that we do not have to
 obtain the explicit results for the higher spin currents
 in terms of WZW affine Kac-Moody currents  for generic
 $N$ (and $k$) manually
 where
 the bosonic spin $1$ affine Kac-Moody current
 has the level $k$,
 contrary to the other cases where
 the complete results are needed
 \cite{GH,GGHR,Ahn1111,Ahn1202,Ahn1211,Ahn1305,AK1308,Ahn1701}
 in the context of \cite{GG1011,GG1205,GG1207}
  except for the overall factor.
That is,  
the lowest $16$ higher spin currents in terms of
affine Kac Moody currents for several $N$-values ($N=3,5,7,9$)
are enough to determine the three-point functions
even at finite $N$ (and $k$).
It was very useful to use the package by Thielemans \cite{Thielemans}
together with the mathematica \cite{mathematica}.

It is natural to ask what 
the three-point functions of the (bosonic) higher spin $2,3,4$ currents
(among
the above next $16$ lowest higher spin currents in (\ref{second}))
with two scalars
are.
One realizes that the (new primary)
higher spin $2$ current, which plays the role of
the `lowest' current inside of the  next $16$  higher spin currents,
occurs in the OPE between the higher spin $1$ current and the
higher spin $3$ current belonging to the lowest $16$ higher spin currents.
However, this naive (and natural) candidate for the
higher spin $2$ current does not provide the $N \leftrightarrow k$
symmetry \cite{GG1305} in the three-point functions.
Therefore, we should look for the more general higher spin 2
current by adding the extra terms to the above
naive higher spin $2$ current.
It turns out that
these extra terms consist of the square of the higher spin $1$
current, the stress energy tensor, 
the product of the spin $1$ currents with $SO(4)$ Kronecker deltas
and the product of spin $1$ currents with $SO(4)$ epsilon tensor.
See the equation (\ref{generalSpintwoother})
\footnote{In the linear version there are additional terms also:
  the quadratic in the spin $1$ current, the product of spin $1$ current
  and the two spin $\frac{1}{2}$ currents with $SO(4)$ Kronecker deltas,
 the product of spin $1$ current
 and the two spin $\frac{1}{2}$ currents with $SO(4)$ epsilon tensor,
 the quartic terms in the spin $\frac{1}{2}$ currents with  $SO(4)$ epsilon tensor and the quadratic terms in the spin $\frac{1}{2}$ currents
 with a derivative. See also the equation (\ref{lineargeneralspintwo}).
 Note that the higher spin $2$ current living in the next $16$ higher
 spin currents in the nonlinear version is the same as the one in the linear
version. They have the same WZW currents.}.
One can easily see that the relative coefficients (which depend on
$N$ and $k$) between
these four terms are determined completely
by the ${\cal N}=4$ primary condition and the regular condition with the
above spin $1$ current and the four spin $\frac{1}{2}$ currents.

Starting from this general higher spin $2$ current
which can be written in terms of
WZW  currents  for generic
$N$ and $k$ (after reading off the general behaviors
for several $N$ values), one can determine other remaining
$15$ higher spin currents with the help of the four spin $\frac{3}{2}$
currents (supersymmetry generators) belonging to the generators
of the large ${\cal N}=4$ nonlinear superconformal algebra.
In order to do this, one should write down the OPEs
between the $11$ currents and
the next $16$ higher spin currents
from those in the linear version as done before.
As emphasized before,
the general $N$ behavior of the general higher spin $2$ current
is not necessary as long as the three-point functions are concerned
(and the determination of the remaining $15$ higher spin currents are
concerned).
By analyzing the zero mode (higher spin)
eigenvalue equations,
the three-point functions of (bosonic) higher spin $2, 3, 4$
currents with two scalars
are obtained for finite $N$ and $k$.
The three-point functions
of (bosonic) higher spin $2, 3, 4$ currents in the linear version
can be obtained.
It turns out that the three-point functions of higher spin $2,3$
currents in the two cases are equal to each other at finite $N$
and $k$.
Under the large $(N,k)$ 't Hooft limit, the two descriptions for the
three-point functions of higher spin $4$ current
coincide with each other.
The $N \leftrightarrow k$ symmetry for the
higher spin $4$ current arises only in the nonlinear version.

According to the observation of \cite{ST,GK},
the condition $k=N$ leads to 
the $SO({\cal N}=4)$ Knizhnik Bershadsky algebra \cite{Knizhnik,Bershadsky}
in the Wolf space coset model.
Then it is straightforward to obtain the
extension of $SO({\cal N}=4)$ Knizhnik Bershadsky algebra
from the previous results by restricting to $k=N$.

In section $2$, the $11$ currents and its large ${\cal N}=4$
nonlinear superconformal algebra are reviewed in $SU(2) \times SU(2)$
basis and $SO(4)$ basis.

In section $3$, the $16$ lowest higher spin currents
in the $SO(4)$ basis are reviewed.

In section $4$, from the known OPEs between the $16$ currents
of the large ${\cal N}=4$ linear superconformal algebra and the
$16$ higher spin currents, the OPEs
between the $11$ currents
of the large ${\cal N}=4$ nonlinear superconformal algebra and the
$16$ lowest higher spin currents are described.

In section $5$,
from the known OPEs between the $16$ lowest higher currents
in the linear version, the OPEs
between  the
$16$ lowest higher spin currents in the nonlinear version are described.
During this calculation, the general higher spin $2$ current
belonging to the next $16$ higher spin currents is found. 

In section $6$, the three-point functions
of the higher spin $2,3,4$ currents inside of the
next $16$ higher spin currents  with two scalars are obtained.

In section $7$, in the Wolf space coset realization of the
$SO(4)$ Knizhnik Bershadsky algebra, its higher spin extension is
described.

In section $8$, we list some future directions.

In Appendices $A$-$E$, some details appearing in previous sections
are provided.

An ancillary (mathematica) file $\tt ancillary.nb$, where
the OPEs appearing in Appendices $A$ and $B$ are given
and some eigenvalue equations (from which the three-point functions
can be determined) are presented, is included.

\section{ The $11$ currents and
  its large ${\cal N}=4$ `nonlinear' superconformal algebra in
the Wolf space: Review}

We describe the $11$ currents
which generate the  large ${\cal N}=4$ nonlinear superconformal algebra
in terms of Wolf space coset currents explicitly.

\subsection{The $11$ currents of the large ${\cal N}=4$
nonlinear superconformal algebra}

Let us consider the $11$ currents, which
generate the large ${\cal N}=4$ nonlinear superconformal algebra,
in terms of ${\cal N}=1$ affine Kac-Moody currents
of spin $1$ and spin $\frac{1}{2}$.
The $G^{\mu}(z)$ currents where $\mu =(0, i)$ with $i=1, 2, 3$
are four supersymmetry currents,
$A^{\pm i}(z)$ are six spin $1$ currents of $SU(2)_k \times SU(2)_N$
and $T(z)$ is the spin $2$ stress energy tensor.
They are given by \cite{VP,GK}
\bea
G^{0}(z) &  = &   \frac{i}{(k+N+2)}  \, Q_{\bar{a}} \, V^{\bar{a}}(z),
\qquad
G^{i}(z)  =  \frac{i}{(k+N+2)} 
\, h^{i}_{\bar{a} \bar{b}} \, Q^{\bar{a}} \, V^{\bar{b}}(z),
\nonu \\
A^{+i}(z) &  = & 
-\frac{1}{4N} \, f^{\bar{a} \bar{b}}_{\,\,\,\,\,\, c} \, h^i_{\bar{a} \bar{b}} \, V^c(z), 
\qquad
A^{-i}(z)  =  
-\frac{1}{4(k+N+2)} \, h^i_{\bar{a} \bar{b}} \, Q^{\bar{a}} \, Q^{\bar{b}}(z),
\nonu \\
T(z)  & = & 
\frac{1}{2(k+N+2)^2} \left[ (k+N+2) \, V_{\bar{a}} \, V^{\bar{a}} 
+k \, Q_{\bar{a}} \, \pa \, Q^{\bar{a}} 
+f_{\bar{a} \bar{b} c} \, Q^{\bar{a}} \, Q^{\bar{b}} \, V^c  \right] (z)
\nonu \\
&&- \frac{1}{(k+N+2)} \sum_{i=1}^3 \left( A^{+i}+A^{-i}  \right)^2 (z).
\label{11currents}
\eea
The spin $2$ stress energy
tensor $T(z)$ can be further written in terms of $V^a(z)$ and $Q^a(z)$
only when the $A^{\pm i}(z)$ is substituted.

Here the adjoint indices $a, b, \cdots$ corresponding to the group $G =SU(N+2)$
with $N$ odd
of Wolf space coset 
run over
\bea
a, b, \cdots = 1, 2, \cdots, \frac{1}{2} [ (N+2)^2-1],
\qquad
1^{\ast}, 2^{\ast}, \cdots, \frac{1}{2} [ (N+2)^2-1]^{\ast}.
\label{Gindex}
\eea
The total number of
generators, $ (N+2)^2-1$,
is divided into two pieces in the complex basis.
By subtracting the number of generators corresponding to
the subgroup $H=SU(N) \times SU(2) \times U(1)$, $(N^2-1)+4$,
we are left with the Wolf space coset indices
$4N
$.

The Wolf space coset indices
$\bar{a}, \bar{b}, \cdots$ corresponding to
the coset $\frac{G}{H}$
run over
\bea
\bar{a}, \bar{b}, \cdots = 1, 2, \cdots,N; N+1, N+2, \cdots, 2N;
1^{\ast}, 2^{\ast}, \cdots, N^{\ast}; (N+1)^{\ast}, (N+2)^{\ast}, \cdots, 2N^{\ast}.
\label{cosetindex}
\eea
One can assign the first $N$ indices of (\ref{cosetindex})
to the generators where each nonzero element $1$ appears
at the matrix elements
$(N+1,1), (N+1,2), \cdots, (N+1,N)$ (other matrix elements vanish).
Similarly, the next $N$ indices of (\ref{cosetindex})
can be assigned to the generators
with matrix elements
$(N+2,1), (N+2,2), \cdots, (N+2,N)$. 
The next $(2N+1)$ index of (\ref{Gindex}) can be assigned to the generator
with matrix element  $(N+2,N+1)$ corresponding to
the one of the generators of $SU(2) \times U(1)$ of $H$.

Among the half of $SU(N)$ indices of $H$,
$1, 2, \cdots, \frac{1}{2}(N^2-1)$, in the complex basis,
one can assign  the first $\frac{1}{2}N(N-1)$
indices
to the generators where each nonzero element $1$ appears
at the matrix elements
$(2,1), (3,1), (3, 2) \cdots, (N,1), (N,2), \cdots (N,N-1)$.
They correspond to the indices $(2N+2), \cdots, (2N+1)+\frac{1}{2}N(N-1)$
of (\ref{Gindex}).
The remaining $\frac{1}{2}(N-1)$ indices 
appear in the diagonal generators which are the combinations of $(N-1)$
Cartan
generators. Then they
correspond to the indices
$(2N+1)+\frac{1}{2}N(N-1)+1, \cdots, \frac{1}{2}(N^2+1) +2N$
of (\ref{Gindex}).
The next element
can be written as $\frac{1}{2}(N^2+1)+2N+1=\frac{1}{2}[(N+2)^2-1]$
which corresponds to the diagonal generator
which is a combination of two remaining Cartan generators
of $SU(N+2)$. This describes
another generator of $SU(2) \times U(1)$ of $H$.
One can describe the remaining half of the $SU(N+2)$
indices in (\ref{Gindex}) with
$\ast$ similarly. See the reference \cite{AK1506} for the nontrivial
diagonal generators with complex elements.

In summary, the half of the
index assignment of $SU(N+2)$ is as follows:
\bea
\mbox{Coset indices} & : & 1, 2, 3, \cdots, 2N, 
\nonu \\
\mbox{Subgroup} \, \, SU(2) \times U(1) \, \, \mbox{index} & : & 2N+1,
\nonu \\
\mbox{Subgroup} \, \, SU(N) \, \, \mbox{indices} & : & 2N+2, 2N+3, \cdots,
(2N+1)+\frac{1}{2}N(N-1),
\nonu \\
\mbox{Subgroup} \, \, SU(N) \, \, \mbox{indices} & : &
(2N+2)+\frac{1}{2}N(N-1), \cdots,
\frac{1}{2}(N^2+1) +2N,
\nonu \\
\mbox{Subgroup} \, \, SU(2) \times U(1) \, \, \mbox{index} & : &
\frac{1}{2}[(N+2)^2-1].
\label{Indexindex}
\eea
With $\ast$, one has the remaining half of the indices of $SU(N+2)$. 
For the off diagonal generators, this operation is equivalent to
take the transpose of the matrix and for the diagonal generators
this operation is to take the complex conjugation of the matrix.
Note that for the large ${\cal N}=4$ `linear' superconformal algebra,
the subgroup $SU(2) \times U(1)$ in (\ref{Indexindex})
is no longer subgroup and plays the role of
coset. 

The spin $1$ current $V^a(z)$ and the spin $\frac{1}{2}$
current $Q^a(z)$ appearing in (\ref{11currents})
satisfy the following OPE \cite{KT}
\bea
V^a(z) \, V^b(w) & = & \frac{1}{(z-w)^2} \, k \, g^{ab}
-\frac{1}{(z-w)} \, f^{ab}_{\,\,\,\,\,\,c} \, V^c(w) 
+\cdots,
\nonu \\
Q^a(z) \, Q^b(w) & = & -\frac{1}{(z-w)} \, (k+N+2) \, g^{ab} + \cdots,
\nonu \\
V^a(z) \, Q^b(w) & = & + \cdots.
\label{opevq}
\eea
The positive integer $k$ is the level of the spin $1$ current.
The metric $g_{ab}$ in (\ref{opevq})
is given by $g_{ab} = \mbox{Tr} (T_a T_b)$ and
the structure constant $f_{abc}$ is given by $f_{abc} = \mbox{Tr}
(T_c [T_a, T_b])$ where $T_a$ is the $SU(N+2)$ generator.
Note that the nonvanishing metric components are given by
$g_{A A^{\ast}} = g_{A^{\ast} A} =1$ where $A = 1, 2, \cdots,
\frac{1}{2}[(N+2)^2-1]$. Then by raising the $SU(N+2)$
adjoint lower index $A$, one has the $SU(N+2)$ adjoint upper
index $A^{\ast}$ and vice versa.
The role of upper index and the lower index
is different from each other.

The three almost complex structures $(h^1, h^2, h^3)$ 
appearing in (\ref{11currents})
are given by the following $4N \times 4N $ matrices (in the notation
of \cite{AK1506})
\bea
h^1_{\bar{a} \bar{b}} = 
\left(
\begin{array}{cccc}
0 & 0  & 0 & -i \\
0 & 0 & -i & 0 \\
0 & i & 0 & 0 \\
i & 0 & 0 & 0 \\
\end{array}
\right), \quad
h^2_{\bar{a} \bar{b}} = 
\left(
\begin{array}{cccc}
0 & 0  & 0 & 1 \\
0 & 0 & -1 & 0 \\
0 & 1 & 0 & 0 \\
-1 & 0 & 0 & 0 \\
\end{array}
\right), \qquad
h^{3}_{\bar{a} \bar{b}}
=
\left(
\begin{array}{cccc}
0 & 0  & i & 0 \\
0 & 0 & 0 & -i \\
-i & 0 & 0 & 0 \\
0 & i & 0 & 0 \\
\end{array}
\right),
\label{himatrix}
\eea
where each element is given by $N \times N$ matrix.
One can classify the above $4N$ indices as
four entries in (\ref{cosetindex}). 
Then the nonzero element $-i$ of $h^1$ in (\ref{himatrix})
appears when the row and the column
are given by the first ($1, 2, \cdots, N$) and last (
$(N+1)^{\ast}, (N+2)^{\ast}, \cdots, 2N^{\ast}$) entries respectively.
The next nonzero element $-i$ arises when the row and the column
are given by the second ($N+1, N+2, \cdots, 2N$) and the third ($1^{\ast},
2^{\ast},
\cdots, N^{\ast}$) entries.
Similarly,   the nonzero element $i$ occurs when the row and the column
are given by the third ($1^{\ast},
2^{\ast},\cdots, N^{\ast}$) and second ($N+1, N+2, \cdots, 2N$) entries.
Finally,
 the nonzero element $i$ can appear when the row and the column
 are given by the last ($(N+1)^{\ast}, (N+2)^{\ast}, \cdots, 2N^{\ast}$)
 and first ($1, 2, \cdots, N$) entries.
One can also analyze the $h^2$ and $h^3$ matrices according to the
coset indices (\ref{cosetindex})
\footnote{
\label{empha}
  Let us emphasize that there is a
summation over index $c$ in the spin $1$
current $A^{+i}(z)$ and this $c$ runs over $c=1, 2, \cdots,
\frac{1}{2}[(N+2)^2-1], 1^{\ast}, 2^{\ast}, \cdots, \frac{1}{2}[(N+2)^2-1]^{\ast}$
along the line of (\ref{Gindex}).
This behavior also occurs in the spin $2$ stress energy tensor $T(z)$.
The remaining summations run over the coset indices
where $\bar{a}, \bar{b}, \cdots=
1, 2, \cdots, 2N,
1^{\ast}, 2^{\ast}, \cdots, 2N^{\ast}$ with (\ref{cosetindex}).
Note that one obtains
$
[V^a_m, V^b_n]  =  k m g^{ab} \delta_{m+n,0} - f^{ab}_{\,\,\,\,\,\,c} V^c_{m+n}
$ from (\ref{opevq}). For the indices $m=0=n$,
this leads to $[V^a_0, V^b_0]  =   -
f^{ab}_{\,\,\,\,\,\,c} V^c_0$ which is related to the commutation relations
of the underlying finite dimensional Lie algebra $SU(N+2)$.}. 

\subsection{The large ${\cal N}=4$ nonlinear superconformal algebra}

The large ${\cal N}=4$ nonlinear superconformal algebra
generated by the above $11$ currents living in the
Wolf space coset
is summarized by \cite{GS}
\bea 
T(z) \, T(w) &=& \frac{1}{(z-w)^4} \, \frac{1}{2}
\Bigg[\frac{3(k+N+2k N)}{(k+N+2)} \Bigg]
+
\frac{1}{(z-w)^2} \, 2 T(w) +\frac{1}{(z-w)} \, \pa T(w)
\nonu \\
& + & \cdots,
\nonu \\
T(z) \,
\left(
\begin{array}{c}
G^{\mu} \\
A^{\pm i} \\
\end{array}
\right)(w) & = & \frac{1}{(z-w)^2}
\left(
\begin{array}{c}
\frac{3}{2} G^{\mu} \\
A^{\pm i} \\
\end{array}
\right)
(w) + \frac{1}{(z-w)} \, 
\left(
\begin{array}{c}
\pa G^{\mu} \\
\pa A^{\pm i} \\
\end{array}
\right)
(w) + \cdots,
\nonu \\
G^{\mu}(z) \, G^{\nu}(w) &=& \frac{1}{(z-w)^3} \,
\frac{2}{3} \delta^{\mu \nu}
\Bigg[ \frac{6 k N}{(2+k+N)}\Bigg] 
\nonu \\
& - & \frac{1}{(z-w)^2} \, \frac{8}{(k+N+2)}
( N \, \alpha^{+i}_{\mu \nu} \, A^{+}_i      
+k \, \alpha^{-i}_{\mu \nu} \, A^{-}_i  )(w)
\nonu \\
& + & \frac{1}{(z-w)} \, \left[ 2 \delta^{\mu \nu} T
-  \frac{4}{(k+N+2)} \pa ( N \, \alpha^{+i}_{\mu \nu} \, A^{+}_i      
+k \, \alpha^{-i}_{\mu \nu} \, A^{-}_i  ) \right.
\nonu \\
&-& \left.  \frac{8}{(k+N+2)}  ( \alpha^{+i} A^{+}_i      
- \alpha^{-i} A^{-}_i  )_{\rho (\mu}
( \alpha^{+j} A^{+}_j      
- \alpha^{-j} A^{-}_j  )_{\nu)}^{\,\,\,\,\, \rho} \right](w) +\cdots,
\nonu \\
A^{\pm i}(z) \, G^{\mu} (w) &=& 
\frac{1}{(z-w)} \, \alpha^{\pm i}_{\mu \nu} \, G^{\nu} (w)
+\cdots,
\nonu \\
A^{\pm i}(z) \, A^{\pm j}(w) &=& 
-\frac{1}{(z-w)^2} \, \frac{1}{2} \, \delta^{ij} \,
\left(
\begin{array}{c}
k \\
N \\
\end{array}
\right)
+ \frac{1}{(z-w)} \, \epsilon^{ijk}  A^{\pm k} (w) + \cdots,
\label{n4sca}
\eea
where
the quantity $\alpha^{\pm i} $ is defined by
\bea
\alpha^{\pm i}_{\mu \nu} & \equiv &
\frac{1}{2} \left( \pm \delta_{i \mu} \delta_{ \nu 0} \mp \delta_{i \nu} \delta_{ \mu 0}
+\ep_{i \mu \nu} \right).
\label{alphadef}
\eea
The quadratic nonlinear structures appear in the OPEs
between the spin $\frac{3}{2}$ currents.
The two levels of the large
${\cal N}=4$ nonlinear superconformal algebra are
identified with the Wolf space quantities
$k$ and $N$ respectively.
The equivalent OPE in different basis is presented in Appendix
$A.1$ \footnote{The explicit relations between the $11$ currents
  in two bases are
  given by
  \bea
  T(z) & \rightarrow &  L(z), \qquad
  G^0(z) \rightarrow G^2(z), \qquad
  G^1(z) \rightarrow G^3(z), \qquad
   G^2(z) \rightarrow -G^4(z), \qquad
   G^3(z) \rightarrow G^1(z), \nonu \\
   A^{\pm 1}(z) & \rightarrow & \frac{i}{2} ( T^{14} \mp T^{23})(z)
   = \frac{i}{2} \, \alpha_{\mu\nu}^{\pm 1} \, T^{\mu\nu}(z),
   \qquad
   A^{\pm 2}(z) \rightarrow \frac{i}{2} ( T^{13} \pm T^{24})(z)
   =\frac{i}{2} \, \alpha_{\mu\nu}^{\pm 2} \, T^{\mu\nu}(z),
   \nonu \\
   A^{\pm 3}(z) & \rightarrow & \pm
   \frac{i}{2} ( T^{12} \mp T^{34})(z)=
   \frac{i}{2} \, \alpha_{\mu\nu}^{\pm 3} \, T^{\mu\nu}(z),
   \label{replacement}
\eea
where the previous quantity in (\ref{alphadef}) with the renaming of
indices is changed into
   \bea
\alpha^{\pm 1}_{\mu\nu}  = 
\left(
\begin{array}{cccc}
0 & 0 & 0 & \frac{1}{2} \\
0 & 0 & \mp \frac{1}{2} & 0 \\
0 & \pm \frac{1}{2} & 0 & 0  \\
-\frac{1}{2} & 0 & 0 & 0 \\
\end{array}
\right),
\alpha^{\pm 2}_{\mu\nu}  = 
\left(
\begin{array}{cccc}
0 & 0 &  \frac{1}{2} & 0 \\
0 & 0 & 0 & \pm \frac{1}{2} \\
- \frac{1}{2} & 0 & 0 & 0  \\
0 &  \mp \frac{1}{2} & 0 & 0 \\
\end{array}
\right),
\alpha^{\pm 3}_{\mu\nu}  = 
\left(
\begin{array}{cccc}
0 &  \pm \frac{1}{2} & 0 & 0 \\
 \mp \frac{1}{2} & 0 & 0 & 0 \\
0 & 0 & 0 & - \frac{1}{2}  \\
0 & 0 &  \frac{1}{2} & 0 \\
\end{array}
\right).
\label{threealpha}
\eea}.
The central term appearing in the fourth order pole in the OPE
between the stress energy tensor and itself is given by
$\frac{c}{2}$ where $c$ is a central charge.
The central term  appearing in the third order pole in the OPE
between the spin $\frac{3}{2}$ currents is given by
$\frac{2c_W}{3}$ where $c_W$ is the Wolf space coset central charge.

\section{ The lowest $16$ higher spin currents  in
the Wolf space: Review}

We describe the lowest $16$ higher spin currents  in terms of
the Wolf space coset currents explicitly.

For fixed spin $s \geq 1$,
the $16$ higher spin currents in $SO(4)$ basis
can be described as \cite{AK1509}
\bea
  \mbox{spin} \, s &:&   \Phi_{0}^{(s)}(z),
\nonu \\
   \mbox{spin} \, (s + \frac{1}{2}) &:& \Phi_{\frac{1}{2}}^{(s),1}(z),
\Phi_{\frac{1}{2}}^{(s),2}(z), \Phi_{\frac{1}{2}}^{(s),3}(z),
\Phi_{\frac{1}{2}}^{(s),4}(z),
\nonu \\
\mbox{spin} \, (s+1) &:&   \Phi_{1}^{(s),12}(z),  \Phi_{1}^{(s),13}(z),
  \Phi_1^{(s),14}(z), \Phi_1^{(s),23}(z),
  \Phi_1^{(s),24}(z), \Phi_1^{(s),34}(z),
  \nonu \\
 \mbox{spin} \, (s+\frac{3}{2}) & : & \Phi_{\frac{3}{2}}^{(s),1}(z),
  \Phi_{\frac{3}{2}}^{(s),2}(z), \Phi_{\frac{3}{2}}^{(s),3}(z),
  \Phi_{\frac{3}{2}}^{(s),4}(z),
  \nonu \\
  \mbox{spin} \, (s+2) &:& \Phi_{2}^{(s)}(z).
\label{phis}
\eea
There are a single $SO(4)$ singlet  $\Phi_{0}^{(s)}(z)$,
four $SO(4)$ vector  $\Phi_{\frac{1}{2}}^{(s), \mu}(z)$,
six $SO(4)$ adjoint  $\Phi_{1}^{(s), \mu \nu}(z)$,
four $SO(4)$ vector $ \Phi_{\frac{3}{2}}^{(s),\mu}(z)$
and a single $SO(4)$ singlet  $\Phi_{2}^{(s)}(z)$.
For $s=0$, the large ${\cal N}=4$ nonlinear superconformal algebra
is generated by
six  $SO(4)$ adjoint  $T^{\mu \nu}(z)$ of spin $1$,
four $SO(4)$ vector $ G^{\mu}(z)$ of spin $\frac{3}{2}$
and a single $SO(4)$ singlet  $L(z)$ of spin $2$ discussed in previous
section \footnote{
  One can consider this $16$ higher spin currents (\ref{phis})
  in $SU(2) \times SU(2)$
  basis \cite{BCG} and their precise relations are given as follows:
  \bea
V_0^{(s)}(z) & = &   -i \, \Phi_0^{(s)}(z), 
\qquad
V_{\frac{1}{2}}^{(s),1}(z)  = 
 i \, \Phi_{\frac{1}{2}}^{(s),1}(z),
\qquad
V_{\frac{1}{2}}^{(s),2}(z)  = 
-i \, \Phi_{\frac{1}{2}}^{(s),2}(z), 
\nonu \\
V_{\frac{1}{2}}^{(s),3}(z)  & = &   
 -i \, \Phi_{\frac{1}{2}}^{(s),3}(z),\qquad
V_{\frac{1}{2}}^{(s),4}(z)  =  
 -i \, \Phi_{\frac{1}{2}}^{(s),4}(z), 
\qquad
V_{1}^{(s), \pm i}=\mp i \alpha^{\pm i}_{\mu \nu}\,\Phi_{1}^{(s),\mu \nu}, 
\nonu \\
V_{\frac{3}{2}}^{(s), 1}(z)   & = &   
-2 i \,
\Phi_{\frac{3}{2}}^{(s), 1}(z),  
\qquad
V_{\frac{3}{2}}^{(s), 2}(z)  =  
2 i \,
\Phi_{\frac{3}{2}}^{(s), 2} (z),   
\nonu \\
V_{\frac{3}{2}}^{(s), 3}(z)  & = &  
2 i \,
\Phi_{\frac{3}{2}}^{(s), 3}(z), 
\qquad
V_{\frac{3}{2}}^{(s), 4}(z)  =  
2 i \,
\Phi_{\frac{3}{2}}^{(s), 4} (z),
\label{VPhinon}
\\
V_{2}^{(s)}(z) & = & 2i\,\Bigg[\,\Phi_{2}^{(s)}
  -
  \frac{24 i (k - N) s (1 + s)}{3 k + 3 N + 6 k N - 20 s - 4 k s - 4 N s + 12 k N s + 32 s^2 + 
 16 k s^2 + 16 N s^2}\,L\Phi_{0}^{(s)}
\nonu\\
& + &  \frac{36 i  (k - N) s (1 + s)}{(1 + 2 s) (3 k + 3 N + 6 k N - 20 s - 4 k s - 4 N s + 12 k N s + 
   32 s^2 + 16 k s^2 + 16 N s^2)}\partial^2 \Phi_{0}^{(s)}\,\Bigg](z),
\nonu
\eea
where the $\alpha_{\mu\nu}^{\pm i}$ tensor can be
obtained from the ones in \cite{BCG} by changing the sign of
index $1$ appearing in the row or column of  $\alpha_{\mu\nu}^{\pm i}$ in
\cite{BCG},
\bea
\alpha^{\pm 1}_{\mu \nu}  = 
\left(
\begin{array}{cccc}
0 & 0 & 0 & \mp \frac{1}{2} \\
0 & 0 &  \frac{1}{2} & 0 \\
0 & - \frac{1}{2} & 0 & 0  \\
\pm \frac{1}{2} & 0 & 0 & 0 \\
\end{array}
\right), 
\alpha^{\pm 2}_{\mu \nu}  = 
\left(
\begin{array}{cccc}
0 & 0 &  \frac{1}{2} & 0 \\
0 & 0 & 0 & \pm \frac{1}{2} \\
- \frac{1}{2} & 0 & 0 & 0  \\
0 &  \mp \frac{1}{2} & 0 & 0 \\
\end{array}
\right),
\alpha^{\pm 3}_{\mu \nu}  = 
\left(
\begin{array}{cccc}
0 &  - \frac{1}{2} & 0 & 0 \\
 \frac{1}{2} & 0 & 0 & 0 \\
0 & 0 & 0 & \pm \frac{1}{2}  \\
0 & 0 & \mp \frac{1}{2} & 0 \\
\end{array}
\right).
\label{otheralpha}
\eea
The locations for the nonzero elements in (\ref{otheralpha})
are the same as previous ones (\ref{threealpha}).
There are $({\bf 1},{\bf 1})$ for $V_0^{(s)}(z)$,
$({\bf 2},{\bf 2})$ for $V_{\frac{1}{2}}^{(s),\mu}(z)$,
$({\bf 3},{\bf 1}) \oplus ({\bf 1},{\bf 3})$ for
$V_1^{(s), \pm i}(z)$,
$({\bf 2},{\bf 2})$ for $V_{\frac{3}{2}}^{(s),\mu}(z)$,
and $({\bf 1},{\bf 1})$ for $V_2^{(s)}(z)$
under
the $SU(2) \times SU(2)$.}.
Furthermore,
there are additional spin $0$ current which is a $SO(4)$
singlet and four spin $\frac{1}{2}$
currents transforming as a $SO(4)$ vector
in the large ${\cal N}=4$ linear superconformal algebra
(In the right hand sides of OPEs, there exist only the derivatives of
above spin $0$ current).



\subsection{Higher spin $1$ current}

It is known that the lowest higher spin $1$ current  
is described as \cite{AK1411}
\bea
\Phi_0^{(1)} (z) &=&
-\frac{1}{2(k+N+2)} \, d^0_{\bar{a} \bar{b}} \,
f^{\bar{a} \bar{b}}_{\,\,\,\,\,\, c}  V^c  (z)
+ \frac{k}{2(k+N+2)^2} \, d^0_{\bar{a} \bar{b}} \, Q^{\bar{a}} \, Q^{\bar{b}} (z),
\label{finalspinone}
\eea
where the antisymmetric  $d$ tensor of rank $2$ is given by
$4N \times 4N$ matrix as follows:
\bea
d^0_{\bar{a} \bar{b}}  = 
\left(
\begin{array}{cccc}
0 & 0 & -1 & 0 \\
0 & 0 & 0 & -1 \\
1 & 0 & 0 & 0  \\
0 & 1 & 0 & 0 \\
\end{array}
\right).
\label{dzero}
\eea
Each element is $N \times N$ matrix. The locations of the nonzero
elements of this matrix are the same as the previous almost complex
structure $h^3_{\bar{a} \bar{b}}$ but numerical values are different from each
other. As emphasized in the footnote \ref{empha}, the summation over
$c$ index in (\ref{finalspinone}) runs over the whole range of $SU(N+2)$
adjoint indices (\ref{Gindex}).
This higher spin $1$ current plays the role of the `generator'
of the next higher spin currents because one can construct them
using the OPEs between the spin $\frac{3}{2}$ currents of the
large ${\cal N}=4$ nonlinear superconformal algebra and  the
higher spin $1$ current. 

\subsection{Other higher spin currents}

Let us define the four higher spin-$\frac{3}{2}$ currents  $G'^{\mu} (z)$
from the first order pole of the following OPE \cite{AK1411}
\bea
G^{\mu} (z) \, \Phi_0^{(1)} (w) 
&=&
\frac{1}{(z-w)} \, G'^{\mu} (w) + \cdots.
\label{Gprime}
\eea
Then  the first order pole in (\ref{Gprime})
can be obtained from (\ref{11currents}) and (\ref{finalspinone})
with the help of (\ref{opevq})
\bea
G'^{\mu}(z) &=&
\frac{i}{(k+N+2)} \, 
d^{\mu}_{\bar{a} \bar{b}} \, Q^{\bar{a}} \, V^{\bar{b}} (z),
\label{gprimemu}
\eea
where  $d^{\mu}_{\bar{a} \bar{b}} \equiv d^{0 \bar{c} }_{ \bar{a} } \, 
h^{\mu }_{\bar{c} \bar{b}}$ and $h^0_{\bar{a} \bar{b}} \equiv g_{\bar{a} \bar{b}}$
with (\ref{dzero}).
These four independent higher spin-$\frac{3}{2}$ currents
(\ref{gprimemu})
also appear in the linear version.
It is easy to see that the exact relations
between the higher spin $\frac{3}{2}$ currents in \cite{AK1411}
and those in (\ref{phis}) are given by
\footnote{
  We have the following relations between the higher spin $\frac{3}{2}$
  currents in the $SU(2) \times SU(2)$ basis of \cite{BCG} and
  those in (\ref{phis})
  as follows:
\bea
  G'^{11}(z) & = &
  -\frac{1}{\sqrt{2}}\,( \Phi_{\frac{1}{2}}^{(1),3} +i\,
  \Phi_{\frac{1}{2}}^{(1),4} )(z),
\qquad
G'^{12}(z)  = 
\frac{1}{\sqrt{2}}\,( \Phi_{\frac{1}{2}}^{(1),1} -i\,\Phi_{\frac{1}{2}}^{(1),2} )(z),
\nonu\\
G'^{21}(z) & = &
\frac{1}{\sqrt{2}}\,( \Phi_{\frac{1}{2}}^{(1),1} +
i\,\Phi_{\frac{1}{2}}^{(1),2} )(z),
\qquad
G'^{22}(z)  = 
-\frac{1}{\sqrt{2}}\,( \Phi_{\frac{1}{2}}^{(1),3} -
i\,\Phi_{\frac{1}{2}}^{(1),4} )(z).
\label{gprime}
\eea
One can read off 
the currents $\Phi_{\frac{1}{2}}^{(1),\mu}(z) $ from (\ref{gprime}).}
\bea
T_{\pm}^{(\frac{3}{2})}(z) & = & 
\frac{1}{2\sqrt{2}}\,(\Phi_{\frac{1}{2}}^{(1),1} \pm i\,
\Phi_{\frac{1}{2}}^{(1),2} \pm G^{1}+i\,G^{2})(z),
\nonu \\
\left(
\begin{array}{c}
U^{(\frac{3}{2})} \\
 V^{(\frac{3}{2})} \\
\end{array}
\right)
(z) & = &
-\frac{1}{2\sqrt{2}}\,(\Phi_{\frac{1}{2}}^{(1),3} \pm i\,
\Phi_{\frac{1}{2}}^{(1),4}+G^{3} \pm i\,G^{4})(z).
\label{spin3halfrelations}
\eea

Then the six higher spin $2$ currents  can be obtained 
from the OPEs between the spin $\frac{3}{2}$ currents of the
large ${\cal N}=4$ nonlinear superconformal algebra and the above
higher spin $\frac{3}{2}$ currents (\ref{spin3halfrelations}).
Then the four higher spin $\frac{5}{2}$ currents
can be determined by the OPEs
 between the spin $\frac{3}{2}$ currents of the
large ${\cal N}=4$ nonlinear superconformal algebra and the newly obtained
higher spin $2$ currents. The final higher spin $3$ current
can be obtained from the OPEs
 between the spin $\frac{3}{2}$ currents of the
large ${\cal N}=4$ nonlinear superconformal algebra and the newly obtained
higher spin $\frac{5}{2}$ currents.
In this way, all the higher spin currents in (\ref{phis})
can be written
in terms of  $\mathcal N =1$ Kac-Moody currents $V^a(z), Q^{\bar{b}}(z)$, 
the three almost complex structures
 $h^i_{\bar{a} \bar{b}} $, antisymmetric tensor 
 $d^{0}_{\bar{a} \bar{b}}$ of rank $2$ and symmetric tensors 
 $d^{i}_{\bar{a} \bar{b}}$  ($ \equiv d^{0 \bar{c} }_{ \bar{a} } \, 
h^{i }_{\bar{c} \bar{b}} $) of rank $2$
\footnote{
  The higher spin $\frac{5}{2}$ currents have the following
  relations
\bea
W_{\pm}^{(\frac{5}{2})}(z) & = &
\frac{1}{2\sqrt{2}}\,( \pm \Phi_{\frac{3}{2}}^{(1),1}
+i\,\Phi_{\frac{3}{2}}^{(1),2})(z)+\frac{4}{3\sqrt{2}(2+k+N)}\,(\partial G^{1}
\pm i\,\partial G^{2})(z) \nonu\\
& + &  \frac{\varepsilon^{1 2 \mu\nu}}{\sqrt{2}(2+k+N)}\,\Bigg[\,(T^{1\mu}
 \pm i\,T^{2\mu})\Phi_{\frac{1}{2}}^{(1),\nu}
+ (\pm \widetilde{T}^{1\mu}+i\,\widetilde{T}^{2\mu})G^{\nu}
+\,T^{\mu\nu}(\Phi_{\frac{1}{2}}^{(1),1}
\pm i\,\Phi_{\frac{1}{2}}^{(1),2})
\nonu\\
&+ &  \widetilde{T}^{\mu\nu}( \pm G^{1}+i\,G^{2})\,\Bigg](z),
\nonu\\
\left(
\begin{array}{c}
U^{(\frac{5}{2})} \\
 V^{(\frac{5}{2})} \\
\end{array}
\right)(z) & = &
-\frac{1}{2\sqrt{2}}\, ( \pm
\Phi_{\frac{3}{2}}^{(1),3}+i\,\Phi_{\frac{3}{2}}^{(1),4})(z)
\nonu\\
&-& \frac{2}{3\sqrt{2}(2+N+k)}\,
(\partial \Phi_{\frac{1}{2}}^{(1),3}
\pm i\,\partial \Phi_{\frac{1}{2}}^{(1),4}
\pm \,\partial G^{3}
+i\,\partial G^{4} )(z) \nonu\\
& - &
\frac{\varepsilon^{34 \mu\nu}}{2\sqrt{2}(2+N+k)}\,
\Bigg[\,
  (\widetilde{T}^{3\mu}\pm i\,\widetilde{T}^{4\mu})
  (\pm \Phi_{\frac{1}{2}}^{(1),\nu}+G^{\nu})
+(T^{3\mu}\pm i\,T^{4\mu})(3\,\Phi_{\frac{1}{2}}^{(1),\nu}\mp G^{\nu})
 \nonu\\
& + &   \widetilde{T}^{\mu\nu}(\pm \Phi_{\frac{1}{2}}^{(1),3}
+i\,\Phi_{\frac{1}{2}}^{(1),4}
+G^{3}
\pm i\,G^{4})
 +T^{\mu\nu}(\Phi_{\frac{1}{2}}^{(1),3}\pm i\,\Phi_{\frac{1}{2}}^{(1),4}
 \pm  G^{3}+i\,G^{4})
 \,\Bigg](z).
\label{5halfexpression}
\eea
One can read off 
the currents $\Phi_{\frac{3}{2}}^{(1),\mu}(z) $ from
(\ref{5halfexpression}).
The explicit relations between the higher spin $2, 3$ currents
will appear in section $6$ later. }.

\section{ The  OPEs between the $11$ currents and the
   $16$ higher spin currents
for generic $N$ and $k$}

We describe
how  the OPEs between the $11$ currents and the
   $16$ higher spin currents
  arise
  and
we  present them in Appendix $A.2$.
  
\subsection{The $11$ currents using the $16$ currents}

Recall that
the explicit relation between the
$11$ currents of the large ${\cal N}=4$ nonlinear superconformal
algebra and the $16$ currents (with boldface notation) of the
large ${\cal N}=4$ linear superconformal algebra
is described by \cite{GS}
\bea
T^{\mu \nu}(z) & = &
{\bf T^{\mu \nu}}(z)-\frac{2i}{(2+k+N)}\,{\bf \Gamma^{\mu} \Gamma^{\nu}}(z),
\nonu\\
G^{\mu}(z) & = & {\bf G^{\mu}}(z)-\frac{2i}{(2+k+N)}\,{\bf U\Gamma^{\mu}}(z)
\nonu \\
& - & \varepsilon^{\mu\nu\rho\si} \Bigg[  
\frac{4i}{3(2+k+N)^2}{\bf \Gamma^{\nu}\Gamma^{\rho}\Gamma^{\si}}
-\frac{1}{(2+k+N)}{\bf T^{\nu\rho} \Gamma^{\si}}
\Bigg](z),
\nonu\\
L(z) & = & 
{\bf L}(z)+\frac{1}{(2+k+N)}\,\Bigg[\,{\bf UU}-{\bf
  \partial \Gamma^{\mu} \Gamma^{\mu}}\,\Bigg](z).
\label{gsformula}
\eea

The four fermionic spin $\frac{1}{2}$ currents are \cite{Saulina,AK1506}
\bea
{\bf \Gamma}^0 (z) =-\frac{i }{4(N+1)}
h^j_{\tilde{a} \tilde{b} }
f^{\tilde{a} \tilde{b}}_{\,\,\,\,\,\, \tilde{c}} h^{j \tilde{c}}_{ \,\,\,\, \tilde{d} } Q^{\tilde{d}}(z),
\qquad
{\bf \Gamma}^j (z) =-\frac{i }{4(N+1)} h^j_{\tilde{a} \tilde{b} }
f^{\tilde{a} \tilde{b}}_{\,\,\,\,\,\, \tilde{c}} Q^{\tilde{c}} (z),
\label{Gamma}
\eea
where $j=1,2,3$ and there is no sum over $j$ in the first equation of
(\ref{Gamma}) \footnote{One should change the index structures
  as follows: ${\bf \Gamma^0} \rightarrow -i {\bf \Gamma^2}$,
  ${\bf \Gamma^1} \rightarrow
  -i {\bf \Gamma^3}$,
  ${\bf \Gamma^2} \rightarrow i {\bf \Gamma^4}$ and
  ${\bf \Gamma^3} \rightarrow -i {\bf \Gamma^1}$ in order to use
  (\ref{gsformula}) from (\ref{Gamma}).
  We introduce the coset $\frac{G}{H}=\frac{SU(N+2)}{SU(N)}$ notation 
  $\tilde{a}=(\bar{a}, \hat{a})$ 
where the $\bar{a}$ index runs over $4N$ values
as before
and the index $\hat{a}$ associates with the  
$2 \times 2$ matrix corresponding to $SU(2)\times U(1)$
and runs over $4$ values. 
  Let us represent the $4 \times 4$ matrices $h^{i}_{\hat{a}  \hat{b}}$
appearing in (\ref{Gamma})
as follows \cite{AK1506}:
\bea
h^1_{\hat{a} \hat{b}} & = & 
\left(
\begin{array}{cccc}
0 & 0  & 0 & -\frac{1}{A} \\
0 & 0 & A & 0 \\
0 & -A & 0 & 0 \\
\frac{1}{A} & 0 & 0 & 0 \\
\end{array}
\right), 
h^2_{\hat{a} \hat{b}} = 
\left(
\begin{array}{cccc}
0 & 0  & 0 & \frac{i}{A} \\
0 & 0 & i A & 0 \\
0 & -i A & 0 & 0 \\
-\frac{i}{A} & 0 & 0 & 0 \\
\end{array}
\right), 
h^{3}_{\hat{a} \hat{b}}
= 
\left(
\begin{array}{cccc}
0 & 0  & -i & 0 \\
0 & 0 & 0 & i \\
i & 0 & 0 & 0 \\
0 & -i & 0 & 0 \\
\end{array}
\right),
\label{lastexp}
\eea
where the quantity $A$ in (\ref{lastexp}) which depends on $N$ is 
given by
\bea
A \equiv -\frac{1}{2(N+1)} \left( 
\sqrt{2N(N+1)} + i  \sqrt{2(N+1)(N+2)}
\right).
\label{adef}
\eea
The $\hat{a}$ indices are given by  $(2N+1)$, $\frac{1}{2}[(N+2)^2-1]$,
$(2N+1)^\ast$ and  
$\frac{1}{2}[(N+2)^2-1]^{\ast}$ of (\ref{Indexindex}).
The $N$ dependence in (\ref{adef}) can be fixed by 
the several $N$ cases.
}.
The bosonic spin $1$ current is given by
\bea
{\bf U} (z) =-\frac{1}{4(N+1)} h^j_{\tilde{a} \tilde{b} }
f^{\tilde{a} \tilde{b}}_{\,\,\,\,\,\, \tilde{c}} h^{j \tilde{c}}_{ \,\,\,\, \tilde{d} } \left[
 V^{\tilde{d}}
-\frac{1}{2(k+N+2)}  
f^{\tilde{d} }_{\,\,\,\, \tilde{e} \tilde{f}} Q^{\tilde{e}} Q^{\tilde{f}}
\right](z),
\label{ulinear}
\eea
where there is no sum over the index $j$.
Of course, the $11$ currents in (\ref{11currents})
or (\ref{gsformula}) are regular in the OPEs between them
and the spin $\frac{1}{2}$ currents
${\bf \Gamma}^{\mu}(z)$ and the spin $1$ current ${\bf U}(z)$.

\subsection{The $16$ higher spin currents in the extension of
the large ${\cal N}=4$ nonlinear superconformal algebra}

Let us introduce the $16$ higher spin currents (with boldface notation)
in the context of an extension of the large ${\cal N}=4$
linear superconformal algebra in $SO(4)$ manifest way \cite{AK1509}
\bea
  \mbox{spin} \, s &:&   {\bf \Phi_{0}^{(s)}}(z),
\nonu \\
   \mbox{spin} \, (s + \frac{1}{2}) &:& {\bf \Phi_{\frac{1}{2}}^{(s),1}}(z),
{\bf \Phi_{\frac{1}{2}}^{(s),2}}(z), {\bf \Phi_{\frac{1}{2}}^{(s),3}}(z),
{\bf \Phi_{\frac{1}{2}}^{(s),4}}(z),
\nonu \\
\mbox{spin} \, (s+1) &:&   {\bf \Phi_{1}^{(s),12}}(z),
     {\bf \Phi_{1}^{(s),13}}(z),
  {\bf \Phi_1^{(s),14}}(z), {\bf \Phi_1^{(s),23}}(z),
  {\bf \Phi_1^{(s),24}}(z), {\bf \Phi_1^{(s),34}}(z),
  \nonu \\
 \mbox{spin} \, (s+\frac{3}{2}) & : & {\bf \Phi_{\frac{3}{2}}^{(s),1}}(z),
  {\bf \Phi_{\frac{3}{2}}^{(s),2}}(z), {\bf \Phi_{\frac{3}{2}}^{(s),3}}(z),
  {\bf \Phi_{\frac{3}{2}}^{(s),4}}(z),
  \nonu \\
  \mbox{spin} \, (s+2) &:& {\bf \Phi_{2}^{(s)}}(z).
\label{16linear}
\eea
Note that in the usual ${\cal N}=4$ superspace approach,
each component current in this ${\cal N}=4$ multiplet for fixed $s$
can be put into the coefficients in the expansion of
fermionic Grassmann coordinate for the ${\cal N}=4$ super primary
current.

As done in (\ref{gsformula}), one can
apply the mechanism in (\ref{gsformula})
to the higher spin currents.
From the OPEs between $16$ currents and the $16$  higher
spin currents, one realizes
that
${\bf \Phi_{0}^{(s)}} (z)$
and ${\bf \Phi_{\frac{1}{2}}^{(s),\mu}} (z)$
are regular with the above currents
${\bf \Gamma^{\mu}}(w)$ in (\ref{Gamma}) and
${\bf U}(w)$ in (\ref{ulinear})
\cite{BCG}.
Then we do not have to add the extra terms to these higher spin currents.
For the other higher spin currents, one should
add possible terms with correct spin and $SO(4)$ indices. 
The coefficients can be fixed by using the above
regularity with ${\bf \Gamma^{\mu}}(w)$ and ${\bf U}(w)$.
Furthermore, by requiring that they should transform as
primary currents under the stress energy tensor $T(z)$,
further undetermined coefficients
can be determined. The final
relations between (\ref{phis}) and (\ref{16linear})
are given by \cite{BCG}
\bea 
\Phi_{0}^{(s)}(z) & = & {\bf \Phi_{0}^{(s)}} (z),
\nonu\\
\Phi_{\frac{1}{2}}^{(s),\mu}(z) & = & {\bf \Phi_{\frac{1}{2}}^{(s),\mu}} (z),
\nonu\\
\Phi_{1}^{(s),\mu\nu}(z) & = & {\bf \Phi_{1}^{(s),\mu\nu}}(z)
+c_{1}\,({\bf \Gamma^{\mu}\Phi_{\frac{1}{2}}^{(s),\nu}}
-{\bf \Gamma^{\nu}\Phi_{\frac{1}{2}}^{(s),\mu}})(z),
\nonu\\
\Phi_{\frac{3}{2}}^{(s),\mu}(z) & = &
    {\bf \Phi_{\frac{3}{2}}^{(s),\mu}}(z)+c_{2}\,
     \partial {\bf \Phi_{\frac{1}{2}}^{(s),\mu}}(z)+c_{3}\,
      \partial {\bf \Gamma^{\mu}\Phi_{0}^{(s)}}(z)+c_{4}\,
    {\bf \Gamma^{\mu}} \partial {\bf \Phi_{0}^{(s)}}(z)+c_{5}\,
    {\bf  U \Phi_{\frac{1}{2}}^{(s),\mu}}(z)
\nonu\\
&+& c_{6}\,  \varepsilon^{\mu\nu\rho\si}\,
    {\bf (\Gamma^{\nu}\Phi_{1}^{(s),\rho\si}+T^{\nu\rho}
      \Phi_{\frac{1}{2}}^{(s),\si})}(z),
\nonu\\
\Phi_{2}^{(s)}(z) & = & {\bf \Phi_{2}^{(s)}}(z)+c_{7}\,
    {\bf L\Phi_{0}^{(s)}}(z)+c_{8}\, \partial^{2} {\bf \Phi_{0}^{(s)}}(z)+
    c_{9}\, \partial {\bf U\Phi_{0}^{(s)}}(z)+c_{10}\,
    {\bf U } \partial {\bf \Phi_{0}^{(s)}}(z)
    \label{Phinonandlin}
    \\
&+& c_{11}\,{\bf U U \Phi_{0}^{(s)}}(z)
+c_{12}\, \partial {\bf \Gamma^{\mu}\Phi_{\frac{1}{2}}^{(s),\mu}}(z)+
c_{13}\,{\bf \Gamma^{\mu}} \partial {\bf \Phi_{\frac{1}{2}}^{(s),\mu}}(z)+
c_{14}\,  \partial {\bf \Gamma^{\mu}\Gamma^{\mu}\Phi_{0}^{(s)}}(z),
\nonu
\eea
where the coefficients are given by
{\small
\bea
c_ {1} & = & \frac{2 i}{(2 + k + N)}, \qquad
c_ {2} = -\frac{(k - N)}{(2 s + 1) (2 + k + N)}, \qquad
c_ {3} = -\frac{4 i\,s}{(2 + k + N)},
\label{coeffcoeff}
\\
c_ {4} & = & \frac{2 i}{(2 + k + N)},  \qquad
c_ {5} = \frac{2}{(2 + k + N)},  \qquad
c_ {6} = -\frac{i}{(2 + k + N)},\nonu\\
c_ {7} & = & \frac{24 (k - N) \,s (1 + s)}{
 (3 k + 3 N + 6 k N - 20 s - 4 k s - 4 N s + 12 k N s + 32 s^2 + 
  16 k s^2 + 16 N s^2)}, \nonu\\
c_ {8} & = & -\frac{(k - N) (3 k + 3 N + 6 k N + 52 s + 26 k s + 26 N s)}{(2 +
       k + N) (3 k + 3 N + 6 k N - 20 s - 4 k s - 4 N s + 12 k N s + 
      32 s^2 + 16 k s^2 + 16 N s^2)},\nonu\\
c_ {9}& = & -\frac{4 s}{(2 + k + N)}, \qquad
c_ {10} =\frac{ 4}{(2 + k + N)}, \nonu\\
c_ {11} & = & \frac{24 (k - N) \,s (1 + s)}{(2 + k + N) (3 k + 3 N + 6 k N - 
      20 s - 4 k s - 4 N s + 12 k N s + 32 s^2 + 16 k s^2 + 
      16 N s^2)}, \nonu\\
c_ {12}& = & \frac{2 i(1 + 2 s)}{(2 + k + N)}, \qquad
c_ {13} = -\frac{2 i}{(2 + k + N)},  \nonu\\
c_ {14} & = & -\frac{24 (k - N) \,s (1 + s)}{(2 + k + N) (3 k + 3 N + 6 k N - 
      20 s - 4 k s - 4 N s + 12 k N s + 32 s^2 + 16 k s^2 + 16 N s^2)}.
\nonu
\eea}
For the condition $k =N$, some of the coefficients vanish.
According to the relations (\ref{VPhinon}), the left hand sides
of (\ref{Phinonandlin}) can be written in terms of
the $16$ higher spin currents in the $SU(2) \times SU(2)$ basis
and from the observation of \cite{BCG}, one can also write down
the right hand side of (\ref{Phinonandlin}) 
in terms of
the $16$ higher spin currents in the $SU(2) \times SU(2)$ basis.
Then one can check that the above relations 
match with the findings of \cite{BCG}.

\subsection{ The  OPEs between the $11$ currents and the
   $16$ higher spin currents}
  
Let us describe how one can obtain the
OPEs 
 between the $11$ currents and the
  $16$ higher spin currents.
 One of the simplest OPE in the OPEs between
 the $16$ currents and the $16$ higher spin currents
in the linear version
 is described by
\bea
    {\bf A^{\pm, i}}(z) \, {\bf V_{\frac{1}{2}}^{(s), \mu}}(w)  & = &
    \frac{1}{(z-w)} \,
\alpha_{\mu \nu}^{\pm, i} \, {\bf V_{\frac{1}{2}}^{(s), \nu}}(w)+\cdots.
\label{one3halfabove}
\eea
One can rewrite this OPE (\ref{one3halfabove})
in the $SO(4)$ manifest way.
Using the relations in \cite{AK1509} \footnote{ The relations 
  between the spin $1$ currents in two different bases
  are given by
  \bea
{\bf T^{12}}(z) & = & i( {\bf A^{+3}}+{\bf A^{-3}})(z),\qquad 
{\bf T^{13}}(z)=-i( {\bf A^{+2}}+{\bf A^{-2}})(z), \nonu \\
{\bf T^{14}}(z) & = & i( {\bf A^{+1}}-{\bf A^{-1}})(z),
\qquad {\bf T^{23}}(z)=-i( {\bf A^{+1}}+{\bf A^{-1}})(z),
\nonu \\
{\bf T^{24}}(z) & = & -i( {\bf A^{+2}}-{\bf A^{-2}})(z),\qquad 
{\bf T^{34}}(z)=-i( {\bf A^{+3}}-{\bf A^{-3}})(z).
\label{spinonerelation}
\eea
Or one can express these in (\ref{spinonerelation})
as ${\bf T^{\mu\nu}}(z) = -2i \, \alpha_{\mu\nu}^{+i}
\, {\bf A^{+i}}-2i \, \alpha_{\mu\nu}^{-i} \,
   {\bf A^{-i}}$ where $\alpha_{\mu\nu}^{\pm i}$ are given in
   (\ref{threealpha}).},
one can reexpress the currents $ {\bf A^{\pm, i}}(z)$
in terms of ${\bf T^{\mu \nu}}(z)$. Furthermore,
the higher spin $\frac{3}{2}$ current
$ {\bf V_{\frac{1}{2}}^{(s), \mu}}(w)$
is the same as the $  V_{\frac{1}{2}}^{(s), \mu}(w)$
which is equivalent to $ \Phi_{\frac{1}{2}}^{(s),\mu}(w)$
from (\ref{VPhinon}) 
up to an overall factor. This leads to
$ {\bf \Phi_{\frac{1}{2}}^{(s),\mu}}(w)$ from (\ref{Phinonandlin}).
Then the above OPE (\ref{one3halfabove})
is equal to
\bea
{\bf T^{\mu \nu}}(z)\; {\bf \Phi_{\frac{1}{2}}^{(s),\rho}}(w)
& = & 
-\frac{1}{(z-w)}\,i\,\Bigg[\,\delta^{\mu \rho}\, {\bf \Phi_{\frac{1}{2}}^{(s),\nu}}
-\delta^{\nu \rho}\, {\bf \Phi_{\frac{1}{2}}^{(s),\mu}} \,\Bigg](w)+\cdots.
\label{one3halfother}
\eea
The next step is to rewrite this OPE
(\ref{one3halfother}) in the context of an extension of the
large ${\cal N}=4$ nonlinear superconformal algebra. 
Then it is easy to see that the following OPE
holds
\bea
T^{\mu \nu}(z)\;\Phi_{\frac{1}{2}}^{(s),\rho}(w)
& = & 
-\frac{1}{(z-w)}\,i\,\Bigg[\,\delta^{\mu \rho}\,\Phi_{\frac{1}{2}}^{(s),\nu}-
  \delta^{\nu \rho}\,\Phi_{\frac{1}{2}}^{(s), \mu}\,\Bigg](w)+\cdots,
\label{one3halfnon}
\eea
where the precise relation between
the spin $1$ currents, ${\bf T^{\mu \nu}}(z)$ and $ T^{\mu \nu}(z)$, 
is given by (\ref{gsformula}) and the 
regularity between the current ${\bf \Gamma^{\mu}}(z)$
and $
 {\bf \Phi_{\frac{1}{2}}^{(s),\nu}}(w)$ is used.
 This OPE (\ref{one3halfnon})
 is one of the OPEs in Appendix $A.2$.
 
In this way,
one can construct the
remaining OPEs between the $11$ currents and the $16$ higher spin currents
in the nonlinear version 
and they are described in Appendix $A.2$.

\section{ The  OPEs between the lowest $16$ higher spin currents
  and itself for generic $N$ and $k$}

We explain
how
we obtain the 
 OPEs between the lowest $16$ higher spin currents
 and itself in the nonlinear version.
 They will appear in Appendix $B$.
  We also write down the most general higher spin $2$ current
  in terms of the Wolf space coset currents explicitly.
  
\subsection{The OPE between the higher spin $1$ current
and itself}

It is known that the
OPE between the higher spin $1$ current
and itself in the linear version is given by \cite{AK1411}
\bea
    {\bf {\Phi}_{0}^{(1)}}(z)\, {\bf {\Phi}_{0}^{(1)}}(w) &
    = & \frac{1}{(z-w)^{2}}\,
\Bigg[ \frac {2k\,N}{(2 + k + N)} \Bigg]
+\cdots.
    \label{11OPE}
\eea
According to (\ref{Phinonandlin}), the
higher spin $1$ current in the nonlinear version
is equal to the one in the linear version.
Then it is obvious that
the
OPE between the higher spin $1$ current
and itself in the nonlinear version
is the same as the right hand side of (\ref{11OPE}).

\subsection{The OPE between the higher spin $1$ current
and other $15$ higher spin currents}

One of the simplest known OPE in the
OPEs between the higher spin $1$ current and the
other higher spin currents in the context of
an extension of the large ${\cal N}=4$ linear superconformal
algebra  is described by the following OPE \cite{AK1509} 
\bea
    {\bf {\Phi}_{0}^{(1)}}(z) \, {\bf {\Phi}_{\frac{1}{2}}^{(1), \mu}}(w)
    & = & \frac{1}{(z-w)}\, \Bigg[ {\bf {G}^{\mu} } -
      \frac{2i}{ (k+N+2)} \, {\bf U} {\bf \Gamma^{\mu}} -
      \frac{8i}{6(k+N+2)} \, \ep^{\mu\nu\rho\si} \, {\bf \Gamma^{\nu}}
           {\bf \Gamma^{\rho}}  {\bf \Gamma^{\si}}
           \nonu \\
           &+& \frac{1}{(k+N+2)} \, \ep^{\mu\nu\rho\si} \,
                 {\bf T^{\nu\rho}} {\bf \Gamma^{\si}}
           \Bigg](w)+\cdots.
    \label{one3half2}
\eea
According to the previous analysis, the left hand side
of this OPE (\ref{one3half2}) can be replaced by
the corresponding higher spin currents in the nonlinear version (in
the context of an extension of the large ${\cal N}=4$
nonlinear superconformal algebra). 
On the other hand, the right hand side of this OPE
should be written in terms of the (higher) spin currents in the nonlinear
version.
By using the relations in (\ref{gsformula}), one can
rewrite the spin $\frac{3}{2}$ currents ${\bf G^{\mu}}(w)$ and
the spin $1$
currents ${\bf T^{\mu\nu}}(w)$ in terms of their
linear expressions $G^{\mu}(w)$ and $T^{\mu\nu}(w)$ respectively.
Note that in the second equation of (\ref{gsformula}),
one should also change ${\bf T^{\nu\rho}}(w)$ by using the first equation
of (\ref{gsformula}).
One expects that there will be no ${\bf \Gamma^{\mu}}(w)$
and ${\bf U}(w)$ dependences because one should have the extension of
the large ${\cal N}=4$ nonlinear superconformal algebra where they
are decoupled in the OPEs.

Then it turns out that the
OPE between the higher spin $1$ current and the higher
spin $\frac{3}{2}$ current in the nonlinear version has very simple form
and is given by
\bea
    {\Phi}_{0}^{(1)}(z)\,{\Phi}_{\frac{1}{2}}^{(1), \mu}(w) & = &
    \frac{1}{(z-w)}\,{G}^{\mu}(w)+\cdots.
    \label{one3half1}
\eea
Or one can see that the first order pole of the right hand side
in (\ref{one3half2}) is exactly same as the spin $\frac{3}{2}$ current
in the nonlinear version according to (\ref{gsformula}).

What happens for the OPEs between the higher spin $1$ current
and the higher spin $2$ currents?
We have the explicit OPEs between these higher spin currents
in the linear version \cite{AK1509}.
Then using the relation of (\ref{Phinonandlin}), one should write down
the higher spin $2$ currents ${\bf {\Phi}_{1}^{(1), \mu\nu}}(w)$
in terms of those in the nonlinear version where
the higher spin $\frac{3}{2}$ current
${\bf {\Phi}_{\frac{1}{2}}^{(1), \mu}}(w)$ is the same as the one in
the nonlinear version according to the second equation of
(\ref{Phinonandlin}).
Then the left hand side of the OPEs
 between the higher spin $1$ current
 and the higher spin $2$ currents in the linear version
 consist of the OPEs between them in the nonlinear version
 plus the OPEs between the higher spin $1$ current
 and $ ({\bf \Gamma^{\mu}} \Phi_{\frac{1}{2}}^{(1),\nu}
 -{\bf \Gamma^{\nu}} \Phi_{\frac{1}{2}}^{(1),\mu})(w)$.
 Note that the extra contributions can be calculated by using
 (\ref{one3half1}) and the fact that the higher spin $1$ current
 is regular in the OPEs with ${\bf \Gamma^{\mu}}(w)$.
The OPEs between  the higher spin $1$ current
and the higher spin $2$ currents
in the linear version contain the spin $1$ current ${\bf T^{\mu\nu}}(w)$
and the spin
$\frac{3}{2}$ currents ${\bf G^{\mu}}(w)$ which can be replaced with
the ones in the nonlinear version according to (\ref{gsformula}).
Then one can read off the OPEs
 between the higher spin $1$ current
 and the higher spin $2$ currents in the nonlinear version
 by simplifying the above singular terms. This is described in
 Appendix (\ref{onetwo}).

 Now one can analyze the OPEs
 between the higher spin $1$ current
 and the higher spin $\frac{5}{2}$ currents.
 For given OPEs between them in the linear version,
 one can rewrite the higher spin $\frac{5}{2}$ currents in the
 linear version in terms of the higher spin currents and the currents
 in the nonlinear version using (\ref{Phinonandlin})
 and (\ref{gsformula}).
One can easily see these OPEs from the previous OPE results.
By simplifying the right hand sides of the OPEs where
all the linear (higher) spin currents are replaced with
the nonlinear (higher) spin currents,
one arrives at the final OPEs and presents them in Appendix
(\ref{one5half}) where the new higher spin $\frac{5}{2}$
current $\Phi_{\frac{1}{2}}^{(2),\mu}(w)$
living in next $16$ higher spin multiplet occurs.

Let us consider the 
 the OPEs
 between the higher spin $1$ current
 and the higher spin $3$ current.
 For given OPEs between them in the linear version,
 one can reexpress the higher spin $3$ current in the
 linear version in terms of the higher spin currents and the currents
 in the nonlinear version using (\ref{Phinonandlin})
 and (\ref{gsformula}).
By simplifying the right hand sides of the OPEs where
all the linear (higher) spin currents are replaced with
the nonlinear (higher) spin currents,
one arrives at the final OPEs and presents them in Appendix
(\ref{onethree}) where the new higher spin $2$
current $\Phi_0^{(2)}(w)$
living in next $16$ higher spin multiplet occurs. 
The fourth, third and first order poles appearing in the
linear version are disappeared in the nonlinear version.

\subsection{
The OPE between the higher spin $1$ current
and higher spin $3$ current
  and next higher $2$ spin current}

In order to calculate the three-point functions
of the higher spin currents living in the
next higher spin multiplet,
one should obtain the higher spin currents 
for several $N$ values  at least.
One realizes that the lowest higher spin $2$ current
in the next higher spin multiplet occurs in the following OPE
from Appendix (\ref{onethree})
\bea
    {\Phi}_{0}^{(1)}(z)\,{\Phi}_{2}^{(1)}(w) & = & \frac{1}{(z-w)^{2}}\,\Bigg[\,
      c_{1}\,{\Phi}_{0}^{(2)}+c_{2}\,{\Phi}_{0}^{(1)}{\Phi}_{0}^{(1)}+c_{3}\,{L}+
      c_{4}\,({T}^{\mu\nu})^{2}+c_5\, \varepsilon^{\mu\nu\rho\si}\,
      {T}^{\mu\nu}{T}^{\rho\si}\,\Bigg](w)
    \nonu \\
    & + & \cdots,
    \label{onethree1}
\eea
where
the new higher spin $2$ current in terms of the Wolf space coset
currents (the spin $1$
current and the spin $\frac{1}{2}$ current) is given by
\bea
\Phi_0^{(2)}(z)  & = & \Bigg( a_1 \, V_{\bar{a}} \, V^{\bar{a}} 
+ a_2 \sum_{a':SU(N)}\, V_{a'} \, V^{a'} +  a_3 \, \sum_{a'':SU(2) \times U(1)} \,
V_{a''} \, V^{a''}
\nonu \\
& + & 
a_4 \, \frac{1}{16N^2} \, h^i_{\bar{a} \bar{b}} \, h^i_{\bar{d} \bar{e}} \, 
f^{\bar{a} \bar{b}}_{\,\,\,\,\,\, c} \,  f^{\bar{d} \bar{e}}_{\,\,\,\,\,\, f} \, V^c \, V^f
+ 
a_5 \, \frac{1}{16(k+N+2)^2} \, h^i_{\bar{a} \bar{b}} \,
h^i_{\bar{c} \bar{d}} \,
(Q^{\bar{a}} \, Q^{\bar{b}})
 \,  (Q^{\bar{c}} \, Q^{\bar{d}})
\nonu \\
 & + & a_6 \, d^{0}_{\bar{a} \bar{b}} \,
d^{0}_{\bar{c} \bar{d}} \,
Q^{\bar{a}} \, Q^{\bar{b}} \, Q^{\bar{c}} \, Q^{\bar{d}}
+ a_7 \, Q_{\bar{a}} \pa Q^{\bar{a}}
+   a_8 \, h^{\mu}_{\bar{a} \bar{b}} \, h^{\mu}_{\bar{c} \bar{d}} \,
f^{\bar{a} \bar{c}}_{\,\,\,\,\,\, e} \, Q^{\bar{b}} \, Q^{\bar{d}} \, V^e
\nonu \\
& + &  a_9 \, 
d^{0}_{\bar{a} \bar{b}} \, d^0_{\bar{c} \bar{d}} \,
f^{\bar{c} \bar{d}}_{\,\,\,\,\,\, e}
\, Q^{\bar{a}} \, Q^{\bar{b}} \, V^e \Bigg)(z), 
\label{Spintwo}
\eea
where the coefficients with the information for
$N=3,5,7,9,11$ and $13$ can be determined as follows
\bea
a_1 & = & -\frac{(k+2) (2 k+3 N+5)}{(k+N+2)^3},
\nonu \\
a_2 & = & -\frac{8}{(k+N+2)^2},
\nonu \\
a_3 & = & \frac{1}{(k+N+2)^4} ( 20 k^2 N^2+18 k^2 N-8 k^2+42 k N^3+107 k N^2+45 k N-32 k+16 N^4 \nonu \\
  & + & 75 N^3+97 N^2+12 N-32 ),
\nonu \\
a_4 & = & \frac{4} {(k+N+2)^4} (10 k^2 N^2+8 k^2 N-4 k^2+21 k N^3+51 k N^2+18 k N-16 k+8 N^4 \nonu \\
& + & 36 N^3+43 N^2+N-16 ),
\nonu \\
a_5 & = & -\frac{2 k
  (6 k^2+11 k N+17 k+4 N^2+14 N+10)}{(N+2) (k+N+2)^3},
\nonu \\
a_6 & = & \frac{k}{8 (N+2) (k+N+2)^6}
(20 k^3 N+22 k^3+42 k^2 N^2+119 k^2 N+85 k^2
\nonu \\
& + & 16 k N^3+87 k N^2+153 k N+92 k+4 N^3+24 N^2+48 N+32 ),
\nonu \\
a_7 & = & -\frac{k}{2 (N+2) (k+N+2)^5}
(-8 k^3 N+2 k^3-8 k^2 N^2+7 k^2 N+31 k^2+14 k N^3
\nonu \\
& + & 79 k N^2+147 k N+84 k+8 N^4
+  56 N^3+140 N^2+144 N+48),
\nonu \\
a_8 & = & -\frac{(6 k^2+11 k N+17 k+4 N^2+14 N+10)}{4 (k+N+2)^4},
\nonu \\
a_9 & = & -\frac{1}{4 (N+2) (k+N+2)^5}
(20 k^3 N+22 k^3+42 k^2 N^2+119 k^2 N+85 k^2 \nonu \\
& + & 16 k N^3+87 k N^2+153 k N+92 k+4 N^3+24 N^2+48 N+32).
\label{Spintwocoeff}
\eea
One has the following identities due to the ordering of the operators 
\cite{BBSSfirst,BBSSsecond,Fuchs}
\bea
h^i_{\bar{a} \bar{b}} \,
h^i_{\bar{c} \bar{d}} \,
(Q^{\bar{a}} \, Q^{\bar{b}})
\,  (Q^{\bar{c}} \, Q^{\bar{d}}) & = &
h^i_{\bar{a} \bar{b}} \,
h^i_{\bar{c} \bar{d}} \,
Q^{\bar{a}} \, Q^{\bar{b}}
\,  Q^{\bar{c}} \, Q^{\bar{d}}  -4 (2+N+k)
Q_{\bar{a}} \pa Q^{\bar{a}},
\nonu \\
d^{0}_{\bar{a} \bar{b}} \,
d^{0}_{\bar{c} \bar{d}} \,
(Q^{\bar{a}} \, Q^{\bar{b}}) \, (Q^{\bar{c}} \, Q^{\bar{d}})
& = & d^{0}_{\bar{a} \bar{b}} \,
d^{0}_{\bar{c} \bar{d}} \,
Q^{\bar{a}} \, Q^{\bar{b}} \, Q^{\bar{c}} \, Q^{\bar{d}}
+  4 (2+N+k) Q_{\bar{a}} \pa Q^{\bar{a}}.
\label{ordering}
\eea
The $a_4$ term in (\ref{Spintwo}) is nothing but
$A^{+i} A^{+i}(z)$
while
$a_5$ term in (\ref{Spintwo}) is equal to
$A^{-i} A^{-i}(z)$ with (\ref{ordering}).
Compared to the higher spin $2$ current in the orthogonal
Wolf space coset \cite{AKP1510},
the $a_6$ and $a_9$ terms are the extra terms.
Note that in the orthogonal case, there is no $d^{0}_{\bar{a} \bar{b}}$
tensor. Compared to the previous higher spin $1$ current in
(\ref{finalspinone}), 
the indices in the quadratic terms in the spin $1$ currents 
contain the coset ones as well as the subgroup indices
and the relative coefficients are different.

In order to see the behavior of the above higher spin $2$ current
in the context of the three-point functions, we try to
obtain the more general higher spin $2$ current by adding
the possible terms to the one in (\ref{Spintwo}).
Let us consider the field contents in (\ref{Spintwo})
 with arbitrary coefficients $\hat{a}_i$ where $i =1,2, \cdots, 9$
and the higher spin $2$ current takes the form of
 \bea
\hat{\Phi}_0^{(2)}(z)  & = & \Bigg( \hat{a}_1 \, V_{\bar{a}} \, V^{\bar{a}} 
+ \hat{a}_2 \sum_{a':SU(N)}\, V_{a'} \, V^{a'} +
\hat{a}_3 \, \sum_{a'':SU(2) \times U(1)} \,
V_{a''} \, V^{a''}
\nonu \\
& + & 
\hat{a}_4 \, \frac{1}{16N^2} \, h^i_{\bar{a} \bar{b}} \, h^i_{\bar{d} \bar{e}} \, 
f^{\bar{a} \bar{b}}_{\,\,\,\,\,\, c} \,  f^{\bar{d} \bar{e}}_{\,\,\,\,\,\, f} \, V^c \, V^f
+ 
\hat{a}_5 \, \frac{1}{16(k+N+2)^2} \, h^i_{\bar{a} \bar{b}} \,
h^i_{\bar{c} \bar{d}} \,
(Q^{\bar{a}} \, Q^{\bar{b}})
 \,  (Q^{\bar{c}} \, Q^{\bar{d}})
\nonu \\
 & + & \hat{a}_6 \, d^{0}_{\bar{a} \bar{b}} \,
d^{0}_{\bar{c} \bar{d}} \,
Q^{\bar{a}} \, Q^{\bar{b}} \, Q^{\bar{c}} \, Q^{\bar{d}}
+ \hat{a}_7 \, Q_{\bar{a}} \pa Q^{\bar{a}}
+   \hat{a}_8 \, h^{\mu}_{\bar{a} \bar{b}} \, h^{\mu}_{\bar{c} \bar{d}} \,
f^{\bar{a} \bar{c}}_{\,\,\,\,\,\, e} \, Q^{\bar{b}} \, Q^{\bar{d}} \, V^e
\nonu \\
& + &  \hat{a}_9 \, 
d^{0}_{\bar{a} \bar{b}} \, d^0_{\bar{c} \bar{d}} \,
f^{\bar{c} \bar{d}}_{\,\,\,\,\,\, e}
\, Q^{\bar{a}} \, Q^{\bar{b}} \, V^e \Bigg)(z).
\label{generalSpintwo}
\eea
%
This higher spin $2$ current (\ref{generalSpintwo})
should be  a primary
field under the stress energy tensor $T(z)$ in (\ref{11currents})
and should have the regular OPE with
the spin $1$ currents $A^{\pm i}(z)$ in (\ref{11currents}).
Furthermore, by definition,
the OPEs between 
the spin $\frac{1}{2}$ currents $ {\bf \Gamma^{\mu}}(z)$
of the large ${\cal N}=4$ linear superconformal algebra
and the above higher spin $2$ current is regular
and similarly the OPE with the spin $1$ current
$ {\bf U}(z)$ (of the large ${\cal N}=4$ linear superconformal algebra)
has no singular terms. This implies that 
\bea
T(z) \, \hat{\Phi}_0^{(2)}(w) & = &
\frac{1}{(z-w)^2} \, 2  \, \hat{\Phi}_0^{(2)}(w)  +
\frac{1}{(z-w)^2} \, \pa \hat{\Phi}_0^{(2)}(w) + \cdots,
\nonu \\
\left(
\begin{array}{c}
  A^{\pm i} \\
  {\bf \Gamma^{\mu}} \\
  {\bf U} \\
\end{array}
\right)(z) \, \hat{\Phi}_0^{(2)}(w) & = & +\cdots.
\label{conditions}
\eea

After using the conditions (\ref{conditions}),
some of the coefficients can be written in terms of
$\hat{a}_1$ and $\hat{a}_2$ as follows:
{\small
\bea
\hat{a}_3 & = & -\frac{1}{(k+2) (k+N+2)}
( \hat{a}_2 k^2 N^2-\hat{a}_2 k^2+
  2 \hat{a}_2 k N^3+6 \hat{a}_1 k N^2+4 \hat{a}_2 k N^2+
  9 \hat{a}_1 k N-2 \hat{a}_2 k N \nonu \\
  & - &
  4 \hat{a}_2 k+4 \hat{a}_2 N^3+9 \hat{a}_1 N^2+4 \hat{a}_2 N^2+
  12 \hat{a}_1 N-
  4 \hat{a}_2 N-4 \hat{a}_2),
\nonu \\
\hat{a}_4 & = & -\frac{2}{(k+2) (k+N+2)} (\hat{a}_2 k^2 N^2-\hat{a}_2 k^2
+2 \hat{a}_2 k N^3
+6 \hat{a}_1 k N^2+4 \hat{a}_2 k N^2+8 \hat{a}_1 k N-2 \hat{a}_2 k N\nonu \\
& - & 4 \hat{a}_2 k+4 \hat{a}_2 N^3+8 \hat{a}_1 N^2+4 \hat{a}_2 N^2
+10 \hat{a}_1 N-4 \hat{a}_2 N-4 \hat{a}_2),
\nonu \\
\hat{a}_5 & = & \frac{k (\hat{a}_2 k+\hat{a}_2 N+2 \hat{a}_1)}{(N+2)},
\nonu \\
\hat{a}_6 & = & -\frac{k}{16 (k+2) (N+2) (k+N+2)^3}
(2 \hat{a}_2 k^3 N+\hat{a}_2 k^3+4 \hat{a}_2 k^2 N^2+12 \hat{a}_1 k^2 N+
10 \hat{a}_2 k^2 N
\nonu \\
& + & 18 \hat{a}_1 k^2+
4 \hat{a}_2 k^2+9 \hat{a}_2 k N^2+18 \hat{a}_1 k N+16 \hat{a}_2 k N+
24 \hat{a}_1 k+8 \hat{a}_2 k+2 \hat{a}_2 N^2+
8 \hat{a}_2 N+8 \hat{a}_2),
\nonu \\
\hat{a}_7 & = & \frac{k}{4 (k+2) (N+2) (k+N+2)^2}
(3 \hat{a}_2 k^3-8 \hat{a}_1 k^2 N+6 \hat{a}_2 k^2 N-10 \hat{a}_1 k^2+
12 \hat{a}_2 k^2
\nonu \\
& + & 2 \hat{a}_2 k N^3+
4 \hat{a}_1 k N^2+7 \hat{a}_2 k N^2+6 \hat{a}_1 k N+
16 \hat{a}_2 k N+8 \hat{a}_1 k+8 \hat{a}_2 k+4 \hat{a}_2 N^3+
8 \hat{a}_1 N^2\nonu \\
& + & 14 \hat{a}_2 N^2+32 \hat{a}_1 N+8 \hat{a}_2 N+32 \hat{a}_1-8 \hat{a}_2),
\nonu \\
\hat{a}_8 & = & \frac{(\hat{a}_2 k+\hat{a}_2 N+2 \hat{a}_1)}{8 (k+N+2)},
\label{coeffgeneralSpintwo} \\
\hat{a}_9 & = & \frac{1}{8 (k+2) (N+2) (k+N+2)^2}
(2 \hat{a}_2 k^3 N+\hat{a}_2 k^3+4 \hat{a}_2 k^2 N^2+
12 \hat{a}_1 k^2 N+10 \hat{a}_2 k^2 N \nonu \\
& + &
18 \hat{a}_1 k^2+4 \hat{a}_2 k^2+9 \hat{a}_2 k N^2+18 \hat{a}_1 k N+
16 \hat{a}_2 k N+24 \hat{a}_1 k+8 \hat{a}_2 k+
2 \hat{a}_2 N^2+8 \hat{a}_2 N+8 \hat{a}_2).
\nonu
\eea}

One can check that the above higher spin $2$ current
can be expressed as the previous higher spin $2$ current
(\ref{Spintwo}) and others 
as follows:
\bea
\hat{\Phi}_0^{(2)}(z) & = & 
-\frac{1}{8} \hat{a}_2 \, (k+N+2)^2 \, \Phi_0^{(2)}(z)
\nonu \\
& + &    
(-2 \hat{a}_2 k^2-3 \hat{a}_2 k N+8 \hat{a}_1 k-9 \hat{a}_2 k+
8 \hat{a}_1
N-6 \hat{a}_2 N+16 \hat{a}_1-10 \hat{a}_2) \nonu \\
& \times & \Bigg[
-\frac{3 (2 k N+3 k+3 N+4)}{16 (k+2) (N+2)}
  \, {\Phi}_{0}^{(1)}{\Phi}_{0}^{(1)}+
  \frac{1}{4} \,{L} \nonu \\
  & - & 
\frac{(k+N+4)}{32 (k+2) (N+2)}
  \,({T}^{\mu\nu})^{2}+
\frac{(N-k)}{64 (k+2) (N+2)}
  \, \varepsilon^{\mu\nu\rho\si} \,{T}^{\mu\nu}{T}^{\rho\si} \Bigg](z).
\label{generalSpintwoother}
\eea
When
$\hat{a}_1 =a_1$ and $\hat{a}_2 =a_2$ together with (\ref{Spintwocoeff}),
it turns out that
$\hat{\Phi}_0^{(2)}(z)  = \Phi_0^{(2)}(z)$.
The coefficient in the second line of (\ref{generalSpintwoother})
vanishes.
So far the parameters $\hat{a}_1$ and $\hat{a}_2$ are completely 
arbitrary.
One can insert $\Phi_0^{(2)}(w)$ by using the relation
(\ref{generalSpintwoother}) into (\ref{onethree1})
and then the second order pole of (\ref{onethree1})
can be written similarly and the structure constants behave differently.
Furthermore, the four kinds of field contents
appearing in the right hand side of the OPE are the same.
One sees that
the expression in the third and fourth lines in (\ref{generalSpintwoother})
is primary field of spin $2$ under the stress energy tensor.
The point is that this is written in terms of the known quantities. 
Furthermore, this expression with bracket  in (\ref{generalSpintwoother})
satisfies the standard ${\cal N}=4$ primary condition described
in Appendix $A.2$.

One can calculate the OPE between the
higher spin $2$ current (\ref{generalSpintwo}) and itself
and it is given by
\bea 
\hat{\Phi}_{0}^{(2)}(z) \, \hat{\Phi}_{0}^{(2)}(w) &=& \frac{1}{(z-w)^4} \, c_{0}
\nonu\\
& + &
\frac{1}{(z-w)^2} \,\Bigg[\,
c_{1}\,\hat{\Phi}_{0}^{(2)}
+c_{2}\,\Phi_{0}^{(1)}\Phi_{0}^{(1)}
+c_{3}\,L
+c_{4}\,T^{\mu\nu}T^{\rho\si}
+c_{5}\,\varepsilon^{\mu\nu\rho\si}T^{\mu\nu}T^{\rho\si}
\,\Bigg](w)
\nonu\\
& + &
\frac{1}{(z-w)} \,\frac{1}{2}\,\partial (\mbox{pole-2})(w)
+ \cdots,
\label{spin2spin2ope}
\eea
where the coefficients are given by
\bea 
c_{0} & = & \frac{k\,N}{(2 + k)^2 (2 + N)^2}(-192 \hat{a}_{1}^2 -
336 k \hat{a}_{1}^2 - 
 192 k^2 \hat{a}_{1}^2 - 36 k^3 \hat{a}_{1}^2 - 
 336 N \hat{a}_{1}^2 - 240 k N \hat{a}_{1}^2
\nonu\\
& + & 168 k^2 N \hat{a}_{1}^2 + 120 k^3 N \hat{a}_{1}^2 - 
 192 N^2 \hat{a}_{1}^2 
+168 k N^2 \hat{a}_{1}^2 + 
 504 k^2 N^2 \hat{a}_{1}^2 + 204 k^3 N^2 \hat{a}_{1}^2
\nonu\\
& - & 36 N^3 \hat{a}_{1}^2 + 120 k N^3 \hat{a}_{1}^2 + 
 204 k^2 N^3 \hat{a}_{1}^2 + 72 k^3 N^3 \hat{a}_{1}^2 
- 192 k \hat{a}_{1} \hat{a}_{2} - 
 336 k^2 \hat{a}_{1} \hat{a}_{2}
\nonu\\
& - & 192 k^3 \hat{a}_{1} \hat{a}_{2} - 
 36 k^4 \hat{a}_{1} \hat{a}_{2} - 
 336 k N \hat{a}_{1} \hat{a}_{2} - 
 480 k^2 N \hat{a}_{1} \hat{a}_{2} - 
 204 k^3 N \hat{a}_{1} \hat{a}_{2} - 
 24 k^4 N \hat{a}_{1} \hat{a}_{2} 
\nonu\\
& + & 48 k N^2 \hat{a}_ {1} \hat{a}_{2} + 
 168 k^2 N^2 \hat{a}_{1} \hat{a}_{2} + 
 144 k^3 N^2 \hat{a}_{1} \hat{a}_{2} + 
 36 k^4 N^2 \hat{a}_{1} \hat{a}_{2} + 
 336 k N^3 \hat{a}_{1} \hat{a}_{2} 
\nonu\\
& + & 
 480 k^2 N^3 \hat{a}_{1} \hat{a}_{2} + 
 204 k^3 N^3 \hat{a}_{1} \hat{a}_{2} + 
 24 k^4 N^3 \hat{a}_{1} \hat{a}_{2} + 
 144 k N^4 \hat{a}_{1} \hat{a}_{2} + 
 168 k^2 N^4 \hat{a}_{1} \hat{a}_{2} 
\nonu\\
& + & 48 k^3 N^4 \hat{a}_{1} \hat{a}_{2} - 32 \hat{a}_{2}^2 - 
 64 k \hat{a}_{2}^2 - 56 k^2 \hat{a}_{2}^2 - 
 28 k^3 \hat{a}_{2}^2 - 8 k^4 \hat{a}_{2}^2 - 
 k^5 \hat{a}_{2}^2 - 64 N \hat{a}_{2}^2 
\nonu\\
& - & 144 k N \hat{a}_{2}^2 - 144 k^2 N \hat{a}_{2}^2 - 
 76 k^3 N \hat{a}_{2}^2 - 20 k^4 N \hat{a}_{2}^2 - 
 2 k^5 N \hat{a}_{2}^2 - 8 N^2 \hat{a}_{2}^2 - 
 60 k N^2 \hat{a}_{2}^2 
\nonu\\
& - & 70 k^2 N^2 \hat{a}_{2}^2 - 
 25 k^3 N^2 \hat{a}_{2}^2 + k^5 N^2 \hat{a}_{2}^2 + 
 56 N^3 \hat{a}_{2}^2 + 104 k N^3 \hat{a}_{2}^2 + 
 110 k^2 N^3 \hat{a}_{2}^2 
\nonu\\
& + & 68 k^3 N^3 \hat{a}_{2}^2 + 
 20 k^4 N^3 \hat{a}_{2}^2 + 2 k^5 N^3 \hat{a}_{2}^2 + 
 40 N^4 \hat{a}_{2}^2 + 124 k N^4 \hat{a}_{2}^2 + 
 126 k^2 N^4 \hat{a}_{2}^2 
\nonu\\
& + & 53 k^3 N^4 \hat{a}_{2}^2 + 
 8 k^4 N^4 \hat{a}_{2}^2 + 8 N^5 \hat{a}_{2}^2 + 
 40 k N^5 \hat{a}_{2}^2 + 34 k^2 N^5 \hat{a}_{2}^2 + 
 8 k^3 N^5 \hat{a}_{2}^2),
\nonu \\
c_ {1} & = &  2 (2 + N + k) (4 \hat{a}_{1} + N \hat{a}_{2} + 
    k \hat{a}_{2}),
\nonu\\
c_ {2} & = & \frac{ 1}{(2 + k)^2 (2 + N)^2}(24 k \hat{a}_{1} +
18 k^2 \hat{a}_{1} + 18 k N \hat{a}_{1} + 
   12 k^2 N \hat{a}_{1} + 8 \hat{a}_{2} + 8 k \hat{a}_{2} + 
   4 k^2 \hat{a}_{2} + k^3 \hat{a}_{2} 
\nonu\\
& + & 8 N \hat{a}_{2} + 
   16 k N \hat{a}_{2} + 10 k^2 N \hat{a}_{2} + 
   2 k^3 N \hat{a}_{2} + 2 N^2 \hat{a}_{2}
+ 9 k N^2 \hat{a}_{2} + 
   4 k^2 N^2 \hat{a}_{2}) 
\nonu\\
& \times &  (12 N \hat{a}_{1} + 
   9 k N \hat{a}_{1} 
+9 N^2 \hat{a}_{1} + 
   6 k N^2 \hat{a}_{1} - 4 \hat{a}_{2} - 4 k \hat{a}_{2} - 
   k^2 \hat{a}_{2} - 4 N \hat{a}_{2} - 2 k N \hat{a}_{2} + 
   4 N^2 \hat{a}_{2} 
\nonu\\
& + &   4 k N^2 \hat{a}_{2} + 
   k^2 N^2 \hat{a}_{2} + 4 N^3 \hat{a}_{2} + 
   2 k N^3 \hat{a}_{2} ),
\nonu\\
c_ {3} & = & -4\hat{a}_{1} (2 + N + k)^2  (2 \hat{a}_{1} + 
    N \hat{a}_{2} + k \hat{a}_{2} ),
\nonu\\
c_ {4} & = & \frac{\hat{a}_{1} (2 + N + k)^2 (4 + N + k)  (2 \hat{a}_{1} + 
    N \hat{a}_{2} + k \hat{a}_{2})}{2 (2 + N) (2 + k)},
\nonu\\
c_ {5}  & = & \frac{\hat{a}_{1} (-N + k) (2 + N + k)^2  (2 \hat{a}_{1} + 
    N \hat{a}_{2} + k \hat{a}_{2} )}{2 (2 + N) (2 + k)}.
\label{Coeffcoeffcoeff}
\eea
Technically, it is more useful to extract the correct $N$ dependence
in the coefficients
for several $N$ values by considering the nine terms in (\ref{generalSpintwo})
separately. In other words, the coefficients are given in 
(\ref{coeffgeneralSpintwo}). For fixed $N$ case, one considers
the singular terms coming from
each contribution in different combinations
between the nine terms rather than taking the whole expressions. 
For example, for the fourth order pole in (\ref{spin2spin2ope}),
there are many contributions, $\hat{a}_1$ term-$\hat{a}_1$ term
OPE, $\hat{a}_1$ term-$\hat{a}_2$ term OPE, $\cdots$. 
It turns out that there are $13$ contributions. Now one considers
each contribution separately and tries to read off the correct
$N$ dependence for several $N$ values. 
Now the coefficients are known from (\ref{coeffgeneralSpintwo})
and then the final result can be obtained by
summing over the possible contributions as in (\ref{Coeffcoeffcoeff}).

\subsection{The remaining OPEs}

So far we have considered
the OPEs
between the higher spin $1$ current
and the $16$ higher spin currents.
We can continue to compute the other OPEs
between the higher spin $\frac{3}{2}$ current and other
higher spin currents from the known corresponding OPEs
in the linear version.
With the help of the Thielemans package \cite{Thielemans}, 
one can easily obtain the whole OPEs in the nonlinear version.
We present them in Appendix $B$ explicitly.
In particular, the OPE between
$\Phi_1^{(1),\mu\nu}(z)$ and
$\Phi_{\frac{3}{2}}^{(1),\rho}(w)$ in Appendix (\ref{two5half})
contains the first order pole
having the quadratic or cubic terms between the
spin $1$ currents and the higher spin currents in the
$c_{25}, c_{28}, c_{40}$ and $c_{42}$ terms.
They do not appear in the corresponding OPE in the linear version.
Furthermore, there is
the $c_{27}$ term which is a quadratic term in the
higher spin currents and one does not see this kind of term
in the linear version.

We have checked that
under  the large $(N,k)$ 't Hooft like limit,
all the structure constants of the right hand sides of the OPEs
in Appendix $B$ appearing in the nonlinear terms
which contain the $11$ currents vanish.

\section{ Three-point functions of higher spin $2,3,4$ currents
among the second lowest $16$ higher spin currents 
  with two scalars}

We describe the three-point
functions of higher spin $2,3,4$ currents
in two different bases.
In order to obtain them, the eigenvalue equations 
for the zero mode of the higher spin currents
acting the states are used.

The large $N$ 't Hooft limit
is described by \cite{GG1305}
\bea
N,k \rightarrow \infty, \quad \lambda 
\equiv \frac{N+1}{N+k+2} \quad \mbox{fixed}.
\label{limit}
\eea
One introduces the following two types of
column vectors \cite{AK1506}
\bea
|(f;0)>_{+}= (0, \cdots, 0, 1, 0)^T, \qquad
|(f;0)>_{-}= (0, \cdots, 0, 0, 1)^T,
\label{eigenvector}
\eea
where $T$ stands for transpose.
They transform as singlets under the $SU(N)$ characterized
by the first $N$ zeros in (\ref{eigenvector}).  
Moreover, they are ${\bf 2}$ representation under the $SU(2)$.

One further introduces
the following states \cite{AK1506}
\bea
|(0;f)>_{+} & : & \frac{1}{\sqrt{k+N+2}} Q_{-\frac{1}{2}}^{1^{\ast}}|0>,
\qquad 
\cdots, 
\qquad \frac{1}{\sqrt{k+N+2}} Q_{-\frac{1}{2}}^{N^{\ast}}|0>,
\nonu \\
|(0;f)>_{-} & : & \frac{1}{\sqrt{k+N+2}} Q_{-\frac{1}{2}}^{(N+1)^{\ast}}|0>,
\qquad
\cdots, 
\qquad
\frac{1}{\sqrt{k+N+2}} Q_{-\frac{1}{2}}^{(2N)^{\ast}}|0>.
\label{plusminus}
\eea
They transform as fundamental representation ${\bf N}$
under the $SU(N)$ respectively and are doublet of the $SU(2)$.
According to (\ref{cosetindex}), they
correspond to the rectangular  $N \times 2$ matrix inside of
$(N+2) \times (N+2)$ matrix.

\subsection{Eigenvalue equation of spin $2$ current}

The eigenvalue
equations of the zero mode of the spin $2$ stress energy tensor
acting on the
state $|(f;0)>_{\pm}$ (\ref{eigenvector}) and the state
$|(0;f)>_{\pm}$ (\ref{plusminus})
are given by \cite{GG1305,AK1506}
\bea
T_0 |(f;0)>_{\pm}  &=& \left[ \frac{(2N+3)}{4(k+N+2)}  \right] |(f;0)>_{\pm}
\rightarrow \frac{\la}{2} |(f;0)>_{\pm},
\nonu \\
T_0 |(0;f)>_{\pm}
&=&  \left[ \frac{(2k+3)}{4(N+k+2)} \right] |(0;f)>_{\pm}
\rightarrow \frac{1}{2} (1-\la)  |(0;f)>_{\pm},
\label{eigenspin2}
\eea
where the large $N$ 't Hooft limit (\ref{limit}) is taken.
Among the $11$ currents of the large ${\cal N}=4$ nonlinear
superconformal algebra, one can write down the eigenvalue
equations for the spin $1$ currents \footnote{
  One has the following relations between the zero modes and
  the states \cite{AK1506}
  \bea
(A^{+1})_0 |(f;0)>_{\pm} &=&
-\frac{i}{2} \,|(f;0)>_{\mp},
\qquad
(A^{+2})_0  |(f;0)>_{\pm} =
\mp \frac{1}{2} \,|(f;0)>_{\mp},
\nonu \\
(A^{+3})_0 |(f;0)>_{\pm} &=&
\pm \frac{i}{2}\,|(f;0)>_{\pm},
\qquad
(A^{-1})_0 |(0;f)>_{\pm} =
\frac{i}{2}\,|(0;f)>_{\mp},
\nonu \\
(A^{-2})_0  |(0;f)>_{\pm} &=&
\mp \frac{1}{2} \,|(0;f)>_{\mp},
\qquad
(A^{-3})_0 |(0;f)>_{\pm} =
\mp \frac{i}{2}\,|(0;f)>_{\pm}.
\label{spin1eigen}
\eea
Among these (\ref{spin1eigen}),
the zero modes of $A^{\pm 3}$ satisfy the eigenvalue equation.
}.

\subsection{Eigenvalue equation of higher spin $1$, $2$, $3$ currents}

The eigenvalue equations of the zero mode of the
higher spin $1$ current (\ref{finalspinone}) acting on the states
(\ref{eigenvector}) and (\ref{plusminus})
can be summarized as 
\bea
\Phi^{(1)}_0 |(f;0)>_{\pm} &=&  - \left[ \frac{N}{(k+N+2)} \right] |(f;0)>_{\pm}
\equiv (\phi_0^{(1)})_{(f;0)} |(f;0)>_{\pm}
\rightarrow   -\la  |(f;0)>_{\pm},
\label{t1f00f}
\\
\Phi^{(1)}_0 |(0;f)>_{\pm} &=&  - \left[ \frac{k}{(k+N+2)} \right] |(0;f)>_{\pm}
\equiv
(\phi_0^{(1)})_{(0;f)} |(0;f)>_{\pm}
\rightarrow   -(1-\la)  |(0;f)>_{\pm},
\nonu
\eea
where one introduces the corresponding two eigenvalues
in (\ref{t1f00f}) at finite $N$ and $k$ and
the large $N$ 't Hooft limit is taken
at the final stage.

One presents the relevant relations of the zero mode of the
higher spin $2$ currents \footnote{ In the description of
  ${\cal N}=2$ superspace multiplets \cite{AK1506},
  one has the following explicit relations between the
  higher spin $2$ currents as follows:
  \bea
T^{(2)} & = & \frac{i}{2}\Phi_{1}^{(1),34}+\frac{(N+k)}{(k+N+2kN)}L-
\frac{1}{2(2+N+k)}\,\Big((T^{3j})^{2}+(T^{4j})^{2}\Big),
\nonu\\
\tilde{W}^{(2)} & = & \frac{i}{2}\Phi_{1}^{(1),12}+\frac{(N+k)}{(k+N+2kN)}L-
\frac{1}{2(2+N+k)}\,\Big((T^{1j})^{2}+(T^{2j})^{2}\Big),
\nonu\\
U^{(2)}_{\pm} & = &
\frac{1}{4}\,( \pm \Phi_{1}^{(1),13} \pm
i\,\Phi_{1}^{(1),14} + i\,\Phi_{1}^{(1),23} - \Phi_{1}^{(1),24})
\nonu\\
&+&\frac{i}{2(2+N+k)}\,(\pm T^{12}T^{13}
\pm i\,T^{12}T^{14}+ i\,T^{12}T^{23} - T^{12}T^{24})
\nonu\\
&+&\frac{i}{2(2+N+k)}\,(- T^{34}T^{13}
- i\,T^{34}T^{14}\mp i\,T^{34}T^{23} \pm T^{34}T^{24}),
\nonu\\
V^{(2)}_{\pm} & = & \frac{1}{4}\,( \mp \Phi_{1}^{(1),13} \pm
i\,\Phi_{1}^{(1),14}-i\,\Phi_{1}^{(1),23}-\Phi_{1}^{(1),24})
\nonu\\
&+&\frac{i}{2(2+N+k)}\,(\mp T^{12}T^{13}
\pm i\,T^{12}T^{14}- i\,T^{12}T^{23} - T^{12}T^{24})
\nonu\\
&+&\frac{i}{2(2+N+k)}\,(- T^{34}T^{13}
+ i\,T^{34}T^{14}\mp i\,T^{34}T^{23} \mp T^{34}T^{24}).
\label{twotwo1}
\eea
It is straightforward to obtain the eigenvalue equations \cite{AK1506}
of the left hand sides of 
(\ref{twotwo1}) by using the expressions of the
right hand sides of (\ref{twotwo1}), the equations (\ref{eigen3-1}) and
(\ref{eigen3-2}) and other relevant relations. 
} acting on the state
(\ref{eigenvector}), together with (\ref{t1f00f}),  as follows:
\bea
(\Phi_1^{(1), 12})_0 |(f;0)>_{\pm} &=&
(\mp  i )(\phi_0^{(1)})_{(f;0)} |(f;0)>_{\pm},
\nonu \\
(\Phi_1^{(1), 13})_0 |(f;0)>_{\pm} &=&
 (\pm   )(\phi_0^{(1)})_{(f;0)} |(f;0)>_{\mp},
\nonu \\
(\Phi_1^{(1), 14})_0 |(f;0)>_{\pm} &=&
 (  i )(\phi_0^{(1)})_{(f;0)} |(f;0)>_{\mp},
\nonu \\
(\Phi_1^{(1), 23})_0 |(f;0)>_{\pm} &=&
 ( - i )(\phi_0^{(1)})_{(f;0)} |(f;0)>_{\mp},
 \nonu \\
(\Phi_1^{(1), 24})_0 |(f;0)>_{\pm} &=&
 (\pm   )(\phi_0^{(1)})_{(f;0)} |(f;0)>_{\mp},
\nonu \\
(\Phi_1^{(1), 34})_0 |(f;0)>_{\pm} &=&
 (\pm  i )(\phi_0^{(1)})_{(f;0)} |(f;0)>_{\pm}.
\label{eigen3-1}
\eea
It is straightforward to obtain the large $N$ 't Hooft limit (\ref{limit}).
The right hand side of (\ref{eigen3-1}) has simple form
and can be written in terms of the multiple of the previous
eigenvalues (\ref{t1f00f}).
The first and the last ones satisfy the eigenvalue equations.

Similarly, 
one describes the following relations
between the zero mode of the higher spin $2$ currents 
and the above two states with (\ref{t1f00f})
\bea
(\Phi_1^{(1), 12})_0 |(0;f)>_{\pm} &=&
 (\pm  i )(\phi_0^{(1)})_{(0;f)} |(0;f)>_{\pm},
\nonu \\
(\Phi_1^{(1), 13})_0 |(0;f)>_{\pm} &=&
 (\mp   )(\phi_0^{(1)})_{(0;f)} |(0;f)>_{\mp},
\nonu \\
(\Phi_1^{(1), 14})_0 |(0;f)>_{\pm} &=&
 (  i )(\phi_0^{(1)})_{(0;f)} |(0;f)>_{\mp},
\nonu \\
(\Phi_1^{(1), 23})_0 |(0;f)>_{\pm} &=&
 (  i )(\phi_0^{(1)})_{(0;f)} |(0;f)>_{\mp},
 \nonu \\
(\Phi_1^{(1), 24})_0 |(0;f)>_{\pm} &=&
 (\pm   )(\phi_0^{(1)})_{(0;f)} |(0;f)>_{\mp},
\nonu \\
(\Phi_1^{(1), 34})_0 |(0;f)>_{\pm} &=&
 (\pm  i )(\phi_0^{(1)})_{(0;f)} |(0;f)>_{\pm}.
\label{eigen3-2}
\eea
One sees that there is $N \leftrightarrow k$ and $0 \leftrightarrow f$
symmetry 
between (\ref{eigen3-1}) and (\ref{eigen3-2}) up to signs.

Finally, one describes the eigenvalue equations of the zero mode
of the higher spin $3$ current \footnote{
One has the following explicit relation between the higher spin $3$ current
\bea
&& W^{(3)}(z)  =  -\frac{1}{2}\,\Phi_2^{(1)}(z)+\frac{1}{(2+N+k)}
\Bigg[\,\frac{i}{2}\,\partial \Phi_{1}^{(1),12}-i\,T^{12}\Phi_{1}^{(1),12}+2i\,T^{34}\Phi_{1}^{(1),34}
\nonu\\
&& +  \frac{i}{2}(\,T^{1j}\Phi_{1}^{(1),1j}+\,T^{2j}\Phi_{1}^{(1),2j})
+\frac{1}{2}(\,T^{1j}\Phi_{1}^{(1),2j}-\,T^{2j}\Phi_{1}^{(1),1j})
-(T^{34}\partial \Phi_{0}^{(1)}-\partial T^{34}\Phi_{0}^{(1)})
\nonu\\
&& -2i\,  G^{1}G^{2}+\frac{16(1+N+k)}{(4+5k+5N+6kN)}\,T^{12}L
-\frac{4(k-N)}{(4+5k+5N+6kN)}\,T^{34}L
\,\Bigg](z)
\nonu\\
&& +  \frac{1}{(2+N+k)^{2}}\,\Bigg[\, 2\,(T^{12})^{3}-\Big(T^{12}(T^{1j})^{2}+
  T^{12}(T^{2j})^{2}\Big)-4\,T^{12}(T^{34})^{2}
+ 2\,\varepsilon^{12ij}\,T^{1i}T^{2j}T^{34}
\nonu\\
&& +  (N-k)\,(T^{12}\partial T^{34}-\partial T^{12}T^{34})
+\frac{i(N-k)}{2}\,(T^{1j}\partial \tilde{T}^{2j}-T^{2j}\partial \tilde{T}^{1j})
\label{threethree1}
\\
&& -\frac{i(-8+N+k)}{2}\,T^{1j}\partial T^{2j}+\frac{i(N+k)}{2}\,\partial T^{1j}T^{2j}
+\frac{(-12+5N+5k)}{6}\,\partial^{2}T^{12} 
+\frac{4(-N+k)}{3}\,\partial^{2}T^{34} \,\Bigg](z).
\nonu
\eea
Moreover, one can check the eigenvalue equations of the higher spin $3$
current $W^{(3)}(z)$ \cite{AK1506} by using the relation
(\ref{threethree1}) and other relevant relations including
(\ref{spin3eigen}). }
acting on the two states together with
(\ref{t1f00f})
\bea
(\Phi_2^{(1)})_0 |(f;0)>_{\pm} &=&
\Bigg[\frac{4(12+28k+5k^{2}+14N+39kN+6k^{2}N+4N{}^{2}+12kN{}^{2})}
  {3(2+k+N)(4+5k+5N+6kN)}
   \Bigg]
 \nonu \\
 & \times & (\phi_0^{(1)})_{(f;0)} |(f;0)>_{\pm} \rightarrow
 -\frac{4}{3} \la (1+\la)  |(f;0)>_{\pm},
\nonu \\
(\Phi_2^{(1)})_0 |(0;f)>_{\pm} &=&-
\Bigg[\frac{4(12+28N+5N^{2}+14k+39kN+6 k N^2+4 k^{2}+12k^2 N)}
  {3(2+k+N)(4+5k+5N+6kN)}
  \Bigg]
\nonu \\
& \times & (\phi_0^{(1)})_{(0;f)} |(0;f)>_{\pm}
 \rightarrow
 \frac{4}{3} (1-\la)(2-\la)  |(0;f)>_{\pm}.
\label{spin3eigen}
\eea
There is $N \leftrightarrow k$ and $0 \leftrightarrow f$
symmetry in (\ref{spin3eigen}) up to sign.
Note that the extra factors inside of the brackets
in (\ref{spin3eigen})
behave as $(1+\la)$ (which is equal to $\la$ plus one)
and $(2-\la)$ (which is equal to $(1-\la)$ plus one) under the large $N$
't Hooft limit respectively. 

From the diagonal modular invariant with 
pairing up identical representations on the left (holomorphic)
and the right (antiholomorphic) 
sectors \cite{CY}, 
one of the primaries is given by 
$(f;0) \otimes (f;0)$
which is denoted by 
${\cal O}_{+}$ and the other 
is given by $(0;f) \otimes (0;f)$
which is denoted by ${\cal O}_{-}$.

The (nonzero) three-point functions of the
higher spin $1, 2, 3$ currents from (\ref{t1f00f}),
(\ref{eigen3-1}), (\ref{eigen3-2})
and (\ref{spin3eigen}) are given by
\bea
<\overline{{\cal O}}_{+ } 
{\cal O}_{+ } \Phi_0^{(1)}> & = &
(\phi_0^{(1)})_{(f;0)},
\qquad
< \overline{{\cal O}}_{- } 
{\cal O}_{- } \Phi_0^{(1)}>  =  (\phi_0^{(1)})_{(0;f)},
\nonu \\
<\overline{{\cal O}}_{+ } 
{\cal O}_{+ } \Phi_1^{(1),12}> & = &
(\mp i)(\phi_0^{(1)})_{(f;0)},
\qquad
< \overline{{\cal O}}_{- } 
{\cal O}_{- } \Phi_1^{(1),12}>  =  (\pm i) (\phi_0^{(1)})_{(0;f)},
\nonu \\
<\overline{{\cal O}}_{+ } 
{\cal O}_{+ } \Phi_1^{(1),34}> & = &
(\pm i)(\phi_0^{(1)})_{(f;0)},
\qquad
< \overline{{\cal O}}_{- } 
{\cal O}_{- } \Phi_1^{(1),34}>  =  (\pm i) (\phi_0^{(1)})_{(0;f)},
\nonu \\
<\overline{{\cal O}}_{+ } 
{\cal O}_{+ } \Phi_2^{(1)}> & = &
\Bigg[\frac{4(12+28k+5k^{2}+14N+39kN+6k^{2}N+4N{}^{2}+12kN{}^{2})}
  {3(2+k+N)(4+5k+5N+6kN)}
   \Bigg]
 \nonu \\
 & \times & 
(\phi_0^{(1)})_{(f;0)},
\nonu \\
< \overline{{\cal O}}_{- } 
{\cal O}_{- } \Phi_2^{(1)}> & = &
-\Bigg[\frac{4(12+28N+5N^{2}+14k+39kN+6 k N^2+4 k^{2}+12k^2 N)}
  {3(2+k+N)(4+5k+5N+6kN)}
  \Bigg]
\nonu \\
& \times &
(\phi_0^{(1)})_{(0;f)}.
\label{threenonlinear}
\eea
Compared to the three-point functions in the basis of ${\cal N}=2$
mutiplets \cite{AK1506}, the cases of the higher spin $2$ currents
in (\ref{threenonlinear}) have very simple form. They
are proportional to the eigenvalues of the higher spin $1$ current
(\ref{t1f00f}) and the overall factors do not depend on $N$ and $k$.
We will see later that if we go to the three-point functions of the
(nonprimary)
higher spin $3$ current in different basis, they will lead to
behave in simple form also.

\subsection{Eigenvalue equation of higher spin $1$, $2$, $3$ currents
in different basis}

One can calculate the eigenvalue equations in an extension
of the large ${\cal N}=4$ linear superconformal algebra
using the relations in (\ref{Phinonandlin}).
It turns out that they are the same as the ones in (\ref{eigen3-1})
and (\ref{eigen3-2}) even at finite $N$ and $k$.
That is,
\bea
{\bf (\Phi_1^{(1), 12})}_0 |(f;0)>_{\pm} &=&
 (\mp  i )(\phi_0^{(1)})_{(f;0)} |(f;0)>_{\pm},
\nonu \\
{\bf (\Phi_1^{(1), 13})}_0 |(f;0)>_{\pm} &=&
 (\pm   )(\phi_0^{(1)})_{(f;0)} |(f;0)>_{\mp},
\nonu \\
{\bf (\Phi_1^{(1), 14})}_0 |(f;0)>_{\pm} &=&
 (  i )(\phi_0^{(1)})_{(f;0)} |(f;0)>_{\mp},
\nonu \\
{\bf (\Phi_1^{(1), 23})}_0 |(f;0)>_{\pm} &=&
 ( - i )(\phi_0^{(1)})_{(f;0)} |(f;0)>_{\mp},
 \nonu \\
{\bf (\Phi_1^{(1), 24})}_0 |(f;0)>_{\pm} &=&
 (\pm   )(\phi_0^{(1)})_{(f;0)} |(f;0)>_{\mp},
\nonu \\
{\bf (\Phi_1^{(1), 34})}_0 |(f;0)>_{\pm} &=&
 (\pm  i )(\phi_0^{(1)})_{(f;0)} |(f;0)>_{\pm}
\label{eigen3-3}
\eea
and
\bea
{\bf (\Phi_1^{(1), 12})}_0 |(0;f)>_{\pm} &=&
 (\pm  i )(\phi_0^{(1)})_{(0;f)} |(0;f)>_{\pm},
\nonu \\
{\bf (\Phi_1^{(1), 13})}_0 |(0;f)>_{\pm} &=&
 (\mp   )(\phi_0^{(1)})_{(0;f)} |(0;f)>_{\mp},
\nonu \\
{\bf (\Phi_1^{(1), 14})}_0 |(0;f)>_{\pm} &=&
 (  i )(\phi_0^{(1)})_{(0;f)} |(0;f)>_{\mp},
\nonu \\
{\bf (\Phi_1^{(1), 23})}_0 |(0;f)>_{\pm} &=&
 (  i )(\phi_0^{(1)})_{(0;f)} |(0;f)>_{\mp},
 \nonu \\
{\bf (\Phi_1^{(1), 24})}_0 |(0;f)>_{\pm} &=&
 (\pm   )(\phi_0^{(1)})_{(0;f)} |(0;f)>_{\mp},
\nonu \\
{\bf (\Phi_1^{(1), 34})}_0 |(0;f)>_{\pm} &=&
 (\pm  i )(\phi_0^{(1)})_{(0;f)} |(0;f)>_{\pm}.
\label{eigen3-4}
\eea
Note that there is no contribution from the extra $c_1$ term in
 (\ref{Phinonandlin}) during this calculation.
The previous relations (\ref{eigen3-1})
and (\ref{eigen3-2}) are used.

Furthermore, one obtains the following eigenvalue equations
for the (nonprimary) higher spin $3$ current
\footnote{
  One can calculate the eigenvalue equations for the extra terms
  in (\ref{Phinonandlin})
  \bea
({\bf L\Phi_{0}^{(1)}})_0 |(f;0)>_{\pm}
& = &
\Bigg[ -\frac{N \left(2 k N+4 k+3 N^2+12 N+11\right)}{2 (N+2) (k+N+2)^2}
  \Bigg] |(f;0)>_{\pm},
\label{c7c11cal-1}
\\
( \partial^{2} {\bf \Phi_{0}^{(1)}})_0 |(f;0)>_{\pm}
& = & 
\Bigg[ -\frac{2 N}{k+N+2}
  \Bigg] |(f;0)>_{\pm},
(\partial {\bf U\Phi_{0}^{(1)}} )_0 |(f;0)>_{\pm}
 =  
\Bigg[-\frac{i N \sqrt{\frac{N}
      {N+2}}}{2 (k+N+2)}
  \Bigg] |(f;0)>_{\pm},
\nonu \\
( {\bf U } \partial {\bf \Phi_{0}^{(1)}} )_0 |(f;0)>_{\pm}
& = &
\Bigg[-\frac{i N \sqrt{\frac{N}{N+2}}}
  {2 (k+N+2)}
  \Bigg] |(f;0)>_{\pm},
\nonu \\
( {\bf U U \Phi_{0}^{(1)}} )_0 |(f;0)>_{\pm}
 & = & 
\Bigg[\frac{N^2}{4 (N+2) (k+N+2)}
  \Bigg] |(f;0)>_{\pm}.
\nonu
\eea
Similarly, one has the following eigenvalue equations
\bea
({\bf L\Phi_{0}^{(1)}})_0 |(0;f)>_{\pm}
& = &
\Bigg[ -\frac{k \left(3 k N+2 N^2+6 N+1\right)}{2 N (k+N+2)^2}
  \Bigg] |(0;f)>_{\pm},
\label{c7c11cal-2}
\\
( \partial^{2} {\bf \Phi_{0}^{(1)}})_0 |(0;f)>_{\pm}
& = & 
\Bigg[ -\frac{2 k}{k+N+2}
  \Bigg] |(0;f)>_{\pm},
(\partial {\bf U\Phi_{0}^{(1)}} )_0 |(0;f)>_{\pm}
 =  
\Bigg[-\frac{i k \sqrt{\frac{N+2}{N}}}{2 (k+N+2)}
  \Bigg] |(0;f)>_{\pm},
\nonu \\
( {\bf U } \partial {\bf \Phi_{0}^{(1)}} )_0 |(0;f)>_{\pm}
& = &
\Bigg[-\frac{i k \sqrt{\frac{N+2}{N}}}{2 (k+N+2)}
  \Bigg] |(0;f)>_{\pm},
\nonu \\
( {\bf U U \Phi_{0}^{(1)}} )_0 |(0;f)>_{\pm}
& = & 
\Bigg[ \frac{k (N+2)}{4 N (k+N+2)}
  \Bigg] |(0;f)>_{\pm}.
\nonu
\eea
}
{\small
\bea
    {\bf (\Phi_2^{(1)})}_0 |(f;0)>_{\pm} &=&
    (\Phi_2^{(1)})_0 |(f;0)>_{\pm}
    \nonu \\
    &- & 
\Bigg[ c_{7}\,
    {\bf L\Phi_{0}^{(1)}}+c_{8}\, (\partial^{2} {\bf \Phi_{0}^{(1)}}+
    c_{9}\, \partial {\bf U\Phi_{0}^{(1)}}+c_{10}\,
    {\bf U } \partial {\bf \Phi_{0}^{(1)}}
    + c_{11}\,{\bf U U \Phi_{0}^{(1)}} \Bigg]_0 |(f;0)>_{\pm}
\nonu \\
& = & (2)(\phi_0^{(1)})_{(f;0)} |(f;0)>_{\pm},
\nonu \\
{\bf (\Phi_2^{(1)})}_0 |(0;f)>_{\pm} &=&
  (\Phi_2^{(1)})_0 |(0;f)>_{\pm}
    \nonu \\
    &- &
\Bigg[ c_{7}\,
    {\bf L\Phi_{0}^{(1)}}+c_{8}\, (\partial^{2} {\bf \Phi_{0}^{(1)}}+
    c_{9}\, \partial {\bf U\Phi_{0}^{(1)}}+c_{10}\,
    {\bf U } \partial {\bf \Phi_{0}^{(1)}}
    + c_{11}\,{\bf U U \Phi_{0}^{(1)}} \Bigg]_0 |(0;f)>_{\pm}
\nonu \\
& = & (-2)(\phi_0^{(1)})_{(0;f)} |(0;f)>_{\pm},
\label{eigen5}
\eea}
where the previous relations (\ref{spin3eigen})
are used. The coefficients in (\ref{coeffcoeff}) with (\ref{c7c11cal-1})
and (\ref{c7c11cal-2}) are also used in this calculation.
Miraculously, all the $N$ and $k$ dependences are disappeared.

Moreover, one can go to the primary basis.
In the linear version, one can construct the primary current
as follows \cite{AK1509}:
\bea
{\bf \widetilde\Phi_{2}^{(s)}}(z)={\bf \Phi_{2}^{(s)}}(z)+
p_{1}\:\partial^{2}{\bf \Phi_{0}^{(s)}}(z)+p_{2}\:{\bf L}{\bf \Phi_{0}^{(s)}}(z),
\label{widetildenot}
\eea
where the two quantities are introduced
{\small
\bea
p_{1} & \equiv & 
-\frac{(k - N) (3 + 3 k + 3 N + 3 k\,N + 26 s + 13 k\,s + 13 N\,s)}{(2 +
     k + N) (3 + 3 k + 3 N + 3 k\,N - 4 s + k\,s + N\,s + 6 k\,N\,s + 
    16 s^2 + 8 k\,s^2 + 8 N\,s^2)}, \nonu \\
p_{2} & \equiv & 
\frac{12 \,(k - N)\, s \,(1 + s)}{(3 + 3 k + 3 N + 3 k\, N - 4 s + k
\, s + N \,s + 
 6 k\, N\,s + 16 s^2 + 8 k \,s^2 + 8 N\, s^2)}.
\label{p1p2}
\eea}
When $k=N$, both $p_1$ and $p_2$ in (\ref{p1p2}) vanish.

Then using the relations in (\ref{c7c11cal-1}), (\ref{c7c11cal-2})
and (\ref{eigen5}),
one obtains
\bea
{\bf (\widetilde\Phi_2^{(1)})}_0 |(f;0)>_{\pm}  & = &
 \Bigg[\,
   \frac{4 }{3 (2 + N) (2 + k + N)
     (5 + 4 k + 4 N + 3 k N)}(30 + 43 k + 8 k^2 + 50 N 
\nonu \\
&+ & 71 k N + 10 k^2 N + 26 N^2 + 36 k N^2 + 
 3 k^2 N^2 + 5 N^3 + 6 k N^3)
 \,\Bigg]
 \nonu \\
 & \times & (\phi_0^{(1)})_{(f;0)}  |(f;0)>_{\pm}
 \rightarrow  -\frac{4}{3}  \la (1+\la) |(f;0)>_{\pm},
\nonu \\
{\bf (\widetilde\Phi_2^{(1)})}_0 |(0;f)>_{\pm}  & = &
-\Bigg[
\,
\frac{4}{3 N (2 + k + N) (5 + 4 k + 4 N + 3 k N)}\,(-3 k + 18 N + 16 k N + 5 k^2 N 
\nonu \\
&+ & 23 N^2 + 24 k N^2 + 6 k^2 N^2 + 
 4 N^3 + 3 k N^3)
 \Bigg] \nonu \\
& \times &  (\phi_0^{(1)})_{(f;0)} |(0;f)>_{\pm} \rightarrow
 \frac{4}{3} (1-\la)(2-\la) |(0;f)>_{\pm}.
\label{primaryeigen}
\eea
At the final stage of (\ref{primaryeigen}), the large $N$ 't Hooft
limit is taken.
Of course, when $k=N$, this reduces to
(\ref{eigen5}).
One can easily see that the extra terms in ${\bf \widetilde\Phi_2^{(1)}}(z)$
do not contribute the eigenvalues under the large $N$ 't Hooft limit.
As before,
the extra factors inside of the brackets
in (\ref{primaryeigen})
behave as $(1+\la)$
and $(2-\la)$ under the large $N$
't Hooft limit respectively. 
Note that one has the relation ${\bf W^{(3)}}(z) = -\frac{1}{2}
{\bf \widetilde\Phi_2^{(1)}}(z)$ in (\ref{widetildenot})
\footnote{
  The precise relations between the
  higher spin currents in two different bases  are given by
  \bea
{\bf T^{(1)}} & = & {\bf \Phi_{0}^{(1)}},\qquad
{\bf T_{\pm}^{(\frac{3}{2})}}  =  -\frac{1}{2\sqrt{2}}\,
({\bf \Phi_{\frac{1}{2}}^{(1),1}} \pm i\,{\bf \Phi_{\frac{1}{2}}^{(1),2}} \pm
{\bf G^{1}}+i\,{\bf  G^{2}}),\nonu \\ 
{\bf T^{(2)}} &  = &  \frac{i}{2}\,{\bf \Phi_{1}^{(1),34}},\qquad
\left(
\begin{array}{c}
  {\bf U^{(\frac{3}{2})}} \\
  {\bf V^{(\frac{3}{2})}}\\
\end{array}
\right)
 =  \frac{1}{2\sqrt{2}}\,({\bf \Phi_{\frac{1}{2}}^{(1),3}} \pm i
\,{\bf \Phi_{\frac{1}{2}}^{(1),4}} \pm {\bf G^{3}}+i\,{\bf  G^{4}}),\nonu\\
{\bf U_{\pm}^{(2)}} & = & \frac{1}{4}\,(\pm { \bf \Phi_{1}^{(1),13}} \pm i\,{\bf \Phi_{1}^{(1),14}}+i\,{\bf \Phi_{1}^{(1),23}}-{\bf \Phi_{1}^{(1),24}}),\nonu\\
\nonu\\ 
\left(
\begin{array}{c}
  {\bf U^{(\frac{5}{2})}} \\
  {\bf U^{(\frac{5}{2})}} \\
\end{array}
\right)& = & \frac{1}{2\sqrt{2}}\,( \pm {\bf \Phi_{\frac{3}{2}}^{(1),3}}+i\,{\bf \Phi_{\frac{3}{2}}^{(1),4}}) \mp \frac{(k-N)}{6\sqrt{2}(2+k+N)}\,\partial({\bf \Phi_{\frac{1}{2}}^{(1),3}} \pm i\,{\bf \Phi_{\frac{1}{2}}^{(1),4}}), \nonu\\
{\bf V_{\pm}^{(2)}} & = & \frac{1}{4}\,( \mp {\bf \Phi_{1}^{(1),13}} \pm i\,
{\bf \Phi_{1}^{(1),14}}-i\,{\bf \Phi_{1}^{(1),23}}-{\bf \Phi_{1}^{(1),24}}), \qquad 
{\bf W^{(2)}}  =  \frac{i}{2}\,{\bf \Phi_{1}^{(1),12}}, \nonu\\
{\bf W_{\pm}^{(\frac{5}{2})}} & = & \mp \frac{1}{2\sqrt{2}}\,({\bf \Phi_{\frac{3}{2}}^{(1),1}} \pm i\,{\bf \Phi_{\frac{3}{2}}^{(1),2}})+\frac{(k-N)}{6\sqrt{2}(2+k+N)}\,\partial(\pm {\bf \Phi_{\frac{1}{2}}^{(1),1}}+i\,{\bf \Phi_{\frac{1}{2}}^{(1),2}}), \nonu\\
{\bf W^{(3)}} & = & -\frac{1}{2}\,{\bf \Phi_{2}^{(1)}}+
\frac{(k-N)(29+16k+16N+3kN)}{6(2+k+N)(5+4k+4N+3kN)}\,\partial^{2}{\bf \Phi_{0}^{(1)}}\nonu \\
& - & \frac{4(k-N)}{(5+4k+4N+3kN)}\,{\bf  L\Phi_{0}^{(1)}}.
\label{somerelation}
\eea
One can obtain the corresponding eigenvalue equations for the
higher spin currents appearing in the left hand sides of
(\ref{somerelation}) by using these relations.}.

Let us summarize the (nonzero) three-point functions of the
higher spin $1, 2, 3$ currents from (\ref{eigen3-3}), (\ref{eigen3-4}),
 (\ref{eigen5}) and (\ref{primaryeigen}) as follows:
{\small
\bea
<\overline{{\cal O}}_{+ } 
{\cal O}_{+ } {\bf \Phi_0^{(1)}}> & = &
(\phi_0^{(1)})_{(f;0)},
\qquad
< \overline{{\cal O}}_{- } 
{\cal O}_{- } {\bf \Phi_0^{(1)}}>  =  (\phi_0^{(1)})_{(0;f)},
\nonu \\
<\overline{{\cal O}}_{+ } 
{\cal O}_{+ } {\bf \Phi_1^{(1),12}}> & = &
(\mp i)(\phi_0^{(1)})_{(f;0)},
\qquad
< \overline{{\cal O}}_{- } 
{\cal O}_{- } {\bf \Phi_1^{(1),12}}>  =  (\pm i) (\phi_0^{(1)})_{(0;f)},
\nonu \\
<\overline{{\cal O}}_{+ } 
{\cal O}_{+ } {\bf \Phi_1^{(1),34}}> & = &
(\pm i)(\phi_0^{(1)})_{(f;0)},
\qquad
< \overline{{\cal O}}_{- } 
{\cal O}_{- } {\bf \Phi_1^{(1),34}}>  =  (\pm i) (\phi_0^{(1)})_{(0;f)},
\nonu \\
<\overline{{\cal O}}_{+ } 
{\cal O}_{+ } {\bf \Phi_2^{(1)}}> & = &
(2)
(\phi_0^{(1)})_{(f;0)},
\qquad
< \overline{{\cal O}}_{- } 
{\cal O}_{- } {\bf \Phi_2^{(1)}}>  = 
(-2)(\phi_0^{(1)})_{(0;f)},
\nonu \\
<\overline{{\cal O}}_{+ } 
{\cal O}_{+ } {\bf \widetilde\Phi_2^{(1)}}> & = &
 \Bigg[\,
   \frac{4 }{3 (2 + N) (2 + k + N)
     (5 + 4 k + 4 N + 3 k N)}(30 + 43 k + 8 k^2 + 50 N 
\nonu \\
&+ & 71 k N + 10 k^2 N + 26 N^2 + 36 k N^2 + 
 3 k^2 N^2 + 5 N^3 + 6 k N^3)
 \,\Bigg] (\phi_0^{(1)})_{(f;0)},
\nonu \\
< \overline{{\cal O}}_{- } 
{\cal O}_{- } {\bf \widetilde\Phi_2^{(1)}}> & = &
-\Bigg[
\,
\frac{4}{3 N (2 + k + N)
  (5 + 4 k + 4 N + 3 k N)}\,(-3 k + 18 N + 16 k N + 5 k^2 N 
\nonu \\
&+ & 23 N^2 + 24 k N^2 + 6 k^2 N^2 + 
 4 N^3 + 3 k N^3)
 \Bigg] 
(\phi_0^{(1)})_{(0;f)}.
\label{finalthreelinear}
\eea}
Therefore, the three-point functions in (\ref{finalthreelinear})
have simple forms which are multiples of the eigenvalues of
the higher spin $1$ current in (\ref{t1f00f}) even at finite
$N$ and $k$ except of the last two case.
The other three-point functions for the higher spin $2$
currents  vanish.
There is no $N \leftrightarrow k$ symmetry
in the last two cases.
Compared to the nonlinear case (\ref{threenonlinear}),
the only last two in (\ref{finalthreelinear}) at finite $N$ and $k$
are different from each other. However, under the large $N$
't Hooft limit, the three-point functions of the primary
higher spin $3$ current coincide with each other
\footnote{It is easy to check that
  the extra terms vanish under the large $N$ 't Hooft limit.
  One obtains the differences of each eigenvalue in (\ref{spin3eigen})
  and (\ref{primaryeigen})
  as follows:
  \bea
  \frac{4 (k-N) (8 k N+9 k+3 N^3+14 N^2+28 N+18)}{
    (N+2) (k+N+2) (3 k N+4 k+4 N+5) (6 k N+5 k+5 N+4)}
   & \rightarrow &
 -\Bigg[\frac{2 \la^2 (2 \la-1)}{3 (\la-1)^2} \Bigg]\frac{1}{ N^2},
  \nonu \\
  \frac{4 (k-N) (3 k^2 N+2 k N-5 k+2 N^2+2 N-4)}{
    N (k+N+2) (3 k N+4 k+4 N+5) (6 k N+5 k+5 N+4)}
  & \rightarrow & - \Bigg[\frac{2 (2 \la-1)}{3}\Bigg] \frac{1}{ N^2}.
\label{limitexp}
\eea
Obviously they (\ref{limitexp})
vanish under the large $N$ 't Hooft limit.}.

\subsection{Eigenvalue equation of higher spin $2$, $3$, $4$ currents}

Let us obtain the eigenvalue equations for the higher spin $2$ current
(\ref{generalSpintwo}) acting on the states $ |(f;0)>_{\pm}$.
The only $Q^a(z)$ independent terms can contribute to the
zero mode eigenvalue equations \cite{GG1305,AK1506}.
The zero mode of spin $1$ current $V^a(z)$ plays the role of
the generator $T_{a^{\ast}}$ of $SU(N+2)$
where $[T_a, T_b]= f_{ab}^{\,\,\,\,\,\,c} T_c$.
See also the footnote \ref{empha}.
We summarize the eigenvalue equations as follows:
\bea
(V_{\bar{a}} \, V^{\bar{a}})_0 |(f;0)>_{\pm}
& = &
\Bigg[ N \Bigg] |(f;0)>_{\pm},
\nonu \\
(\sum_{a':SU(N)} V_{a'} \, V^{a'})_0 |(f;0)>_{\pm} & = &  0, \nonu \\
 (\sum_{a'':SU(2) \times U(1)} V_{a''} \, V^{a''})_0 |(f;0)>_{\pm} & = & 
\Bigg[ \frac{(2 N+3)}{(N+2)} \Bigg] 
|(f;0)>_{\pm},
\nonu \\
\frac{1}{16N^2} \,  h^i_{\bar{a} \bar{b}} \, h^i_{\bar{d} \bar{e}} \, 
 f^{\bar{a} \bar{b}}_{\,\,\,\,\,\, c} \,  f^{\bar{d} \bar{e}}_{\,\,\,\,\,\, f} \,
 (V^c \, V^f)_0 |(f;0)>_{\pm}
 & = & \Bigg[ -\frac{3}{4} \Bigg]  |(f;0)>_{\pm},
 \nonu \\
 h^{\mu}_{\bar{a} \bar{b}} \, h^{\mu}_{\bar{c} \bar{d}} \,
f^{\bar{a} \bar{c}}_{\,\,\,\,\,\, e} \, (Q^{\bar{b}} \, Q^{\bar{d}} \, V^e)_0
 |(f;0)>_{\pm} & = & 0,
\nonu \\ 
d^{0}_{\bar{a} \bar{b}} \, d^0_{\bar{c} \bar{d}} \,
f^{\bar{c} \bar{d}}_{\,\,\,\,\,\, e}
\, (Q^{\bar{a}} \, Q^{\bar{b}} \, V^e)_0   |(f;0)>_{\pm} & = & 0,
\nonu \\
(Q_{\bar{a}} \pa Q^{\bar{a}})_0  |(f;0)>_{\pm} & = & 0,
\nonu \\
\frac{1}{16(k+N+2)^2}
\,  h^i_{\bar{a} \bar{b}}\, h^i_{\bar{c} \bar{d}} \,
((Q^{\bar{a}} \, Q^{\bar{b}})
 \,  (Q^{\bar{c}} \, Q^{\bar{d}}))_0
 |(f;0)>_{\pm} & = & 0,
 \nonu \\
  d^{0}_{\bar{a} \bar{b}} \,
d^{0}_{\bar{c} \bar{d}} \,
(Q^{\bar{a}} \, Q^{\bar{b}} \, Q^{\bar{c}} \, Q^{\bar{d}})_0
|(f;0)>_{\pm} & = & 0.
\label{eigen1}
 \eea

 For the second simplest representation,
 the eigenvalue equations can be obtained
 by calculating the OPE between the higher spin current
 and the spin $\frac{1}{2}$ current $Q^{\bar{A}\ast}$
 in (\ref{plusminus})
 and reading off the highest order pole \cite{GG1305,AK1506}.
 The terms  containing the spin $1$ current
 $V^a(z)$ do not  contribute to the highest order pole. 
We summarize the eigenvalue equations as follows:
 \bea
 (V_{\bar{a}} \, V^{\bar{a}})_0 |(0;f)>_{\pm}
& = & 0,
\nonu \\
( \sum_{a':SU(N)} V_{a'} \, V^{a'})_0 |(0;f)>_{\pm} & = &  0, \nonu \\
 ( \sum_{a'':SU(2) \times U(1)} V_{a''} \, V^{a''})_0 |(0;f)>_{\pm} & = & 0,
\nonu \\
 h^i_{\bar{a} \bar{b}} \, h^i_{\bar{d} \bar{e}} \, 
 f^{\bar{a} \bar{b}}_{\,\,\,\,\,\, c} \,  f^{\bar{d} \bar{e}}_{\,\,\,\,\,\, f} \,
 (V^c \, V^f)_0 |(0;f)>_{\pm}
 & = & 0,
 \nonu \\
 h^{\mu}_{\bar{a} \bar{b}} \, h^{\mu}_{\bar{c} \bar{d}} \,
f^{\bar{a} \bar{c}}_{\,\,\,\,\,\, e} \, (Q^{\bar{b}} \, Q^{\bar{d}} \, V^e)_0
 |(0;f)>_{\pm} & = & 0,
\nonu \\ 
d^{0}_{\bar{a} \bar{b}} \, d^0_{\bar{c} \bar{d}} \,
f^{\bar{c} \bar{d}}_{\,\,\,\,\,\, e}
\, (Q^{\bar{a}} \, Q^{\bar{b}} \, V^e)_0   |(0;f)>_{\pm} & = & 0,
\nonu \\
(Q_{\bar{a}} \pa Q^{\bar{a}})_0  |(0;f)>_{\pm} & = & \Bigg[(N+k+2) \Bigg]
|(0;f)>_{\pm},
\nonu \\
\frac{1}{16(k+N+2)^2}\, h^i_{\bar{a} \bar{b}} \,
h^i_{\bar{c} \bar{d}} \,
((Q^{\bar{a}} \, Q^{\bar{b}})
 \,  (Q^{\bar{c}} \, Q^{\bar{d}}))_0
 |(0;f)>_{\pm} & = & \Bigg[ -\frac{3}{4} \Bigg]  |(0;f)>_{\pm},
 \nonu \\
  d^{0}_{\bar{a} \bar{b}} \,
d^{0}_{\bar{c} \bar{d}} \,
(Q^{\bar{a}} \, Q^{\bar{b}} \, Q^{\bar{c}} \, Q^{\bar{d}})_0
|(0;f)>_{\pm} & = &  0.
\label{eigen2}
\eea

Therefore, by using (\ref{eigen1}), (\ref{generalSpintwo}) and
(\ref{coeffgeneralSpintwo}),
the eigenvalue equations of the zero mode of
the higher spin $2$ current (the more general expression)
acting on the states 
$ |(f;0)>_{\pm}$
are given by
\bea
(\hat{\Phi}_0^{(2)})_0 |(f;0)>_{\pm} &=&
\Bigg[ \hat{a}_1 \, N  + \hat{a}_3 \left( \frac{(2 N+3)}{(N+2)} \right)  +
  \hat{a}_4 \left(-\frac{3}{4} \right) \Bigg] |(f;0)>_{\pm}\nonu \\
&=& \Bigg[ -\frac{N}{2 (N+2) (k+N+2)}
  (\hat{a}_2 k N^2-2 \hat{a}_1 k N-4 \hat{a}_1 k-
    \hat{a}_2 k+
    2 \hat{a}_2 N^3\nonu \\
    & + & 4 \hat{a}_1 N^2
    +  2 \hat{a}_2 N^2+4 \hat{a}_1 N-
    2 \hat{a}_2 N-2 \hat{a}_1-2 \hat{a}_2 )\Bigg]  |(f;0)>_{\pm}
\nonu \\
& \equiv &  (\hat{\phi}_0^{(2)})_{(f;0)} |(f;0)>_{\pm}.
\label{spin2f0}
\eea

Similarly,
the eigenvalue equations of the zero mode of
the higher spin $2$ current acting on the states 
$ |(0;f)>_{\pm}$ from (\ref{generalSpintwo}), (\ref{coeffgeneralSpintwo})
and (\ref{eigen2}),
are given by
\bea
(\hat{\Phi}_0^{(2)})_0 |(0;f)>_{\pm} &=&
\Bigg[ \hat{a}_7 (N+k+2) + \hat{a}_5 \left(-\frac{3}{4} \right)
  \Bigg] |(0;f)>_{\pm}
\nonu \\
& = &  \Bigg[
  -\frac{k}{2 (k+2) (k+N+2)}
  (4 \hat{a}_1 k^2-\hat{a}_2 k N^2-2 \hat{a}_1 k N+4 \hat{a}_1 k+
  \hat{a}_2 k \nonu \\
  & - & 2 \hat{a}_2 N^2-
    4 \hat{a}_1 N-2 \hat{a}_1+2 \hat{a}_2 )
  \Bigg] |(0;f)>_{\pm}
\nonu \\
& \equiv &  (\hat{\phi}_0^{(2)})_{(0;f)} |(0;f)>_{\pm}.
\label{spin20f}
\eea
The overall factors inside of the bracket
have $N \leftrightarrow k$ symmetry
in (\ref{spin2f0}) and (\ref{spin20f}).

By assuming that the two eigenvalues 
have the $N \leftrightarrow k$ symmetry \footnote{For the choice of
  \bea
\hat{a}_1  & = & \frac{(k+2) (k+3 N+4)}{2 (k-1) (k+1) (k+N+2)}, 
\qquad
\hat{a}_2  =  \frac{3 (k-N) (k N+2 k+2 N+3)}{(k-1) (k+1) (N-1) (N+1) (k+N+2)},
\label{twovalues}
\eea
the eigenvalues are given (after substituting (\ref{twovalues})
into (\ref{spin2f0}) and (\ref{spin20f}) respectively) by
\bea
(\hat{\phi}_0^{(2)})_{(f;0)}  & = &
-\frac{N}{ (k+N+2)} = (\phi_0^{(1)})_{(f;0)} \rightarrow 
-\la,
\nonu \\
(\hat{\phi}_0^{(2)})_{(0;f)}  &  = & 
-\frac{k}{ (k+N+2)} = (\phi_0^{(1)})_{(0;f)} \rightarrow 
-(1-\la),
\label{choice}
\eea
where the large $N$ 't Hooft limit is taken in (\ref{choice}).
}
\bea
(\hat{\phi}_0^{(2)})_{(f;0)}  & = &
\frac{N^2 (k+2 N+3)}{(N+2) (k+N+2)^2} \rightarrow 
\la (1+\la),
\nonu \\
(\hat{\phi}_0^{(2)})_{(0;f)}  & = &
\frac{k^2 (2 k+N+3)}{(k+2) (k+N+2)^2}
\rightarrow (1-\la) (2-\la),
\label{limitspin2}
\eea
where the large $N$ 't Hooft limit is taken,
one can determine the two undetermined coefficients as follows
\footnote{ Instead of (\ref{a1a2hat}),
  for the choice of
  \bea
  \hat{a}_1 & = & -\frac{(2 k^3+4 k^2 N+7 k^2+3 k N+6 k-4 N^2-6 N)}
  {2 (k-1) (k+1) (k+N+2)^2},
  \nonu \\
  \hat{a}_2 & = & -\frac{(4 k^3-6 k^2 N^2-7 k^2 N+8 k^2-7 k N^2-10 k N+
    3 k+4 N^3+8 N^2+3 N)}{(k-1) (k+1) (N-1) (N+1) (k+N+2)^2},
  \label{otherchoice}
  \eea
there is a minus sign in the first equation of (\ref{limitspin2}).}:
\bea
\hat{a}_1  & = &
-\frac{(2 k^3+6 k^2 N+7 k^2+4 k N^2+13 k N+6 k+4 N^2+6 N)}
{2 (k-1) (k+1) (k+N+2)^2},
\nonu \\
\hat{a}_2  & = & -\frac{(k-N)(4 k^2 N+4 k^2+4 k N^2+13 k N+8 k+4 N^2+8 N+3)}
{(k-1) (k+1) (N-1) (N+1) (k+N+2)^2}.
\label{a1a2hat}
\eea


The relations of the zero modes of the higher spin $3$ currents
(which are not described explicitly in this paper but we do have
them for several $N$ values explicitly)
acting on the states $|(f;0)>_{\pm}$
are given by, together with (\ref{spin2f0}),
\bea
(\hat{\Phi}_1^{(2), 12})_0 |(f;0)>_{\pm} &=&
 (\mp 2 i )(\hat{\phi}_0^{(2)})_{(f;0)} |(f;0)>_{\pm},
\nonu \\
(\hat{\Phi}_1^{(2), 13})_0 |(f;0)>_{\pm} &=&
 (\pm 2  )(\hat{\phi}_0^{(2)})_{(f;0)} |(f;0)>_{\mp},
\nonu \\
(\hat{\Phi}_1^{(2), 14})_0 |(f;0)>_{\pm} &=&
 ( 2 i )(\hat{\phi}_0^{(2)})_{(f;0)} |(f;0)>_{\mp},
\nonu \\
(\hat{\Phi}_1^{(2), 23})_0 |(f;0)>_{\pm} &=&
 ( -2 i )(\hat{\phi}_0^{(2)})_{(f;0)} |(f;0)>_{\mp},
 \nonu \\
(\hat{\Phi}_1^{(2), 24})_0 |(f;0)>_{\pm} &=&
 (\pm 2  )(\hat{\phi}_0^{(2)})_{(f;0)} |(f;0)>_{\mp},
\nonu \\
(\hat{\Phi}_1^{(2), 34})_0 |(f;0)>_{\pm} &=&
 (\pm 2 i )(\hat{\phi}_0^{(2)})_{(f;0)} |(f;0)>_{\pm}.
\label{eigen3}
\eea
The first and last one in (\ref{eigen3})
satisfy eigenvalue equations.
Furthermore, 
 the zero modes of the higher spin $3$ currents
 acting on the states $|(0;f)>_{\pm}$, together with (\ref{spin20f}),
 satisfy 
\bea
(\hat{\Phi}_1^{(2), 12})_0 |(0;f)>_{\pm} &=&
 (\pm 2 i )(\hat{\phi}_0^{(2)})_{(0;f)} |(0;f)>_{\pm},
\nonu \\
(\hat{\Phi}_1^{(2), 13})_0 |(0;f)>_{\pm} &=&
 (\mp 2  )(\hat{\phi}_0^{(2)})_{(0;f)} |(0;f)>_{\mp},
\nonu \\
(\hat{\Phi}_1^{(2), 14})_0 |(0;f)>_{\pm} &=&
 ( 2 i )(\hat{\phi}_0^{(2)})_{(0;f)} |(0;f)>_{\mp},
\nonu \\
(\hat{\Phi}_1^{(2), 23})_0 |(0;f)>_{\pm} &=&
 ( 2 i )(\hat{\phi}_0^{(2)})_{(0;f)} |(0;f)>_{\mp},
 \nonu \\
(\hat{\Phi}_1^{(2), 24})_0 |(0;f)>_{\pm} &=&
 (\pm 2  )(\hat{\phi}_0^{(2)})_{(0;f)} |(0;f)>_{\mp},
\nonu \\
(\hat{\Phi}_1^{(2), 34})_0 |(0;f)>_{\pm} &=&
 (\pm 2 i )(\hat{\phi}_0^{(2)})_{(0;f)} |(0;f)>_{\pm}.
\label{eigen4}
\eea
In this case, the first and the last one in (\ref{eigen4})
satisfy the eigenvalue equations.

The behaviors of (\ref{eigen3}) and (\ref{eigen4})
look similar to (\ref{eigen3-1}) and (\ref{eigen3-2}) respectively.
The signs are the same.

Finally, 
the zero modes of the higher spin $4$ current
(which is not described explicitly in this paper but we do have
this for several $N$ values explicitly)
acting on the states $|(f;0)>_{\pm}$ and  $|(0;f)>_{\pm}$
are given by
\bea
(\hat{\Phi}_2^{(2)})_0 |(f;0)>_{\pm} &=&
 \Bigg[
\frac{12 }{(2+k+N)(88 + 59 k + 59 N + 30 k N)}\,(88 + 108 k + 26 k^2 + 98 N 
\nonu \\
& + & 95 k N + 12 k^2 N + 27 N^2 + 
   18 k N^2)
\Bigg](\hat{\phi}_0^{(2)})_{(f;0)} |(f;0)>_{\pm},
 \nonu \\
 & \rightarrow & \frac{12}{5} \la (1+\la) (2+\la) |(f;0)>_{\pm},
 \nonu \\
(\hat{\Phi}_2^{(2)})_0 |(0;f)>_{\pm} &=&
-\Bigg[ \frac{12 }{(2+k+N)(88 + 59 k + 59 N + 30 k N)}\,
(88 + 108 N + 26 N^2 + 98 k 
\nonu \\
& + & 95 N k  + 12 N^2 k + 27 k^2 + 
   18 N k^2)
   \Bigg](\hat{\phi}_0^{(2)})_{(0;f)} |(0;f)>_{\pm}
\nonu \\
 & \rightarrow & -\frac{12}{5} (1-\la) (2-\la)(3-\la) |(0;f)>_{\pm}.
\label{Eigenfour}
\eea
There is $N \leftrightarrow k$ symmetry.
Note that the extra factors in the brackets
in (\ref{Eigenfour})
behave as $\frac{12}{5}(2+\la)$
(which is proportional to $(1+\la)$ plus one)
and $-\frac{12}{5}(3-\la)$ (which is proportional to $(2-\la)$
plus one) under the large $N$
't Hooft limit respectively.
During this calculation,
the contributions come from the coefficients of
in $k$, $N$ in the first factor of denominator, and
$k N$ in the second factor of denominator
and the coefficients of
$k^2 N$, and $k N^2$ in the numerator (they are given by
$1, 1, 30, 12$ and $18$).
This implies that the above large $N$ behavior in
(\ref{limitspin2}) is reasonable because
one observes $(2+\la)$ and $(3-\la)$ respectively.
The above eigenvalues are very similar to the ones
in the orthogonal Wolf space coset case in \cite{AKP1510} under the
large $N$ 't Hooft limit.
For the different choice in (\ref{otherchoice}),
the signs are coincident with each other.


Then the three point functions with these two scalars  
from (\ref{eigen3}), (\ref{eigen4}) and (\ref{Eigenfour})
can be written as
\bea
<\overline{{\cal O}}_{+ } 
{\cal O}_{+ } \hat{\Phi}_0^{(2)}> & = &
(\hat{\phi}_0^{(2)})_{(f;0)},
\qquad
< \overline{{\cal O}}_{- } 
{\cal O}_{- } \hat{\Phi}_0^{(2)}>  =  (\hat{\phi}_0^{(2)})_{(0;f)},
\nonu \\
<\overline{{\cal O}}_{+ } 
{\cal O}_{+ } \hat{\Phi}_1^{(2),12}> & = &
(\mp 2i) (\hat{\phi}_0^{(2)})_{(f;0)},
\qquad
< \overline{{\cal O}}_{- } 
{\cal O}_{- } \hat{\Phi}_1^{(2),12}>  = 
(\pm 2i)
(\hat{\phi}_0^{(2)})_{(0;f)},
\nonu \\
 <\overline{{\cal O}}_{+ } 
 {\cal O}_{+ } \hat{\Phi}_1^{(2),34}>
  & = & (\pm 2i) (\hat{\phi}_0^{(2)})_{(f;0)},
\qquad
 <\overline{{\cal O}}_{- } 
 {\cal O}_{-} \hat{\Phi}_1^{(2),34}>   = 
 (\pm 2i)
(\hat{\phi}_0^{(2)})_{(0;f)},
\nonu \\
<\overline{{\cal O}}_{+ } 
{\cal O}_{+ } \hat{\Phi}_2^{(2)}> & = &
 \Bigg[
\frac{12 }{(2+k+N)(88 + 59 k + 59 N + 30 k N)}\,(88 + 108 k + 26 k^2 + 98 N 
\nonu \\
& + & 95 k N + 12 k^2 N + 27 N^2 + 
   18 k N^2)
\Bigg]
(\hat{\phi}_0^{(2)})_{(f;0)},
 \nonu \\
 <\overline{{\cal O}}_{- } 
 {\cal O}_{- } \hat{\Phi}_2^{(2)}> & = &
-\Bigg[ \frac{12 }{(2+k+N)(88 + 59 k + 59 N + 30 k N)}\,
(88 + 108 N + 26 N^2 + 98 k 
\nonu \\
& + & 95 N k  + 12 N^2 k + 27 k^2 + 
   18 N k^2)
   \Bigg](\hat{\phi}_0^{(2)})_{(0;f)}.
 \label{twothreenon}
\eea
The overall behavior of (\ref{twothreenon})
looks similar to the previous ones in (\ref{threenonlinear}) in the
sense that the three-point functions of the higher spin $3$ currents
have simple form where the overall coefficients do not depend on
$N$ and $k$ explicitly. Of course, the other three-point functions
for the higher spin $3$ currents vanish.

\subsection{Eigenvalue equation of higher spin $2$, $3$, $4$ currents
in the extension of the large ${\cal N}=4$ linear superconformal algebra}

In order to obtain the three-point functions
of the higher spin currents in the linear version,
one should determine the lowest higher spin $2$ current in terms of
the currents of ${\cal N}=4$ linear superconformal algebra
and the corresponding higher spin currents only.
Now one describes the previous higher spin $2$ current
(\ref{generalSpintwoother}) by substituting $L(z)$ and $T^{\mu\nu}(z)$
into the linear ones using the relations
(\ref{gsformula}) (the higher spin currents $\Phi_0^{(1)}(z)$
and $\Phi_0^{(2)}(z)$ remain the same) 
\bea
{\bf \hat{\Phi}_0^{(2)}}(z) & = & 
-\frac{1}{8} \hat{a}_2 \, (k+N+2)^2 \, {\bf \Phi_0^{(2)}}(z)
\nonu \\
& + &    
(-2 \hat{a}_2 k^2-3 \hat{a}_2 k N+8 \hat{a}_1 k-9 \hat{a}_2 k+
8 \hat{a}_1
N-6 \hat{a}_2 N+16 \hat{a}_1-10 \hat{a}_2) \nonu \\
& \times & \Bigg[
-\frac{3 (2 k N+3 k+3 N+4)}{16 (k+2) (N+2)}
  \, {\bf {\Phi}_{0}^{(1)}} {\bf {\Phi}_{0}^{(1)}}+
  \frac{1}{4} \,{\bf L} \nonu \\
  & - & 
\frac{(k+N+4)}{32 (k+2) (N+2)}
  \,({\bf {T}^{\mu\nu}})^{2}+
\frac{(N-k)}{64 (k+2) (N+2)}
\, \varepsilon^{\mu\nu
\rho\si} \,{\bf {T}^{\mu\nu}} {\bf {T}^{\rho\si}} 
\nonu \\
& + & \frac{1}{4(2+k+N)}\,{\bf U}^{2}
+  \frac{i(4+N+k)}{8(2+N)(2+k)(2+k+N)}\,{\bf {T}^{\mu\nu}}
{\bf {\Gamma}^{\mu}}{\bf {\Gamma}^{\nu}}
\nonu \\
& + & 
\frac{(-N+k)}{16(2+N)(2+k)(2+k+N)}\,\varepsilon^{\mu\nu\rho\si} \,
\Bigg(\,i\,
     {\bf {T}^{\mu\nu}}{\bf {\Gamma}^{\rho}}{\bf {\Gamma}^{\si}}
     +\frac{1}{(2+k+N)}\,{\bf {\Gamma}^{\mu}}{\bf {\Gamma}^{\nu}}
     {\bf {\Gamma}^{\rho}}{\bf {\Gamma}^{\si}}\Bigg) 
\nonu \\
& + &  \frac{(20+7N+7k+2Nk)}{8(2+N)(2+k)(2+k+N)}\,
      {\bf {\Gamma}^{\mu}} \partial{\bf {\Gamma}^{\mu}}\, \Bigg](z).
\label{lineargeneralspintwo}
\eea
Let us emphasize that the field contents in 
$\hat{\Phi}_0^{(2)}(z)$ in the nonlinear version
is the same as the ones ${\bf \hat{\Phi}_0^{(2)}}(z)$ in the linear version.
At the level of $V^a(z)$ and $Q^a(z)$, they are exactly the same
quantities.
We have checked that starting from the most general spin
$2$ current in (\ref{lineargeneralspintwo}),
we can construct the remaining $15$ higher spin currents
and the resulting $16$ higher spin currents satisfy the ${\cal N}=4$
primary conditions characterized by Appendix A.2 for generic $\hat{a}_1$
and $\hat{a}_2$. Then one has the relations (\ref{Phinonandlin})
with hat notation.

One can evaluate the relevant equations for the
zero mode of the higher spin $3$ currents acting on the state
(again the $c_1$ term in (\ref{Phinonandlin}) does not contribute the
eigenvalue equations)
\bea
({\bf \hat{\Phi}_1^{(2), 12}})_0 |(f;0)>_{\pm} &=&
 (\mp 2 i )( \hat{\phi}_0^{(2)})_{(f;0)} |(f;0)>_{\pm},
\nonu \\
({\bf \hat{\Phi}_1^{(2), 13}})_0 |(f;0)>_{\pm} &=&
 (\pm 2  )( \hat{\phi}_0^{(2)})_{(f;0)} |(f;0)>_{\mp},
\nonu \\
({\bf \hat{\Phi}_1^{(2), 14}})_0 |(f;0)>_{\pm} &=&
 ( 2 i )( \hat{\phi}_0^{(2)})_{(f;0)} |(f;0)>_{\mp},
\nonu \\
({\bf \hat{\Phi}_1^{(2), 23}})_0 |(f;0)>_{\pm} &=&
 ( -2 i )( \hat{\phi}_0^{(2)})_{(f;0)} |(f;0)>_{\mp},
 \nonu \\
({\bf \hat{\Phi}_1^{(2), 24}})_0 |(f;0)>_{\pm} &=&
 (\pm 2  )( \hat{\phi}_0^{(2)})_{(f;0)} |(f;0)>_{\mp},
\nonu \\
({\bf \hat{\Phi}_1^{(2), 34}})_0 |(f;0)>_{\pm} &=&
 (\pm 2 i )( \hat{\phi}_0^{(2)})_{(f;0)} |(f;0)>_{\pm}.
\label{lineareigeneigen}
\eea
Furthermore, the other relevant equations lead to
the following results
\bea
({\bf \hat{\Phi}_1^{(2), 12}})_0 |(0;f)>_{\pm} &=&
 (\mp 2 i )( \hat{\phi}_0^{(2)})_{(0;f)} |(0;f)>_{\pm},
\nonu \\
({\bf \hat{\Phi}_1^{(2), 13}})_0 |(0;f)>_{\pm} &=&
 (\mp 2  )( \hat{\phi}_0^{(2)})_{(0;f)} |(0;f)>_{\mp},
\nonu \\
({\bf \hat{\Phi}_1^{(2), 14}})_0 |(0;f)>_{\pm} &=&
 ( -2 i )( \hat{\phi}_0^{(2)})_{(0;f)} |(0;f)>_{\mp},
\nonu \\
({\bf \hat{\Phi}_1^{(2), 23}})_0 |(0;f)>_{\pm} &=&
 ( -2 i )( \hat{\phi}_0^{(2)})_{(0;f)} |(0;f)>_{\mp},
 \nonu \\
({\bf \hat{\Phi}_1^{(2), 24}})_0 |(0;f)>_{\pm} &=&
 (\mp 2  )( \hat{\phi}_0^{(2)})_{(0;f)} |(0;f)>_{\mp},
\nonu \\
({\bf \hat{\Phi}_1^{(2), 34}})_0 |(0;f)>_{\pm} &=&
 (\mp 2 i )( \hat{\phi}_0^{(2)})_{(0;f)} |(0;f)>_{\pm}.
\label{eigen4-1}
\eea
In this case, the first and the last one in (\ref{eigen4-1})
satisfy the eigenvalue equations.
They are the same as (\ref{eigen3}) and (\ref{eigen4}) respectively.

Finally, 
 the zero modes of the `nonprimary' higher spin $4$ currents
acting on the states $|(f;0)>_{\pm}$ and  $|(0;f)>_{\pm}$
are given by
\footnote{
  One can calculate the eigenvalue equations for the extra terms
  in (\ref{Phinonandlin})
  \bea
({\bf L \hat{\Phi}_{0}^{(2)}})_0 |(f;0)>_{\pm}
& = &
\Bigg[ \frac{(4 k N+8 k+5 N^2+20 N+19)}{2 (N+2) (k+N+2)}
  \Bigg]( \hat{\phi}_0^{(2)})_{(f;0)} |(f;0)>_{\pm},
\label{c7c11cal-3}
\\
( \partial^{2} {\bf \hat{\Phi}_{0}^{(2)}})_0 |(f;0)>_{\pm}
& = & 
\Bigg[ 6
  \Bigg] ( \hat{\phi}_0^{(2)})_{(f;0)} |(f;0)>_{\pm},
\nonu \\
(\partial {\bf U \hat{\Phi}_{0}^{(2)}} )_0 |(f;0)>_{\pm}
& = & 
\Bigg[\frac{1}{2} i \sqrt{\frac{N}{(N+2)}}
  \Bigg] ( \hat{\phi}_0^{(2)})_{(f;0)} |(f;0)>_{\pm},
\nonu \\
( {\bf U } \partial {\bf \hat{\Phi}_{0}^{(2)}} )_0 |(f;0)>_{\pm}
& = &
\Bigg[i \sqrt{\frac{N}{(N+2)}}
  \Bigg] ( \hat{\phi}_0^{(2)})_{(f;0)} |(f;0)>_{\pm},
\nonu \\
( {\bf U U \hat{\Phi}_{0}^{(2)}} )_0 |(f;0)>_{\pm}
&  = & 
\Bigg[-\frac{N}{4 (N+2)}
  \Bigg] ( \hat{\phi}_0^{(2)})_{(f;0)} |(f;0)>_{\pm}.
\nonu
\eea
Similarly, one has the following eigenvalue equations
\bea
({\bf L \hat{\Phi}_{0}^{(2)}})_0 |(0;f)>_{\pm}
& = &
\Bigg[ \frac{(5 k N+4 N^2+10 N+1)}{2 N (k+N+2)}
  \Bigg] ( \hat{\phi}_0^{(2)})_{(0;f)}  |(0;f)>_{\pm},
\label{c7c11cal-4}
\\
( \partial^{2} {\bf \hat{\Phi}_{0}^{(2)}})_0 |(0;f)>_{\pm}
& = & 
\Bigg[ 6
  \Bigg] ( \hat{\phi}_0^{(2)})_{(0;f)} |(0;f)>_{\pm},
\nonu \\
(\partial {\bf U \hat{\Phi}_{0}^{(2)}} )_0 |(0;f)>_{\pm}
& = &  
\Bigg[ \frac{1}{2} i \sqrt{\frac{(N+2)}{N}}
  \Bigg] ( \hat{\phi}_0^{(2)})_{(0;f)} |(0;f)>_{\pm},
\nonu \\
( {\bf U } \partial {\bf \hat{\Phi}_{0}^{(2)}} )_0 |(0;f)>_{\pm}
& = &
\Bigg[ i \sqrt{\frac{(N+2)}{N}}
  \Bigg] ( \hat{\phi}_0^{(2)})_{(0;f)} |(0;f)>_{\pm},
\nonu \\
( {\bf U U \hat{\Phi}_{0}^{(2)}} )_0 |(0;f)>_{\pm}
& = &
\Bigg[ -\frac{(N+2)}{4 N}
  \Bigg] ( \hat{\phi}_0^{(2)})_{(0;f)} |(0;f)>_{\pm}.
\nonu
\eea
}
{\small
\bea
({\bf \hat{\Phi}_2^{(2)}})_0 |(f;0)>_{\pm} &=&
 (\Phi_2^{(2)})_0 |(f;0)>_{\pm}
    \nonu \\
    &- & 
\Bigg[ c_{7}\,
  {\bf L \hat{\Phi}_{0}^{(2)}}+c_{8}\, \partial^{2}
  {\bf \hat{\Phi}_{0}^{(2)}}+
    c_{9}\, \partial {\bf U \hat{\Phi}_{0}^{(2)}}+c_{10}\,
    {\bf U } \partial {\bf \hat{\Phi}_{0}^{(2)}}
    + c_{11}\,{\bf U U \hat{\Phi}_{0}^{(2)}} \Bigg]_0 |(f;0)>_{\pm}
\nonu \\
& = &
(6)( \hat{\phi}_0^{(2)})_{(f;0)} |(f;0)>_{\pm},
\nonu \\
({\bf \hat{\Phi}_2^{(2)}})_0 |(0;f)>_{\pm} &=&
(\Phi_2^{(2)})_0 |(0;f)>_{\pm}
    \nonu \\
    &- & 
\Bigg[ c_{7}\,
  {\bf L \hat{\Phi}_{0}^{(2)}}+c_{8}\, \partial^{2}
  {\bf \hat{\Phi}_{0}^{(2)}}+
    c_{9}\, \partial {\bf U \hat{\Phi}_{0}^{(2)}}+c_{10}\,
    {\bf U } \partial {\bf \hat{\Phi}_{0}^{(2)}}
    + c_{11}\,{\bf U U \hat{\Phi}_{0}^{(2)}} \Bigg]_0 |(0;f)>_{\pm}
\nonu \\
& = &
(-6)( \hat{\phi}_0^{(2)})_{(0;f)} |(0;f)>_{\pm}.
\label{linearspin4eigen}
\eea}
As in (\ref{eigen5}), all the $N$ and $k$ dependence in the combinations
of the coefficients and the eigenvalues in (\ref{c7c11cal-3})
and (\ref{c7c11cal-4}) disappear.

On the other hand, in the primary basis, using (\ref{widetildenot})
with hat,
(\ref{linearspin4eigen}),
(\ref{c7c11cal-3}) and (\ref{c7c11cal-4}),
the following relations hold
{\small
\bea
({\bf \widetilde{\hat{\Phi}}_2^{(2)}})_0 |(f;0)>_{\pm} & = &
 \Bigg[
\frac{12 }{(2 + N) (2 + k + N) (59 + 37 k + 37 N + 15 k N)}\,(118 + 135 k + 32 k^2 
\nonu \\
&+ & 190 N + 179 k N + 28 k^2 N + 100 N^2 + 
 73 k N^2 + 6 k^2 N^2 + 18 N^3 + 9 k N^3)
 \Bigg] \nonu \\
 & \times & (\hat{\phi}_0^{(2)})_{(f;0)} |(f;0)>_{\pm}
 \rightarrow  \frac{12}{5} \la (1+\la)(2+\la) |(f;0)>_{\pm},
\nonu \\
({\bf \widetilde{\hat{\Phi}}_{2}^{(2)}})_0 |(0;f)>_{\pm}  & = &
\Bigg[ \frac{-12 }{N (2 + k + N) (59 + 37 k + 37 N + 15 k N)}\, (-3 k + 62 N + 64 k N 
\nonu \\
&+ & 18 k^2 N + 69 N^2 + 55 k N^2 + 9 k^2 N^2 + 
 16 N^3 + 6 k N^3) 
 \Bigg]\nonu \\
& \times & (\hat{\phi}_0^{(2)})_{(0;f)} |(0;f)>_{\pm} \rightarrow
-\frac{12}{5} (1-\la)(2-\la)(3-\la) |(0;f)>_{\pm}.
\label{beforerel}
\eea}
As described before, this reduces to (\ref{linearspin4eigen}) for $k=N$.
In this case, the results in (\ref{beforerel})
coincide with the ones in (\ref{Eigenfour}) under the large
$N$ 't Hooft limit \footnote{It is easy to check that
  the extra terms vanish under the large $N$ 't Hooft limit.
  One obtains the differences of each eigenvalue as follows:
  \bea
  && \frac{36 (k-N) (6 k^2 N+12 k^2+12 k N^2+71 k N+93 k+21 N^3+
    89 N^2+178 N+138)}{(N+2) (k+N+2)
    (15 k N+37 k+37 N+59) (30 k N+59 k+59 N+88)}
  \rightarrow   \nonu \\
  & & -\frac{6 (2 \la-1) (5 \la^2+2)}
              {25 (\la-1)^2 } \frac{1}{N^2},
              \nonu \\
            &&  \frac{36 (k-N) (21 k^2 N+12 k N^2+17 k N-59 k+6 N^3+
                17 N^2-34 N-88)}{
                N (k+N+2) (15 k N+37 k+37 N+59) (30 k N+59 k+59 N+88)}  
               \rightarrow  \nonu \\
              & & -  \frac{6 (2 \la-1) (5 \la^2-10 \la+7)}{25
                (\la-1)^2} \frac{1}{N^2}.
\label{limitexpexp}
\eea
Then in the large $N$ 't Hooft limit, these (\ref{limitexpexp})
vanish.
}.
Again the large $N$ 't Hooft limit behavior of (\ref{beforerel})
looks similar to
the ones in the orthogonal coset case in \cite{AKP1510}.

Let us summarize the three-point functions of the higher spin
$2, 3, 4$ currents from (\ref{lineareigeneigen}),
(\ref{eigen4-1}), (\ref{linearspin4eigen}) and (\ref{beforerel})
as follows:
{\small
\bea
<\overline{{\cal O}}_{+ } 
{\cal O}_{+ } {\bf \hat{\Phi}_0^{(2)}}> & = &
(\hat{\phi}_0^{(2)})_{(f;0)},
\qquad
< \overline{{\cal O}}_{- } 
{\cal O}_{- } {\bf \hat{\Phi}_0^{(2)}}>  =  (\hat{\phi}_0^{(2)})_{(0;f)},
\nonu \\
<\overline{{\cal O}}_{+ } 
{\cal O}_{+ } {\bf \hat{\Phi}_1^{(2),12}}> & = &
(\mp 2i) (\hat{\phi}_0^{(2)})_{(f;0)}, 
\qquad
< \overline{{\cal O}}_{- } 
{\cal O}_{- } {\bf \hat{\Phi}_1^{(2),12}}>  = 
(\pm 2i)
(\hat{\phi}_0^{(2)})_{(0;f)},
\nonu \\
<\overline{{\cal O}}_{+ } 
{\cal O}_{+ } {\bf \hat{\Phi}_1^{(2),34}}> & = &
(\pm 2i) (\hat{\phi}_0^{(2)})_{(f;0)},
\qquad
<\overline{{\cal O}}_{- } 
{\cal O}_{-} {\bf \hat{\Phi}_1^{(2),34}}>  = 
(\pm 2i)
(\hat{\phi}_0^{(2)})_{(0;f)},
\nonu \\
<\overline{{\cal O}}_{+ } 
{\cal O}_{+ } {\bf \hat{\Phi}_2^{(2)}}> & = &
(6)
(\hat{\phi}_0^{(2)})_{(f;0)},
 \qquad
 <\overline{{\cal O}}_{- } 
 {\cal O}_{- } {\bf \hat{\Phi}_2^{(2)}}>  = 
 (-6)
 (\hat{\phi}_0^{(2)})_{(0;f)},
 \nonu \\
 <\overline{{\cal O}}_{+ } 
{\cal O}_{+ } {\bf \widetilde{\hat{\Phi}}_2^{(2)}}> & = &
 \Bigg[
\frac{12 }{(2 + N) (2 + k + N) (59 + 37 k + 37 N + 15 k N)}\,(118 + 135 k + 32 k^2 
\nonu \\
&+ & 190 N + 179 k N + 28 k^2 N + 100 N^2 + 
 73 k N^2 + 6 k^2 N^2 + 18 N^3 + 9 k N^3)
 \Bigg] \nonu \\
 & \times &
(\hat{\phi}_0^{(2)})_{(f;0)},
 \nonu \\
 <\overline{{\cal O}}_{- } 
 {\cal O}_{- } {\bf \widetilde{\hat{\Phi}}_2^{(2)}}> & = &
\Bigg[ \frac{-12 }{N (2 + k + N) (59 + 37 k + 37 N + 15 k N)}\, (-3 k + 62 N + 64 k N 
\nonu \\
&+ & 18 k^2 N + 69 N^2 + 55 k N^2 + 9 k^2 N^2 + 
 16 N^3 + 6 k N^3) 
 \Bigg]
 (\hat{\phi}_0^{(2)})_{(0;f)}.
 \label{spinfinallinear}
\eea}
Therefore, the three-point functions in (\ref{spinfinallinear})
have simple forms which are multiples of the eigenvalues of
the higher spin $2$ current in (\ref{spin2f0}) and (\ref{spin20f})
even at finite
$N$ and $k$
except of the last two.
The other three-point functions for the higher spin $3$
currents  vanish.
Compared to the nonlinear case (\ref{twothreenon}),
the only last two in (\ref{spinfinallinear})
are different from each other at finite $N$ and $k$.

\section{ Higher spin
  extension of $SO({\cal N}=4)$ Knizhnik Bershadsky algebra in the Wolf
space}

As an application of the OPEs
between the $16$ currents described in previous sections,
the extension of  $SO({\cal N}=4)$ Knizhnik Bershadsky algebra
is analyzed.

\subsection{$SO({\cal N}=4)$ Knizhnik Bershadsky algebra
  from the large ${\cal N}=4$ nonlinear superconformal algebra with the
  condition
$k=N$}

The  Knizhnik Bershadsky algebra \cite{Knizhnik,Bershadsky}
generated by the stress energy tensor $T_{KB}(z)$ of spin $2$,
the four spin $\frac{3}{2}$ currents $G^{\mu}_{KB}(z)$
where the vector index $\mu=1,2, \cdots, {\cal N}=4$,
and the spin $1$ currents $J_{KB}^a(z)$
where the adjoint index
$a=1,2, \cdots, \frac{1}{2}{\cal N}({\cal N}-1)=6$
can be described as  
\bea
J_{KB}^a(z) \, J_{KB}^b(w) & = & -\frac{1}{(z-w)^2} \, 
k \, \delta^{ab} +
\frac{1}{(z-w)} \,  f^{abc} \, J_{KB}^c(w) +\cdots,
\nonu \\
 J_{KB}^a(z) \, G_{KB}^{\mu}(w) & = & 
\frac{1}{(z-w)} \, T^a_{\nu \mu} \, G_{KB}^{\nu}(w) +\cdots,
\nonu \\
G_{KB}^{\mu}(z) \, G_{KB}^{\nu}(w) & = & 
\frac{1}{(z-w)^3} \,  \frac{2k^2}{(1+k)} \, \delta^{\mu \nu}
+ \frac{1}{(z-w)^2} \,  \frac{2k}{(1+k)} \, T^a_{\mu \nu} \,
J_{KB}^a(w) \nonu \\
& + &  \frac{1}{(z-w)}
\Bigg[  2 \, \delta^{\mu \nu} \, T_{KB} +
  \frac{k}{(1+k)} \, T^a_{\mu \nu} \,
  \pa J_{KB}^a 
  \nonu \\
  & +& \frac{1}{2(1+k)} \left( T^a_{\mu \rho} \, T^b_{\rho \nu} +
  T^b_{\mu \rho} \, T^a_{\rho \nu} +
  2 \delta^{ab} \, \delta_{\mu \nu} \right) J_{KB}^a \, J_{KB}^b
\Bigg](w) +\cdots,
\nonu \\
T_{KB}(z) \, J_{KB}^i(w) & = &  
\frac{1}{(z-w)^2} \,  J_{KB}^i(w) + \frac{1}{(z-w)} \, \pa J_{KB}^i(w) + \cdots,
\nonu \\
T_{KB}(z) \, G_{KB}^i(w) & = &  
\frac{1}{(z-w)^2} \, \frac{3}{2} \, G_{KB}^i(w) + 
\frac{1}{(z-w)} \, \pa G_{KB}^i(w) + \cdots,
\nonu \\
T_{KB}(z) \, T_{KB}(w) & = & \frac{1}{(z-w)^4} \, \frac{3k}{2}  + 
\frac{1}{(z-w)^2} \, 2 \, T_{KB}(w) + \frac{1}{(z-w)} \, \pa T_{KB}(w)
+ \cdots.
\label{nonKB}
\eea
The explicit $SO({\cal N}=4)$ generators $T_{\mu\nu}^a$ are given in
Appendix $C$.
The central charge is given by $c=3k$ (or $c=3N$) where $k =1, 2,
\cdots$. 
The Wolf space central charge $c_W=\frac{3k^2}{(k+1)}$.
Then it is straightforward to check that  
by identifying the following relations
\bea
T_{KB}(z) & = & T(z),
\qquad
G_{KB}^1(z)  =  -G^0(z), 
\nonu \\
G_{KB}^2(z) & = & G^1(z), 
\qquad
G_{KB}^3(z)  =  -G^3(z),
\nonu \\
G_{KB}^4(z) & = & G^2(z), 
\qquad
J_{KB}^1(z)  =  -A^{+1}(z) +A^{-1}(z), 
\nonu \\
J_{KB}^2(z) & = & A^{+3}(z) -A^{-3}(z), 
\qquad
J_{KB}^3(z)  =  -A^{+2}(z) +A^{-2}(z), 
\nonu \\
J_{KB}^4(z) & = & -A^{+2}(z) -A^{-2}(z), 
\qquad
J_{KB}^5(z)  =  -A^{+3}(z) -A^{-3}(z), 
\nonu \\
J_{KB}^6(z) & = & -A^{+1}(z) -A^{-1}(z),
\label{11currentsinKB}
\eea
one sees that the above 
Knizhnik Bershadsky algebra (\ref{nonKB}) leads to
the previous large ${\cal N}=4$ nonlinear superconformal algebra. 
In Appendix $D$, we present the corresponding Appendix $A$ in this case.

\subsection{ Higher spin
  extension of $SO({\cal N}=4)$ Knizhnik Bershadsky algebra
in the Wolf space}

Then one obtains
 the higher spin
  extension of $SO({\cal N}=4)$ Knizhnik Bershadsky algebra
  in the Wolf space and they
  are described in Appendix $B$ together with
  (\ref{11currentsinKB}) and $k=N$.
  In Appendix $E$, the various coefficients in terms of $k$
  are given explicitly.

\subsection{ The $SO({\cal N}=3)$ Knizhnik Bershadsky algebra
and its extension}

By taking the following choices from the currents in the linear version
\bea
\hat{J}^{1}(z) & = & {\bf T}^{23}(z), \qquad
\hat{J}^{2}(z)  =  {\bf T}^{13}(z), \qquad
\hat{J}^{3}(z)  =  {\bf T}^{12}(z), \nonu\\
\hat{G}^{1}(z) & = &
    {\bf G}^{1}(z)-\frac{i(k-N)}{(2+k+N)}\,{\bf \partial \Gamma}^{1}(z)
    +\frac{2}{(2+k+N)}\,{\bf T}^{23}{\bf \Gamma}^{4}(z), \nonu\\
    \hat{G}^{2}(z) & = & -{\bf G}^{2}(z)+\frac{i(k-N)}{(2+k+N)}\,
        {\bf \partial \Gamma}^{2}(z)+
        \frac{2}{(2+k+N)}\,{\bf T}^{13}{\bf \Gamma}^{4}(z), \nonu\\
        \hat{G}^{3}(z) & = & {\bf G}^{3}(z)-\frac{i(k-N)}{(2+k+N)}\,
            {\bf \partial \Gamma}^{3}(z)+
            \frac{2}{(2+k+N)}\,{\bf T}^{12}{\bf \Gamma}^{4}(z), \nonu\\
            \hat{T}(z) & = & {\bf L}(z)-\frac{(k-N)}{2(2+k+N)}\,{\bf \partial U}(z)-
            \frac{1}{(2+k+N)}\,{\bf \partial \Gamma}^{4}{\bf \Gamma}^{4}(z),
\label{so3kb}
\eea
one obtains the  
$SO({\cal N}=3)$ Knizhnik Bershadsky algebra \cite{Knizhnik,Bershadsky}.
By adding the fermion ${\bf \Gamma^4}(z)$ to (\ref{so3kb}), one obtains
the ${\cal N}=3$ linear superconformal algebra \cite{CK,Schoutens,AK1607}.
The central term coming from the fourth order term in the OPE between the
$\hat{T}(z)$ and $\hat{T}(w)$ is given by
$\frac{1}{4}(5+3k+3N)$.

Then one can see that the previous results in \cite{AK1509}
provide the higher spin extension of
 the  
$SO({\cal N}=3)$ Knizhnik Bershadsky algebra.

\section{ Conclusions and outlook}

First of all, we have obtained the OPEs in Appendix (\ref{1116})
which were necessary to calculate the higher spin $\frac{5}{2}, 3,
\frac{7}{2}, 4$ currents in the next $16$ higher spin currents.
For the bosonic higher spin $3, 4$ currents, their expressions
in terms of WZW currents for several $N$ can be read off from the defining
relations in Appendix (\ref{1116}).

Secondly, we also have determined the OPEs
between the lowest $16$ higher spin currents
in Appendix $B$.
One of the motivations to obtain these is to 
describe the marginal deformation which breaks the higher spin symmetry 
and determine the mass for the higher spin currents
in the large $(N, k)$ 't Hooft limit.
The coset model we consider here has ${\cal N}=4$ supersymmetry and there 
should be a marginal deformation \cite{GMMS}.
It would be interesting to determine the mass for the higher spin currents
with the help of the explicit symmetry algebra found in this paper. 
According to the observation of  \cite{CHR1406,HR1503,CH1506},
the $SO(2)_R$ doublet rather than $SO(2)_R$ singlet can have nonzero 
mass contribution.
See also the relevant work in \cite{GPZ1506}.
It is known that the divergence of higher spin-$s$
  currents
  contains some operator which breaks the higher spin symmetry \cite{GJL}.
  Although the explicit form for the
  perturbing marginal operator which transforms as a nontrivial
  primary state from the
  denominator of the coset is not determined yet,
  one can analyze further by assuming that there exist the eigenvalue
  equations for the zeromodes of the spins $s=1,2,3$ acting on the
  state associated with the perturbing operator \cite{Ahn1604}.
  The anormalous dimension of the higher spin
  current at the leading order corresponds to the norm of the above
  divergence of the corresponding higher spin current \cite{CHR1406}.
  Then from our result, one obtains
  the commutation relations
  $[\Phi_{1,m}^{(1),\mu\nu},\Phi_{1,n}^{(1),\rho\sigma}]$
  and   $[\Phi_{1,m}^{(2)},\Phi_{1,n}^{(2)}]$
  as well as the standard commutators
  $[L_m,L_n]$, $[L_m, \Phi_{1,n}^{(1),\mu\nu}]$ and  $[L_m, \Phi_{1,n}^{(2)}]$.
  Then one can write down the various 
modes 
  between the states
 from the higher spin symmetry
 breaking terms.
 By moving the positive modes to the right
  in the corresponding terms, and using
  the commutators and the eigenvalue equations, we will obtain
  the coefficient of the norm of
  perturbing marginal operator which depends on the 
  eigenvalues. It would be interesting to find the perturbing marginal
  operator explicitly
  and describe the mass terms with the explicit eigenvalues
  further.

Thirdly, the lowest higher spin $2$ current (\ref{generalSpintwo})
or (\ref{generalSpintwoother})
in the next $16$ higher spin currents was found explicitly
in terms of WZW currents.
The sixth and ninth terms having the quadratic $d^{0}_{\bar{a}\bar{b}}$
tensors (\ref{dzero}) are
new compared to the orthogonal case \cite{AKP1510}.
They come from the
quadratic higher spin $1$ currents in (\ref{finalspinone}). 
It would be interesting to
obtain the remaining higher spin
$\frac{5}{2}, 3, \frac{7}{2}, 4$ currents in (\ref{second})
in terms of spin $1$ and spin $\frac{1}{2}$ currents
for generic $N$ and $k$.
This can be done manually by using Appendix (\ref{1116}) along the line of
\cite{AK1411} or
one can read off the relative coefficients appearing in
the possible terms from the explicit results for several $N$ cases.
For the latter, it is nontrivial to obtain the possible terms
completely.

Fourthly,
the three-point functions in (\ref{twothreenon})
and (\ref{spinfinallinear})
were obtained.
For the $SO(4)$ adjoint higher spin $3$ currents,
the three-point functions
are the same in nonlinear and linear versions.
For the spin $4$ current, the $N \leftrightarrow k$ symmetry
appears in the nonlinear version as in the stress energy tensor of spin
$2$ in (\ref{eigenspin2}). 
As described before, these three-point functions
look similar to the ones in \cite{AKP1510} in the sense that
the $\la$ dependence appearing in the three-point functions
coincides with each other.

It is straightforward to apply
the work of \cite{AK1509} and the presentation of this paper
to the orthogonal case \cite{AP1410} in the context of
\cite{Ahn1106,GV1106,AP1301,AP1310,CGKV}. 
The lowest $16$ higher spin currents contains
the higher spin $2$ current as its lowest component.
The complete OPEs between
the lowest higher spin $2$ currents
can be constructed and it turns out that
there are next $16$ higher spin currents
containing the higher spin $3$ current as its
lowest component in the right hand sides of the OPEs.
Moreover,  
there are three different 
next $16$ higher spin currents.
They transform as $SO(3)$ triplet
inside of $SO(4)$
rather than a singlet.
Furthermore,
 the third $16$ higher spin currents
containing the higher spin $4$ current as its
lowest component appear in the right hand sides of the
OPEs.
It would be interesting to describe 
this orthogonal case in details.

Recently,  the other type of large ${\cal N}=4$ holography
is studied in \cite{Ferreira}.
It is an open problem to study this holography
in the context of ${\cal N}=4$ coset model.

One can also consider 
i)
the fermionic higher spin current-scalar-spinor three-point functions
as well as
the three-point functions, given by
ii) the bosonic higher spin current-scalar-spinor,
iii) the bosonic higher spin current-spinor-spinor, 
iv) the fermionic higher spin current-scalar-scalar,
and
v) the fermionic higher spin current-spinor-spinor.
The states corresponding to the spinors can be
obtained by acting the spin $\frac{3}{2}$ currents $G^{\mu}$
on the states $|(f;0)>$ or $|(0;f)>$ \cite{mz,chr}.
For example, one should calculate the four-point functions
$<{\overline{\cal O}}_{\pm} {\cal O}_{\pm}
\, G^{\mu} \, \Phi^{(1),\nu}_{\frac{1}{2}}>$ for the case i) when we take
the simplest fermionic higher spin-$\frac{3}{2}$ currents.
The corresponding
OPE $G^{\mu}(z_1) \, \Phi^{(1),\nu}_{\frac{1}{2}}(z_2)$
in this four-point function is given 
in Appendix (\ref{1116}). The normal ordered product
$(G^{\mu} \, \Phi^{(1),\nu}_{\frac{1}{2}})(z_2)$
appears in the zeroth order \cite{BBSSfirst} and
the positive power of $(z_1-z_2)$ has the form of
$ \sum_{n=1}^{\infty} \frac{1}{n!} \, (z_1-z_2)^n \,
(\pa^n G^{\mu} \, \Phi^{(1),\nu}_{\frac{1}{2}})(z_2)$.
By analyzing the contributions of zero modes in the field dependent
parts (appearing in the singular and regular terms) in the states,
the above
four-point functions can be determined together with $z_1$ and $z_2$
dependences as in section $6$.

For the case ii), one can use the OPE
$G^{\mu}(z_1) \, \Phi^{(1)}_{0}(z_2)$, for example, when one considers
the simplest bosonic higher spin current.
The previous analysis for the case i) can be done.

For the case iii), one should analyze
 the five-point functions
$<{\overline{\cal O}}_{\pm} {\cal O}_{\pm}
 \, G^{\mu} \, \Phi^{(1)}_{0} \, G^{\nu}>$ where
 the $\mu$ and $\nu$ indices are complex conjugated to each other.
We take the simplest higher spin-$1$ current here.
 The operator $
G^{\mu}(z_1) \, \Phi^{(1)}_{0}(z_2) \, G^{\nu}(z_3) $
consists of four pieces \cite{rt}.
That is,
\bea
G^{\mu}(z_1) \, \Phi^{(1)}_{0}(z_2) \, G^{\nu}(z_3) & = &
: G^{\mu}(z_1) \, \Phi^{(1)}_{0}(z_2) \, G^{\nu}(z_3) :
+: (\mbox{sing. part of } G^{\mu}(z_1) \, \Phi^{(1)}_{0}(z_2)) \, G^{\nu}(z_3) :
\nonu \\
& + &
G^{\mu}(z_1) \, (\mbox{sing. part of } \Phi^{(1)}_{0}(z_2) \, G^{\nu}(z_3))
\nonu \\
& + &
\Phi^{(1)}_{0}(z_2) \, (\mbox{sing. part of }
G^{\mu}(z_1) \,  G^{\nu}(z_3)),
\label{four}
\eea
where the notation
$:A(z_1) \, B(z_2):$ means all the regular (or nonsingular)
terms of the OPE
$A(z_1) \, B(z_2)$ \cite{rt}. One has all the three OPEs,
$ G^{\mu}(z_1) \, \Phi^{(1)}_{0}(z_2)$, $\Phi^{(1)}_{0}(z_2) \, G^{\nu}(z_3)$
and $G^{\mu}(z_1) \,  G^{\nu}(z_3)$ from Appendices (\ref{nonalgebra})
and (\ref{1116}).
In the first term of (\ref{four}),
one should take the regular terms in the OPE $ G^{\mu}(z_1)$ and
the $z_3$ dependent fields
in the OPE of  $
\Phi^{(1)}_{0}(z_2) \, G^{\nu}(z_3)$.
Similarly, for the second term,
one should take the regular terms in the OPE between
the $z_2$ dependent fields appearing in the OPE
$G^{\mu}(z_1) \, \Phi^{(1)}_{0}(z_2)$
and $G^{\nu}(z_3)$. 
One can also extract the nontrivial
$z_3$ dependent terms in the remaining third and fourth terms of
(\ref{four}).
After that, one should calculate the OPEs with the first factors.
By analyzing the contributions of zero modes in the field dependent
parts appearing in the singular and regular terms in the states,
one can calculate  the five-point functions including $z_1, z_2$ and
$z_3$ dependences.

For the case iv), one should examine the zero mode contributions
in the fermionic higher spin currents as in section $6$.

For the case v), one should analyze
 the five-point functions
$<{\overline{\cal O}}_{\pm} {\cal O}_{\pm}
 \, G^{\mu} \, \Phi^{(1), \nu}_{\frac{1}{2}} \, G^{\rho}>$
 with simplest higher spin-$\frac{3}{2}$ currents.
This can be done as in previous case iii).

What happens for the three-point and four-point functions
between the higher spin currents with vacuum $|0>$?
For example, one can easily see that the two-point function
in the higher spin-$1$ current is
\bea
<0| \Phi_0^{(1)}(z_1) \, \Phi_0^{(1)}(z_2)|0> = \frac{1}{z_{12}^2} \,
\frac{2 k N}{(N+k+2)}
\nonu
\eea
from the result of Appendix (\ref{oneone}).
Note that $z_{12} \equiv (z_1-z_2)$.
One sees that the three-point function
in the higher spin-$1$ current
$
<0| \Phi_0^{(1)}(z_1) \, \Phi_0^{(1)}(z_2) \,  \Phi_0^{(1)}(z_3)|0>
$ vanishes by analyzing the four contributions in the vacuum
as in (\ref{four}).
For the four-point function, one can use the equation $(2.64)$ of
\cite{rt} where the difference between the OPE and the vacuum expectation
value of the OPE is introduced in $(2.59)$ of \cite{rt}.
Then it is easy to see that this quantity corresponding to
the above OPE  $\Phi_0^{(1)}(z_1) \, \Phi_0^{(1)}(z_2)$ vanishes.
Then the four-point function can be reduced to three contributions
from the product of two point-functions.  
It turns out that
\bea
<0| \Phi_0^{(1)}(z_1) \, \Phi_0^{(1)}(z_2) \,  \Phi_0^{(1)}(z_3) \,
\Phi_0^{(1)}(z_4) |0>  & = &
<0| \Phi_0^{(1)}(z_1) \, \Phi_0^{(1)}(z_4)|0>
<0| \Phi_0^{(1)}(z_2) \, \Phi_0^{(1)}(z_3)|0>
\nonu \\
&+& <0| \Phi_0^{(1)}(z_1) \, \Phi_0^{(1)}(z_2)|0>
<0| \Phi_0^{(1)}(z_3) \, \Phi_0^{(1)}(z_4)|0>
\nonu \\
&+&
<0| \Phi_0^{(1)}(z_1) \, \Phi_0^{(1)}(z_3)|0>
<0| \Phi_0^{(1)}(z_2) \, \Phi_0^{(1)}(z_4)|0>
\nonu \\
& = &
\left( \frac{1}{z_{14}^2 z_{23}^2}  +  \frac{1}{z_{12}^2 z_{34}^2} +
 \frac{1}{z_{13}^2 z_{24}^2} 
\right)
\frac{4 k^2 N^2}{(N+k+2)^2}.
\nonu
\eea
For the four-point functions having other higher spin currents, 
the similar analysis (after further OPEs can be found) can be done.

Recall that the state $| (0;f)>$ contains the vacuum $|0>$
in (\ref{plusminus}).
The four-point function
$<{\overline{\cal O}}_{-} {\cal O}_{-}
\, \Phi_0^{(1)} \, \Phi^{(1),\mu \nu}_{1}>$
can be related to $<0| Q^{a}(z_1)
\, \Phi_0^{(1)}(z_2) \, \Phi^{(1),\mu \nu}_{1}(z_3) \, Q^{\bar{b}}(z_4) |0>$.
Because the OPEs between the higher spin currents and
spin-$\frac{1}{2}$ currents are known explicitly,
one can calculate the four-point functions. 




\vspace{.7cm}

\centerline{\bf Acknowledgments}

This research was supported by Kyungpook National University
Bokhyeon Research Fund, 2015.
Part of the results (Appendices $B$ and $E$) in this paper is
based on the Master thesis of Dong-gyu Kim.
CA acknowledges warm hospitality from 
the School of  Liberal Arts (and Institute of Convergence Fundamental
Studies), Seoul National University of Science and Technology.

\newpage

\appendix

\renewcommand{\theequation}{\Alph{section}\mbox{.}\arabic{equation}}

\section{ The  OPEs between the $11$ currents and the 
  $16$ higher spin currents}

Let us
present  the ${\cal N}=4$ nonlinear superconformal algebra
and  the OPEs between the $11$ currents and the 
  $16$ higher spin currents.

\subsection{The ${\cal N}=4$ nonlinear superconformal algebra}

The large ${\cal N}=4$ nonlinear superconformal algebra
appearing in (\ref{n4sca}) from the $11$ currents
can be written in $SO(4)$ manifest way 
and using the relations in the  (\ref{replacement})
the corresponding OPEs 
are described by
{\small
\bea
L(z)\,L(w) & = & \frac{1}{(z-w)^{4}}\, \Bigg[\frac{3(k+N+2\,kN)}{(2+k+N)}
  \Bigg]
+\frac{1}{(z-w)^{2}}\,2\, L(w)+\frac{1}{(z-w)}\,\partial L(w)+\cdots,  \nonu \\ 
L(z)\,G^{\mu}(w) & = & \frac{1}{(z-w)^{2}}\,\frac{3}{2}\, G^{\mu}(w)+
\frac{1}{(z-w)}\,\partial G^{\mu}(w)+\cdots,  \nonu \\ 
L(z)\,T^{\mu \nu}(w) & = & \frac{1}{(z-w)^{2}}\,T^{\mu \nu}(w)+
\frac{1}{(z-w)}\,\partial T^{\mu \nu}(w)+\cdots,  \nonu \\ 
\nonu\\
G^{\mu}(z)\,G^{\nu}(w) & = & \frac{1}{(z-w)^{3}} \Bigg[
  \frac{4\,kN}{(2+k+N)} \Bigg] \,\delta^{\mu \nu}
-\frac{1}{(z-w)^{2}}\Bigg[
  \frac{2i(k+N)}{(2+k+N)}\,T^{\mu \nu}+\frac{2i(k-N)}{(2+k+N)}\,
  \widetilde{T}^{\mu \nu}
  \Bigg]
\nonu \\ 
& + &\frac{1}{(z-w)}\,\Bigg[\,\delta^{\mu \nu}\,\Bigg( 2\, L
  -\frac{2}{(2+k+N)}\,(T^{\mu \rho}T^{\mu \rho}+
  \widetilde{T}^{\mu \rho}\widetilde{T}^{\mu \rho})\,
  \Bigg)+\frac{2}{(2+k+N)}\, T^{\mu \rho}T^{\nu \rho}
\nonu \\ 
& + &\frac{i(2-k-N)}{(2+k+N)}\,\partial T^{\mu \nu}-\frac{i(k-N)}{(2+k+N)}\,
\partial\widetilde{T}^{\mu \nu}\Bigg](w)+\cdots,
\nonu \\ 
G^{\mu}(z)\,T^{\nu \rho}(w) & = & \frac{1}{(z-w)}\,i\, \Bigg[\,
  \delta^{\mu \nu}\,G^{\rho}-\delta^{\mu \rho}\,G^{\nu}\, \Bigg](w)+\cdots,
\nonu \\ 
T^{\mu \nu}(z)\,T^{\rho \si}(w) & = & \frac{1}{(z-w)^{2}}\,
\Bigg[-\frac{(k-N)}{2} \, \varepsilon^{\mu \nu \rho \si}\, +
  \frac{(k+N)}{2}\, (\delta^{\mu \rho}\delta^{\nu \si}-
  \delta^{\mu \si}\delta^{\nu \rho})\Bigg]
\nonu \\ 
& - &\frac{1}{(z-w)}\,i\, \Bigg[
  \delta^{\mu \rho}T^{\nu \si}-\delta^{\mu \si}T^{\nu \rho}-
  \delta^{\nu \rho}T^{\mu \si}+\delta^{\nu \si} T^{\mu \rho}
  \Bigg](w)+
  \cdots.
\label{nonalgebra}
\eea}
The nonlinearity from the OPEs between the spin $\frac{3}{2}$
currents occurs. We introduce the notation $\widetilde{T}^{\mu \nu}(z) \equiv
\frac{1}{2} \varepsilon^{\mu \nu \rho \si} T^{\rho \si}(z)$ in
(\ref{nonalgebra}).

\subsection{The OPEs between the
  generators of ${\cal N}=4$ nonlinear superconformal algebra
and the  $16$ higher spin currents}

One describes the OPEs between the
$11$ currents and the $16$  higher spin currents
(as far as we know, there is no explicit form in the literature so far)
as follows:
{\small
\bea
L(z)\,
\left(
\begin{array}{c}
 \Phi_0^{(s)} \\
\Phi_{\frac{1}{2}}^{(s),\mu}  \\
\Phi_1^{(s),\mu\nu}  \\
\Phi_{\frac{3}{2}}^{(s),\mu}  \\
\Phi_{2}^{(s)} \\
\end{array}
\right)
(w) & = &
\frac{1}{(z-w)^{2}}\, \left(
\begin{array}{c}
s  \,\Phi_0^{(s)} \\
(s+\frac{1}{2})\, \Phi_{\frac{1}{2}}^{(s),\mu}  \\
(s+1) \, \Phi_1^{(s),\mu\nu}  \\
(s+\frac{3}{2}) \, \Phi_{\frac{3}{2}}^{(s),\mu}  \\
(s+2) \, \Phi_{2}^{(s)} \\
\end{array}
\right)(w)+\frac{1}{(z-w)}\,\partial
\left(
\begin{array}{c}
 \Phi_0^{(s)} \\
\Phi_{\frac{1}{2}}^{(s),\mu}  \\
\Phi_1^{(s),\mu\nu}  \\
\Phi_{\frac{3}{2}}^{(s),\mu}  \\
\Phi_{2}^{(s)} \\
\end{array}
\right)
(w)+\cdots,
\nonu \\
\nonu \\  
G^{\mu}(z)\,\Phi_{2}^{(s)}(w) & = & 
\frac{1}{(z-w)^{3}}\, \Bigg[
  \frac{4  (6 - 3 k - 3 N - 12 k N - 32 s - 16 k s -
   16 N s)}{(2 + k + N)}\,c_{s}
  \Bigg] \,\Phi_{\frac{1}{2}}^{(s),\mu}(w)
\nonu \\ 
& + & 
\frac{1}{(z-w)^{2}}\,\Bigg[-(2\,s+3)\,\Phi_{\frac{3}{2}}^{(s),\mu}
+36\,c_{s}\Bigg(\, G^{\mu}\Phi_{0}^{(s)}
+\frac{2}{(1 + 2 s)}\,\partial\Phi_{\frac{1}{2}}^{(s),\mu}\Bigg)
\nonu \\ 
& - & \frac{4\, i(1+s)}{(2+k+N)}\,\varepsilon^{\mu \nu' \rho' \si'}\, T^{\nu' \rho'}\Phi_{\frac{1}{2}}^{(s),\si'}\,\Bigg](w)
\nonu \\ 
& + &\frac{1}{(z-w)}\,\Bigg[
-\partial\Phi_{\frac{3}{2}}^{(s),\mu}
-\frac{2\, i(s+1)}{(2+k+N)}\, \varepsilon^{\mu \nu' \rho' \si'}\,
\partial T^{\nu' \rho'}\Phi_{\frac{1}{2}}^{(s),\si'}
\nonu \\ 
& + & 12\,c_{s}\,
\Bigg(\,\frac{3}{(2\,s+1)}\,\partial^{2}\Phi_{\frac{1}{2}}^{(s),\mu}-
2\, L\Phi_{\frac{1}{2}}^{(s),\mu}+\partial
G^{\mu}\Phi_{0}^{(s)}\,\Bigg)\,\Bigg](w)+\cdots,
\nonu \\
G^{\mu}(z)\,\Phi_{\frac{3}{2}}^{(s),\nu}(w) & = & 
\frac{1}{(z-w)^{3}}\Bigg[\frac{8(s^{2}+s)(k-N)}{(2\,s+1)(2+k+N)}
  \Bigg] \delta^{\mu \nu} \, \Phi_{0}^{(s)}(w)
\nonu \\
& - & \frac{1}{(z-w)^{2}}\,\Bigg[\,2(s+1)\,\Phi_{1}^{(s),\mu \nu}
+\frac{2(s+1)(k-N)}{(2\,s+1)(2+k+N)}\,\widetilde{\Phi}_{1}^{(s),\mu \nu}\,\Bigg](w)\nonu
\nonu \\ 
& + & \frac{1}{(z-w)}\,\Bigg[\,\delta^{\mu \nu}\,\Bigg(-\Phi_{2}^{(s)}
+24\,c_{s}\,( L\Phi_{0}^{(s)}
-\frac{3 }{2(1 + 2 s) }\,\partial^{2}\Phi_{0}^{(s)})\,\Bigg)
\nonu \\ 
& - &
\partial\Phi_{1}^{(s),\mu \nu}
-\frac{(k-N)}{(2\,s+1)(2+k+N)}\,\partial\widetilde{\Phi}_{1}^{(s),\mu \nu}
+\frac{4\,s\, i}{(2+k+N)}\,\partial\widetilde{T}^{\mu \nu}\Phi_{0}^{(s)}
\nonu \\ 
& - & \frac{4\, i}{(2+k+N)}\,\widetilde{T}^{\mu \nu}\partial\Phi_{0}^{(s)}
+ \frac{2\, i}{(2+k+N)}\,
\Bigg(T^{\mu \rho'}\Phi_{1}^{(s),\nu \rho'}-T^{\nu \rho'}\Phi_{1}^{(s),\mu \rho'}\Bigg)
\nonu \\ 
& - &\frac{2 }{(2+k+N)}\,\varepsilon^{\mu \nu \rho' \si'}\, G^{\rho'}\Phi_{\frac{1}{2}}^{(s),\si'}\,\Bigg](w)
+ \cdots,
\nonu \\
G^{\mu}(z)\,\Phi_{1}^{(s),\nu \rho}(w) & = & 
\frac{1}{(z-w)^{2}}\,\Bigg[-\,
  \frac{(k-N)}{(2+k+N)}\delta^{\mu \nu} \,\Phi_{\frac{1}{2}}^{(s),\rho}
\nonu \\
& + &
\frac{(-2 + k + N + 4 s + 2 k s + 2 N s)}{(2+k+N)}
\varepsilon^{\mu \nu \rho \si'} \Phi_{\frac{1}{2}}^{(s),\si'}\Bigg](w)
\nonu \\
& + & 
\frac{1}{(z-w)}\,\Bigg[\,
\frac{2\, i}{(2+k+N)}\,
\Bigg(\widetilde{T}^{\mu \rho}\Phi_{\frac{1}{2}}^{(s),\nu}-\widetilde{T}^{\mu \nu}\Phi_{\frac{1}{2}}^{(s),\rho}\Bigg)
+\varepsilon^{\mu \nu \rho \si'}\,\partial\Phi_{\frac{1}{2}}^{(s),\si'}
\nonu \\
& - & \delta^{\mu \nu}(
  \Phi_{\frac{3}{2}}^{(s),\rho}+\frac{(k-N)}{(2\,s+1)(2+k+N)}\partial
  \Phi_{\frac{1}{2}}^{(s),\rho}+\frac{2\, i}{(2+k+N)}
  \widetilde{T}^{\rho \si'}\Phi_{\frac{1}{2}}^{(s),\si'})
\Bigg]
\nonu \\
& - & \delta^{\mu \rho}\, \sum_{n=2}^1 \, \frac{1}{(z-w)^n} \,
\left(\nu \;\leftrightarrow \;\rho\right)(w)+\cdots,
\nonu \\
G^{\mu}(z)\,\Phi_{\frac{1}{2}}^{(s),\nu}(w) & = & 
-\frac{1}{(z-w)^{2}}\,2\,s\:\delta^{\mu\nu}\,\Phi_{0}^{(s)}(w)
\nonu \\
& + & 
\frac{1}{(z-w)}\,\Bigg[-\delta^{\mu \nu}\,\partial\Phi_{0}^{(s)}+\widetilde{\Phi}_{1}^{(s),\mu \nu}\,\Bigg](w)+\cdots,
\nonu \\
G^{\mu}(z) \, \Phi_{0}^{(s)}(w) & = & 
-\frac{1}{(z-w)} \, \Phi_{\frac{1}{2}}^{(s),\mu}(w)
+\cdots,
\nonu \\
\nonu \\ 
T^{\mu \nu}(z)\,\Phi_{2}^{(s)}(w) & = & 
\frac{1}{(z-w)^{2}}\,\Bigg[\,2\,i(s+1)\,\Phi_{1}^{(s),\mu \nu}+
  24\,c_{s}\, T^{\mu \nu}
  \Phi_{0}^{(s)}\,\Bigg](w)+\cdots,
\nonu \\
T^{\mu \nu}(z)\,\Phi_{\frac{3}{2}}^{(s),\rho}(w) & = &
\frac{1}{(z-w)^{2}}\Bigg[\,\frac{2i(s+1)(k-N)}{(2\,s+1)(2+k+N)}\,
  \Bigg(\delta^{\mu \rho}\,\Phi_{\frac{1}{2}}^{(s),\nu}-\delta^{\nu \rho}\,
  \Phi_{\frac{1}{2}}^{(s)\mu}\Bigg)
\nonu\\
& - &
  2i(s+1)\, \varepsilon^{\mu \nu \rho \si'}\, \Phi_{\frac{1}{2}}^{(s),\si'}
\,\Bigg](w)
-\frac{1}{(z-w)}\,i\,\Big(\delta^{\mu \rho}\,\Phi_{\frac{3}{2}}^{(s),\nu}-\delta^{\nu \rho}\,\Phi_{\frac{3}{2}}^{(s),\mu}\Big)
(w)\nonu \\
& + & \cdots,
\nonu \\
T^{\mu \nu}(z)\,\Phi_{1}^{(s), \rho \si}(w)
& = & 
\frac{1}{(z-w)^{2}}\, 2\,i\,s\,\varepsilon^{\mu \nu \rho \si}\,\Phi_{0}^{(s)}(w)
\nonu \\
& - &
\frac{1}{(z-w)}\,i\,\Bigg[\,\delta^{\mu \rho}\,\Phi_{1}^{(s),\nu \si}-\delta^{\mu \si}\,\Phi_{1}^{(s),\nu \rho}-\delta^{\nu \rho}\,\Phi_{1}^{(s),\mu \si}+\delta^{\nu \si}\,\Phi_{1}^{(s),\mu \rho}\,\Bigg](w)\nonu \\
& + & \cdots,
\nonu \\
T^{\mu \nu}(z)\;\Phi_{\frac{1}{2}}^{(s),\rho}(w)
& = & 
-\frac{1}{(z-w)}\,i\,\Bigg(\,\delta^{\mu \rho}\,\Phi_{\frac{1}{2}}^{(s),\nu}-\delta^{\nu \rho}\,\Phi_{\frac{1}{2}}^{(s),\mu}\,\Bigg)(w)+\cdots,
\nonu \\
T^{\mu\nu}(z)\;\Phi_{0}^{(s)}(w) & = & + \cdots,
\label{1116}
\eea}
where we introduce
\bea
c_{s} \equiv
\frac{(k - N)  (s^2 + s)}
     {(3 k + 3 N + 6 k N - 20 s - 4 k s - 4 N s + 12 k N s + 32 s^2 + 
 16 k s^2 + 16 N s^2)}
\nonu
\eea
and 
$\widetilde{\Phi}_1^{(s),\mu \nu}(z) \equiv
\frac{1}{2} \varepsilon^{\mu \nu \rho \si} \Phi_1^{(s),\rho \si}(z)$.
Compared to the linear version, the higher spin currents
are primary under the stress energy tensor from the first equation
of (\ref{1116}).

\section{ The  OPEs between the lowest $16$ higher spin currents
  and itself for generic $N$ and $k$}

In the right hand sides of these OPEs,
there are indices summed over $1,2,3,4$ denoted by
primed ones, $\mu', \nu', \rho', \cdots$.
The unprimed indices $\mu, \nu, \rho, \cdots$
are not summed.

\subsection{The OPEs between the higher spin $1$ current and the
$16$ higher spin currents}

The OPE between the higher spin $1$ current and itself is given by
\bea
    {\Phi}_{0}^{(1)}(z)\,{\Phi}_{0}^{(1)}(w) & = & \frac{1}{(z-w)^{2}}\, c_{1}
    +\cdots,
    \label{oneone}
\eea
where
\bea
c_ {1} = \frac {2k\,N}{(2 + k + N)}. \nonu
\eea

The OPE between the higher spin $1$ current and the higher
spin $\frac{3}{2}$ currents is given by
\bea
    {\Phi}_{0}^{(1)}(z)\,{\Phi}_{\frac{1}{2}}^{(1),\mu}(w) & = & \frac{1}{(z-w)}\,{G}^{\mu}+\cdots.
    \label{one3half}
\eea
The OPE (\ref{one3half}) also appeared in (\ref{one3half1}) before.

The OPE between the higher spin $1$ current and the higher spin $2$
currents is given by
\bea
    {\Phi}_{0}^{(1)}(z)\,{\Phi}_{1}^{(1),\mu \nu}(w)=\frac{1}{(z-w)^{2}}\,\Bigg[\, c_{1}\,{T}^{\mu \nu}+c_{2}\,\widetilde{T}^{\mu \nu}\,\Bigg](w)+\cdots,
    \label{onetwo}
\eea
where
\bea
c_ {1} = \frac {2i(k-N)}{(2 + k + N)},\qquad c_ {2} =
\frac {2i(k+N)}{(2 + k + N)}.
\nonu
\eea

The OPE between the higher spin $1$ current and the higher spin
$\frac{5}{2}$
currents is given by
\bea
{\Phi}_{0}^{(1)}(z)\,{\Phi}_{\frac{3}{2}}^{(1),\mu}(w) & = & \frac{1}{(z-w)^{2}}\, c_{1}\,{G}^{\mu}(w)
+\frac{1}{(z-w)}\,\Bigg[\, c_{2}\,{\Phi}_{\frac{1}{2}}^{(2),\mu}+c_{3}\,{\Phi}_{0}^{(1)}{\Phi}_{\frac{1}{2}}^{(1),\mu}+c_{4}\,{G}^{\nu'}{T}^{\mu \nu'}
\nonu\\ & + & 
c_{5}\,\varepsilon^{\mu \nu' \rho' \si'}\, {G}^{\nu'}{T}^{\rho' \si'}+c_{6}\,\partial{G}^{\mu}\,\Bigg](w)+\cdots,
\label{one5half}
\eea
where
\bea
c_ {1} & = & \frac {8(k-N)}{3(2 + k + N)},\qquad c_ {2} = -\frac {1}{2}, \nonu\\ 
c_ {3} & = & \frac {(60 + 77 k + 22 k^2 + 121 N + 115 k N + 20 k^2\, N
  + 79 N^2 + 
 42 k\, N^2 + 16 N^3)}{2 (2 + N) (2 + k + N)^2}, \nonu\\
c_ {4} & = & -\frac {i(20 + 21 k + 6 k^2 + 25 N + 13 k\, N + 7 N^2)}{2 (2 + N) (2 + k + N)^2}, \nonu\\
c_ {5} & = & -\frac {i(32 + 29 k + 6 k^2 + 35 N + 17 k\, N + 9 N^2)}{4 (2 + N) (2 + k + N)^2}, \nonu\\
c_ {6}  & = & -\frac {i(-60 - 49 k - 14 k^2 - 29 N + 37 k\, N + 20 k^2\, N + 37 N^2 + 
  42 k\, N^2 + 16 N^3)}{24 (32 + 29 k + 6 k^2 + 35 N + 17 k\, N + 9 N^2)}.
\nonu
\eea

The OPE between the higher spin $1$ current and the higher spin $3$
current is given by
{\small
\bea
    {\Phi}_{0}^{(1)}(z)\,{\Phi}_{2}^{(1)}(w) & = &
    \frac{1}{(z-w)^{2}}\,\Bigg[\, c_{1}\,{\Phi}_{0}^{(2)}+c_{2}\,{\Phi}_{0}^{(1)}{\Phi}_{0}^{(1)}+c_{3}\,{L}+c_{4}\,({T}^{\mu' \nu'})^{2}+c_{5}\,\varepsilon^{\mu' \nu' \rho' \si'}{T}^{\mu' \nu'}{T}^{ \rho' \si'}\,\Bigg](w)
    \nonu \\
    & + & \cdots,
    \label{onethree}
\eea}
where
\bea
c_{1} & = & 2, \nonu\\
c_{2} & = & -\frac{3}{(2+N)(2+k+N)^{2}(4+5k+5N+6kN)}(80+160k+115k^{2}+26k^{3}+304N  \nonu\\
& + &
582kN+377k^{2}N+72k^{3}N+371N^{2}+632kN^{2}+328k^{2}N^{2}+40k^{3}N^{2}+185N^{3} \nonu\\
& + & 
260kN^{3}+84k^{2}N^{3}+32N^{4}+32kN^{4}), \nonu\\
c_{3} & = & \frac{4}{(2+k+N)^{2}(4+5k+5N+6kN)}(40+86k+53k^{2}+10k^{3}+106N+203kN \nonu\\
& + & 115k^{2}N+20k^{3}N+86N^{2}+123kN^{2}+42k^{2}N^{2}+20N^{3}+16kN^{3}), \nonu\\
c_{4} & = & -\frac{(20+21k+6k^{2}+25N+13kN+7N^{2})}{2(2+N)(2+k+N)^{2}},
c_{5} = -\frac{(13k+6k^{2}+3N+9kN+N^{2})}{4(2+N)(2+k+N)^{2}}.
\nonu
\eea

\subsection{The OPEs between the higher spin $\frac{3}{2}$ currents and the
remaining $15$ higher spin currents}

The OPE between the higher spin $\frac{3}{2}$ currents and
the higher spin $\frac{3}{2}$
currents is given by
\bea
{\Phi}_{\frac{1}{2}}^{(1),\mu}(z)\,{\Phi}_{\frac{1}{2}}^{(1),\nu}(w) & = &
\frac{1}{(z-w)^{3}}\,\delta^{\mu \nu}\, c_{1}+\frac{1}{(z-w)^{2}}\,\Bigg[\, c_{2}\,{T}^{\mu \nu}+c_{3}\,\widetilde{T}^{\mu \nu}\,\Bigg](w)
\nonu\\
& + & \frac{1}{(z-w)}\,\Bigg[\,\delta^{\mu \nu}\, c_{4}\,{L}
  +c_{5}\,\widetilde{T}^{\mu \rho'}\widetilde{T}^{\nu \rho'}
  +c_{6}\,\partial{T}^{\mu \nu}+c_{7}\,\partial\widetilde{T}^{\mu \nu}\,\Bigg](w)
\nonu \\
& + & \cdots,
\label{3half3half}
\eea
where
\bea
c_{1} & = & -\frac{4kN}{(2+k+N)},\qquad
c_{2} = \frac{2i(k+N)}{(2+k+N)},\qquad
c_{3} = \frac{2i(k-N)}{(2+k+N)},\qquad c_{4} = -2,\nonu\\
c_{5} & = & \frac{2}{(2+k+N)},\qquad c_{6}= i,\qquad
c_{7} = \frac{i(k-N)}{(2+k+N)}.\nonu
\eea

The OPE between the higher spin $\frac{3}{2}$ currents and
the higher spin $2$
currents is given by
\bea
{\Phi}_{\frac{1}{2}}^{(1),\mu}(z)\,{\Phi}_{1}^{(1),\nu \rho}(w) & = & 
\frac{1}{(z-w)^{2}}\,\Bigg[\,\delta^{\mu \nu}\, c_{1}\,{G}^{\rho}+\varepsilon^{\mu \nu \rho \si'}\, c_{2}\,{G}^{\si'}\,\Bigg](w)
\nonu\\ & + & 
\frac{1}{(z-w)}\,\Bigg[\,\delta^{\mu \nu}\,\Bigg(\, c_{3}\,{\Phi}_{\frac{1}{2}}^{(2),\rho}
+c_{4}\,{\Phi}_{0}^{(1)}{\Phi}_{\frac{1}{2}}^{(1),\rho}+c_{5}\,\partial{G}^{\rho}+c_{6}\,{G}^{\si'}\,{T}^{\rho \si'}
\nonu\\ & + & 
c_{7}\,{G}^{\si'}\,\widetilde{T}^{\rho \si'}\,\Bigg)+c_{8}\,({G}^{\nu}\widetilde{T}^{\mu \rho}-{G}^{\rho}\widetilde{T}^{\mu \nu})
+c_{9}\,\varepsilon^{\mu \nu \rho \si'}\, \partial{G}^{\si'}\,\Bigg](w)
\nonu\\ & + & 
\delta^{ik}\,\sum_{n=2}^{1}\,\frac{1}{(z-w)^{n}}\,\left(j\;\leftrightarrow\; k\right)(w)+\cdots,
\label{3halftwo}
\eea
where
\bea
c_{1} & = & -\frac{(k-N)}{(2+k+N)}, \qquad 
c_{2}=\frac{(2+3k+3N)}{(2+k+N)},\qquad 
c_{3}=\frac{1}{2},\nonu\\
c_{4} & = & -\frac{(60+77k+22k^{2}+121N+115kN+20k^{2}N+79N^{2}+
  42kN^{2}+16N^{3})}{2(2+N)(2+k+N)^{2}},\nonu\\
c_{5} & = & \frac{(-20-19k-6k^{2}-7N+11kN+6k^{2}N+15N^{2}+14kN^{2}+6N^{3})}{2(2+N)(2+k+N)^{2}},\nonu\\
c_{6} & = & \frac{i(20+21k+6k^{2}+25N+13kN+7N^{2})}{2(2+N)(2+k+N)^{2}},\nonu\\
c_{7} & = & \frac{i(16+21k+6k^{2}+19N+13kN+5N^{2})}{2(2+N)(2+k+N)^{2}},\nonu\\
c_{8} & = & \frac{2i}{(2+k+N)}, \qquad 
c_{9}=\frac{(-2+k+N)}{(2+k+N)}.\nonu
\eea

The OPE between the higher spin $\frac{3}{2}$ currents and
the higher spin $\frac{5}{2}$
currents is given by
{\small
\bea
{\Phi}_{\frac{1}{2}}^{(1),\mu}(z)\,{\Phi}_{\frac{3}{2}}^{(1),\nu}(w) & = & 
\frac{1}{(z-w)^{3}}\,\Bigg[\, c_{1}\,{T}^{\mu \nu}+c_{2}\,\widetilde{T}^{\mu \nu}\,\Bigg](w)
+\frac{1}{(z-w)^{2}}\,\Bigg[\,\delta^{\mu \nu}\Bigg(\, c_{3}\,{\Phi}_{0}^{(2)}+c_{4}\,{\Phi}_{0}^{(1)}{\Phi}_{0}^{(1)}
\nonu\\ 
& + & c_{5}\,{L}+c_{6}\,({T}^{\mu \rho'})^{2}+c_{7}\,(\widetilde{T}^{\mu \rho'})^{2}+c_{8}\,{T}^{\mu \rho'}\widetilde{T}^{\mu \rho'}\,\Bigg)
+c_{9}\,\partial{T}^{\mu \nu}+c_{10}\,{T}^{\mu \rho'}{T}^{\nu \rho'}\,\Bigg](w)
\nonu\\ 
& + & \frac{1}{(z-w)}\,\Bigg[\,\frac{1}{4}\,\partial\Big(\mbox{pole-2}\Big)+c_{27}\,\widetilde{\Phi}_{1}^{(2),\mu \nu}
+ c_{14}\,{\Phi}_{\frac{1}{2}}^{(1),\mu}{\Phi}_{\frac{1}{2}}^{(1),\nu}+c_{28}\,{\Phi}_{0}^{(1)}\widetilde{\Phi}_{1}^{(1),\mu \nu}
\nonu\\ 
& + &c_{29}\, \varepsilon^{\mu \nu \rho' \si'}\,{\Phi}_{\frac{1}{2}}^{(1),\rho'}{\Phi}_{\frac{1}{2}}^{(1),\si'}
+\delta^{\mu \nu}\Bigg(\,
c_{11}\,\partial{L}
+c_{12}\,\partial({T}^{\mu \rho'}\widetilde{T}^{\mu \rho'})
+c_{13}\,(\partial{T}^{\mu \rho'}{T}^{\mu \rho'}
\nonu\\ 
& + & \partial\widetilde{T}^{\mu \rho'}\widetilde{T}^{\mu \rho'})\Bigg)
+c_{15}\,{L}{T}^{\mu \nu}+c_{16}\,{L}\widetilde{T}^{\mu \nu}
+c_{17}\,{G}^{\mu}{G}^{\nu}+c_{18}\,\partial^{2}{T}^{\mu \nu}
+c_{19}\,\partial^{2}\widetilde{T}^{\mu \nu}
\nonu\\ 
& + & c_{20}\,{T}^{\mu \nu}{T}^{\mu \nu} \widetilde{T}^{\mu \nu}
+c_{21}\,(\widetilde{T}^{\mu \nu})^{3}
+c_{22}\,\Bigg({T}^{\mu \rho'}{T}^{\mu \rho'}\widetilde{T}^{\mu \nu}-{T}^{\nu \rho'}{T}^{\nu \rho'} \widetilde{T}^{\nu \mu}\Bigg)
\nonu\\ 
& + & c_{23}\,\partial{T}^{\mu \rho'}{T}^{\nu \rho'}
+c_{24}\,{T}^{\mu \rho'}\partial{T}^{\nu \rho'}
+c_{25}\,{T}^{\mu \rho'}\partial\widetilde{T}^{\nu \rho'}
+c_{26}\,{T}^{\nu \rho'}\partial\widetilde{T}^{\mu \rho'}
\nonu\\ 
& + & \varepsilon^{\mu \nu \rho' \si'}\,\Bigg(
c_{30}\,{G}^{\rho'}{G}^{\si'}+c_{31}\,\widetilde{T}^{\rho' \si'}{T}^{\alpha' \rho'}{T}^{\alpha' \rho'}+c_{32}\,{T}^{\alpha' \rho'}\widetilde{T}^{\alpha' \rho'}T^{\rho' \si'}\,
\Bigg)\,\Bigg](w)
\nonu \\
& + & \cdots,
\label{3half5half}
\eea}
where
{\small
\bea
c_{1} & = & -\frac{16i(k-N)(3+k+N)}{3(2+k+N)^{2})}, \qquad  c_{2}=-\frac{16i(3k+k^{2}+3N+4kN+N^{2})}{3(2+k+N)^{2}},  \qquad c_{3}=-2, \nonu\\
c_{4} & = & \frac{(60+77k+22k^{2}+121N+115kN+20k^{2}N+79N^{2}+42kN^{2}+16N^{3})}
{(2+N)(2+k+N)^{2}}, \nonu\\
c_{5} & = & -\frac{4(3+2k+N)(10+5k+8N)}{3(2+k+N)^{2}}, \qquad 
c_{6}= \frac{(60+95k+18k^{2}+43N+55kN+5N^{2})}{12(2+N)(2+k+N)^{2}}, \nonu\\
c_{7} & = & \frac{(60+95k+18k^{2}+43N+55kN+5N^{2})}{3(2+N)(2+k+N)^{2}}, \qquad 
c_{8}=\frac{2(13k+6k^{2}+3N+9kN+N^{2})}{(2+N)(2+k+N)^{2}}, \nonu\\
c_{9} & = & -\frac{16i(k-N)}{3(2+k+N)^{2}}, \qquad 
c_{10}=-\frac{16(k-N)}{3(2+k+N)^{2}},  \nonu\\
c_{11} & = & \frac{(40+49k+14k^{2}+73N+64kN+10k^{2}N+43N^{2}+21kN^{2}+8N^{3})}
{(2+N)(2+k+N)^{2}},  \nonu\\
c_{12} & = & \frac{4}{(2+k+N)},  \nonu\\
c_{13} & = & -\frac{2(40+49k+14k^{2}+73N+64kN+10k^{2}N+43N^{2}+21kN^{2}+8N^{3})}{(2+N)(2+k+N)^{3}},  \nonu\\
c_{14} & = &\frac{(60+77k+22k^{2}+121N+115kN+20k^{2}N+79N^{2}+42kN^{2}+16N^{3})}{2(2+N)(2+k+N)^{2}}, \nonu\\
c_{15} & = & -\frac{i(20+21k+6k^{2}+25N+13kN+7N^{2})}{(2+N)(2+k+N)^{2}},\nonu\\
c_{16} & = & -\frac{i(32+29k+6k^{2}+35N+17kN+9N^{2})}{(2+N)(2+k+N)^{2}},\nonu\\
c_{17} & = & -\frac{(20+21k+6k^{2}+25N+13kN+7N^{2})}{2(2+N)(2+k+N)^{2}},\nonu\\
c_{18} & = & \frac{1}{3(2+N)(2+k+N)^{3}}i(-180-219k-59k^{2}+2k^{3}-255N-171kN+16k^{2}N
\nonu\\ & + & 10k^{3}N-82N^{2}+25kN^{2}+31k^{2}N^{2}+17N^{3}+29kN^{3}+8N^{4}),\nonu\\
c_{19} & = &
\frac{1}{6(2+N)(2+k+N)^{3}}i(576+498k+55k^{2}-22k^{3}+846N+442kN-45k^{2}N
\nonu\\ & - & 20k^{3}N+463N^{2}+132kN^{2}-22k^{2}N^{2}+127N^{3}+26kN^{3}+16N^{4}),\nonu\\
c_{20} & = & -\frac{i(32+29k+6k^{2}+35N+17kN+9N^{2})}{(2+N)(2+k+N)^{3}},\nonu\\
c_{21} & = &\frac{i(32+29k+6k^{2}+35N+17kN+9N^{2})}{(2+N)(2+k+N)^{3}},\nonu\\
c_{22} & = &\frac{i(32+29k+6k^{2}+35N+17kN+9N^{2})}{(2+N)(2+k+N)^{3}},\nonu\\
c_{23} & = &\frac{(40+82k+41k^{2}+6k^{3}+90N+128kN+35k^{2}N+
  67N^{2}+48kN^{2}+15N^{3})}{2(2+N)(2+k+N)^{3}},\nonu\\
c_{24} & = &\frac{(120+114k+15k^{2}-6k^{3}+202N+128kN+5k^{2}N+105N^{2}+36kN^{2}+17N^{3})}{2(2+N)(2+k+N)^{3}},\nonu\\
c_{25} & = &-\frac{(-2+k+N)(32+29k+6k^{2}+35N+17kN+9N^{2})}{2(2+N)(2+k+N)^{3}},\nonu\\
c_{26} & = &-\frac{(192+154k+15k^{2}-6k^{3}+230N+98kN-7k^{2}N+79N^{2}+6kN^{2}+
  7N^{3})}{2(2+N)(2+k+N)^{3}},\nonu\\
c_{27} & = & \frac{1}{4}, \nonu\\
c_{28} & = & -\frac{(60+77k+22k^{2}+121N+115kN+20k^{2}N+79N^{2}+42kN^{2}+16N^{3})}{4(2+N)(2+k+N)^{2}},\nonu\\
c_{29} & = & -\frac{2}{(2+k+N)}, \qquad 
c_{30}=\frac{(16+21k+6k^{2}+19N+13kN+5N^{2})}{2(2+N)(2+k+N)^{2}},\nonu\\
c_{31} & = &-\frac{i(32+29k+6k^{2}+35N+17kN+9N^{2})}{2(2+N)(2+k+N)^{3}},\nonu\\
c_{32} & = & \frac{i(20+21k+6k^{2}+25N+13kN+7N^{2})}{2(2+N)(2+k+N)^{3}}.\nonu
\eea}

The OPE between the higher spin $\frac{3}{2}$ currents and
the higher spin $3$
current is given by
\bea
{\Phi}_{\frac{1}{2}}^{(1),\mu}(z)\,{\Phi}_{2}^{(1)}(w) & = & 
\frac{1}{(z-w)^{3}}\, c_{1}\,{G}^{\mu}(w)
+\frac{1}{(z-w)^{2}}\,\Bigg[\, c_{2}\,{\Phi}_{\frac{1}{2}}^{(2),\mu}+c_{3}\,{\Phi}_{0}^{(1)}{\Phi}_{\frac{1}{2}}^{(1),\mu}+c_{4}\,{G}^{\nu'}{T}^{\mu \nu'}
\nonu\\
& + & c_{5}\,{G}^{\nu'}\widetilde{T}^{\mu \nu'}+c_{6}\,\partial{G}^{\mu}\,\Bigg](w)
+\frac{1}{(z-w)}\,\Bigg[\frac{1}{5}\,\partial\Big(\mbox{pole-2}\Big)+c_{7}\,{\Phi}_{0}^{(1)}\partial{\Phi}_{\frac{1}{2}}^{(1),\mu}
\nonu\\
& + & c_{8}\,\partial\hat{\Phi}_{0}^{(1)}{\Phi}_{\frac{1}{2}}^{(1),\mu}+c_{9}\,{L}\hat{G}^{\mu}+c_{10}\,{G}^{\nu'}\partial\widetilde{T}^{\mu \nu'}+c_{11}\,\partial{G}^{\nu'}\widetilde{T}^{\mu \nu'}+c_{12}\,\partial^{2}{G}^{\mu}\,\Bigg](w)
\nonu \\
& + & \cdots, 
\label{3halfthree}
\eea
where
\bea
c_ {1} & = & -\frac {8 (k - N) (26 + 19 k + 19 N + 12 kN)} {3 (2 + k + 
      N) (4 + 5 k + 5 N + 6 kN)}, \qquad 
c_ {2} = \frac {5} {2}, \nonu\\
c_ {3} & = & \frac {1} {2 (2 + N) (2 + k + N)^{2} (4 + 5 k + 5 N +  
       6 kN)} (-1200 - 2656 k - 1981 k^{2} - 454 k^{3} 
\nonu\\
& - &  4304 N - 
    8858 kN - 5847 k^{2} N - 1112 k^{3} N - 5181 N^{2} - 
    9416 kN^{2} - 4952 k^{2} N^{2} 
\nonu\\
& - &  600 k^{3} N^{2} - 2583 N^{3} - 
    3868 kN^{3} - 1260 k^{2} N^{3} - 448 N^{4} - 480 kN^{4}), \nonu\\
c_ {4}  & = & \frac {5 i (20 + 21 k + 6 k^{2} + 25 N + 13 kN + 
       7 N^{2})} {2 (2 + N) (2 + k + N)^{2}},  \nonu\\
c_ {5}  & = &  \frac {i (32 + 81 k + 30 k^{2} + 47 N + 53 kN + 
      13 N^{2})} {2 (2 + N) (2 + k + N)^{2}},  \nonu\\
c_ {6} & = &  \frac {1} {6 (2 + N) (2 + k + N)^{2} (4 + 5 k + 5 N + 
       6 kN)} (-1200 - 2864 k - 1889 k^{2} - 446 k^{3} 
\nonu\\
& - &  1696 N - 3202 kN - 783 k^{2} N + 32 k^{3} N + 591 N^{2} + 1916 kN^{2} + 
    2612 k^{2} N^{2} + 600 k^{3} N^{2}
\nonu\\
& + &  1533 N^{3} + 2608 kN^{3} + 
    1260 k^{2} N^{3} + 448 N^{4} + 480 kN^{4}),  \nonu\\
c_ {7} & = & \frac {16 (k - N)} {5 (4 + 5 k + 5 N + 6 kN)}, \qquad 
c_ {8} = -\frac {24 (k - N)} {5 (4 + 5 k + 5 N + 6 kN)},  \nonu\\
c_ {9} & = & -\frac {16 (k - N)} {(4 + 5 k + 5 N + 6 kN)},  \qquad 
c_ {10} = -\frac {24 i} {5 (2 + k + N)},  \qquad 
c_ {11} = \frac {16 i} {5 (2 + k + N)},   \nonu\\
c_ {12} & = & \frac {28 (k - N)} {5 (4 + 5 k + 5 N + 6 kN)}.\nonu
\eea

\subsection{The OPEs between the higher spin $2$ currents and the
other $11$ higher spin currents}

The OPE between the higher spin $2$ currents and
the higher spin $2$
currents is given by
{\small
\bea
&&{\Phi}_{1}^{(1),\mu \nu}(z)\,{\Phi}_{1}^{(1),\rho \si}(w)  = 
\frac{1}{(z-w)^{4}}\,\Bigg[\,(\delta^{\mu \rho}\delta^{\nu \si}-\delta^{\mu \si}\delta^{\nu \rho})\, c_{1}+\varepsilon^{\mu \nu \rho \si}\, c_{2}\Bigg]
\nonu\\
&& + \frac{1}{(z-w)^{3}}\,\delta^{\mu \rho}\,\Bigg[\, c_{3}\,{T}^{\nu \si}+c_{4}\,\widetilde{{T}}^{\nu \si}\,\Bigg](w)
+\frac{1}{(z-w)^{2}}\,\Bigg[\,\varepsilon^{\mu \nu \rho \si}\,\Bigg(\, c_{17}\,{\Phi}_{0}^{(2)}+c_{18}\,{\Phi}_{0}^{(1)}{\Phi}_{0}^{(1)}\,\Bigg)
\nonu\\
&& +(\delta^{\mu \rho}\delta^{\nu \si}-\delta^{\mu \si}\delta^{\nu \rho})\,\Bigg(\, c_{5}\,{L}
+c_{6}\,({T}^{\mu \nu})^{2}+c_{7}\,(\widetilde{T}^{\mu \nu})^{2}+c_{8}\,({T}^{\mu \alpha'})^{2}+c_{9}\,({T}^{\nu \alpha'})^{2}
\nonu\\
&&+c_{10}\,\varepsilon^{\mu \nu \alpha' \beta'}\, T^{\mu \alpha'}T^{\nu \beta'}
+c_{11}\,{T}^{\mu \nu}\widetilde{T}^{\mu \nu}\,\Bigg)
+\delta^{\mu \rho}\,\Bigg(\, c_{12}\,{T}^{\mu \nu}{T}^{\mu \si}
+c_{13}\,({T}^{\mu \nu}\widetilde{T}^{\mu \si}+{T}^{\mu \si}\widetilde{T}^{\mu \nu})
\nonu\\
&&+c_{14}\,{T}^{\mu \si}\widetilde{T}^{\mu \nu}
+c_{15}\,\partial{T}^{\nu \si}
+c_{16}\,\partial\widetilde{T}^{\nu \si}\,\Bigg)
+\varepsilon^{\mu \nu \rho \si}\,\Bigg(\,c_{19}\,{L}
+c_{20}\,\Big(({T}^{\mu \nu})^{2}+({T}^{\rho \si})^{2}\Big)
+c_{21}\,({T}^{\alpha' \beta'})^{2}
\nonu\\
&&+c_{22}\,{T}^{\alpha' \beta'}\widetilde{T}^{\alpha' \beta'}\,\Bigg)
+c_{23}\,{T}^{\mu \nu}{T}^{\rho \si}\,\Bigg](w)
\nonu \\
& + &
\frac{1}{(z-w)}\,\Bigg[\,\frac{1}{2}\,\partial\Big(\mbox{pole-2}\Big)
+\delta^{\mu \rho}\,\Bigg(\, c_{27}\,{\Phi}_{1}^{(2),\nu \si}+c_{28}\,{\Phi}_{0}^{(1)}{\Phi}_{1}^{(1),\nu \si}
\nonu\\
&&+c_{29}\,{\Phi}_{\frac{1}{2}}^{(1),\nu}{\Phi}_{\frac{1}{2}}^{(1),\si}
+c_{32}\,\varepsilon^{\mu \nu \alpha' \beta'}\,{\Phi}_{\frac{1}{2}}^{(1),\alpha'}{\Phi}_{\frac{1}{2}}^{(1), \beta'}\Bigg)
+(\delta^{\mu \rho}\delta^{\nu \si}-\delta^{\mu \si}\delta^{\nu \rho})\,\Bigg(
c_{24}\,\partial{L}
+c_{25}\,\partial{T}^{\mu \nu}{T}^{\mu \nu}
\nonu\\
&&+c_{26}\,(\partial T^{\mu \alpha'}{T}^{\mu \alpha'}+\partial{T}^{\nu \alpha'}{T}^{\nu \alpha'})\Bigg)
+\delta^{\mu \rho}\,\Bigg(\, c_{30}\,{L}\widetilde{T}^{\nu \si}+c_{31}\,{L}{T}^{\nu \si}
+c_{33}\,\varepsilon^{\mu \nu \alpha' \beta'}\,{G}^{\alpha'}{G}^{\beta'}
+c_{34}\,{G}^{\nu}{G}^{\si}
\nonu\\
&&+ c_{35}\,({T}^{\mu \nu}\widetilde{T}^{\mu \nu}\widetilde{T}^{\nu \si}+{T}^{\mu \si}\widetilde{T}^{\nu \si}\widetilde{T}^{\mu \si})
+c_{36}\,{T}^{\mu \alpha'}\widetilde{T}^{\mu \alpha'}{T}^{\nu \si}
+c_{37}\,{T}^{\alpha' \beta'}{T}^{\alpha' \beta'}{T}^{\nu \si}
+c_{38}\,\widetilde{T}^{\nu \si}\widetilde{T}^{\nu \si}{T}^{\nu \si}
\nonu\\
&&+c_{39}\,({T}^{\mu \nu}\partial\widetilde{T}^{\mu \si}-{T}^{\mu \si}\partial\widetilde{T}^{\mu \nu})
+c_{40}\,(\partial{T}^{\mu \nu}\widetilde{T}^{\mu \si}-\partial{T}^{\mu \si}\widetilde{T}^{\mu \nu})
+c_{41}\,{T}^{\mu \nu}\partial{T}^{\mu \si}
+c_{42}\,\partial{T}^{\mu \nu}{T}^{\mu \si}
\nonu\\
&&+c_{43}\,\widetilde{T}^{\mu \nu}\partial\widetilde{T}^{\mu \si}
+c_{44}\,\partial\widetilde{T}^{\mu \nu}\widetilde{T}^{\mu \si}
+c_{45}\,\partial^{2}{T}^{\nu \si}
+c_{46}\,\partial^{2}\widetilde{T}^{\nu \si}\,\Bigg)
+ c_{47}\,({T}^{\mu \nu}\partial{T}^{\rho \si}-\partial{T}^{\mu \nu}{T}^{\rho \si})\,\Bigg](w)
\nonu\\
&&-\delta^{\mu \si}\,
\sum_{n=3}^1 \, \frac{1}{(z-w)^n} \, \left(\rho \;\leftrightarrow \;\si\right) (w)
 -\delta^{\nu \rho}\, \sum_{n=3}^1 \, \frac{1}{(z-w)^n} \,
\left(\mu\;\leftrightarrow \;\nu\right)(w)
\nonu\\
&&+ \delta^{\nu \si}\, 
\sum_{n=3}^1 \, \frac{1}{(z-w)^n} \,
\left(\mu \;\leftrightarrow \;\nu,\;\;\rho\;\leftrightarrow \;\si\right)(w)
+\cdots,
\label{twotwo}
\eea}
where
{\small
\bea
c_{1} & = &- \frac{4kN(2+3k+3N)}{(2+k+N)^{2}}, \qquad 
c_{2}=\frac{4k(k-N)N}{(2+k+N)^{2}}, \qquad 
c_{3}=\frac{4i(k+k^{2}+N+4kN+N^{2})}{(2+k+N)^{2}},\nonu\\
c_{4} & = & \frac{4i(k-N)(1+k+N)}{(2+k+N)^{2}}, \qquad 
c_{5}=-\frac{8(k+N)}{(2+k+N)}, \qquad 
c_{6} =- \frac{4}{(2+k+N)}, \nonu\\
c_{7} & = & \frac{4}{(2+k+N)}, \qquad 
c_{8}=\frac{2(k+N)}{(2+k+N)^{2}},  \qquad 
c_{9}=\frac{2(k+N)}{(2+k+N)^{2}},  \qquad 
c_{10}= \frac{4(k-N)}{(2+k+N)^{2}}, \nonu\\
c_{11} & = & -\frac{2(k-N)}{(2+k+N)^{2}},  \qquad
c_{12}=\frac{2(4+k+N)}{(2+k+N)^{2}},  \qquad
c_{13}=\frac{2(k-N)}{(2+k+N)^{2}}, \nonu\\
c_{14} & = &-\frac{2(4+k+N)}{(2+k+N)^{2}}, \qquad
c_{15}=\frac{2i(4+3k+k^{2}+3N+4kN+N^{2})}{(2+k+N)^{2}}, \nonu\\
c_{16} & = & \frac{2i(k-N)(1+k+N)}{(2+k+N)^{2}}, \qquad
c_{17} =2, \nonu\\
c_{18} & = & -\frac{(60+77k+22k^{2}+121N+115kN+20k^{2}N+79N^{2}+
  42kN^{2}+16N^{3})}{(2+N)(2+k+N)^{2}}, \nonu\\
c_{19} & = &\frac{4(10+13k+4k^{2}+10N+7kN+2N^{2})}{(2+k+N)^{2}}, \qquad
c_{20}  = \frac{2(k-N)}{(2+k+N)^{2}}, \nonu\\
c_{21} & = & -\frac{(20+25k+6k^{2}+21N+15kN+5N^{2})}{2(2+N)(2+k+N)^{2}},  \qquad
c_{22}  = -\frac{(9k+6k^{2}-N+7kN-N^{2})}{2(2+N)(2+k+N)^{2}}, \nonu\\
c_{23} & = &\frac{4(k+N)}{(2+k+N)^{2}},  \qquad
c_{24} =\frac{2(13k+6k^{2}+3N+9kN+N^{2})}{(2+N)(2+k+N)^{2}}, \nonu\\
c_{25} & = &-\frac{2(13k+6k^{2}+3N+9kN+N^{2})}{(2+N)(2+k+N)^{3}}, \qquad
c_{26}  = -\frac{2(13k+6k^{2}+3N+9kN+N^{2})}{(2+N)(2+k+N)^{3}}, \nonu\\
c_{27} & = & -\frac{1}{2}, \nonu\\
c_{28} & = & \frac{(60+77k+22k^{2}+121N+115kN+20k^{2}N+79N^{2}+42kN^{2}
  +16N^{3})}{2(2+N)(2+k+N)^{2}},\nonu\\
c_{29} & = & -\frac{2}{(2+k+N)}, \qquad
c_{30}=\frac{i(20+21k+6k^{2}+25N+13kN+7N^{2})}{(2+N)(2+k+N)^{2}},  \nonu\\
c_{31} & = & \frac{i(32+29k+6k^{2}+35N+17kN+9N^{2})}{(2+N)(2+k+N)^{2}}, \nonu\\
c_{32} & = & -\frac{(60+77k+22k^{2}+121N+115kN+20k^{2}N+79N^{2}+42kN^{2}+
  16N^{3})}{4(2+N)(2+k+N)^{2}}, \nonu\\
c_{33} & = & \frac{(20+21k+6k^{2}+25N+13kN+7N^{2})}{4(2+N)(2+k+N)^{2}}, \nonu\\
c_{34} & = & -\frac{(8+17k+6k^{2}+11N+11kN+3N^{2})}{(2+N)(2+k+N)^{2}}, \nonu\\
c_{35} & = & \frac{i(13k+6k^{2}+3N+9kN+N^{2})}{(2+N)(2+k+N)^{3}}, \nonu\\
c_{36} & = & -\frac{i(20+21k+6k^{2}+25N+13kN+7N^{2})}{(2+N)(2+k+N)^{3}}, \nonu\\
c_{37} & = & -\frac{i(16+21k+6k^{2}+19N+13kN+5N^{2})}{2(2+N)(2+k+N)^{3}}, \nonu\\
c_{38} & = & \frac{i(13k+6k^{2}+3N+9kN+N^{2})}{(2+N)(2+k+N)^{3}}, \nonu\\
c_{39} & = & \frac{(-40-24k+9k^{2}+6k^{3}-48N-4kN+13k^{2}N-11N^{2}+6kN^{2}+N^{3})}{2(2+N)(2+k+N)^{3}}, \nonu\\
c_{40} & = & -\frac{(-40-24k+9k^{2}+6k^{3}-48N-4kN+13k^{2}N-11N^{2}+6kN^{2}+N^{3})}{2(2+N)(2+k+N)^{3}}, \nonu\\
c_{41} & = & \frac{(-16+17k+6k^{2}-N+11kN+3N^{2})}{2(2+N)(2+k+N)^{2}}, \nonu\\
c_{42} & = & -\frac{(-4+k+N)(8+17k+6k^{2}+11N+11kN+3N^{2})}{2(2+N)(2+k+N)^{3}}, \nonu\\
c_{43} & = & \frac{(-32-34k+5k^{2}+6k^{3}-30N+2kN+17k^{2}N+N^{2}+14kN^{2}+3N^{3})}{2(2+N)(2+k+N)^{3}}, \nonu\\
c_{44} & = & -\frac{(-16+17k+6k^{2}-N+11kN+3N^{2})}{2(2+N)(2+k+N)^{2}}, \nonu\\
c_{45} & = & -\frac{1}{2(2+N)(2+k+N)^{3}}i(64+47k+8k^{2}+129N+91kN+4k^{2}N-6k^{3}N+109N^{2}
 \nonu\\
 & + & 58kN^{2}-4k^{2}N^{2}+42N^{3}+12kN^{3}+6N^{4}), \nonu\\
c_{46} & = & \frac{1}{(2+N)(2+k+N)^{3}}i(-20-16k-3k^{2}-20N+11kN+16k^{2}N+3k^{3}N+9N^{2}
 \nonu\\
 & + & 30kN^{2}+10k^{2}N^{2}+13N^{3}+10kN^{3}+3N^{4}), \nonu\\
c_{47} & = & -\frac{4}{(2+k+N)}. \nonu
\eea}

The OPE between the higher spin $2$ currents and
the higher spin $\frac{5}{2}$
currents is given by
{\small
\bea
&& \Phi_{1}^{(1),\mu \nu}(z)\,\Phi_{\frac{3}{2}}^{(1),\rho}(w) =
\frac{1}{(z-w)^{3}}\,\Bigg[\,\delta^{\mu \rho}\, c_{1}\, G^{\nu}+\varepsilon^{\mu \nu \rho \si'}\, c_{2}\, G^{\si'}\,\Bigg](w)
\nonu\\
&&+ \frac{1}{(z-w)^{2}}\,\Bigg[\,\delta^{\mu \rho}\,\Bigg(\, c_{3}\,\Phi_{\frac{1}{2}}^{(2),\nu}
+c_{4}\,\Phi_{0}^{(1)}\Phi_{\frac{1}{2}}^{(1),\nu}\,\Bigg)+
\varepsilon^{\mu \nu \rho \si'}\,\Bigg(\, c_{9}\,\Phi_{\frac{1}{2}}^{(2),\si'}+c_{10}\,\Phi_{0}^{(1)}\Phi_{\frac{1}{2}}^{(1),\si'}\,\Bigg)
\nonu\\
&&+ \delta^{\mu \rho}\Bigg(\, c_{5}\,\partial G^{\nu}
+c_{6}\, G^{\si'}T^{\nu \si'}
+c_{7}\, G^{\si'}\widetilde{T}^{\nu \si'}\,\Bigg)
+\varepsilon^{\mu \nu \rho \si'}
\Bigg(c_{11} G^{\alpha'}\widetilde{T}^{\alpha' \si}
+c_{12} G^{\alpha'}T^{\alpha' \si'}
+c_{13} \partial G^{\si'}\Bigg)\Bigg]
\nonu\\
&&+ \frac{1}{(z-w)}\,\Bigg[\,\frac{2}{5}\,\partial\Big(\mbox{pole-2}\Big)+\delta^{\mu \rho}\,\Bigg(\, 
c_{14}\,\Phi_{\frac{3}{2}}^{(2),\nu}
+c_{15}\,\partial\Phi_{\frac{1}{2}}^{(2),\nu}
+c_{16}\,\Phi_{0}^{(1)}\Phi_{\frac{3}{2}}^{(1),\nu}
+c_{17}\,\partial\Phi_{0}^{(1)}\Phi_{\frac{1}{2}}^{(1),\nu}
\nonu\\
&&+ c_{18}\,\Phi_{0}^{(1)}\partial\Phi_{\frac{1}{2}}^{(1),\nu}
+c_{26}\,\Phi_{\frac{1}{2}}^{(1),\si'}\Phi_{1}^{(1),\nu \si'}
+c_{27}\,\Phi_{\frac{1}{2}}^{(1),\si'}\widetilde{\Phi}_{1}^{(1),\nu \si'}\Bigg)
\nonu \\
&& +\varepsilon^{\mu \nu \rho \si'}\,\Bigg(
c_{55}\,\partial\Phi_{\frac{1}{2}}^{(2),\si'}
+c_{56}\,\partial\Phi_{0}^{(1)}\Phi_{\frac{1}{2}}^{(1),\si'}
\nonu\\
&&+ c_{57}\,\Phi_{0}^{(1)}\partial\Phi_{\frac{1}{2}}^{(1),\si'}\,\Bigg)
+ c_{41}\,(\Phi_{\frac{1}{2}}^{(1),\mu}\widetilde{\Phi}_{1}^{(1),\nu \rho}-\Phi_{\frac{1}{2}}^{(1),\nu}\widetilde{\Phi}_{1}^{(1),\mu \rho})
\nonu \\
&& +\delta^{\mu \rho}\,\Bigg(
c_{25}\,\widetilde{T}^{\nu \si'}\Phi_{\frac{1}{2}}^{(2),\si'}
+c_{28}\,\widetilde{T}^{\nu \si'}\Phi_{0}^{(1)}\Phi_{\frac{1}{2}}^{(1),\si'}\Bigg)
\nonu\\
&&+ c_{40}\,(\widetilde{T}^{\mu \rho}\Phi_{\frac{1}{2}}^{(2),\nu}-\widetilde{T}^{\nu \rho}\Phi_{\frac{1}{2}}^{(2),\mu})
+ c_{42}\,(\widetilde{T}^{\nu \rho}\Phi_{0}^{(1)}\Phi_{\frac{1}{2}}^{(1),\mu}-\widetilde{T}^{\mu \rho}\Phi_{0}^{(1)}\Phi_{\frac{1}{2}}^{(1),\nu})
\nonu\\
&&+\delta^{\mu \rho}\,\Bigg(\,c_{19}\, LG^{\nu}+c_{20}\, G^{\mu}\partial T^{\mu \nu}+c_{21}\,\partial^{2}G^{\nu}
+c_{22}\, G^{\nu}(T^{\mu \nu})^{2}
+c_{23}\, G^{\nu}T^{\mu \nu}\widetilde{T}^{\mu \nu}+c_{24}\, G^{\nu}(\widetilde{T}^{\mu \nu})^{2}
\nonu\\
&&+ c_{29}\,\partial G^{\si'}T^{\nu \si'}+c_{30}\, G^{\si'}\partial T^{\nu \si'}+c_{31}\,\partial G^{\si'}\widetilde{T}^{\nu \si'}
+c_{32}\, G^{\si'}\partial\widetilde{T}^{\nu \si'}+c_{33}\, G^{\nu}(T^{\mu \si'})^{2}+c_{34}\, G^{\nu}(T^{\nu \si'})^{2}
\nonu\\
&&+c_{35}\, G^{\nu}T^{\mu \si'}\widetilde{T}^{\mu \si'}
+c_{36}\, G^{\mu}T^{\mu \si'}T^{\nu \si'}+c_{37}\, G^{\si'}T^{\mu \si'}\widetilde{T}^{\mu \nu}+c_{38}\, G^{\si'}\widetilde{T}^{\mu \si'}\widetilde{T}^{\mu \nu}
+c_{39}\, G^{\si'}T^{\mu \nu}T^{\mu \si'}\,\Bigg)
\nonu\\
&&+c_{43}\,(\widetilde{G}^{\mu \nu \rho}T^{\mu \rho}\widetilde{T}^{\mu \rho}-\widetilde{G}^{\nu \mu \rho}T^{\nu \rho}\widetilde{T}^{\nu \rho})
+c_{44}\,(\widetilde{G}^{\mu \nu \rho}\widetilde{T}^{\mu \rho}\widetilde{T}^{\mu \rho}-\widetilde{G}^{\nu \mu \rho}\widetilde{T}^{\nu \rho}\widetilde{T}^{\nu \rho})
\nonu\\
&&+ c_{45}\,(G^{\mu}T^{\mu \nu}\widetilde{T}^{\mu \rho}-G^{\nu}T^{\nu \mu}\widetilde{T}^{\nu \rho}
+G^{\rho}T^{\mu \rho}\widetilde{T}^{\nu \rho}
+G^{\rho}T^{\nu \rho}\widetilde{T}^{\mu \rho})+c_{46}\,(G^{\mu}T^{\mu \nu}T^{\mu \rho}-G^{\nu}T^{\nu \mu}T^{\nu \rho})
\nonu\\
&&+c_{47}\,(G^{\mu}\widetilde{T}^{\mu \rho}\widetilde{T}^{\mu \nu}-G^{\nu}\widetilde{T}^{\nu \rho}\widetilde{T}^{\nu \mu})
+ c_{48}\,(G^{\mu}\partial\widetilde{T}^{\nu \rho}-G^{\nu}\partial\widetilde{T}^{\mu \rho})
+c_{49}\,(\partial G^{\mu}\widetilde{T}^{\nu \rho}-\partial G^{\nu}\widetilde{T}^{\mu \rho})
\nonu\\
&&+c_{50}\,\partial G^{\rho}T^{\mu \nu}
+c_{51}\, G^{\rho}\partial T^{\mu \nu}
+c_{52}\,\partial G^{\rho}\widetilde{T}^{\mu \nu}+c_{53}\,(\partial G^{\mu}T^{\nu \rho}-\partial G^{\nu}T^{\mu \rho})
\nonu\\
&&+c_{54}\,(G^{\mu}\partial T^{\nu \rho}-G^{\nu}\partial T^{\mu \rho})+c_{58}\,\varepsilon^{\mu \nu \rho \si'}\,\partial^{2}G^{\si'}\,\Bigg](w)
-\delta^{\nu \rho}\, \sum_{n=3}^1 \, \frac{1}{(z-w)^n} \,
\left(\mu \;\leftrightarrow \;\nu\right)(w)\nonu \\
& & +  \cdots,
\label{two5half}
\eea}
where
{\small
\bea
c_ {1} & = &  \frac {16 (3 + 6 k + 2 k^{2} + 6 N + 5 kN + 
      2 N^{2})} {3 (2 + k + N)^{2}}, \qquad
c_ {2} = - \frac {16 (k - N)} {3 (2 + k + N)^{2}}, \nonu\\
c_ {3} & = &  \frac {(k - N)} {6 (2 + k + N)}, \nonu\\
c_ {4} & = &  \frac {1}{6 (2 + N) (2 + k + N)^{3}}
(384 + 324 k + 19 k^{2} - 22 k^{3} + 636 N + 
     340 kN - 45 k^{2} N - 20 k^{3} N 
\nonu\\ & + & 409 N^{2} + 132 kN^{2} - 
     22 k^{2} N^{2} + 127 N^{3} + 26 kN^{3} + 
     16 N^{4}), \nonu\\
c_ {5} & = &  \frac {1}{6 (2 + N) (2 + k + N)^{3}}
(-212 k - 27 k^{2} + 38 k^{3} - 172 N - 68 kN + 
     161 k^{2} N + 28 k^{3} N 
\nonu\\ & - & 97 N^{2} + 160 kN^{2} + 
     82 k^{2} N^{2} + 25 N^{3} + 66 kN^{3} + 
     16 N^{4}) , \nonu\\
c_ {6} & = &  \frac {i (212 k + 117 k^{2} + 6 k^{3} + 172 N + 
      292 kN + 55 k^{2} N + 167 N^{2} + 90 kN^{2} + 41 N^{3})} {6 (2 +
       N) (2 + k + N)^{3}}, \nonu\\
c_ {7} & = &  \frac {i (k - N) (-32 - 3 k + 6 k^{2} - 29 N + kN - 
      7 N^{2})} {6 (2 + N) (2 + k + N)^{3}}, \nonu\\
c_ {8} & = &  - \frac {16 i (k - N)} {3 (2 + k + N)^{2}},\qquad
c_ {9}  =   - \frac {5} {2}, \nonu\\
c_ {10} & = &  \frac {5 (60 + 77 k + 22 k^{2} + 121 N + 115 kN + 
      20 k^{2} N + 79 N^{2} + 42 kN^{2} + 16 N^{3})} {2 (2 + 
      N) (2 + k + N)^{2}}, \nonu\\
c_ {11} & = &  \frac {i (32 + 81 k + 30 k^{2} + 47 N + 53 kN + 
      13 N^{2})} {2 (2 + N) (2 + k + N)^{2}}, \nonu\\
c_ {12} & = &  \frac {i (300 + 379 k + 90 k^{2} + 311 N + 227 kN + 
      73 N^{2})} {6 (2 + N) (2 + k + N)^{2}}, \nonu\\
c_ {13} & = &  - \frac {(-300 - 309 k - 70 k^{2} - 81 N + 153 kN + 
      100 k^{2} N + 217 N^{2} + 210 kN^{2} + 
      80 N^{3})} {6 (2 + N) (2 + k + N)^{2}}, \nonu\\
c_ {14} & = &  - \frac {1} {2}, \nonu\\
c_ {15} & = &  - \frac {(60 + 77 k + 22 k^{2} + 121 N + 115 kN + 
      20 k^{2} N + 79 N^{2} + 42 kN^{2} + 
      16 N^{3})} {4 (2 + N) (2 + k + N)^{2}},\nonu\\ 
c_ {16} & = &  \frac {(60 + 77 k + 22 k^{2} + 121 N + 115 kN + 
     20 k^{2} N + 79 N^{2} + 42 kN^{2} + 
     16 N^{3})} {2 (2 + N) (2 + k + N)^{2}},\nonu\\ 
c_ {17} & = &  \frac {1} {20 (2 + N)^{2} (2 + k + N)^{4}} (17232 + 
    44568 k + 41413 k^{2} + 16360 k^{3} + 2332 k^{4} + 71160 N 
\nonu\\ & + & 
    159274 kN + 124650 k^{2} N + 39742 k^{3} N + 4276 k^{4} N + 
    120133 N^{2} + 223872 kN^{2} 
\nonu\\ & + & 138595 k^{2} N^{2} + 
    31738 k^{3} N^{2} + 1960 k^{4} N^{2} + 105950 N^{3} + 
    154380 kN^{3} + 67362 k^{2} N^{3} 
\nonu\\ & + & 8316 k^{3} N^{3} + 
    51327 N^{4} + 52010 kN^{4} + 12028 k^{2} N^{4} + 12902 N^{5} + 
    6804 kN^{5} + 1312 N^{6}), \nonu\\
c_ {18} & = &  \frac {1} {60 (2 + N)^{2} (2 + k + N)^{4}} (55536 + 
    141864 k + 131399 k^{2} + 51980 k^{3} + 7436 k^{4} 
\nonu\\ & + & 220680 N + 492302 kN + 386550 k^{2} N + 124016 k^{3} N + 13448 k^{4} N + 
    362759 N^{2} 
\nonu\\ & + & 678156 kN^{2} + 423035 k^{2} N^{2} + 
    97724 k^{3} N^{2} + 6080 k^{4} N^{2} + 314050 N^{3} + 
    461190 kN^{3} 
\nonu\\ & + & 203376 k^{2} N^{3} + 25368 k^{3} N^{3} + 
    150171 N^{4} + 153880 kN^{4} + 36044 k^{2} N^{4} + 37396 N^{5}
\nonu\\ & + & 
    19992 kN^{5} + 3776 N^{6}), \nonu\\
c_ {19} & = &  \frac {3 (16 + 21 k + 6 k^{2} + 19 N + 13 kN + 
      5 N^{2})} {(2 + N) (2 + k + N)^{2}},\nonu\\ 
c_ {20} & = &  - \frac {4 i (24 + 25 k + 6 k^{2} + 27 N + 15 kN + 
       7 N^{2})} {(2 + N) (2 + k + N)^{3}},\nonu\\ 
c_ {21} & = &  - \frac {1} {60 (2 + N)^{2} (2 + k + N)^{4}} (8880 + 
       12048 k - 133 k^{2} - 4828 k^{3} - 1396 k^{4} + 19968 N 
\nonu\\ & + &  
       27554 kN + 11052 k^{2} N + 2216 k^{3} N + 656 k^{4} N + 
       32111 N^{2} + 63660 kN^{2} + 49433 k^{2} N^{2}
\nonu\\ & + & 
       16790 k^{3} N^{2} + 1892 k^{4} N^{2} + 41980 N^{3} + 
       78546 kN^{3} + 46464 k^{2} N^{3} + 8250 k^{3} N^{3} + 
       30993 N^{4} 
\nonu\\ & + &  40534 kN^{4} + 12272 k^{2} N^{4} + 10876 N^{5} + 
       7158 kN^{5} + 1424 N^{6}), \nonu\\
c_ {22} & = &  \frac {(32 + 55 k + 18 k^{2} + 41 N + 35 kN + 
     11 N^{2})} {(2 + N) (2 + k + N)^{3}}, \nonu\\
c_ {23} & = &  - \frac {(20 + 21 k + 6 k^{2} + 25 N + 13 kN + 
      7 N^{2})} {(2 + N) (2 + k + N)^{3}}, \nonu\\
c_ {24} & = &  - \frac {4 (16 + 21 k + 6 k^{2} + 19 N + 13 kN + 
       5 N^{2})} {(2 + N) (2 + k + N)^{3}},\nonu\\ 
c_ {25} & = &  - \frac {i} {(2 + k + N)}, \nonu\\
c_ {26} & = &  - \frac {(60 + 77 k + 22 k^{2} + 121 N + 115 kN + 
      20 k^{2} N + 79 N^{2} + 42 kN^{2} + 
      16 N^{3})} {2 (2 + N) (2 + k + N)^{2}}, \nonu\\ 
c_ {27} & = &  - \frac {2} {(2 + k + N)}, \nonu\\
c_ {28} & = &  \frac {i (60 + 77 k + 22 k^{2} + 121 N + 115 kN + 
      20 k^{2} N + 79 N^{2} + 42 kN^{2} + 16 N^{3})} {(2 + 
      N) (2 + k + N)^{3}}, \nonu\\
c_ {29} & = &  - \frac {1} {20 (2 + N)^{2} (2 + k + 
        N)^{4}} i (-6800 - 7776 k - 419 k^{2} + 1924 k^{3} + 
     516 k^{4} - 9056 N 
\nonu\\ & + &
 10 kN + 10152 k^{2} N + 4968 k^{3} N + 
     528 k^{4} N + 853 N^{2} + 14678 kN^{2} + 11917 k^{2} N^{2} +
\nonu\\  & + &
     2228 k^{3} N^{2} + 6162 N^{3} + 10678 kN^{3} + 
     3370 k^{2} N^{3} + 3093 N^{4} + 2138 kN^{4} + 468 N^{5}),\nonu\\ 
c_ {30} & = &  - \frac {1} {20 (2 + N)^{2} (2 + k + N)^{4}} i (9840 + 
     22464 k + 18501 k^{2} + 6624 k^{3} + 876 k^{4} + 29824 N 
\nonu\\  & + & 
     55290 kN + 35232 k^{2} N + 8978 k^{3} N + 708 k^{4} N + 
     35773 N^{2} + 50858 kN^{2} + 22427 k^{2} N^{2} 
\nonu\\  & + & 
     3058 k^{3} N^{2} + 21162 N^{3} + 20588 kN^{3} + 
     4720 k^{2} N^{3} + 6143 N^{4} + 3068 kN^{4} + 698 N^{5}), \nonu\\
c_ {31} & = &  - \frac {1} {20 (2 + N)^{2} (2 + k + 
          N)^{4}} i (3676 k + 5877 k^{2} + 3044 k^{3} + 516 k^{4} + 
       2724 N + 14616 kN 
\nonu\\  & + &15372 k^{2} N + 5368 k^{3} N + 
       528 k^{4} N + 6227 N^{2} + 17946 kN^{2} + 12025 k^{2} N^{2} + 
       2148 k^{3} N^{2} 
\nonu\\  & + & 5158 N^{3} + 8740 kN^{3} + 
       2906 k^{2} N^{3} + 1831 N^{4} + 1462 kN^{4} + 236 N^{5}), \nonu\\
c_ {32} & = &  - \frac {1} {20 (2 + N)^{2} (2 + k + N)^{4}} i (4800 + 
     13356 k + 13477 k^{2} + 5744 k^{3} + 876 k^{4} + 14484 N 
\nonu\\  & + &
     33296 kN + 26312 k^{2} N + 7978 k^{3} N + 708 k^{4} N + 
     17307 N^{2} + 30966 kN^{2} + 17135 k^{2} N^{2} 
\nonu\\  & + & 
     2778 k^{3} N^{2} + 10178 N^{3} + 12610 kN^{3} + 
     3676 k^{2} N^{3} + 2921 N^{4} + 1872 kN^{4} + 326 N^{5}), \nonu\\
c_ {33} & = &  - \frac {(16 + 21 k + 6 k^{2} + 19 N + 13 kN + 
      5 N^{2})} {(2 + N) (2 + k + N)^{3}}, \nonu\\
c_ {34} & = &  - \frac {(64 + 71 k + 18 k^{2} + 73 N + 43 kN + 
      19 N^{2})} {(2 + N) (2 + k + N)^{3}}, \nonu\\
c_ {35} & = &  - \frac {(20 + 21 k + 6 k^{2} + 25 N + 13 kN + 
      7 N^{2})} {(2 + N) (2 + k + N)^{3}}, \nonu\\
c_ {36} & = &  - \frac {2 (24 + 25 k + 6 k^{2} + 27 N + 15 kN + 
       7 N^{2})} {(2 + N) (2 + k + N)^{3}}, \nonu\\
c_ {37} & = &  \frac {(20 + 21 k + 6 k^{2} + 25 N + 13 kN + 
     7 N^{2})} {(2 + N) (2 + k + N)^{3}}, \nonu\\
c_ {38} & = &  \frac {3 (16 + 21 k + 6 k^{2} + 19 N + 13 kN + 
      5 N^{2})} {(2 + N) (2 + k + N)^{3}}, \nonu\\
c_ {39} & = &  - \frac {2 (8 + 17 k + 6 k^{2} + 11 N + 11 kN + 
       3 N^{2})} {(2 + N) (2 + k + N)^{3}}, \nonu\\
c_ {40} & = &  - \frac {i} {(2 + k + N)}, \qquad
c_ {41} =  \frac {4} {(2 + k + N)}, \nonu\\
c_ {42} & = &  - \frac {i (60 + 77 k + 22 k^{2} + 121 N + 115 kN + 
       20 k^{2} N + 79 N^{2} + 42 kN^{2} + 16 N^{3})} {(2 + 
       N) (2 + k + N)^{3}}, \nonu\\
c_ {43} & = &  - \frac {(32 + 29 k + 6 k^{2} + 35 N + 17 kN + 
      9 N^{2})} {(2 + N) (2 + k + N)^{3}}, \nonu\\
c_ {44} & = &  - \frac {(20 + 21 k + 6 k^{2} + 25 N + 13 kN + 
      7 N^{2})} {(2 + N) (2 + k + N)^{3}}, \nonu\\
c_ {45} & = &  - \frac {(20 + 21 k + 6 k^{2} + 25 N + 13 kN + 
      7 N^{2})} {(2 + N) (2 + k + N)^{3}}, \nonu\\
c_ {46} & = &  - \frac {8} {(2 + k + N)^{2}}, \qquad
c_ {47}  =   - \frac {(13 k + 6 k^{2} + 3 N + 9 kN + 
      N^{2})} {(2 + N) (2 + k + N)^{3}}, \nonu\\
c_ {48} & = &  \frac {i (100 + 89 k + 22 k^{2} + 141 N + 57 kN - 
      4 k^{2} N + 51 N^{2} + 4 N^{3})} {5 (2 + N) (2 + k + N)^{3}}, \nonu\\ 
c_ {49} & = &  \frac {2 i (17 k + 6 k^{2} + 33 N + 66 kN + 
      18 k^{2} N + 43 N^{2} + 35 kN^{2} + 12 N^{3})} {5 (2 + 
      N) (2 + k + N)^{3}}, \nonu\\
c_ {50} & = &  \frac {2 i (288 + 273 k + 62 k^{2} + 367 N + 213 kN + 
      16 k^{2} N + 141 N^{2} + 32 kN^{2} + 16 N^{3})} {5 (2 + 
      N) (2 + k + N)^{3}}, \nonu\\
c_ {51} & = &  - \frac {i (224 + 239 k + 66 k^{2} + 401 N + 299 kN + 
       48 k^{2} N + 243 N^{2} + 96 kN^{2} + 48 N^{3})} {5 (2 + 
       N) (2 + k + N)^{3}}, \nonu\\
c_ {52} & = &  \frac {2 i (20 + 21 k + 6 k^{2} + 25 N + 13 kN + 
      7 N^{2})} {(2 + N) (2 + k + N)^{3}}, \nonu\\
c_ {53} & = &  - \frac {i (224 + 79 k - 14 k^{2} + 241 N + 59 kN + 
       8 k^{2} N + 83 N^{2} + 16 kN^{2} + 8 N^{3})} {5 (2 + 
       N) (2 + k + N)^{3}}, \nonu\\
c_ {54} & = &  \frac {2 i (48 + 113 k + 42 k^{2} + 87 N + 93 kN + 
      6 k^{2} N + 41 N^{2} + 12 kN^{2} + 6 N^{3})} {5 (2 + 
      N) (2 + k + N)^{3}}, \nonu\\
c_ {55} & = &  - \frac {2} {(2 + k + N)}, \nonu\\ 
c_ {56} & = &  \frac {2 (60 + 77 k + 22 k^{2} + 121 N + 115 kN + 
      20 k^{2} N + 79 N^{2} + 42 kN^{2} + 16 N^{3})} {(2 + 
      N) (2 + k + N)^{3}}, \nonu\\
c_ {57} & = &  \frac {2 (60 + 77 k + 22 k^{2} + 121 N + 115 kN + 
      20 k^{2} N + 79 N^{2} + 42 kN^{2} + 16 N^{3})} {(2 + 
      N) (2 + k + N)^{3}}, \nonu\\
c_ {58} & = &  - \frac {(-300 - 277 k - 86 k^{2} - 113 N + 169 kN + 
      92 k^{2} N + 217 N^{2} + 210 kN^{2} + 
      88 N^{3})} {15 (2 + N) (2 + k + N)^{3}}.
\nonu
\eea}

The OPE between the higher spin $2$ currents and
the higher spin $3$
current is given by
{\small
\bea
&&\Phi_{1}^{(1),\mu \nu}(z)\,\Phi_{2}^{(1)}(w)  = 
\frac{1}{(z-w)^{4}}\,\Bigg[\, c_{1}\, T^{\mu \nu}+c_{2}\,\widetilde{T}^{\mu \nu}\,\Bigg](w)
+\frac{1}{(z-w)^{2}}\,\Bigg[\, c_{3}\,\Phi_{1}^{(2),\mu \nu}+c_{4}\,\Phi_{0}^{(1)}\Phi_{1}^{(1),\mu \nu}
\nonu\\
&& +  c_{5}\,(\Phi_{\frac{1}{2}}^{(1),\mu}\Phi_{\frac{1}{2}}^{(1),\nu}-\Phi_{\frac{1}{2}}^{(1),\nu}\Phi_{\frac{1}{2}}^{(1),\mu})
+c_{6}\, LT^{\mu \nu}
+c_{16}\,\varepsilon^{\mu \nu \rho' \si'}\,\Phi_{\frac{1}{2}}^{(1),\rho'}\Phi_{\frac{1}{2}}^{(1),\si'}
+c_{7}\, L\widetilde{T}^{\mu \nu}
\nonu\\
&&+c_{8}\,(G^{\mu}G^{\nu}-G^{\nu}G^{\mu})
+  c_{9}\,(T^{\mu \nu})^{3}
+c_{10}\,\partial^{2}T^{\mu \nu}
+c_{11}\,\partial^{2}\widetilde{T}^{\mu \nu}
+c_{12}\,T^{\mu \nu}(\widetilde{T}^{\mu \nu})^{2}
\nonu\\
&& +  c_{13}\,\Big(T^{\mu \nu}(T^{\mu \rho'})^{2}-T^{\nu \mu}(T^{\nu \rho'})^{2}\Big)
+c_{14}\,(T^{\mu \rho'}\partial T^{\nu \rho'}-T^{\nu \rho'}\partial T^{\mu \rho'})
+  c_{15}\,(T^{\mu \rho'}\partial\widetilde{T}^{\nu \rho'}-T^{\nu \rho'}\partial\widetilde{T}^{\mu \rho'})
\nonu\\
&& +\varepsilon^{\mu \nu \rho' \si'}\,\Bigg(\, 
c_{17}\, G^{\rho'}G^{\si'}
+c_{18}\, T^{\alpha' \rho'}\widetilde{T}^{\alpha' \rho'}T^{\rho' \si'}
+  c_{19}\,\widetilde{T}^{\rho' \si'}T^{\alpha' \rho'}\widetilde{T}^{\alpha' \rho'}\,
\Bigg)\,\Bigg](w)
\nonu\\
&& +\frac{1}{(z-w)}\,\Bigg[\,\frac{1}{3}\,\partial\Big(\mbox{pole-2}\Big)+c_{20}\,\Phi_{0}^{(1)}\partial\Phi_{1}^{(1),\mu \nu}
+  c_{21}\,\partial\Phi_{0}^{(1)}\Phi_{1}^{(1),\mu \nu}+c_{22}\, L\partial T^{\mu \nu}+c_{23}\,\partial LT^{\mu \nu}
\nonu\\
&& +c_{24}\, L\partial\widetilde{T}^{\mu \nu}+c_{25}\,\partial L\widetilde{T}^{\mu \nu}+c_{26}\,\partial^{3}T^{\mu \nu}
+  c_{27}\,\partial^{3}\widetilde{T}^{\mu \nu}
+c_{28}\,\Big(T^{\mu \nu}\partial\widetilde{T}^{\mu \nu}\widetilde{T}^{\mu \nu}-\partial T^{\mu \nu}(\widetilde{T}^{\mu \nu})^{2}\Big)
\nonu\\
&& +c_{29}\,(T^{\mu \nu}\partial T^{\mu \rho'}T^{\mu \rho'}-T^{\nu \mu}\partial T^{\nu \rho'}T^{\nu \rho'}
-  \partial T^{\mu \nu}T^{\mu \rho'}T^{\mu \rho'}+\partial T^{\nu \mu}T^{\nu \rho'}T^{\nu \rho'})
\nonu\\
&&+c_{30}\,(T^{\mu \rho'}\partial^{2}T^{\nu \rho'}-T^{\nu \rho'}\partial^{2}T^{\mu \rho'})
 +  \varepsilon^{\mu \nu \rho' \si'}\,\Bigg( c_{31}\, T^{\alpha' \rho'}\widetilde{T}^{\alpha' \rho'}\partial T^{\rho' \si'}+c_{32}\,\partial(T^{\alpha' \rho'}\widetilde{T}^{\alpha' \rho'})T^{\rho' \si'}\Bigg)\,\Bigg](w)
\nonu \\
&& +  \cdots,
\label{twothree}
\eea}
where
{\small
\bea
c_ {1} & = & -\frac{1}{(2 + k + N)^{2} (4 + 5 k + 5 N + 6 kN)} 16 i (4 k + 5 k^{2} + 3 k^{3} + 4 N + 34 kN + 
      33 k^{2} N 
\nonu\\
& + & 6 k^{3} N + 5 N^{2} + 33 kN^{2} + 24 k^{2} N^{2} + 
      3 N^{3} + 6 kN^{3}), \nonu\\
c_ {2} & = & -\frac {16 i (k - N) (4 + 5 k + 3 k^{2} + 5 N + 12 kN + 
      6 k^{2} N + 3 N^{2} + 6 kN^{2})} {(2 + k + N)^{2} (4 + 5 k + 
      5 N + 6 kN)}, \nonu\\
c_ {3} & = & 3,  \nonu\\
c_ {4} & = & -\frac {1}{(2 + N) (2 + k + N)^{2} (4 + 5 k + 5 N + 6 kN)}(720 + 1568 k + 1163 k^{2} + 266 k^{3} + 2608 N 
\nonu\\
& + & 5302 kN + 3489 k^{2} N + 664 k^{3} N + 3147 N^{2} + 
     5656 kN^{2} + 2968 k^{2} N^{2} + 360 k^{3} N^{2} + 1569 N^{3}
\nonu\\
& + & 2324 kN^{3} + 756 k^{2} N^{3} + 272 N^{4} + 
     288 kN^{4}),  \nonu\\
c_ {5} & = & \frac {6} {(2 + k + N)},  \nonu\\
c_ {6} & = &  \frac{1} {(2 + N) (2 + k + N)^{2} (4 + 5 k + 5 N + 6 kN)}2 i (-128 - 380 k - 283 k^{2} - 58 k^{3} - 324 N 
\nonu\\
& - & 940 kN - 595 k^{2} N - 92 k^{3} N - 185 N^{2} - 572 kN^{2} - 
      226 k^{2} N^{2} + 9 N^{3} - 82 kN^{3} + 16 N^{4}),  \nonu\\
c_ {7} & = & -\frac {2 i }{(2 + N) (2 + k + N)^{2} (4 + 5 k + 5 N + 6 kN)}(240 + 552 k + 323 k^{2} + 58 k^{3} + 600 N 
\nonu\\
& + & 1206 kN + 599 k^{2} N + 92 k^{3} N + 523 N^{2} + 782 kN^{2} + 
      218 k^{2} N^{2} + 169 N^{3} + 142 kN^{3} + 16 N^{4}) ,  \nonu\\
c_ {8} & = & \frac {3 (8 + 17 k + 6 k^{2} + 11 N + 11 kN + 3 N^{2})} {(2 +
       N) (2 + k + N)^{2}},  \nonu\\
c_ {9} & = & -\frac {2 i (32 + 55 k + 18 k^{2} + 41 N + 35 kN + 
      11 N^{2})} {(2 + N) (2 + k + N)^{3}},  \nonu\\
c_ {10} & = & \frac{1}{(2 + N) (2 + k + N)^{3} (4 + 5 k + 
      5 N + 6 kN)}i (-608 k - 1356 k^{2} - 801 k^{3} - 142 k^{4} + 
      352 N 
\nonu\\
& - & 288 kN - 2021 k^{2} N - 1341 k^{3} N - 248 k^{4} N + 
      1004 N^{2} + 1641 kN^{2} - 341 k^{2} N^{2} - 672 k^{3} N^{2} 
\nonu\\
& - & 120 k^{4} N^{2} + 1053 N^{3} + 2021 kN^{3} + 460 k^{2} N^{3} - 
      132 k^{3} N^{3} + 443 N^{4} + 780 kN^{4} + 156 k^{2} N^{4} 
\nonu\\
& + & 64 N^{5} + 96 kN^{5}),  \nonu\\
c_ {11} & = & -\frac{}{(2 + N) (2 + k + N)^{3} (4 + 5 k + 5 N + 6 kN)} i (-480 - 608 k + 410 k^{2} + 581 k^{3} + 
      142 k^{4} 
\nonu\\
& - & 1216 N - 892 kN + 1917 k^{2} N + 1521 k^{3} N + 
      248 k^{4} N - 838 N^{2} + 559 kN^{2} + 2959 k^{2} N^{2}
\nonu\\
& + & 1360 k^{3} N^{2} + 120 k^{4} N^{2} + 135 N^{3} + 1479 kN^{3} + 
      1764 k^{2} N^{3} + 372 k^{3} N^{3} + 283 N^{4} + 716 kN^{4}
\nonu\\
& + & 348 k^{2} N^{4} + 64 N^{5} + 96 kN^{5}),  \nonu\\
c_ {12} & = & \frac {2 i (32 + 55 k + 18 k^{2} + 41 N + 35 kN + 
      11 N^{2})} {(2 + N) (2 + k + N)^{3}},  \nonu\\
c_ {13} & = & \frac {2 i (32 + 55 k + 18 k^{2} + 41 N + 35 kN + 
      11 N^{2})} {(2 + N) (2 + k + N)^{3}},  \nonu\\
c_ {14} & = & -\frac {(64 + 188 k + 111 k^{2} + 18 k^{3} + 100 N + 
     172 kN + 45 k^{2} N + 37 N^{2} + 30 kN^{2} + 
     3 N^{3})} {(2 + N) (2 + k + N)^{3}},  \nonu\\
c_ {15} & = & -\frac {3 (26 k + 25 k^{2} + 6 k^{3} - 6 N + 26 kN + 
      15 k^{2} N - 5 N^{2} + 6 kN^{2} - N^{3})} {(2 + 
      N) (2 + k + N)^{3}},  \nonu\\
c_ {16} & = & \frac {3 (60 + 77 k + 22 k^{2} + 121 N + 115 kN + 
      20 k^{2} N + 79 N^{2} + 42 kN^{2} + 
      16 N^{3})} {2 (2 + N) (2 + k + N)^{2}},  \nonu\\
c_ {17} & = & -\frac {3 (20 + 21 k + 6 k^{2} + 25 N + 13 kN + 
      7 N^{2})} {2 (2 + N) (2 + k + N)^{2}},  \nonu\\
c_ {18} & = &  -\frac {i (32 + 55 k + 18 k^{2} + 41 N + 35 kN + 
      11 N^{2})} {(2 + N) (2 + k + N)^{3}},  \nonu\\
c_ {19} & = & \frac {3 i (20 + 21 k + 6 k^{2} + 25 N + 13 kN + 
      7 N^{2})} {(2 + N) (2 + k + N)^{3}},  \nonu\\
c_ {20} & = & \frac {16 (k - N)} {3 (4 + 5 k + 5 N + 6 kN)},  \qquad 
c_ {21}  =  -\frac {32 (k - N)} {3 (4 + 5 k + 5 N + 6 kN)},  \nonu\\
c_ {22} & = & \frac {64 i (4 + 5 k + k^{2} + 5 N + 4 kN + N^{2})} {3 (2 + 
      k + N) (4 + 5 k + 5 N + 6 kN)},  \nonu\\
c_ {23} & = &  -\frac {32 i (4 + 5 k + k^{2} + 5 N + 4 kN + 
      N^{2})} {3 (2 + k + N) (4 + 5 k + 5 N + 6 kN)},  \nonu\\
c_ {24} & = & \frac {64 i (k - N) (k + N)} {3 (2 + k + N) (4 + 5 k + 
      5 N + 6 kN)},  \nonu\\
c_ {25} & = & -\frac {32 i (k - N) (k + N)} {3 (2 + k + N) (4 + 5 k + 5 N + 6 kN)},  \nonu\\
c_ {26} & = & -\frac {16 i (28 + 47 k + 21 k^{2} + 3 k^{3} + 47 N + 
      60 kN + 15 k^{2} N + 21 N^{2} + 15 kN^{2} + 
      3 N^{3})} {9 (2 + k + N)^{2} (4 + 5 k + 5 N + 6 kN)},  \nonu\\
c_ {27} & = & -\frac {16 i (k - N) (k + N)} {3 (2 + k + N) (4 + 5 k + 5 N + 6 kN)},  \qquad 
c_ {28}  =  \frac {16 i} {3 (2 + k + N)^{2}},  \nonu\\
c_ {29} & = & \frac {16 i} {3 (2 + k + N)^{2}},  \qquad 
c_ {30}  =  -\frac {8} {3 (2 + k + N)^{2}},  \qquad 
c_ {31}  =  -\frac {16 i} {3 (2 + k + N)^{2}},  \nonu\\
c_ {32} & = & \frac {8 i} {3 (2 + k + N)^{2}}.
\nonu
\eea}

\subsection{The OPEs between the higher spin $\frac{5}{2}$ currents and the
remaining $5$ higher spin currents}

The OPE between the higher spin $\frac{5}{2}$ currents and
the higher spin $\frac{5}{2}$
currents is given by
\bea
&& \Phi_{\frac{3}{2}}^{(1),\mu}(z)\,\Phi_{\frac{3}{2}}^{(1),\nu}(w) =
\frac{1}{(z-w)^{5}}\,\delta^{\mu \nu}\, c_{1}+\frac{1}{(z-w)^{4}}\,\Bigg[\, c_{2}\,T^{\mu \nu}+c_{3}\,\widetilde{T}^{\mu \nu}\,\Bigg](w)
\nonu\\ 
&& +\frac{1}{(z-w)^{3}}\,\Bigg[\,\delta^{\mu \nu}\,\Bigg(\, c_{4}\,\Phi_{0}^{(2)}+c_{5}\,\Phi_{0}^{(1)}\Phi_{0}^{(1)}+c_{6}\,L+c_{7}\,(T^{\mu \rho'}T^{\mu \rho'}+\widetilde{T}^{\mu \rho'}\widetilde{T}^{\mu \rho'})\,\Bigg)
\nonu\\ 
&& +c_{8}\,\varepsilon^{\mu \nu \rho' \si'}\, T^{\mu \nu}T^{\rho' \si'}
+c_{9}\,T^{\mu \rho'}T^{\nu \rho'}+c_{10}\,\partial T^{\mu \nu}+c_{11}\,\partial\widetilde{T}^{\mu \nu}\,\Big](w)
\nonu\\ 
&& +\frac{1}{(z-w)^{2}}\,\Bigg[\,\frac{1}{2}\,\partial\Big(\mbox{pole-3}\Big)
+c_{14}\,\Phi_{1}^{(2),\mu \nu}
+c_{15}\,\widetilde{\Phi}_{1}^{(2),\mu \nu}
+c_{16}\,\Phi_{0}^{(1)}\Phi_{1}^{(1),\mu \nu}
+c_{17}\,\Phi_{0}^{(1)}\widetilde{\Phi}_{1}^{(1),\mu \nu}
\nonu\\ 
&& +c_{18}\,\Phi_{\frac{1}{2}}^{(1),\mu}\Phi_{\frac{1}{2}}^{(1),\nu}
+ c_{33}\, \varepsilon^{\mu \nu \rho' \si'}\,
\Phi_{\frac{1}{2}}^{(1),\rho'}\Phi_{\frac{1}{2}}^{(1),\si'}
+c_{19}\,\widetilde{T}_{\mu \nu}\Phi_{0}^{(2)}
+c_{20}\,\widetilde{T}^{\mu \nu}\Phi_{0}^{(1)}\Phi_{0}^{(1)}
+\delta^{\mu \nu}\,\Bigg(\, c_{12}\,\partial L
\nonu\\ 
&& + c_{13}\,(\partial T^{\mu \rho'}T^{\mu \rho'}+\partial\widetilde{T}^{\mu \rho'}\widetilde{T}^{\mu \rho'})\,\Bigg)
+c_{21\,}LT^{\mu \nu}+c_{22\,}L\widetilde{T}^{\mu \nu}
+c_{23}\,G^{\mu }G^{\nu}+c_{24}\,\partial^{2}T^{\mu \nu}
+c_{25}\,\partial^{2}\widetilde{T}^{\mu \nu}
\nonu\\ 
&& +c_{26}\,\Big((T^{\mu \nu})^{3}-T^{\mu \nu}(\widetilde{T}^{\mu \nu})^{2}\Big)
+c_{27}\,\Big(T^{\mu \nu}T^{\mu \nu}\widetilde{T}^{\mu \nu}
-(\widetilde{T}^{\mu \nu})^{3}\Big)
+c_{28}\,\Big(T^{\mu \nu}(T^{\mu \rho'})^{2}-T^{\nu \mu}(T^{\nu \rho'})^{2}\Big)
\nonu\\ 
&& +c_{29}\,\partial T^{\mu \rho'}T^{\nu \rho'}+c_{30}\,T^{\mu \rho'}\partial T^{\nu \rho'}
+c_{31}\,\Big(T^{\mu \rho'}T^{\mu \rho'}\widetilde{T}^{\mu \nu}-T^{\nu \rho'}T^{\nu \rho'}\widetilde{T}^{\nu \mu}\Big)
\nonu\\ 
&&+c_{32}\,\Big(\partial T^{\mu \rho'}\widetilde{T}^{\nu \rho'}-\partial T^{\nu \rho'}\widetilde{T}^{\mu \rho'}\Big)
+c_{34}\,\varepsilon^{\mu \nu \rho' \si'}\,G^{\rho'}G^{\si'}
+\Big(c_{35}\,T^{\mu \nu}+c_{36}\,\widetilde{T}^{\mu \nu}\Big)T^{\rho' \si'}\widetilde{T}^{\rho' \si'}\,\Bigg](w)
\nonu\\ 
&& + \frac{1}{(z-w)^{1}}\,\Bigg[\frac{1}{2}\,\partial\Big(\mbox{pole-2}\Big)+\delta^{\mu \nu}\,\Bigg(\, c_{37}\,\Phi_{2}^{(2)}+c_{38}\,\partial\Phi_{0}^{(2)}
+c_{39}\,\Phi_{0}^{(1)}\Phi_{2}^{(1)}
+ c_{40}\,\partial^{2}\Phi_{0}^{(1)}\Phi_{0}^{(1)}
\nonu\\ 
&& +c_{41}\,\partial\Phi_{0}^{(1)}\partial\Phi_{0}^{(1)}
+ c_{44}\,(\Phi_{1}^{(1),\mu \rho'}\Phi_{1}^{(1),\mu \rho'}+\widetilde{\Phi}_{1}^{(1),\mu \rho'}\widetilde{\Phi}_{1}^{(1),\mu \rho'})
+c_{45}\,\Phi_{\frac{1}{2}}^{(1),\rho'}\Phi_{\frac{3}{2}}^{(1),\rho'}
+c_{46}\,\partial\Phi_{\frac{1}{2}}^{(1),\rho'}\Phi_{\frac{1}{2}}^{(1),\rho'}
\nonu\\ 
&& +c_{69}\,\varepsilon^{\mu \alpha' \beta' \gamma'}\, \Phi_{1}^{(1),\mu \alpha'}\Phi_{1}^{(1),\beta' \gamma'}\,\Bigg)
+c_{75}\,\partial\Phi_{1}^{(2),\mu \nu}
+ c_{76}\,\partial(\Phi_{0}^{(1)}\Phi_{1}^{(1),\mu \nu})
+c_{77}\,\partial\Phi_{\frac{1}{2}}^{(1),\mu}\Phi_{\frac{1}{2}}^{(1),\nu}
\nonu\\ 
&&+c_{78}\,\Phi_{\frac{1}{2}}^{(1),\mu}\partial\Phi_{\frac{1}{2}}^{(1),\nu}
+c_{91}\,\Phi_{1}^{(1),\mu \rho'}\Phi_{1}^{(1),\nu \rho'}
+\varepsilon^{\mu \nu \rho' \si'}\, c_{116}\,\Phi_{\frac{1}{2}}^{(1),\rho'}\partial\Phi_{\frac{1}{2}}^{(1),\si'}
+\delta^{\mu \nu}\, \Bigg(
c_{42}\, L\Phi_{0}^{(2)}
\nonu\\ 
&&+c_{43}\, L\Phi_{0}^{(1)}\Phi_{0}^{(1)}
+ c_{47}\,(T^{\mu \rho'}\Phi_{1}^{(2),\mu \rho'}+\widetilde{T}^{\mu \rho'}\widetilde{\Phi}_{1}^{(2),\mu \rho'})
+c_{48}\,(T^{\mu \rho'}\Phi_{0}^{(1)}\Phi_{1}^{(1),\mu \rho'}+\widetilde{T}^{\mu \rho'}\Phi_{0}^{(1)}\widetilde{\Phi}_{1}^{(1),\mu \rho'})
\nonu\\ 
&& + c_{49}\, T^{\mu \rho'}\Phi_{\frac{1}{2}}^{(1),\mu}\Phi_{\frac{1}{2}}^{(1),\rho'}
+c_{50}\,\widetilde{T}^{\mu \rho'}\Phi_{\frac{1}{2}}^{(1),\mu}\Phi_{\frac{1}{2}}^{(1),\rho'}
+\varepsilon^{\mu \alpha' \beta' \gamma'}\,\Big(c_{70}\, T^{\mu \alpha' }
+c_{71}\,\widetilde{T}^{\mu \alpha' }\Big)\Phi_{\frac{1}{2}}^{(1),\beta'}\Phi_{\frac{1}{2}}^{(1),\gamma'}
\,\Bigg)
\nonu\\ 
&& + c_{90}\,(T^{\mu \rho'}\Phi_{1}^{(2),\nu \rho'}+T^{\nu \rho'}\Phi_{1}^{(2),\mu \rho'})
+ c_{92}\,\Big(T^{\mu \rho'}\Phi_{0}^{(1)}\Phi_{1}^{(1),\nu \rho'}
+T^{\nu \rho'}\Phi_{0}^{(1)}\Phi_{1}^{(1),\mu \rho'}
+\widetilde{T}^{\mu \rho'}\Phi_{\frac{1}{2}}^{(1),\nu}\Phi_{\frac{1}{2}}^{(1),\rho'}
\nonu\\ 
&&+\widetilde{T}^{\nu \rho'}\Phi_{\frac{1}{2}}^{(1),\mu}\Phi_{\frac{1}{2}}^{(1),\rho'}\Big)
+ c_{93}\,(T^{\mu \rho'}\Phi_{\frac{1}{2}}^{(1),\nu}\Phi_{\frac{1}{2}}^{(1),\rho'}+T^{\nu \rho'}\Phi_{\frac{1}{2}}^{(1),\mu}\Phi_{\frac{1}{2}}^{(1),\rho'})
+c_{51}\,\Big(L(T^{\mu \rho'})^{2}+L(\widetilde{T}^{\mu \rho'})^{2}\Big)
\nonu\\ 
&& + c_{52}\, LT^{\mu \rho'}\widetilde{T}^{\mu \rho'}
+c_{53}\,\partial G^{\rho'}G^{\rho'}
+c_{54}\, G^{\mu}G^{\rho'}T^{\mu \rho'}
+c_{55}\,\Big((T^{\mu \rho'})^{4}+T^{\mu \rho'}T^{\mu \rho'}\widetilde{T}^{\mu \rho'}\widetilde{T}^{\mu \rho'} \Big)
\nonu\\ 
&& + c_{56}\,T^{\mu \rho'}T^{\mu \rho'}T^{\mu \rho'}\widetilde{T}^{\mu \rho'}
+c_{57}\,\partial^{2}T^{\mu \rho'}T^{\mu \rho'}
+c_{58}\,(\partial T^{\mu \rho'})^{2}
+c_{59}\,\partial^{2}\widetilde{T}^{\mu \rho'}\widetilde{T}^{\mu \rho'}
+c_{60}(\partial\widetilde{T}^{\mu \rho'})^{2}
\nonu\\ 
&& + c_{61}\, T^{\mu \rho'}\partial^{2}\widetilde{T}^{\mu \rho'}
+c_{62}\,\partial T^{\mu \rho'}\partial\widetilde{T}^{\mu \rho'}
+c_{63}\,\partial^{2}T^{\mu \rho'}\widetilde{T}^{\mu \rho'}+c_{64}\,(L)^{2}+c_{65}\,\partial^{2}L
\nonu\\ 
&& + c_{66}\,\Big(T^{\mu \rho'}T^{\mu \rho'}T^{\mu \si'}T^{\mu \si'}+T^{\mu \rho'}T^{\mu \rho'}\widetilde{T}^{\mu \si'}\widetilde{T}^{\mu \si'}
+\widetilde{T}^{\mu \nu}\widetilde{T}^{\mu \nu}\widetilde{T}^{\mu \si'}\widetilde{T}^{\mu \si'} \Big)
+c_{67}\, T^{\mu \rho'}T^{\mu \si'}\widetilde{T}^{\mu \rho'}\widetilde{T}^{\mu \si'}
\nonu\\ 
&& + c_{68}\, T^{\mu \rho'}\widetilde{T}^{\mu \rho'}(\widetilde{T}^{\mu \si'})^{2}
+\varepsilon^{\mu  \alpha' \beta' \gamma'}\,\Big(\, c_{72}\, G^{ \alpha'}G^{ \beta'}\widetilde{T}^{ \gamma' \mu}
+c_{73}\, T^{\mu \alpha'}\partial T^{\mu \beta'}\widetilde{T}^{\mu \gamma'}
+c_{74}\, T^{\mu \alpha'}\partial\widetilde{T}^{\mu \beta'}\widetilde{T}^{\mu \gamma'}\Big)
\nonu\\ 
&& +c_{79}\, L\partial T^{\mu \nu}
+ c_{80}\,\partial( L\widetilde{T}^{\mu \nu})
+c_{81}\,\partial LT^{\mu \nu}+c_{82}\,\partial G^{\mu}G^{\nu}+c_{83}\, G^{\mu}\partial G^{\nu}+c_{84}\,\partial T^{\mu \nu}(T^{\mu \nu})^{2}
\nonu\\ 
&& + c_{85}\,\partial T^{\mu \nu}(\widetilde{T}^{\mu \nu})^{2}
+c_{86}\, T^{\mu \nu}\partial\widetilde{T}^{\mu \nu}\widetilde{T}^{\mu \nu}
+c_{87}\,T^{\mu \nu}T^{\mu \nu}\partial\widetilde{T}^{\mu \nu}
+c_{88}\,\partial^{3}T^{\mu \nu}
+c_{89}\,\partial^{3}\widetilde{T}^{\mu \nu}
\nonu\\ 
&& +c_{94}\, LT^{\mu \rho'}T^{\nu \rho'}
+ c_{95}\,(G^{\mu}G^{\rho'}\widetilde{T}^{\nu \rho'}+G^{\nu}G^{\rho'}\widetilde{T}^{\mu \rho'})
+c_{96}\,(G^{\mu}G^{\rho'}T^{\nu \rho'}+G^{\nu}G^{\rho'}T^{\mu \rho'})
\nonu\\ 
&& + c_{97}\,\Big(T^{\mu \rho'}(T^{\nu \rho'})^{3}+T^{\mu \rho'}T^{\nu \rho'}\widetilde{T}^{\mu \nu}\widetilde{T}^{\mu \nu}+T^{\mu \rho'}T^{\nu \rho'}\widetilde{T}^{\mu \rho'}\widetilde{T}^{\mu \rho'}\Big)
\nonu\\ 
&& + c_{98}\,\Big(T^{\mu \nu}T^{\mu \rho'}T^{\nu \rho'}\widetilde{T}^{\mu \nu}+T^{\mu \rho'}\widetilde{T}^{\nu \rho'}T^{\nu \rho'}T^{\nu \rho'}+T^{\nu \rho'}\widetilde{T}^{\mu \rho'}T^{\mu \rho'}T^{\mu \rho'}\Big)
+c_{99}\, T^{\mu \nu}\partial T^{\mu \rho'}T^{\mu \rho'}
\nonu\\ 
&& + c_{100}\, T^{\mu \nu}\partial T^{\nu \rho'}T^{\nu \rho'}
+c_{101}\,\partial T^{\mu \nu}T^{\mu \rho'}T^{\mu \rho'}
+c_{102}\,\partial T^{\mu \nu}T^{\nu \rho'}T^{\nu \rho'}
+c_{103}\, T^{\mu \nu}T^{\mu \rho'}\partial\widetilde{T}^{\mu \rho'}
\nonu\\ 
&& + c_{104}\, T^{\mu \nu}T^{\nu \rho'}\partial\widetilde{T}^{\nu \rho'}
+c_{105}\,\Big(T^{\mu \rho'}T^{\mu \rho'}\partial\widetilde{T}^{\mu \nu}+T^{\nu \rho'}T^{\nu \rho'}\partial\widetilde{T}^{\nu \mu}\Big)
+c_{106}\, T^{\mu \rho'}\widetilde{T}^{\mu \rho'}\partial\widetilde{T}^{\mu \nu}
\nonu\\ 
&& + c_{107}\, T^{\mu \rho'}\partial\widetilde{T}^{\mu \rho'}\widetilde{T}^{\mu \nu}
+c_{108}\, T^{\nu \rho'}\partial\widetilde{T}^{\nu \rho'}\widetilde{T}^{\nu \mu}
+c_{109}\,(\partial T^{\mu \rho'}T^{\mu \rho'}\widetilde{T}^{\mu \nu}+\partial T^{\nu \rho'}T^{\nu \rho'}\widetilde{T}^{\nu \mu})
\nonu\\ 
&& + c_{110}\, T^{\mu \rho'}\partial^{2}T^{\nu \rho'}+c_{111}\, T^{\nu \rho'}\partial^{2}T^{\mu \rho'}
+c_{112}\, T^{\mu \rho'}\partial^{2}\widetilde{T}^{\nu \rho'}
+c_{113}\, T^{\nu \rho'}\partial^{2}\widetilde{T}^{\mu \rho'}
+ c_{114}\,\partial T^{\mu \rho'}\partial T^{\nu \rho'}
\nonu\\ 
&& +c_{115}\,T^{\mu \rho'}T^{\mu \rho'}T^{\mu \si'}T^{\nu \si'}+\varepsilon^{\mu \nu \rho' \si'}\,\Bigg(\,
c_{117}\,\partial G^{\rho'}G^{\si'}
+ c_{118}\,\partial T^{\mu \nu}T^{\mu \rho'}T^{\nu \si'}\,\Bigg)\,\Bigg](w)+\cdots,
\label{5half5half}
\eea
where
{\small
\bea
c_ {1} & = & \frac {64 kN (15 + 12 k + 2 k^{2} + 12 N + 5 kN + 
      2 N^{2})} {3 (2 + k + N)^{3}},  \nonu\\
c_ {2} & = & - \frac {32 i (k + N) (15 + 12 k + 2 k^{2} + 12 N + 
       5 kN + 2 N^{2})} {3 (2 + k + N)^{3}},  \nonu\\
c_ {3} & = & - \frac {32 i (k - N) (15 + 12 k + 2 k^{2} + 12 N + 
       5 kN + 2 N^{2})} {3 (2 + k + N)^{3}},  \nonu\\
c_ {4} & = & - \frac {8 (k - N)} {3 (2 + k + N)},  \nonu\\
c_ {5} & = & - \frac{1}{3 (2 + N) (2 + k + N)^{3}} 4 (96 + 36 k - 53 k^{2} - 22 k^{3} + 204 N + 
       52 kN - 81 k^{2} N - 20 k^{3} N 
\nonu\\
 & + & 193 N^{2} + 60 kN^{2} - 
       22 k^{2} N^{2} + 91 N^{3} + 26 kN^{3} + 
       16 N^{4}),  \nonu\\
c_ {6} & = & \frac {1} {9 (2 + k + N)^{3}} 16 (180 + 204 k + 
      77 k^{2} + 10 k^{3} + 264 N + 245 kN + 59 k^{2} N + 146 N^{2}
\nonu\\
 & + & 83 kN^{2} + 28 N^{3}),  \nonu\\
c_ {7} & = & \frac {4 (-144 + 204 k + 151 k^{2} + 18 k^{3} + 12 N + 
      340 kN + 65 k^{2} N + 85 N^{2} + 110 kN^{2} + 
      23 N^{3})} {9 (2 + N) (2 + k + N)^{3}},  \nonu\\
c_ {8} & = & - \frac {4 (-k + N) (48 + 37 k + 6 k^{2} + 51 N + 
       21 kN + 13 N^{2})} {3 (2 + N) (2 + k + N)^{3}},  \nonu\\
c_ {9} & = & - \frac {32 (27 + 36 k + 10 k^{2} + 36 N + 25 kN + 
       10 N^{2})} {9 (2 + k + N)^{3}},  \nonu\\
c_ {10} & = & - \frac {1} {9 (2 + k + N)^{3}} 16 i (54 + 117 k + 
       56 k^{2} + 6 k^{3} + 117 N + 122 kN +21 k^{2} N +56 N^{2} 
\nonu\\
 & + & 21 kN^{2} + 6 N^{3}),  \nonu\\
c_ {11} & = & - \frac {16 i (k - N) (15 + 12 k + 2 k^{2} + 12 N + 
       5 kN + 2 N^{2})} {3 (2 + k + N)^{3}},  \nonu\\
c_ {12} & = & - \frac{1}{3 (2 + N) (2 + k + N)^{3}}2 (480 + 608 k + 260 k^{2} + 40 k^{3} + 
       688 N + 600 kN + 109 k^{2} N 
\nonu\\
 & - &  10 k^{3} N + 340 N^{2} + 
       159 kN^{2} - 11 k^{2} N^{2} + 76 N^{3} + 13 kN^{3} + 
       8 N^{4}),  \nonu\\
c_ {13} & = & \frac{1}{3 (2 + N) (2 + k + N)^{4}}4 (480 + 608 k + 260 k^{2} + 40 k^{3} + 688 N + 
      600 kN + 109 k^{2} N 
\nonu\\
 & - &  10 k^{3} N + 340 N^{2} + 159 kN^{2} - 
      11 k^{2} N^{2} + 76 N^{3} + 13 kN^{3} + 
      8 N^{4}) ,  \nonu\\
c_ {14} & = & \frac {(8 + 3 k + 3 N)} {(2 + k + N)},  \nonu\\
c_ {15} & = &\frac {(k - N)} {3 (2 + k + N)},  \nonu\\
c_ {16} & = & - \frac {(8 + 3 k + 3 N) (60 + 77 k + 22 k^{2} + 
       121 N + 115 kN + 20 k^{2} N + 79 N^{2} + 42 kN^{2} + 
       16 N^{3})} {(2 + N) (2 + k + N)^{3}},  \nonu\\
c_ {17} & = & \frac {1}{3 (2 + N) (2 + k + N)^{3}}(192 + 132 k - 29 k^{2} - 22 k^{3} + 348 N + 
     148 kN - 69 k^{2} N - 20 k^{3} N 
\nonu\\
 & + &  265 N^{2} + 84 kN^{2} - 
     22 k^{2} N^{2} + 103 N^{3} + 26 kN^{3} + 
     16 N^{4}) ,  \nonu\\
     c_ {18} & = & - \frac {1}{3 (2 + N) (2 + k + N)^{3}}
     (480 + 324 k - 5 k^{2} - 22 k^{3} + 684 N + 
     292 kN - 57 k^{2} N - 20 k^{3} N 
\nonu\\
 & + & 385 N^{2} + 108 kN^{2} - 
     22 k^{2} N^{2} + 115 N^{3} + 26 kN^{3} + 
     16 N^{4}),  \nonu\\
c_ {19} & = & \frac {8 i} {(2 + k + N)},  \nonu\\
c_ {20} & = & - \frac {4 i (60 + 77 k + 22 k^{2} + 121 N + 115 kN + 
       20 k^{2} N + 79 N^{2} + 42 kN^{2} + 16 N^{3})} {(2 + 
       N) (2 + k + N)^{3}},  \nonu\\
c_ {21} & = & - \frac {4 i (192 + 358 k + 189 k^{2} + 30 k^{3} + 
       314 N + 374 kN + 95 k^{2} N + 157 N^{2} + 90 kN^{2} + 
       25 N^{3})} {3 (2 + N) (2 + k + N)^{3}},  \nonu\\
c_ {22} & = & - \frac {4 i (94 k + 109 k^{2} + 30 k^{3} - 34 N + 
       66 kN + 55 k^{2} N - 37 N^{2} + 2 kN^{2} - 
       9 N^{3})} {3 (2 + N) (2 + k + N)^{3}},  \nonu\\
c_ {23} & = & \frac {(480 + 892 k + 525 k^{2} + 102 k^{3} + 692 N + 
     908 kN + 275 k^{2} N + 295 N^{2} + 210 kN^{2} + 
     37 N^{3})} {3 (2 + N) (2 + k + N)^{3}},  \nonu\\
c_ {24} & = & \frac{1}{9 (2 + N) (2 + k + N)^{4}} 4 i (-144 + 126 k + 159 k^{2} + 23 k^{3} - 
      2 k^{4} + 954 N + 1494 kN 
\nonu\\
 & + &  427 k^{2} N - 116 k^{3} N - 
      46 k^{4} N + 1947 N^{2} + 1834 kN^{2} + 228 k^{2} N^{2} - 
      87 k^{3} N^{2} + 1316 N^{3} 
\nonu\\
 & + & 797 kN^{3} + 38 k^{2} N^{3} + 
      389 N^{4} + 123 kN^{4} + 
      44 N^{5}),  \nonu\\
c_ {25} & = & - \frac{1}{9 (2 + N) (2 + k + N)^{4}} 2 i (-1440 - 3756 k - 2706 k^{2} - 769 k^{3} - 
       86 k^{4} + 84 N 
\nonu\\ & - & 852 kN + 147 k^{2} N + 401 k^{3} N + 
       92 k^{4} N + 3018 N^{2} + 3633 kN^{2} + 1905 k^{2} N^{2} + 
       350 k^{3} N^{2} 
\nonu\\ & + & 2569 N^{3} + 2267 kN^{3} + 516 k^{2} N^{3} + 
       805 N^{4} + 358 kN^{4} + 
       88 N^{5}),  \nonu\\
c_ {26} & = & - \frac {2 i (128 + 200 k + 103 k^{2} + 18 k^{3} + 
       184 N + 200 kN + 53 k^{2} N + 81 N^{2} + 46 kN^{2} + 
       11 N^{3})} {(2 + N) (2 + k + N)^{4}},  \nonu\\
c_ {27} & = & \frac{1}{3 (2 + N) (2 + k + N)^{4}} 2 i (240 + 436 k + 217 k^{2} + 30 k^{3} + 
      356 N + 474 kN + 127 k^{2} N 
\nonu\\
 & + &  173 N^{2} + 128 kN^{2} + 
      27 N^{3}),  \nonu\\
c_ {28} & = & \frac {2 i (128 + 200 k + 103 k^{2} + 18 k^{3} + 
      184 N + 200 kN + 53 k^{2} N + 81 N^{2} + 46 kN^{2} + 
      11 N^{3})} {(2 + N) (2 + k + N)^{4}},  \nonu\\
c_ {29} & = & \frac{1}{3 (2 + N) (2 + k + N)^{4}} 2 (-480 - 200 k + 266 k^{2} + 181 k^{3} + 
      30 k^{4} - 520 N + 272 kN 
\nonu\\ & + & 505 k^{2} N + 121 k^{3} N - 
      106 N^{2} + 385 kN^{2} + 179 k^{2} N^{2} + 33 N^{3} + 
      93 kN^{3} + 9 N^{4}),  \nonu\\
c_ {30} & = & - \frac{1}{3 (2 + N) (2 + k + N)^{4}} 2 (480 + 1016 k + 786 k^{2} + 261 k^{3} + 
       30 k^{4} + 856 N + 1472 kN 
\nonu\\ & + & 723 k^{2} N + 101 k^{3} N + 
       574 N^{2} + 703 kN^{2} + 157 k^{2} N^{2} + 185 N^{3} + 
       119 kN^{3} + 25 N^{4}),  \nonu\\
c_ {31} & = & - \frac {1}{3 (2 + N) (2 + k + N)^{4}} 2 i (240 + 436 k + 217 k^{2} + 30 k^{3} + 
       356 N + 474 kN + 127 k^{2} N 
\nonu\\ & + & 173 N^{2} + 128 kN^{2} + 
       27 N^{3}),  \nonu\\
c_ {32} & = & - \frac{1}{3 (2 + N) (2 + k + N)^{4}} 2 (-480 - 1280 k - 960 k^{2} - 281 k^{3} - 
       30 k^{4} - 544 N - 1388 kN 
\nonu\\ & - & 740 k^{2} N - 111 k^{3} N - 
       112 N^{2} - 384 kN^{2} - 104 k^{2} N^{2} + 55 N^{3} - 
       6 kN^{3} + 17 N^{4}),  \nonu\\
c_ {33} & = & \frac {1}{6 (2 + N) (2 + k + N)^{3}} (1440 + 2308 k + 1181 k^{2} + 198 k^{3} + 
     3524 N + 4502 kN + 1693 k^{2} N 
\nonu\\ & + & 180 k^{3} N + 3065 N^{2} + 
     2754 kN^{2} + 558 k^{2} N^{2} + 1115 N^{3} + 522 kN^{3} + 
     144 N^{4}),  \nonu\\
c_ {34} & = & - \frac {(480 + 700 k + 315 k^{2} + 42 k^{3} + 764 N + 
     754 kN + 169 k^{2} N + 383 N^{2} + 196 kN^{2} + 
     61 N^{3})} {6 (2 + N) (2 + k + N)^{3}},  \nonu\\
c_ {35} & = & \frac {i (120 + 187 k + 88 k^{2} + 12 k^{3} + 179 N + 
      192 kN + 46 k^{2} N + 83 N^{2} + 47 kN^{2} + 
      12 N^{3})} {3 (2 + N) (2 + k + N)^{4}},  \nonu\\
c_ {36} & = & - \frac {i (96 + 199 k + 135 k^{2} + 30 k^{3} + 137 N + 
       215 kN + 77 k^{2} N + 58 N^{2} + 54 kN^{2} + 
       7 N^{3})} {3 (2 + N) (2 + k + N)^{4}}, \nonu\\
c_ {37} & = & \frac {1} {2}, \nonu\\
c_ {38} & = & \frac{1}{12 (2 + N) (2 + k + N)^{2} (88 + 59 k + 
             59 N + 30 kN)}(15840 + 34148 k + 22829 k^{2} + 
            4790 k^{3} 
\nonu\\ & + & 39364 N + 72406 kN + 39435 k^{2} N + 
            6160 k^{3} N + 37281 N^{2} + 55036 kN^{2} + 
            21964 k^{2} N^{2} 
\nonu\\ & + & 1896 k^{3} N^{2} + 15615 N^{3} + 
            16544 kN^{3} + 3780 k^{2} N^{3} + 2384 N^{4} + 
            1344 kN^{4}),  \nonu\\
c_ {39} & = & - \frac {(60 + 77 k + 22 k^{2} + 121 N + 
             115 kN + 20 k^{2} N + 79 N^{2} + 42 kN^{2} + 
             16 N^{3})} {2 (2 + N) (2 + k + N)^{2}},  \nonu\\
c_ {40} & = & - \frac{1}{12 (2 + N)^{2} (2 + k + N)^{4} (4 + 5 k + 5 N + 6 kN)}
(49344 + 231696 k + 374420 k^{2} 
\nonu\\ & + & 275703 k^{3} + 94812 k^{4} + 12364 k^{5} + 201360 N + 
930344 kN + 1398077 k^{2} N + 925572 k^{3} N
\nonu\\ & + & 276820 k^{4} N + 30160 k^{5} N + 363380 N^{2} + 
1630133 kN^{2} + 2218658 k^{2} N^{2} + 
1271887 k^{3} N^{2} 
\nonu\\ & + & 312052 k^{4} N^{2} + 
25648 k^{5} N^{2} + 378287 N^{3} + 1607432 kN^{3} + 
1903753 k^{2} N^{3} 
\nonu\\ & + & 889510 k^{3} N^{3} + 
160480 k^{4} N^{3} + 7680 k^{5} N^{3} + 244478 N^{4} + 
949405 kN^{4} + 918412 k^{2} N^{4} 
\nonu\\ & + & 312196 k^{3} N^{4} + 
31248 k^{4} N^{4} + 96763 N^{5} + 328198 kN^{5} + 
231412 k^{2} N^{5} + 43176 k^{3} N^{5} 
\nonu\\ & + & 21316 N^{6} + 
59632 kN^{6} + 23184 k^{2} N^{6} + 1984 N^{7} + 
4224 kN^{7}),  \nonu\\
c_ {41} & = & - \frac {1}{12 (2 + N)^{2} (2 + k + N)^{4}}
 (5424 + 20136 k + 22531 k^{2} + 9976 k^{3} + 
1540 k^{4} + 29640 N 
\nonu\\ & + & 81766 kN + 71910 k^{2} N + 
24610 k^{3} N + 2764 k^{4} N + 57475 N^{2} + 
122592 kN^{2} 
\nonu\\ & + & 82021 k^{2} N^{2} + 19654 k^{3} N^{2} + 
1240 k^{4} N^{2} + 54674 N^{3} + 87396 kN^{3} + 
40254 k^{2} N^{3} 
\nonu\\ & + & 5124 k^{3} N^{3} + 27657 N^{4} + 
29942 kN^{4} + 7204 k^{2} N^{4} + 7130 N^{5} + 
3948 kN^{5} + 736 N^{6}),  \nonu\\
c_ {42} & = & - \frac {72 (k - N)} {(88 + 59 k + 59 N + 30 kN)},  \nonu\\
c_ {43} & = & - \frac {8 (-k + N) (60 + 77 k + 22 k^{2} + 
            121 N + 115 kN + 20 k^{2} N + 79 N^{2} + 42 kN^{2} + 
            16 N^{3})} {(2 + N) (2 + k + N)^{2} (4 + 5 k + 5 N + 
            6 kN)}, \nonu\\
c_ {44} & = & \frac {2} {(2 + k + N)},  \nonu\\
c_ {45} & = & - \frac {(60 + 77 k + 22 k^{2} + 121 N + 115 kN + 
           20 k^{2} N + 79 N^{2} + 42 kN^{2} + 
           16 N^{3})} {2 (2 + N) (2 + k + N)^{2}},  \nonu\\
c_ {46} & = & - \frac {1} {6 (2 + N) (2 + k + N)^{3}} (480 + 
    516 k + 91 k^{2} - 22 k^{3} + 876 N + 580 kN - 9 k^{2} N 
\nonu\\ & - &
    20 k^{3} N + 577 N^{2} + 204 kN^{2} - 22 k^{2} N^{2} + 
    163 N^{3} + 26 kN^{3} + 16 N^{4}),  \nonu\\
c_ {47} & = & \frac {2 i} {(2 + k + N)},  \nonu\\
c_ {48} & = & - \frac {2 i (60 + 77 k + 22 k^{2} + 121 N + 115 kN + 
       20 k^{2} N + 79 N^{2} + 42 kN^{2} + 16 N^{3})} {(2 + 
       N) (2 + k + N)^{3}},  \nonu\\
c_ {49} & = & - \frac {8 i} {(2 + k + N)^{2}},  \nonu\\
c_ {50} & = & - \frac {2 i (60 + 77 k + 22 k^{2} + 121 N + 115 kN + 
       20 k^{2} N + 79 N^{2} + 42 kN^{2} + 16 N^{3})} {(2 + 
       N) (2 + k + N)^{3}},  \nonu\\
c_ {51} & = & - \frac {2 (13 k + 6 k^{2} + 3 N + 9 kN + N^{2})} {(2 + 
       N) (2 + k + N)^{3}},  \qquad
c_ {52}  =  \frac {4 (20 + 21 k + 6 k^{2} + 25 N + 13 kN + 
      7 N^{2})} {(2 + N) (2 + k + N)^{3}},  \nonu\\
c_ {53} & = & \frac {608 + 872 k + 303 k^{2} + 18 k^{3} + 776 N + 
     632 kN + 57 k^{2} N + 249 N^{2} + 54 kN^{2} + 
     15 N^{3}} {2 (2 + N) (2 + k + N)^{3}},  \nonu\\
c_ {54} & = & \frac {10 i (16 + 21 k + 6 k^{2} + 19 N + 13 kN + 
      5 N^{2})} {(2 + N) (2 + k + N)^{3}},  \nonu\\
c_ {55} & = & - \frac {4 (24 + 25 k + 6 k^{2} + 27 N + 15 kN + 
       7 N^{2})} {(2 + N) (2 + k + N)^{4}},  \nonu\\
c_ {56} & = & - \frac {8 (20 + 21 k + 6 k^{2} + 25 N + 13 kN + 
       7 N^{2})} {(2 + N) (2 + k + N)^{4}},  \nonu\\
c_ {57} & = & - \frac {1} {36 (2 + N)^{2} (2 + k + N)^{4}} (-26064 - 
       25440 k + 3401 k^{2} + 6436 k^{3} + 1044 k^{4} - 31584 N 
\nonu\\ & + & 
       1106 kN + 29380 k^{2} N + 10768 k^{3} N + 1008 k^{4} N + 
       2513 N^{2} + 38014 kN^{2} + 26125 k^{2} N^{2} 
\nonu\\ & + & 
       4180 k^{3} N^{2} + 16062 N^{3} + 23206 kN^{3} + 
       6146 k^{2} N^{3} + 6717 N^{4} + 3794 kN^{4} + 820 N^{5}),  \nonu\\
c_ {58} & = & - \frac {1} {36 (2 + N)^{2} (2 + k + N)^{4}} (-32976 - 
       12048 k + 23201 k^{2} + 13456 k^{3} + 1692 k^{4} 
\nonu\\ & - & 38064 N + 
       40778 kN + 64948 k^{2} N + 18706 k^{3} N + 1332 k^{4} N + 
       4673 N^{2} + 76570 kN^{2}
\nonu\\ & + & 47275 k^{2} N^{2} + 
       6394 k^{3} N^{2} + 20742 N^{3} + 38740 kN^{3} + 
       10304 k^{2} N^{3} + 8679 N^{4} 
\nonu\\ & + & 6044 kN^{4} + 1090 N^{5}),  \nonu\\
c_ {59} & = & - \frac {1} {36 (2 + N)^{2} (2 + k + N)^{4}} (-42192 - 
       54384 k - 7039 k^{2} + 6004 k^{3} + 1044 k^{4} - 57936 N
\nonu\\ & - &
       30214 kN + 23944 k^{2} N + 10552 k^{3} N + 1008 k^{4} N - 
       9295 N^{2} + 30598 kN^{2} + 26017 k^{2} N^{2} 
\nonu\\ & + & 
       4180 k^{3} N^{2} + 15522 N^{3} + 23710 kN^{3} + 
       6146 k^{2} N^{3} + 7113 N^{4} + 3794 kN^{4} + 820 N^{5}),  \nonu\\
c_ {60} & = & - \frac {1} {36 (2 + N)^{2} (2 + k + N)^{4}} (-51408 - 
       41856 k + 13697 k^{2} + 13456 k^{3} + 1692 k^{4} - 70464 N 
\nonu\\ & + & 
       7010 kN + 60196 k^{2} N + 18706 k^{3} N + 1332 k^{4} N - 
       13111 N^{2} + 67138 kN^{2} + 47275 k^{2} N^{2} 
\nonu\\ & + & 
       6394 k^{3} N^{2} + 17646 N^{3} + 38740 kN^{3} + 
       10304 k^{2} N^{3} + 8679 N^{4} + 6044 kN^{4} + 1090 N^{5}), \nonu\\
c_ {61} & = & \frac {1} {12 (2 + N)^{2} (2 + k + N)^{4}} (-5760 - 
         6628 k - 4419 k^{2} - 1916 k^{3} - 348 k^{4} - 11420 N 
\nonu\\ & - & 
         10136 kN - 6924 k^{2} N - 2856 k^{3} N - 336 k^{4} N - 
         7333 N^{2} - 5574 kN^{2} - 4335 k^{2} N^{2}
\nonu\\ & - &
         1084 k^{3} N^{2} - 1474 N^{3} - 1396 kN^{3} - 
         990 k^{2} N^{3} + 151 N^{4} - 146 kN^{4} + 60 N^{5}),  \nonu\\
c_ {62} & = & - \frac {1} {6 (2 + N)^{2} (2 + k + N)^{4}} (-2400 + 
    2508 k + 6227 k^{2} + 3376 k^{3} + 564 k^{4} - 6348 N 
\nonu\\ & + & 5620 kN + 
    10936 k^{2} N + 4262 k^{3} N + 444 k^{4} N - 6783 N^{2} + 
    3966 kN^{2} + 6229 k^{2} N^{2} 
\nonu\\ & + & 1422 k^{3} N^{2} - 3998 N^{3} + 
    842 kN^{3} + 1160 k^{2} N^{3} - 1313 N^{4} - 36 kN^{4} - 
    182 N^{5}),  \nonu\\
c_ {63} & = & \frac {1} {12 (2 + N)^{2} (2 + k + N)^{4}} (-960 + 
      1292 k + 213 k^{2} - 1004 k^{3} - 348 k^{4} + 3700 N 
\nonu\\ & + & 
      11104 kN + 2952 k^{2} N - 1440 k^{3} N - 336 k^{4} N + 
      11195 N^{2} + 15258 kN^{2} + 2421 k^{2} N^{2} 
\nonu\\ & - & 
      604 k^{3} N^{2} + 9674 N^{3} + 7484 kN^{3} + 498 k^{2} N^{3} + 
      3451 N^{4} + 1246 kN^{4} + 444 N^{5}),  \nonu\\
c_ {64} & = & \frac {6 (16 + 21 k + 6 k^{2} + 19 N + 13 kN + 
      5 N^{2})} {(2 + N) (2 + k + N)^{2}},  \nonu\\
c_ {65} & = & \frac {1} {18 (2 + N) (2 + k + N)^{4}} (-28008 - 
      46002 k - 22063 k^{2} - 1984 k^{3} + 524 k^{4} - 46446 N 
\nonu\\ & - & 
      44575 kN - 652 k^{2} N + 6508 k^{3} N + 856 k^{4} N - 
      23740 N^{2} - 1816 kN^{2} + 12469 k^{2} N^{2} 
\nonu\\ & + & 
      2904 k^{3} N^{2} - 3018 N^{3} + 6449 kN^{3} + 
      3094 k^{2} N^{3} + 492 N^{4} + 1056 kN^{4} + 64 N^{5}),  \nonu\\
c_ {66} & = & - \frac {4 (24 + 25 k + 6 k^{2} + 27 N + 15 kN + 
       7 N^{2})} {(2 + N) (2 + k + N)^{4}},  \nonu\\
c_ {67} & = & \frac {10 (16 + 21 k + 6 k^{2} + 19 N + 13 kN + 
      5 N^{2})} {(2 + N) (2 + k + N)^{4}},  \nonu\\
c_ {68} & = & - \frac {4 (20 + 21 k + 6 k^{2} + 25 N + 13 kN + 
       7 N^{2})} {(2 + N) (2 + k + N)^{4}},  \nonu\\
c_ {69} & = & - \frac {(60 + 77 k + 22 k^{2} + 121 N + 115 kN + 
      20 k^{2} N + 79 N^{2} + 42 kN^{2} + 
      16 N^{3})} {4 (2 + N) (2 + k + N)^{2}},  \nonu\\
c_ {70} & = & - \frac {i (60 + 77 k + 22 k^{2} + 121 N + 115 kN + 
       20 k^{2} N + 79 N^{2} + 42 kN^{2} + 16 N^{3})} {(2 + 
       N) (2 + k + N)^{3}},  \nonu\\
c_ {71} & = & - \frac {4 i} {(2 + k + N)^{2}},  \qquad
c_ {72}  = - \frac {5 i (16 + 21 k + 6 k^{2} + 19 N + 13 kN + 
       5 N^{2})} {(2 + N) (2 + k + N)^{3}},  \nonu\\
c_ {73} & = & \frac {2 i (304 + 304 k + 59 k^{2} - 6 k^{3} + 376 N + 
      204 kN - 3 k^{2} N + 129 N^{2} + 14 kN^{2} + 11 N^{3})} {(2 + 
      N) (2 + k + N)^{4}},  \nonu\\
c_ {74} & = & - \frac {2 i (-100 - 63 k + 3 k^{2} + 6 k^{3} - 147 N - 
       31 kN + 19 k^{2} N - 56 N^{2} + 6 kN^{2} - 5 N^{3})} {(2 + 
       N) (2 + k + N)^{4}},  \nonu\\ 
c_ {75} & = & - \frac {1} {(2 + k + N)},  \nonu\\
c_ {76} & = & \frac {(60 + 77 k + 22 k^{2} + 121 N + 115 kN + 
     20 k^{2} N + 79 N^{2} + 42 kN^{2} + 
     16 N^{3})} {(2 + N) (2 + k + N)^{3}},  \nonu\\
c_ {77} & = & \frac {8 (1 + k + N)} {(2 + k + N)^{2}}, \qquad 
c_ {78}  =  - \frac {8} {(2 + k + N)},  \nonu\\
c_ {79} & = & - \frac {2 i (13 k + 6 k^{2} + 3 N + 9 kN + 
       N^{2})} {(2 + N) (2 + k + N)^{3}},  \nonu\\
c_ {80} & =  & \frac {2 i (20 + 21 k + 6 k^{2} + 25 N + 13 kN + 
      7 N^{2})} {(2 + N) (2 + k + N)^{3}},  \nonu\\
c_ {81} & = & \frac {2 i (32 + 29 k + 6 k^{2} + 35 N + 17 kN + 
      9 N^{2})} {(2 + N) (2 + k + N)^{3}},  \nonu\\
c_ {82} & = & - \frac {2 (80 + 95 k + 26 k^{2} + 105 N + 71 kN + 
       4 k^{2} N + 39 N^{2} + 8 kN^{2} + 4 N^{3})} {(2 + 
       N) (2 + k + N)^{3}},  \nonu\\
c_ {83} & = & \frac {2 (64 + 61 k + 14 k^{2} + 83 N + 49 kN + 
      4 k^{2} N + 33 N^{2} + 8 kN^{2} + 4 N^{3})} {(2 + 
      N) (2 + k + N)^{3}},  \nonu\\
c_ {84} & = & - \frac {6 i (32 + 29 k + 6 k^{2} + 35 N + 17 kN + 
       9 N^{2})} {(2 + N) (2 + k + N)^{4}},  \nonu\\
c_ {85} & = & \frac {4 i (16 + 21 k + 6 k^{2} + 19 N + 13 kN + 
      5 N^{2})} {(2 + N) (2 + k + N)^{4}},  \nonu\\
c_ {86} & = & \frac {2 i (64 + 45 k + 6 k^{2} + 67 N + 25 kN + 
      17 N^{2})} {(2 + N) (2 + k + N)^{4}},  \nonu\\
c_ {87} & = & \frac {2 i (20 + 21 k + 6 k^{2} + 25 N + 13 kN + 
      7 N^{2})} {(2 + N) (2 + k + N)^{4}},  \nonu\\
c_ {88} & = & \frac {1} {27 (2 + N) (2 + k + N)^{4}} i (-576 + 
      3294 k + 3701 k^{2} + 1042 k^{3} + 48 k^{4} + 1026 N 
\nonu\\ & + & 
      7214 kN + 4980 k^{2} N + 872 k^{3} N + 24 k^{4} N + 
      1901 N^{2} + 4689 kN^{2} + 1903 k^{2} N^{2}
\nonu\\ & + & 108 k^{3} N^{2} + 
      953 N^{3} + 1135 kN^{3} + 168 k^{2} N^{3} + 218 N^{4} + 
      108 kN^{4} + 24 N^{5}),  \nonu\\
c_ {89} & = & - \frac {1} {9 (2 + N) (2 + k + N)^{4}} i (360 - 456 k - 
       858 k^{2} - 296 k^{3} - 16 k^{4} + 864 N - 786 kN 
\nonu\\ & - & 
       1089 k^{2} N - 224 k^{3} N - 8 k^{4} N + 606 N^{2} - 
       573 kN^{2} - 381 k^{2} N^{2} - 20 k^{3} N^{2} + 170 N^{3} 
\nonu\\ & - &
       119 kN^{3} + 38 N^{4} + 20 kN^{4} + 8 N^{5}),  \nonu\\   
c_ {90} & = & - \frac {i} {(2 + k + N)}, \qquad
c_ {91}  =  - \frac {2} {(2 + k + N)},   \nonu\\   
c_ {92} & = & \frac {i (60 + 77 k + 22 k^{2} + 121 N + 115 kN + 
      20 k^{2} N + 79 N^{2} + 42 kN^{2} + 16 N^{3})} {(2 + 
      N) (2 + k + N)^{3}},   \nonu\\   
c_ {93} & = & \frac {4 i} {(2 + k + N)^{2}},   \qquad
c_ {94}  =  - \frac {4 (16 + 21 k + 6 k^{2} + 19 N + 13 kN + 
       5 N^{2})} {(2 + N) (2 + k + N)^{3}},   \nonu\\   
c_ {95} & = & - \frac {i (20 + 21 k + 6 k^{2} + 25 N + 13 kN + 
       7 N^{2})} {(2 + N) (2 + k + N)^{3}},   \nonu\\   
c_ {96} & = & - \frac {2 i (24 + 25 k + 6 k^{2} + 27 N + 15 kN + 
       7 N^{2})} {(2 + N) (2 + k + N)^{3}},   \nonu\\   
c_ {97} & = & \frac {4 (24 + 25 k + 6 k^{2} + 27 N + 15 kN + 
      7 N^{2})} {(2 + N) (2 + k + N)^{4}},   \nonu\\   
c_ {98} & = & \frac {4 (20 + 21 k + 6 k^{2} + 25 N + 13 kN + 
      7 N^{2})} {(2 + N) (2 + k + N)^{4}},   \nonu\\   
c_ {99} & = & - \frac {i (384 + 396 k + 83 k^{2} - 6 k^{3} + 468 N + 
       260 kN - 3 k^{2} N + 153 N^{2} + 14 kN^{2} + 11 N^{3})} {(2 + 
       N) (2 + k + N)^{4}},   \nonu\\   
c_ {100} & = & \frac {i (256 + 228 k + 35 k^{2} - 6 k^{3} + 316 N + 
      156 kN - 3 k^{2} N + 113 N^{2} + 14 kN^{2} + 11 N^{3})} {(2 + 
      N) (2 + k + N)^{4}},   \nonu\\   
c_ {101} & = & \frac {i (384 + 370 k + 71 k^{2} - 6 k^{3} + 462 N + 
      242 kN - 3 k^{2} N + 151 N^{2} + 14 kN^{2} + 11 N^{3})} {(2 + 
      N) (2 + k + N)^{4}},   \nonu\\   
c_ {102} & = & - \frac {i (32 + 11 k - 6 k^{2} + 37 N + 3 kN + 
       11 N^{2})} {(2 + N) (2 + k + N)^{3}},   \nonu\\   
c_ {103} & = & - \frac {i (120 + 98 k + 7 k^{2} - 6 k^{3} + 218 N + 
       120 kN + k^{2} N + 121 N^{2} + 36 kN^{2} + 21 N^{3})} {(2 + 
       N) (2 + k + N)^{4}},   \nonu\\   
c_ {104} & = & \frac {i (40 + 14 k - 17 k^{2} - 6 k^{3} + 118 N + 
      68 kN + k^{2} N + 93 N^{2} + 36 kN^{2} + 21 N^{3})} {(2 + 
      N) (2 + k + N)^{4}},   \nonu\\   
c_ {105} & = & \frac {i (80 + 56 k - 5 k^{2} - 6 k^{3} + 168 N + 
      94 kN + k^{2} N + 107 N^{2} + 36 kN^{2} + 21 N^{3})} {(2 + 
      N) (2 + k + N)^{4}},   \nonu\\   
c_ {106} & = & - \frac {6 i (32 + 29 k + 6 k^{2} + 35 N + 17 kN + 
       9 N^{2})} {(2 + N) (2 + k + N)^{4}},   \nonu\\   
c_ {107} & = & - \frac {i (64 + 11 k - 6 k^{2} + 53 N + 3 kN + 
       11 N^{2})} {(2 + N) (2 + k + N)^{3}},   \nonu\\   
c_ {108} & = & - \frac {i (128 + 138 k + 23 k^{2} - 6 k^{3} + 182 N + 
       106 kN - 3 k^{2} N + 79 N^{2} + 14 kN^{2} + 11 N^{3})} {(2 + 
       N) (2 + k + N)^{4}},   \nonu\\   
c_ {109} & = & - \frac {i (160 + 140 k + 19 k^{2} - 6 k^{3} + 268 N + 
       146 kN + k^{2} N + 135 N^{2} + 36 kN^{2} + 21 N^{3})} {(2 + 
       N) (2 + k + N)^{4}},   \nonu\\   
c_ {110} & = & \frac {1} {18 (2 + N) (2 + k + N)^{4}} (-1728 + 180 k + 
      877 k^{2} + 182 k^{3} - 1044 N + 964 kN + 99 k^{2} N 
\nonu\\ & - &
      152 k^{3} N + 463 N^{2} + 462 kN^{2} - 244 k^{2} N^{2} + 
      409 N^{3} + 44 kN^{3} + 64 N^{4}),   \nonu\\   
c_ {111} & = & \frac {1} {18 (2 + N) (2 + k + N)^{4}}(3456 + 3996 k + 
      985 k^{2} - 34 k^{3} + 4644 N + 2764 kN - 369 k^{2} N 
\nonu\\ & - & 
      152 k^{3} N + 2011 N^{2} + 246 kN^{2} - 244 k^{2} N^{2} + 
      445 N^{3} + 44 kN^{3} + 64 N^{4}),   \nonu\\   
c_ {112} & = & - \frac {1} {6 (2 + N) (2 + k + N)^{4}} (240 + 428 k + 
       203 k^{2} + 26 k^{3} + 604 N + 886 kN + 345 k^{2} N 
\nonu\\ & + & 
       40 k^{3} N + 567 N^{2} + 580 kN^{2} + 124 k^{2} N^{2} + 
       225 N^{3} + 116 kN^{3} + 32 N^{4}),   \nonu\\   
c_ {113} & = & - \frac {1} {6 (2 + N) (2 + k + N)^{4}} (240 + 644 k + 
       455 k^{2} + 98 k^{3} + 628 N + 1150 kN + 501 k^{2} N 
\nonu\\ & + & 
       40 k^{3} N + 603 N^{2} + 652 kN^{2} + 124 k^{2} N^{2} + 
       237 N^{3} + 116 kN^{3} + 32 N^{4}),   \nonu\\   
c_ {114} & = & \frac {1} {9 (2 + N) (2 + k + N)^{4}} 2 (3744 + 
      4914 k + 1775 k^{2} + 154 k^{3} + 5886 N + 5450 kN 
\nonu\\ & + &
      1269 k^{2} N + 104 k^{3} N + 3287 N^{2} + 2094 kN^{2} + 
      364 k^{2} N^{2} + 875 N^{3} + 364 kN^{3} + 104 N^{4}),   \nonu\\   
c_ {115} & = & \frac {4 (24 + 25 k + 6 k^{2} + 27 N + 15 kN + 
      7 N^{2})} {(2 + N) (2 + k + N)^{4}},   \nonu\\   
c_ {116} & = & - \frac {(60 + 77 k + 22 k^{2} + 121 N + 115 kN + 
      20 k^{2} N + 79 N^{2} + 42 kN^{2} + 
      16 N^{3})} {(2 + N) (2 + k + N)^{3}},   \nonu\\   
c_ {117} & = & \frac {(20 + 21 k + 6 k^{2} + 25 N + 13 kN + 
     7 N^{2})} {(2 + N) (2 + k + N)^{3}},   \nonu\\   
c_ {118} & = & \frac {2 i (20 + 21 k + 6 k^{2} + 25 N + 13 kN + 
  7 N^{2})} {(2 + N) (2 + k + N)^{4}}.
\nonu   
\eea}

The OPE between the higher spin $\frac{5}{2}$ currents and
the higher spin $3$
current is given by
{\small
\bea
&&{\Phi}_{\frac{3}{2}}^{(1),\mu}(z)\,{\Phi}_{2}^{(1)}(w)  = 
\frac{1}{(z-w)^{4}}\, c_{1}\,{G}^{\mu}(w)+\frac{1}{(z-w)^{3}}\,\Bigg[\, c_{2}\,{\Phi}_{\frac{1}{2}}^{(2),\mu}+c_{3}\,{\Phi}_{0}^{(1)}{\Phi}_{\frac{1}{2}}^{(1),\mu}+c_{4}\,{G}^{\nu'}{T}^{\mu \nu'}
\nonu\\
&& + \varepsilon^{\mu \nu' \rho' \si'}\, c_{5}\,{G}^{\nu'}{T}^{\rho' \si'}
+c_{6}\,\partial{G}^{\mu}\,\Bigg](w)
+\frac{1}{(z-w)^{2}}\,\Bigg[\, c_{7}\,{\Phi}_{\frac{3}{2}}^{(2),\mu}
+c_{8}\,\partial{\Phi}_{\frac{1}{2}}^{(2),\mu}+c_{9}\,{\Phi}_{0}^{(1)}{\Phi}_{\frac{3}{2}}^{(1),\mu}
\nonu\\ 
&& + c_{10}\,{\Phi}_{0}^{(1)}\partial{\Phi}_{\frac{1}{2}}^{(1),\mu}+c_{11}\,\partial{\Phi}_{0}^{(1)}{\Phi}_{\frac{1}{2}}^{(1),\mu}
+c_{12}\,{\Phi}_{\frac{1}{2}}^{(1),\nu'}{\Phi}_{1}^{(1),\mu \nu'}+\varepsilon^{\mu \nu' \rho' \si'}\,\Bigg(\, 
c_{17}\,{\Phi}_{\frac{1}{2}}^{(1),\nu'}{\Phi}_{1}^{(1),\rho' \si'}
\nonu\\ 
&& +  c_{18}\,{T}^{\nu' \rho'}{\Phi}_{\frac{1}{2}}^{(1),\si'}
+c_{19}\,{T}^{\nu' \rho'}{\Phi}_{0}^{(1)}{\Phi}_{\frac{1}{2}}^{(1),\si'}\Bigg)
+c_{13}\,({G}^{\mu}{T}^{\mu \nu'}{T}^{\mu \nu'}+{G}^{\mu}\widetilde{T}^{\mu \nu'}\widetilde{T}^{\mu \nu'})
+c_{14}\,{G}^{\nu'}\partial{T}^{\mu \nu'}
\nonu\\ 
&& + c_{15}\,\partial{G}^{\nu'}{T}^{\mu \nu'}
+c_{16}\,{G}^{\nu'}{T}^{\mu \rho'}{T}^{\nu' \rho'}
+\varepsilon^{\mu \nu' \rho' \si'}\Big(c_{20}\,{G}^{\mu}{T}^{\mu \nu'}{T}^{\rho' \si'}
+c_{21}\,{G}^{\nu'}\partial{T}^{\rho' \si'}\,
+c_{22}\,\partial{G}^{\nu'}{T}^{\rho' \si'}\,\Big)
\nonu\\ 
&& +  c_{23}\,{L}{G}^{\mu}+c_{24}\,\partial^{2}{G}^{\mu}\,\Bigg](w)
+\frac{1}{(z-w)}\,\Bigg[\,\frac{3}{7}\,\partial\Big(\mbox{pole-2}\Big)
+c_{26}\,\partial^{2}{\Phi}_{\frac{1}{2}}^{(2),\mu}
+ c_{27}\,{\Phi}_{0}^{(1)}\partial{\Phi}_{\frac{3}{2}}^{(1),\mu}
\nonu\\ 
&& +c_{28}\,\partial{\Phi}_{0}^{(1)}{\Phi}_{\frac{3}{2}}^{(1),\mu}
+c_{29}\,\partial{\Phi}_{0}^{(1)}\partial{\Phi}_{\frac{1}{2}}^{(1),\mu}
+c_{30}\,\partial^{2}{\Phi}_{0}^{(1)}{\Phi}_{\frac{1}{2}}^{(1),\mu}
+ c_{31}\,{\Phi}_{0}^{(1)}\partial^{2}{\Phi}_{\frac{1}{2}}^{(1),\mu}
\nonu\\
&& +\varepsilon^{\mu \nu' \rho' \si'}\Bigg(\, 
c_{47}\,\partial{\Phi}_{\frac{1}{2}}^{(1),\nu'}{\Phi}_{1}^{(1),\rho' \si'}
+c_{48}\,{\Phi}_{\frac{1}{2}}^{(1),\nu'}\partial{\Phi}_{1}^{(1),\rho' \si'}
+c_{49}\,\partial{T}^{\nu' \rho'}{\Phi}_{\frac{1}{2}}^{(2),\si'}
+ c_{50}\,{T}^{\nu' \rho'}\partial{\Phi}_{\frac{1}{2}}^{(2),\si'}
\nonu\\
&& +c_{51}\,\Big(\partial({T}^{\nu' \rho'}{\Phi}_{0}^{(1)}{\Phi}_{\frac{1}{2}}^{(1),\si'})
-\frac{7}{2}\,\partial{T}^{\nu' \rho'}{\Phi}_{0}^{(1)}{\Phi}_{\frac{1}{2}}^{(1),\si'}
\Big)\Bigg)
+c_{32}\,{L}{\Phi}_{\frac{1}{2}}^{(2),\mu}
+c_{33}\,{L}{\Phi}_{0}^{(1)}{\Phi}_{\frac{1}{2}}^{(1),\mu}
+c_{34}\,{L}\partial{G}^{\mu}
\nonu\\
&& +c_{35}\,\partial{L}{G}^{\mu}
+c_{36}\,\partial^{3}{G}^{\mu}
+c_{37}\,{L}{G}^{\nu'}{T}^{\mu \nu'}
+ c_{38}\,\partial^{2}{G}^{\nu'}{T}^{\mu \nu'}
+c_{39}\,\partial{G}^{\nu'}\partial{T}^{\mu \nu'}+c_{40}\,{G}^{\nu'}\partial^{2}{T}^{\mu \nu'}
\nonu\\
&& +c_{41}\,(\,{G}^{\mu}\partial{T}^{\mu \nu'}{T}^{\mu \nu'}+{G}^{\mu}\partial\widetilde{T}^{\mu \nu'}\widetilde{T}^{\mu \nu'})
+  c_{42}\,\partial{G}^{\mu}{T}^{\mu \nu'}{T}^{\mu \nu'}
+c_{43}\,\partial{G}^{\mu}\widetilde{T}^{\mu \nu'}\widetilde{T}^{\mu \nu'}
\nonu\\
&&+c_{44}\,\partial{G}^{\nu'}{T}^{\mu \rho'}{T}^{\nu' \rho'}
+c_{45}\,{G}^{\nu'}\partial{T}^{\mu \rho'}{T}^{\nu' \rho'}
+c_{46}\,{G}^{\nu'}{T}^{\mu \rho'}\partial{T}^{\nu' \rho'}
+\varepsilon^{\mu \nu' \rho' \si'}\,\Bigg(\,
c_{52}\,{L}{G}^{\nu'}{T}^{\rho' \si'}
\nonu\\
&&+c_{53}\,\Big(\partial({G}^{\mu}{T}^{\mu \nu'}){T}^{\rho' \si'}-\frac{5}{2}\,{G}^{\mu}{T}^{\mu \nu'}\partial{T}^{\rho' \si'}\Big)
+  c_{54}\,\partial^{2}{G}^{\nu'}{T}^{\rho' \si'}
+c_{55}\,\partial{G}^{\nu'}\partial{T}^{\rho' \si'}
\nonu\\ 
&& +c_{56}\,{G}^{\nu'}\partial{T}^{\nu' \rho'}{T}^{\nu' \si'}
+c_{57}\,{G}^{\nu'}\partial^{2}{T}^{\rho' \si'}\,\Bigg)\,\Bigg](w)+\cdots,
\label{5halfthree}
\eea}
where
{\small
\bea
c_ {1} & = & - \frac {1}{3 (2 + k + N)^{2} (4 + 5 k + 5 N + 6 kN)}16 (12 + 111 k + 136 k^{2} + 38 k^{3} + 111 N 
\nonu\\  & + &
       406 kN + 331 k^{2} N + 60 k^{3} N +136 N^{2} + 331 kN^{2} + 
       150 k^{2} N^{2} + 38 N^{3} + 
       60 kN^{3}), \nonu\\
c_ {2} & = & \frac {8 (k - N) (17 + 10 k + 10 N + 3 kN)} {3 (2 + k + 
      N) (4 + 5 k + 5 N + 6 kN)}, \nonu\\
c_ {3} & = & \frac {1} {3 (2 + 
      N) (2 + k + N)^{3} (4 + 5 k + 5 N + 6 kN)} 8 (192 - 588 k - 1621 k^{2} - 1084 k^{3} - 
      220 k^{4} 
\nonu\\  & + & 1548 N + 332 kN - 2359 k^{2} N - 1619 k^{3} N - 
      266 k^{4} N + 3161 N^{2} + 2462 kN^{2} - 698 k^{2} N^{2} 
\nonu\\  & - & 
      663 k^{3} N^{2} - 60 k^{4} N^{2} + 2757 N^{3} + 2261 kN^{3} + 
      220 k^{2} N^{3} - 66 k^{3} N^{3} + 1092 N^{4} + 693 kN^{4}
\nonu\\  & + & 
      78 k^{2} N^{4} + 160 N^{5} + 48 kN^{5}), \nonu\\
c_ {4} & = & - \frac{1}{3 (2 + N) (2 + k + N)^{3} (4 + 5 k + 
       5 N + 6 kN)}8 i (384 + 428 k - 149 k^{2} - 252 k^{3} - 
       60 k^{4} 
\nonu\\  & + & 1300 N + 1708 kN + 387 k^{2} N - 91 k^{3} N - 
       18 k^{4} N + 1417 N^{2} + 1680 kN^{2} + 428 k^{2} N^{2} 
\nonu\\  & + &
       15 k^{3} N^{2} + 633 N^{3} + 583 kN^{3} + 90 k^{2} N^{3} + 
       100 N^{4} + 57 kN^{4}), \nonu\\
c_ {5} & = & - \frac{1}{3 (2 + N) (2 + k + N)^{3} (4 + 5 k + 5 N + 6 kN)}
       4 i (-k + N) (576 + 869 k + 412 k^{2} + 
       60 k^{3}
\nonu\\  & + & 987 N + 1141 kN + 351 k^{2} N + 18 k^{3} N + 
       551 N^{2} + 433 kN^{2} + 63 k^{2} N^{2} + 100 N^{3} + 
       39 kN^{3}), \nonu\\
 c_ {6} & = & - \frac{1}{9 (2 + N) (2 + k + N)^{3} (4 + 5 k + 5 N + 
       6 kN)}8 (-2016 - 2220 k + 805 k^{2} + 1276 k^{3} 
\nonu\\  & + & 
       292 k^{4} - 5268 N - 5312 kN + 901 k^{2} N + 1307 k^{3} N + 
       158 k^{4} N - 3341 N^{2} - 734 kN^{2} 
\nonu\\ & + & 2990 k^{2} N^{2} + 
       1065 k^{3} N^{2} + 60 k^{4} N^{2} + 573 N^{3} + 3127 kN^{3} + 
       2444 k^{2} N^{3} + 354 k^{3} N^{3} 
\nonu\\ & + & 1068 N^{4} + 
       1569 kN^{4} + 498 k^{2} N^{4} + 236 N^{5} + 
       168 kN^{5}), \nonu\\
c_ {7} & = & \frac {7} {2}, \nonu\\
c_ {8} & = & \frac {1} {60 (2 + N) (2 + k + N)^{2} (4 + 5 k + 5 N + 
      6 kN)} (25200 + 70112 k + 56321 k^{2} + 13310 k^{3} 
\nonu\\  & + & 76048 N + 
    193186 kN + 134691 k^{2} N + 25624 k^{3} N + 86913 N^{2} + 
    193192 kN^{2}
\nonu\\ & + & 106264 k^{2} N^{2} + 12792 k^{3} N^{2} + 
    43107 N^{3} + 78572 kN^{3} + 26460 k^{2} N^{3} + 7520 N^{4} + 
    9888 kN^{4}), \nonu\\
c_ {9} & = & - \frac {1} {2 (2 + N) (2 + k + N)^{2} (4 + 5 k + 5 N + 
       6 kN)} (1680 + 3616 k + 2671 k^{2} + 610 k^{3}
\nonu\\ & + & 6128 N + 
     12350 kN + 8109 k^{2} N + 1544 k^{3} N + 7407 N^{2} + 
     13208 kN^{2} + 6920 k^{2} N^{2} 
\nonu\\ & + & 840 k^{3} N^{2} + 3693 N^{3} + 
     5428 kN^{3} + 1764 k^{2} N^{3} + 640 N^{4} + 672 kN^{4}), \nonu\\
c_ {10} & = & - \frac {1} {4 (2 + N)^{2} (2 + k + N)^{4} (4 + 5 k + 
       5 N + 6 kN)} 3 (35648 + 142832 k + 214572 k^{2} 
\nonu\\ & + & 
     152505 k^{3} + 51492 k^{4} + 6644 k^{5} + 176240 N + 660824 kN + 
     910867 k^{2} N + 581724 k^{3} N 
\nonu\\ & + & 171980 k^{4} N + 
     18800 k^{5} N + 373580 N^{2} + 1290331 kN^{2} + 
     1592030 k^{2} N^{2} + 880129 k^{3} N^{2} 
\nonu\\ & + & 214668 k^{4} N^{2} + 
     17936 k^{5} N^{2} + 438209 N^{3} + 1366104 kN^{3} + 
     1455655 k^{2} N^{3} + 656026 k^{3} N^{3} 
\nonu\\ & + & 117952 k^{4} N^{3} + 
     5760 k^{5} N^{3} + 305474 N^{4} + 836307 kN^{4} + 
     726932 k^{2} N^{4} + 238876 k^{3} N^{4} 
\nonu\\ & + & 23856 k^{4} N^{4} + 
     125749 N^{5} + 290746 kN^{5} + 185388 k^{2} N^{5} + 
     33624 k^{3} N^{5} + 28124 N^{6} 
\nonu\\ & + & 51696 kN^{6} + 
     18480 k^{2} N^{6} + 2624 N^{7} + 3456 kN^{7}), \nonu\\
c_ {11} & = & - \frac {1} {12 (2 + N)^{2} (2 + k + N)^{4} (4 + 5 k + 
         5 N + 6 kN)} (250176 + 1077744 k + 1704908 k^{2} 
\nonu\\ & + &
       1253961 k^{3} + 432984 k^{4} + 56716 k^{5} + 1334640 N + 
       5259608 kN + 7510931 k^{2} N 
\nonu\\ & + & 4901808 k^{3} N + 
       1466482 k^{4} N + 161164 k^{5} N + 3022124 N^{2} + 
       10765163 kN^{2} 
\nonu\\ & + & 13575866 k^{2} N^{2} + 7590265 k^{3} N^{2} + 
       1857610 k^{4} N^{2} + 154816 k^{5} N^{2} + 3733601 N^{3} 
\nonu\\ & + & 
       11830628 kN^{3} + 12750499 k^{2} N^{3} + 5765230 k^{3} N^{3} + 
       1033840 k^{4} N^{3} + 50160 k^{5} N^{3} 
\nonu\\ & + & 2705042 N^{4} + 
       7448995 kN^{4} + 6498196 k^{2} N^{4} + 2129476 k^{3} N^{4} + 
       211176 k^{4} N^{4} 
\nonu\\ & + & 1145011 N^{5} + 2644570 kN^{5} + 
       1682860 k^{2} N^{5} + 302952 k^{3} N^{5} + 261262 N^{6} + 
       478024 kN^{6} 
\nonu\\ & + & 169848 k^{2} N^{6} + 24736 N^{7} + 
       32448 kN^{7}), \nonu\\
c_ {12} & = & \frac {7 (60 + 77 k + 22 k^{2} + 121 N + 115 kN + 
      20 k^{2} N + 79 N^{2} + 42 kN^{2} + 16 N^{3})} {2 (2 + 
      N) (2 + k + N)^{2}}, \nonu\\
c_ {13} & = & \frac {2 (24 + 25 k + 6 k^{2} + 27 N + 15 kN + 
      7 N^{2})} {(2 + N) (2 + k + N)^{3}}, \nonu\\
c_ {14} & = & \frac {1} {12 (2 + N)^{2} (2 + k + N)^{4} (4 + 5 k + 
      5 N + 6 kN)} i (103872 + 445808 k + 686660 k^{2} 
\nonu\\ & + & 
      485009 k^{3} + 161184 k^{4} + 20508 k^{5} + 452080 N + 
      1723688 kN + 2320763 k^{2} N 
\nonu\\ & + &1398642 k^{3} N + 
      383950 k^{4} N + 38460 k^{5} N + 809636 N^{2} + 
      2694531 kN^{2} + 3058174 k^{2} N^{2} 
\nonu\\ & + & 1475205 k^{3} N^{2} + 
      297482 k^{4} N^{2} + 18072 k^{5} N^{2} + 756377 N^{3} + 
      2150502 kN^{3} + 1947571 k^{2} N^{3} 
\nonu\\ & + & 670996 k^{3} N^{3} + 
      74508 k^{4} N^{3} + 386762 N^{4} + 909111 kN^{4} + 
      592308 k^{2} N^{4} + 109824 k^{3} N^{4} 
\nonu\\ & + &  102199 N^{5} + 
      188192 kN^{5} + 67656 k^{2} N^{5} + 10882 N^{6} + 
      14268 kN^{6}), \nonu\\
c_ {15} & = & \frac {1} {12 (2 + N)^{2} (2 + k + N)^{4} (4 + 5 k + 
      5 N + 6 kN)} i (-9792 - 7312 k + 53156 k^{2} + 108233 k^{3}
\nonu\\ & + &
    66516 k^{4} + 12948 k^{5} + 31216 N + 247208 kN + 
    522371 k^{2} N + 491550 k^{3} N + 195052 k^{4} N 
\nonu\\ & + &
    25608 k^{5} N + 210884 N^{2} + 891915 kN^{2} + 
    1235938 k^{2} N^{2} + 764013 k^{3} N^{2} + 193508 k^{4} N^{2} 
\nonu\\ & + & 
    13536 k^{5} N^{2} + 341585 N^{3} + 1105722 kN^{3} + 
    1123111 k^{2} N^{3} + 452956 k^{3} N^{3} + 57912 k^{4} N^{3} 
\nonu\\ & + &  
    241202 N^{4} + 612759 kN^{4} + 429480 k^{2} N^{4} + 
    88764 k^{3} N^{4} + 78565 N^{5} + 151052 kN^{5} 
\nonu\\ & + &  
    57180 k^{2} N^{5} + 9652 N^{6} + 12792 kN^{6}), \nonu\\
c_ {16} & = & \frac {(176 + 335 k + 114 k^{2} + 233 N + 215 kN + 
    63 N^{2})} {(2 + N) (2 + k + N)^{3}}, \nonu\\
c_ {17} & = & \frac {1} {(2 + k + N)}, \qquad
c_ {18}  =  \frac {i} {(2 + k + N)}, \nonu\\
c_ {19} & = & - \frac {i (60 + 77 k + 22 k^{2} + 121 N + 115 kN + 
       20 k^{2} N + 79 N^{2} + 42 kN^{2} + 16 N^{3})} {(2 + 
       N) (2 + k + N)^{3}}, \nonu\\
c_ {20} & = & \frac {(20 + 21 k + 6 k^{2} + 25 N + 13 kN + 
     7 N^{2})} {(2 + N) (2 + k + N)^{3}},\nonu\\
c_ {21} & = & \frac {1} {24 (2 + N)^{2} (2 + k + N)^{4} (4 + 5 k + 
       5 N + 6 kN)} i (72960 + 354672 k + 589072 k^{2} 
\nonu\\ & + &
     444121 k^{3} + 155488 k^{4} + 20508 k^{5} + 266256 N + 
     1235648 kN + 1876577 k^{2} N 
\nonu\\ & + &  1246902 k^{3} N 
+ 367934 k^{4} N + 38460 k^{5} N + 417264 N^{2} + 1777831 kN^{2} + 
     2360298 k^{2} N^{2} 
\nonu\\ & + &  1292541 k^{3} N^{2} 
+285234 k^{4} N^{2} + 
     18072 k^{5} N^{2} + 355775 N^{3} + 1327946 kN^{3} + 
     1445137 k^{2} N^{3} 
\nonu\\ & + &  580376 k^{3} N^{3}
+ 71676 k^{4} N^{3} + 
     171126 N^{4} + 528219 kN^{4} + 421832 k^{2} N^{4} + 
     93768 k^{3} N^{4} 
\nonu\\ & + &  43197 N^{5}
+ 101620 kN^{5} + 
     45504 k^{2} N^{5} + 4414 N^{6} + 6852 kN^{6}), \nonu\\
 c_ {22} & = & \frac {1} {24 (2 + N)^{2} (2 + k + N)^{4} (4 + 5 k + 
       5 N + 6 kN)} i (7680 + 114416 k + 256864 k^{2}
\nonu\\ & + & 
     229225 k^{3} + 90020 k^{4} + 12948 k^{5} + 22288 N + 410656 kN + 
     853673 k^{2} N + 673694 k^{3} N 
\nonu\\ & + &  223476 k^{4} N + 
     25608 k^{5} N + 54016 N^{2} + 677607 kN^{2} + 
     1187970 k^{2} N^{2} + 764597 k^{3} N^{2} 
\nonu\\ & + &  189940 k^{4} N^{2} + 
     13536 k^{5} N^{2} + 76935 N^{3} + 594018 kN^{3} + 
     810853 k^{2} N^{3} + 376276 k^{3} N^{3} 
\nonu\\ & + & 51960 k^{4} N^{3} + 
     53978 N^{4} + 271051 kN^{4} + 258968 k^{2} N^{4} + 
     65484 k^{3} N^{4} + 17555 N^{5} 
\nonu\\ & + &  57548 kN^{5} + 
     29652 k^{2} N^{5} + 2124 N^{6} + 4104 kN^{6}), \nonu\\
 c_ {23} & = & \frac {1}{(3 (2 + N) (2 + k + N)^{2} (4 + 5 k + 5 N + 6 kN)}
 (-2496 - 7644 k - 6655 k^{2} - 1634 k^{3} 
\nonu\\ & - &  
      6372 N - 17380 kN - 12291 k^{2} N - 2140 k^{3} N - 4429 N^{2} - 
      10356 kN^{2} - 4466 k^{2} N^{2} 
\nonu\\ & - &  583 N^{3} - 1442 kN^{3} + 
      128 N^{4}),\nonu\\
 c_ {24} & = & \frac {1} {36 (2 + N)^{2} (2 + k + N)^{4} (4 + 5 k + 
      5 N + 6 kN)} (93888 + 148656 k - 95876 k^{2} 
\nonu\\  & - & 326745 k^{3} - 
      210876 k^{4} - 41668 k^{5} + 376944 N + 727672 kN + 
      276145 k^{2} N - 279246 k^{3} N 
\nonu\\ & + &  210100 k^{4} N - 
      32392 k^{5} N + 728572 N^{2} + 1930645 kN^{2} + 
      2148328 k^{2} N^{2} + 1223477 k^{3} N^{2} 
\nonu\\ & + & 
      374414 k^{4} N^{2} + 51860 k^{5} N^{2} + 1065547 N^{3} + 
      3440458 kN^{3} + 4303151 k^{2} N^{3} + 2449004 k^{3} N^{3}
\nonu\\ & + &  
      602042 k^{4} N^{3} + 46680 k^{5} N^{3} + 1109392 N^{4} + 
      3439007 kN^{4} + 3627866 k^{2} N^{4} + 1510130 k^{3} N^{4} 
\nonu\\ & + &  
      201564 k^{4} N^{4} + 677381 N^{5} + 1756088 kN^{5} + 
      1330658 k^{2} N^{5} + 296736 k^{3} N^{5} + 209564 N^{6} 
\nonu\\ & + & 
      413798 kN^{6} + 170820 k^{2} N^{6} + 25232 N^{7} + 
      33504 kN^{7}), \nonu\\
c_ {26} & = & - \frac {4 (k - N) (35 + 19 k + 19 N + 
       3 kN)} {21 (2 + k + N) (4 + 5 k + 5 N + 6 kN)}, \nonu\\
c_ {27} & = & \frac {48 (k - N)} {7 (4 + 5 k + 5 N + 6 kN)}, \qquad
c_ {28}  =  - \frac {120 (k - N)} {7 (4 + 5 k + 5 N + 6 kN)}, \nonu\\
c_ {29} & = & - \frac {1}{21 (2 + N) (2 + k + N)^{3} (4 + 5 k + 5 N + 6 kN)}
8 (960 + 60 k - 2395 k^{2} - 1933 k^{3} 
\nonu\\  & - & 
       418 k^{4} + 4740 N + 3860 kN - 2674 k^{2} N - 2606 k^{3} N - 
       446 k^{4} N + 7895 N^{2} + 7403 kN^{2} 
\nonu\\ & - &  95 k^{2} N^{2} - 
       897 k^{3} N^{2} - 60 k^{4} N^{2} + 6084 N^{3} + 4988 kN^{3} + 
       544 k^{2} N^{3} - 66 k^{3} N^{3} + 2211 N^{4}
\nonu\\ & + &  1215 kN^{4} + 
       78 k^{2} N^{4} + 304 N^{5} + 48 kN^{5}) , \nonu\\
c_ {30} & = & - \frac{1}{21 (2 + N) (2 + k + N)^{3} (4 + 5 k + 5 N + 6 kN)}
4 (-576 - 3396 k - 4699 k^{2} - 2413 k^{3} 
\nonu\\  & - & 
       418 k^{4} + 516 N - 4780 kN - 7570 k^{2} N - 3422 k^{3} N - 
       446 k^{4} N + 3863 N^{2} + 203 kN^{2} 
\nonu\\ & - &  3119 k^{2} N^{2} - 
       1185 k^{3} N^{2} - 60 k^{4} N^{2} + 4452 N^{3} + 2540 kN^{3} - 
       32 k^{2} N^{3} - 66 k^{3} N^{3} 
\nonu\\ & + &  1971 N^{4} + 927 kN^{4} + 
       78 k^{2} N^{4} + 304 N^{5} + 48 kN^{5}),\nonu\\
 c_ {31} & = & - \frac{1} {21 (2 + N) (2 + k + N)^{3} (4 + 5 k + 5 N + 6 kN)} 4 (-192 - 2532 k - 4123 k^{2} - 2293 k^{3} 
\nonu\\  & - & 
        418 k^{4} + 1572 N - 2620 kN - 6346 k^{2} N - 3218 k^{3} N - 
        446 k^{4} N + 4871 N^{2} + 2003 kN^{2}
\nonu\\ & - &  2363 k^{2} N^{2} - 
        1113 k^{3} N^{2} - 60 k^{4} N^{2} + 4860 N^{3} + 
        3152 kN^{3} + 112 k^{2} N^{3} - 66 k^{3} N^{3} 
\nonu\\ & + &  2031 N^{4} + 
        999 kN^{4} + 78 k^{2} N^{4} + 304 N^{5} + 48 kN^{5}),  \nonu\\
 c_ {32} & = & \frac {8 (k - N)} {(4 + 5 k + 5 N + 6 k N)},  \nonu\\
 c_ {33} & = & - \frac {8 (k - N) (60 + 77 k + 22 k^{2} + 121 N + 
        115 kN + 20 k^{2} N + 79 N^{2} + 42 kN^{2} + 16 N^{3})} {(2 + 
        N) (2 + k + N)^{2} (4 + 5 k + 5 N + 6 kN)},  \nonu\\
 c_ {34} & = & - \frac {1}{7 (2 + N) (2 + k + N)^{2} (4 + 5 k + 5 N + 6 kN)}
 8 (-256 - 308 k - 131 k^{2} - 10 k^{3} 
\nonu\\ & - &  
        716 N - 804 kN - 391 k^{2} N - 68 k^{3} N - 537 N^{2} - 
        416 kN^{2} - 126 k^{2} N^{2} - 79 N^{3} 
\nonu\\ & - &  14 kN^{3} + 
        16 N^{4}), \nonu\\
c_ {35} & = & - \frac {32 (12 + 15 k + 4 k^{2} + 15 N + 10 kN + 
        4 N^{2})} {7 (2 + k + N) (4 + 5 k + 5 N + 6 kN)},  \nonu\\
 c_ {36} & = & \frac{1}{21 (2 + N) (2 + k + N)^{3} (4 + 5 k + 5 N + 6 kN)}
 4 (-704 - 884 k - 275 k^{2} + 147 k^{3} + 
       78 k^{4} 
\nonu\\ & - &  2636 N - 3468 kN - 1902 k^{2} N - 478 k^{3} N - 
       54 k^{4} N - 2737 N^{2} - 2529 kN^{2} - 775 k^{2} N^{2} 
\nonu\\ & - & 
       41 k^{3} N^{2} + 12 k^{4} N^{2} - 692 N^{3} + 144 kN^{3} + 
       396 k^{2} N^{3} + 90 k^{3} N^{3} + 247 N^{4} + 419 kN^{4} 
\nonu\\ & + &  
       138 k^{2} N^{4} + 96 N^{5} + 48 kN^{5}),  \nonu\\
 c_ {37} & = & \frac {8 i (k - N) (20 + 21 k + 6 k^{2} + 25 N + 
       13 kN + 7 N^{2})} {(2 + N) (2 + k + N)^{2} (4 + 5 k + 5 N + 
       6 kN)},  \nonu\\
c_ {38} & = & - \frac {1} {21 (2 +
         N) (2 + k + N)^{3} (4 + 5 k + 5 N + 6 kN)} 4 i (1536 + 3952 k + 
        3044 k^{2} + 939 k^{3} 
\nonu\\ & + &  114 k^{4} + 2960 N + 7136 kN + 
        4197 k^{2} N + 736 k^{3} N + 18 k^{4} N + 1436 N^{2} + 
        3699 kN^{2} 
\nonu\\ & + &  1399 k^{2} N^{2} + 57 k^{3} N^{2} - 51 N^{3} + 
        542 kN^{3} + 54 k^{2} N^{3} - 103 N^{4} + 15 kN^{4}),  \nonu\\
 c_ {39} & = & \frac{1}{21 (2 + N) (2 + k + N)^{3} (4 + 
       5 k + 5 N + 6 kN)}8 i (2688 + 4484 k + 1405 k^{2} - 309 k^{3}
\nonu\\ & - &  
       114 k^{4} + 7228 N + 11524 kN + 4116 k^{2} N + 200 k^{3} N -
       18 k^{4} N + 6367 N^{2} + 8841 kN^{2} 
\nonu\\ & + &  2507 k^{2} N^{2} + 
       159 k^{3} N^{2} + 2280 N^{3} + 2500 kN^{3} + 378 k^{2} N^{3} + 
       283 N^{4} + 201 kN^{4}),  \nonu\\
 c_ {40} & = & - \frac {1}{21 (2 + N) (2 + k + N)^{3} (4 + 5 k + 
        5 N + 6 kN)} 4 i (1152 + 2536 k + 1538 k^{2} + 519 k^{3} 
\nonu\\ & + &  
        114 k^{4} + 2072 N + 4424 kN + 1851 k^{2} N + 292 k^{3} N + 
        18 k^{4} N + 950 N^{2} + 2643 kN^{2} 
\nonu\\ & + &  895 k^{2} N^{2} + 
        129 k^{3} N^{2} + 27 N^{3} + 758 kN^{3} + 198 k^{2} N^{3} - 
        43 N^{4} + 87 kN^{4}),  \nonu\\
 c_ {41} & = & \frac {12 (32 + 3 k - 6 k^{2} + 29 N - kN + 
       7 N^{2})} {7 (2 + N) (2 + k + N)^{3}},  \nonu\\
 c_ {42} & = & - \frac {8 (32 + 3 k - 6 k^{2} + 29 N - kN + 
        7 N^{2})} {7 (2 + N) (2 + k + N)^{3}},  \nonu\\
 c_ {43} & = & - \frac {8 (32 + 3 k - 6 k^{2} + 29 N - kN + 
        7 N^{2})} {7 (2 + N) (2 + k + N)^{3}},  \nonu\\
 c_ {44} & = & \frac {8 (32 + 3 k - 6 k^{2} + 29 N - kN + 
       7 N^{2})} {7 (2 + N) (2 + k + N)^{3}},  \nonu\\
 c_ {45} & = & - \frac {8 (80 + 53 k + 6 k^{2} + 83 N + 29 kN + 
        21 N^{2})} {7 (2 + N) (2 + k + N)^{3}},  \nonu\\
 c_ {46} & = & \frac {4 (64 + 97 k + 30 k^{2} + 79 N + 61 kN + 
       21 N^{2})} {7 (2 + N) (2 + k + N)^{3}},  \nonu\\
 c_ {47} & = & \frac {32} {7 (2 + k + N)},  \qquad
 c_ {48}  =  - \frac {24} {7 (2 + k + N)},  \qquad
 c_ {49}  =  - \frac {10 i} {7 (2 + k + N)},  \nonu\\
 c_ {50} & = & \frac {4 i} {7 (2 + k + N)},  \nonu\\
 c_ {51} & = & - \frac {4 i (60 + 77 k + 22 k^{2} + 121 N + 115 kN + 
        20 k^{2} N + 79 N^{2} + 42 kN^{2} + 
        16 N^{3})} {7 (2 + N) (2 + k + N)^{3}},  \nonu\\
 c_ {52} & = & \frac {4 i (k - N) (32 + 29 k + 6 k^{2} + 35 N + 
       17 kN + 9 N^{2})} {(2 + N) (2 + k + N)^{2} (4 + 5 k + 5 N + 
       6 kN)}, \nonu\\ 
c_ {53} & = & \frac {4 (20 + 21 k + 6 k^{2} + 25 N + 
       13 kN + 7 N^{2})} {7 (2 + N) (2 + k + N)^{3}},  \nonu\\
 c_ {54} & = & \frac {1}{21 (2 + N) (2 + k + N)^{3} (4 + 5 k + 5 N + 
        6 kN)}2 i (-1200 - 3736 k - 3330 k^{2} - 1151 k^{3} 
\nonu\\ & - &
       114 k^{4} - 1544 N - 5016 kN - 3237 k^{2} N - 668 k^{3} N - 
       18 k^{4} N + 390 N^{2} - 663 kN^{2}
\nonu\\ & + & 321 k^{2} N^{2} + 
       219 k^{3} N^{2} + 1115 N^{3} + 1016 kN^{3} + 528 k^{2} N^{3} + 
       321 N^{4} + 
       207 kN^{4}),  \nonu\\
c_ {55} & = & - \frac{1}{21 (2 + N) (2 + k + N)^{3} (4 + 5 k + 5 N + 
       6 kN)} 2 i (-1680 - 1024 k + 
       1527 k^{2} + 1192 k^{3} 
\nonu\\ & + & 228 k^{4} - 5840 N - 4842 kN + 
       1431 k^{2} N + 1024 k^{3} N + 36 k^{4} N - 5289 N^{2} - 
       2214 kN^{2} 
\nonu\\ & + & 2682 k^{2} N^{2} + 786 k^{3} N^{2} - 
       1405 N^{3} + 872 kN^{3} + 1212 k^{2} N^{3} - 42 N^{4} + 
       306 kN^{4}),  \nonu\\
c_ {56} & = & - \frac {4 (20 + 21 k + 6 k^{2} + 25 N + 
       13 kN + 7 N^{2})} {(2 + N) (2 + k + N)^{3}},  \nonu\\
c_ {57} & = & - \frac{1}{21 (2 + N) (2 + k + N)^{3} (4 + 5 k + 
       5 N + 6 kN)} 2 i (480 + 2128 k + 2229 k^{2} + 881 k^{3}
\nonu\\ & + &
       114 k^{4} + 176 N + 2574 kN + 2184 k^{2} N + 524 k^{3} N + 
       18 k^{4} N - 699 N^{2} + 777 kN^{2} 
\nonu\\ & + & 441 k^{2} N^{2} - 
       3 k^{3} N^{2} - 626 N^{3} - 2 kN^{3} - 24 k^{2} N^{3} - 
       141 N^{4} + 9 kN^{4}).
       \nonu
\eea}

\subsection{The OPE between the higher spin $3$ current and
 itself}

The OPE between the higher spin $3$ current and
 itself is given by
\bea
&&{\Phi}_{2}^{(1)}(z)\,{\Phi}_{2}^{(1)}(w)  =  
\frac{1}{(z-w)^{6}}\, c_{1}+
\frac{1}{(z-w)^{4}}\,\Bigg[\, c_{2}\,{\Phi}_{0}^{(2)}+c_{3}\,{\Phi}_{0}^{(1)}{\Phi}_{0}^{(1)}
+c_{4}\,{L}+c_{5}\,{T}^{\mu' \nu'}{T}^{\mu' \nu'}
\nonu\\ && +
\varepsilon^{\mu' \nu' \rho' \si'}\, c_{6}\,{T}^{\mu' \nu'}{T}^{\rho' \si'}\,\Bigg](w)+\frac{1}{(z-w)^{3}}\,\frac{1}{2}\,\partial\Big(\mbox{pole-4}\Big)
\nonu\\  && +
\frac{1}{(z-w)^{2}}\,\Bigg[\, c_{7}\,{\Phi}_{2}^{(2)}
+c_{8}\,\partial{\Phi}_{0}^{(2)}+c_{9}\,{\Phi}_{0}^{(1)}{\Phi}_{2}^{(1)}
+c_{10}\,{\Phi}_{\frac{1}{2}}^{(1),\mu'}{\Phi}_{\frac{3}{2}}^{(1),\mu'}
+c_{12}\,\partial \Phi_{\frac{1}{2}}^{(1),\mu'}{\Phi}_{\frac{1}{2}}^{(1),\mu'}
\nonu\\ && +
c_{13}\,\partial^{2}{\Phi}_{0}^{(1)}{\Phi}_{0}^{(1)}
+c_{14}\,\partial{\Phi}_{0}^{(1)}\partial{\Phi}_{0}^{(1)}
+c_{11}\, \varepsilon^{\mu' \nu' \rho' \si'}\,\Phi_{1}^{(1),\mu' \nu'} \Phi_{1}^{(1),\rho' \si'}
+c_{15}\,{L}{\Phi}_{0}^{(2)}
\nonu\\ && + 
c_{16}\,{L}{\Phi}_{0}^{(1)}{\Phi}_{0}^{(1)}
+c_{17}\,\varepsilon^{\mu' \nu' \rho' \si'}\,
{T}^{\mu' \nu'}{\Phi}_{\frac{1}{2}}^{(1),\rho'}{\Phi}_{\frac{1}{2}}^{(1),\si'}
+c_{23}\,\partial{G}^{\mu'}{G}^{\mu'}
+c_{18}\,(L)^{2}
+c_{21}\,\partial^{2}{L}
\nonu\\ && +
c_{19}\,{L}{T}^{\mu' \nu'}{T}^{\mu' \nu'}
+c_{22}\,{G}^{\mu'}{G}^{\nu'}{T}^{\mu' \nu'}
+c_{25}\,\partial^{2}{T}^{\mu' \nu'}{T}^{\mu' \nu'}
+c_{26}\,\partial{T}^{\mu' \nu'}\partial{T}^{\mu' \nu'}
+c_{27}\,\partial^{2}{T}^{\mu' \nu'}\widetilde{T}^{\mu' \nu'}
\nonu\\ && + c_{28}\,\partial{T}^{\mu' \nu'}\partial\widetilde{T}^{\mu' \nu'}
+c_{24}\,{T}^{\mu' \nu'}{T}^{\rho' \si'}\widetilde{T}^{\mu' \nu'}\widetilde{T}^{\rho' \si'}
+c_{20}\,\varepsilon^{\mu' \nu' \rho' \si'}\,{L}{T}^{\mu' \nu'}{T}^{\rho' \si'}\,\Bigg](w)
\nonu\\ && +
\frac{1}{(z-w)}\,\Bigg[\,\frac{1}{2}\,\partial\Big(\mbox{pole-2}\Big)
+c_{29}\,\partial^{3}{\Phi}_{0}^{(2)}
+c_{30}\,\partial^{3}({\Phi}_{0}^{(1)}{\Phi}_{0}^{(1)})
+c_{31}\,\partial^{3}{L}
\nonu\\ && +
c_{32}\,\partial^{3}({T}^{\mu' \nu'}{T}^{\mu' \nu'})
+c_{33}\,\partial^{3}({T}^{\mu' \nu'}\widetilde{T}^{\mu' \nu'})\,\Bigg](w)+\cdots,
\label{threethree}
\eea
where
{\small
\bea
c_ {1} & = & \frac {64 k\,N}{3 (k+N+2)^3 (6 k\,N+5 k+5 N+4)} (60 k^3 N+38 k^3+150 k^2\, N^2+331 k^2\, N   \nonu\\
 & + & 136 k^2+60 k\, N^3+331 k\, N^2+406 k\, N+111 k+38 N^3+136 N^2+111 N+12), \nonu\\
c_ {2} & = & -\frac{8 (k-N) (6 k\, N-11 k-11 N-28)}{(k+N+2) (6 k\, N+5 k+5 N+4)}, \nonu\\
c_ {3} & = & -\frac{4}{(N+2) (k+N+2)^3 (6 k\, N+5 k+5 N+4)^2} (-720 k^5\, N^3+120 k^5\, N^2+2308 k^5\, N  \nonu\\
 & + & 1274 k^5-792 k^4\, N^4-1404 k^4\, N^3+10042 k^4\, N^2+20495 k^4\, N+8883 k^4+936 k^3\, N^5   \nonu\\
 & + & 1632 k^3\, N^4+9742 k^3\, N^3+44312 k^3\, N^2+59682 k^3\, N+22524 k^3+576 k^2\, N^6+1764 k^2\, N^5 \nonu\\
 & - & 2138 k^2\, N^4+5294 k^2\, N^3+50124 k^2\, N^2+68636 k^2\, N+25264 k^2-384 k N^6-6434 k\, N^5 \nonu\\
 & - & 25666 k\, N^4-33090 k\, N^3+1028 k\, N^2+27520 k\, N+12096 k-848 N^6-6925 N^5 \nonu\\
 & - & 21087 N^4-29020 N^3-16304 N^2-576 N+1536), \nonu\\
c_ {4} & = & \frac {16}{(k+N+2)^3 (6 k\, N+5 k+5 N+4)^2} (360 k^5\, N^2+856 k^5\, N+410 k^5+1548 k^4\, N^3  \nonu\\
 & + & 4968 k^4\, N^2+6139 k^4\, N+2343 k^4+1836 k^3\, N^4+9268 k^3\, N^3+16523 k^3\, N^2+14345 k^3\, N \nonu\\
 & + & 4622 k^3+576 k^2\, N^5+5292 k^2\, N^4+14493 k^2\, N^3+17651 k^2\, N^2+11702 k^2\, N+3328 k^2 \nonu\\
 & + & 640 k\, N^5+3347 k\, N^4+5207 k\, N^3+2594 k\, N^2+368 k\, N+96 k+80 N^5-26 N^4 \nonu\\
 & - & 1382 N^3-2928 N^2-2144 N-512), \nonu\\
c_ {5} & = & -\frac {5}{(N+2) (k+N+2)^3 (6 k\, N+5 k+5 N+4)} (-36 k^4 N+66 k^4+30 k^3 N^2+221 k^3 N \nonu\\
& + & 519 k^3+180 k^2 N^3+503 k^2 N^2+895 k^2 N+984 k^2+114 k N^4+379 k N^3+161 k N^2 \nonu\\
& - & 48 k N+304 k-17 N^4-263 N^3-872 N^2-944 N-256), \nonu\\
c_ {6} & = & \frac{(k-N)}{(N+2) (k+N+2)^3 (6 k\, N+5 k+5 N+4)} (36 k^3\, N-66 k^3+126 k^2\, N^2+117 k^2\, N \nonu\\
& - & 191 k^2+78 k\, N^3+316 k\, N^2+268 k\, N-28 k+49 N^3+227 N^2+348 N+192), \nonu\\
c_ {7} & = & 4, \nonu\\
c_ {8} & = & \frac{2}{(N+2) (k+N+2)^2 (30 k\, N+59 k+59 N+88)} (582 k^3\, N^2+1855 k^3\, N+1400 k^3 \nonu\\
& + & 1260 k^2\, N^3+7123 k^2\, N^2+12405 k^2\, N+6923 k^2+498 k\, N^4+5813 k\, N^3+18742 k\, N^2 \nonu\\
& + & 23842 k\, N+10796 k+893 N^4+5745 N^3+13407 N^2+13708 N+5280), \nonu\\
c_ {9} & = & -\frac{4}{(N+2) (k+N+2)^2 (6 k\, N+5 k+5 N+4)} (120 k^3\, N^2+208 k^3\, N+62 k^3 \nonu\\
& + & 252 k^2\, N^3+976 k^2\, N^2+1083 k^2\, N+281 k^2+96 k\, N^4+788 k\, N^3+1912 k\, N^2 \nonu\\
& + & 1714 k\, N+416 k+104 N^4+603 N^3+1209 N^2+976 N+240), \nonu\\
c_ {10} & = & -\frac{4 (20 k^2\, N+22 k^2+42 k\, N^2+115 k\, N+77 k+16 N^3+79 N^2+121 N+60)}{(N+2) (k+N+2)^2}, \nonu\\
c_ {11} & = & -\frac{60 + 77 k + 22 k^2 + 121 N + 115 k N + 20 k^2 N + 79 N^2 + 42 k N^2 + 
 16 N^3}{2 (2 + N) (2 + k + N)^2}, \nonu\\
c_ {12} & = & -\frac{4}{3(N+2) (k+N+2)^3} (-288 - 348 k - 149 k^2 - 22 k^3 - 372 N - 332 k N - 129 k^2 N 
\nonu\\ &- & 20 k^3 N - 95 N^2 - 36 k N^2 - 22 k^2 N^2 + 43 N^3 + 26 k N^3 + 
 16 N^4), \nonu\\
c_ {13} & = & -\frac{2}{3 (N+2)^2 (k+N+2)^4 (6 k\, N+5 k+5 N+4)^2} (42480 k^6\, N^4+175704 k^6\, N^3 \nonu\\
& + & 281812 k^6\, N^2+203650 k^6\, N+53632 k^6+179928 k^5\, N^5+1096908 k^5\, N^4 \nonu\\
& + & 2657606 k^5\, N^3+3233965 k^5\, N^2+1963717 k^5\, N+461858 k^5+259776 k^4\, N^6 \nonu\\
& + & 2217300 k^4\, N^5+7618352 k^4\, N^4+13733635 k^4\, N^3+13841716 k^4\, N^2+7373186 k^4\, N \nonu\\
& + & 1583127 k^4+146664 k^3\, N^7+1779396 k^3\, N^6+8543372 k^3\, N^5+21738802 k^3\, N^4 \nonu\\
& + & 32355238 k^3\, N^3+28512928 k^3\, N^2+13766828 k^3\, N+2752976 k^3+28224 k^2\, N^8 \nonu\\
& + & 538980 k^2\, N^7+3758780 k^2\, N^6+13500040 k^2\, N^5+28510252 k^2\, N^4+37249800 k^2\, N^3 \nonu\\
& + & 29848874 k^2\, N^2+13427888 k^2\, N+2549728 k^2+38400 k\, N^8+513998 k\, N^7 \nonu\\
& + & 2851009 k\, N^6+8723101 k\, N^5+16402446 k\, N^4+19696172 k\, N^3+14853680 k\, N^2 \nonu\\
& + & 6399424 k\, N+1178880 k+11248 N^8+127251 N^7+618984 N^6+1715014 N^5 \nonu\\
& + & 3009687 N^4+3468624 N^3+2568160 N^2+1102080 N+203520), \nonu\\
c_ {14} & = & -\frac{2}{(N+2)^2 (k+N+2)^4} (380 k^4\, N^2+818 k^4\, N+440 k^4+1638 k^3\, N^3+6197 k^3\, N^2 \nonu\\
& + & 7661 k^3\, N+3098 k^3+2408 k^2\, N^4+13395 k^2\, N^3+27284 k^2\, N^2+24174 k^2\, N \nonu\\
& + & 7853 k^2+1386 k\, N^5+10531 k\, N^4+30993 k\, N^3+44382 k\, N^2+30986 k\, N+8424 k \nonu\\
& + & 272 N^6+2659 N^5+10494 N^4+21418 N^3+23885 N^2+13800 N+3216), \nonu\\
c_ {15} & = & -\frac{128 (k-N) (12 k\, N-7 k-7 N-26)}{(6 k\, N+5 k+5 N+4) (30 k\, N+59 k+59 N+88)}, \nonu\\
c_ {16} & = & \frac{96(k-N)}{(N+2) (k+N+2)^2 (6 k\, N+5 k+5 N+4)^2} (40 k^3\, N^2+72 k^3\, N+26 k^3+84 k^2\, N^3 \nonu\\
& + & 328 k^2\, N^2+377 k^2\, N+115 k^2+32 k\, N^4+260 k\, N^3+632 k\, N^2+582 k\, N+160 k \nonu\\
& + & 32 N^4+185 N^3+371 N^2+304 N+80), \nonu\\
c_ {17} & = & -\frac{4 i (60 + 77 k + 22 k^2 + 121 N + 115 k N + 20 k^2 N + 79 N^2 + 
   42 k N^2 + 16 N^3)}{(2 + N) (2 + k + N)^3}), \nonu\\
c_ {18} & = & -\frac{16}{(2 + N) (2 + k + N)^2(4 + 5 k + 5 N + 6 k N)^2} (-512 - 2800 k - 4664 k^2 - 2943 k^3 
\nonu\\ &-& 610 k^4 - 1296 N - 8176 k N - 
 13543 k^2 N - 7915 k^3 N - 1416 k^4 N - 600 N^2 - 7441 k N^2 
\nonu\\ &-&
 12651 k^2 N^2 - 6352 k^3 N^2 - 776 k^4 N^2 + 759 N^3 - 1701 k N^3 - 
 4076 k^2 N^3 - 1468 k^3 N^3
\nonu\\ &-& 733 N^4 + 548 k N^4 - 220 k^2 N^4 + 
 160 N^5 + 160 k N^5), \nonu\\
c_ {19} & = & \frac{8}{(N+2) (k+N+2)^3 (6 k\, N+5 k+5 N+4)} (-12 k^4-134 k^3\, N-156 k^3-212 k^2\, N^2 \nonu\\
& - & 673 k^2\, N-471 k^2-40 k\, N^3-410 k\, N^2-828 k\, N-460 k+14 N^4+23 N^3-109 N^2 \nonu\\
& - & 244 N-128), \nonu\\
c_ {20} & = & \frac{-8 (k - N) (13 k + 6 k^2 + 3 N + 9 k N + N^2)}{(2 + N) (2 + k + N)^2 (4 + 5 k + 5 N + 6 k N)}, \nonu\\
c_ {21} & = & \frac{4}{3 (2 + N) (2 + k + N)^4} (-1656 - 3366 k - 1001 k^2 + 598 k^3 + 232 k^4 - 954 N 
\nonu\\ & + & 2251 k N + 
   5980 k^2 N + 2885 k^3 N + 314 k^4 N + 2512 N^2 + 8116 k N^2 + 
   6203 k^2 N^2 
\nonu\\ & + &1137 k^3 N^2 + 2640 N^3 + 4228 k N^3 + 
   1352 k^2 N^3 + 834 N^4 + 591 k N^4 + 80 N^5), \nonu\\
c_ {22} & = & \frac{8 i (32 + 55 k + 18 k^2 + 41 N + 35 k N + 11 N^2)}{(2 + N) (2 + k + N)^3} , \nonu\\
c_ {23} & = & \frac{4}{(2 + N) (2 + k + N)^3} (224 + 440 k + 191 k^2 + 18 k^3 + 312 N + 344 k N + 49 k^2 N
\nonu\\ & + &
 105 N^2 + 38 k N^2 + 7 N^3), \nonu\\
 c_ {24} & = & \frac{(32 + 55 k + 18 k^2 + 41 N + 35 k N + 11 N^2)}{
   (2 + N) (2 + k + N)^4}, \nonu\\
c_ {25} & = & -\frac{1}{3 (2 + N)^2 (2 + k + N)^4}(-5616 - 9096 k + 47 k^2 + 1954 k^3 + 288 k^4 - 6072 N 
\nonu\\ & + & 554 k N + 
  10564 k^2 N + 3793 k^3 N + 306 k^4 N + 3083 N^2 + 13888 k N^2 + 
  10078 k^2 N^2 
\nonu\\ & + & 1543 k^3 N^2 + 5574 N^3 + 8305 k N^3 + 
  2405 k^2 N^3 + 2076 N^4 + 1307 k N^4 + 235 N^5), \nonu\\
c_ {26} & = & -\frac{1}{3 (2 + N)^2 (2 + k + N)^4}
(-2544 - 1944 k + 4511 k^2 + 3286 k^3 + 504 k^4 - 168 N 
\nonu\\ & + & 11090 k N + 
  16000 k^2 N + 5215 k^3 N + 414 k^4 N + 6899 N^2 + 19492 k N^2 + 
  12580 k^2 N^2 
\nonu\\ & + & 1921 k^3 N^2 + 6642 N^3 + 9763 k N^3 + 
  2855 k^2 N^3 + 2238 N^4 + 1505 k N^4 + 253 N^5)
, \nonu\\
c_ {27} & = & \frac{1}{(2 + N)^2 (2 + k + N)^4}
(-236 k - 709 k^2 - 486 k^3 - 96 k^4 + 236 N - 336 k N - 1400 k^2 N 
\nonu\\ & - &
  763 k^3 N - 102 k^4 N + 565 N^2 - 180 k N^2 - 1004 k^2 N^2 - 
  305 k^3 N^2 + 482 N^3 - 45 k N^3 
\nonu\\ & - & 241 k^2 N^3 + 180 N^4 - k N^4 + 
  25 N^5), 
\nonu\\
c_ {28} & = & \frac{1}{(2 + N)^2 (2 + k + N)^4}
(-604 k - 1181 k^2 - 786 k^3 - 168 k^4 + 124 N - 1496 k N 
\nonu\\ & - & 2472 k^2 N - 1165 k^3 N - 138 k^4 N + 373 N^2 - 1256 k N^2 - 
 1674 k^2 N^2 - 431 k^3 N^2 
\nonu\\ & + & 410 N^3 - 399 k N^3 - 367 k^2 N^3 + 
 190 N^4 - 31 k N^4 + 31 N^5), 
\nonu\\
c_ {29} & = & \frac{(k-N) (6 k\, N-11 k-11 N-28)}{3 (k+N+2) (6 k\, N+5 k+5 N+4)}, \nonu\\
c_ {30} & = & \frac{1}{6 (N+2) (k+N+2)^3 (6 k\, N+5 k+5 N+4)^2} (-720 k^5\, N^3+120 k^5\, N^2+2308 k^5\, N \nonu\\
& + & 1274 k^5-792 k^4\, N^4-1404 k^4\, N^3+10042 k^4\, N^2+20495 k^4\, N+8883 k^4+936 k^3\, N^5 \nonu\\
& + & 1632 k^3\, N^4+9742 k^3\, N^3+44312 k^3\, N^2+59682 k^3\, N+22524 k^3+576 k^2\, N^6 \nonu\\
& + & 1764 k^2\, N^5-2138 k^2\, N^4+5294 k^2\, N^3+50124 k^2\, N^2+68636 k^2\, N+25264 k^2 \nonu\\
& - & 384 k\, N^6-6434 k\, N^5-25666 k\, N^4-33090 k\, N^3+1028 k\, N^2+27520 k\, N+12096 k \nonu\\
& - & 848 N^6-6925 N^5-21087 N^4-29020 N^3-16304 N^2-576 N+1536), \nonu\\
c_ {31} & = & -\frac{2}{3 (k+N+2)^3 (6 k\, N+5 k+5 N+4)^2} (360 k^5\, N^2+856 k^5\, N+410 k^5+1548 k^4\, N^3 \nonu\\
& + & 4968 k^4\, N^2+6139 k^4\, N+2343 k^4+1836 k^3\, N^4+9268 k^3\, N^3+16523 k^3\, N^2+14345 k^3\, N \nonu\\
& + & 4622 k^3+576 k^2\, N^5+5292 k^2\, N^4+14493 k^2\, N^3+17651 k^2\, N^2+11702 k^2\, N+3328 k^2 \nonu\\
& + & 640 k\, N^5+3347 k\, N^4+5207 k\, N^3+2594 k\, N^2+368 k\, N+96 k+80 N^5-26 N^4 \nonu\\
& - & 1382 N^3-2928 N^2-2144 N-512), \nonu\\
c_ {32} & = & \frac{1}{12 (N+2) (k+N+2)^3 (6 k\, N+5 k+5 N+4)} (-36 k^4\, N+66 k^4+30 k^3\, N^2+221 k^3\, N \nonu\\
& + & 519 k^3+180 k^2\, N^3+503 k^2\, N^2+895 k^2\, N+984 k^2+114 k\, N^4+379 k\, N^3+161 k\, N^2 \nonu\\
& - & 48 k\, N+304 k-17 N^4-263 N^3-872 N^2-944 N-256), \nonu\\
c_ {33} & = & \frac{(N-k)}{12 (N+2) (k+N+2)^3 (6 k\, N+5 k+5 N+4)} (36 k^3\, N-66 k^3+126 k^2\, N^2+117 k^2\, N \nonu\\
& - & 191 k^2+78 k\, N^3+316 k\, N^2+268 k\, N-28 k+49 N^3+227 N^2+348 N+192). \nonu
\eea}

\section{ The $SO(4)$ generators in the description of 
 $SO({\cal N}=4)$ Knizhnik Bershadsky algebra
}

The $SO(4)$ generators are given by
\bea
T^1 & = &
\left(
\begin{array}{cccc}
0 & 1 &0 &0   \\
-1 & 0 &0 &0  \\
0 & 0 &0 &0 \\
0 & 0 &0 &0\\
\end{array} \right),
\qquad
T^2=
\left(
\begin{array}{cccc}
0 & 0 &1 &0   \\
0 & 0 &0 &0  \\
-1 & 0 &0 &0 \\
0 & 0 &0 &0\\
\end{array} \right),
\nonu \\
T^3 & = &
\left(
\begin{array}{cccc}
0 & 0 &0 & 1   \\
0 & 0 &0 &0  \\
0 & 0 &0 &0 \\
-1 & 0 &0 &0\\
\end{array} \right),
\qquad
T^4=
\left(
\begin{array}{cccc}
0 & 0 &0 &0   \\
0 & 0 & 1 &0  \\
0 & -1 &0 &0 \\
0 & 0 &0 &0\\
\end{array} \right),
\nonu \\
T^5 & = &
\left(
\begin{array}{cccc}
0 & 0 &0 &0   \\
0 & 0 &0 & 1  \\
0 & 0 &0 &0 \\
0 & -1 &0 &0\\
\end{array} \right),
\qquad
T^6=
\left(
\begin{array}{cccc}
0 & 0 &0 &0   \\
0 & 0 &0 &0  \\
0 & 0 &0 & 1 \\
0 & 0 & -1 &0\\
\end{array} \right).
\label{genso4}
\eea
The structure constant is
given by $f^{abc} = -\frac{1}{2} \mbox{Tr} (T^c [T^a, T^b])$
with (\ref{genso4}).
Furthermore, one has $\mbox{Tr} (T^a T^b) =-2 \delta^{ab}$,
$f^{abc} f^{abd} = 4 \delta^{cd}$, $[T^a, T^b] =f^{abc} T^c$
and $ T^a_{\mu \nu} T^a_{\rho \sigma} = \delta_{\mu \rho} \delta_{\nu \sigma}-
\delta_{\mu \sigma} \delta_{\nu \rho}$.

\section{
The OPEs between the
  generators of  $SO({\cal N}=4)$ Knizhnik Bershadsky algebra
and the  $16$ higher spin currents
}

We present the OPEs in Appendix $(A.2)$ for $N=k$ condition 
as follows:
{\small
\bea
L(z)\,
\left(
\begin{array}{c}
 \Phi_0^{(s)} \\
\Phi_{\frac{1}{2}}^{(s),\mu}  \\
\Phi_1^{(s),\mu\nu}  \\
\Phi_{\frac{3}{2}}^{(s),\mu}  \\
\Phi_{2}^{(s)} \\
\end{array}
\right)
(w) & = &
\frac{1}{(z-w)^{2}}\, \left(
\begin{array}{c}
s  \,\Phi_0^{(s)} \\
(s+\frac{1}{2})\, \Phi_{\frac{1}{2}}^{(s),\mu}  \\
(s+1) \, \Phi_1^{(s),\mu\nu}  \\
(s+\frac{3}{2}) \, \Phi_{\frac{3}{2}}^{(s),\mu}  \\
(s+2) \, \Phi_{2}^{(s)} \\
\end{array}
\right)(w)+\frac{1}{(z-w)}\,\partial
\left(
\begin{array}{c}
 \Phi_0^{(s)} \\
\Phi_{\frac{1}{2}}^{(s),\mu}  \\
\Phi_1^{(s),\mu\nu}  \\
\Phi_{\frac{3}{2}}^{(s),\mu}  \\
\Phi_{2}^{(s)} \\
\end{array}
\right)
(w)+\cdots,
\nonu \\
\nonu \\  
G^{\mu}(z)\,\Phi_{2}^{(s)}(w) & = & 
\frac{1}{(z-w)^{2}}\,\Bigg[-(2\,s+3)\,\Phi_{\frac{3}{2}}^{(s),\mu}
-\frac{2\, i(s+1)}{(1+k)}\,\varepsilon^{\mu \nu' \rho' \si'}\, T^{\nu' \rho'}\Phi_{\frac{1}{2}}^{(s),\si'}\,\Bigg](w)
\nonu \\ 
& - &\frac{1}{(z-w)}\,\Bigg[
\,\partial\Phi_{\frac{3}{2}}^{(s),\mu}
+\frac{ i(s+1)}{(1+k)}\, \varepsilon^{\mu \nu' \rho' \si'}\,
\partial T^{\nu' \rho'}\Phi_{\frac{1}{2}}^{(s),\si'}\,\Bigg](w)+\cdots,
\nonu \\
G^{\mu}(z)\,\Phi_{\frac{3}{2}}^{(s),\nu}(w) & = & 
 \frac{1}{(z-w)^{2}}\,\Big(-2(s+1)\,\Phi_{1}^{(s),\mu \nu}
\,\Big)(w)\nonu
\nonu \\ 
& + & \frac{1}{(z-w)}\,\Bigg[-\delta^{\mu \nu}\,\Phi_{2}^{(s)}
-\partial\Phi_{1}^{(s),\mu \nu}
+\frac{2\,s\, i}{(1+k)}\,\partial\widetilde{T}^{\mu \nu}\Phi_{0}^{(s)}
\nonu \\ 
& - & \frac{2\, i}{(1+k)}\,\widetilde{T}^{\mu \nu}\partial\Phi_{0}^{(s)}
+ \frac{ i}{(1+k)}\,
\Big(T^{\mu \rho'}\Phi_{1}^{(s),\nu \rho'}-T^{\nu \rho'}\Phi_{1}^{(s),\mu \rho'}\Big)
\nonu \\ 
& - &\frac{1 }{(1+k)}\,\varepsilon^{\mu \nu \rho' \si'}\, G^{\rho'}\Phi_{\frac{1}{2}}^{(s),\si'}\,\Bigg](w)
+ \cdots,
\nonu \\
G^{\mu}(z)\,\Phi_{1}^{(s),\nu \rho}(w) & = & 
\frac{1}{(z-w)^{2}}\,\Bigg[\,
\frac{(-1 + k + 2 s + 2 k s)}{(1+k)}\,
\varepsilon^{\mu \nu \rho \si'} \,\Phi_{\frac{1}{2}}^{(s),\si'}
\,\Bigg](w)
\nonu \\
& + & 
\frac{1}{(z-w)}\,\Bigg[\,
\frac{i}{(1+k)}\,
\Big(\widetilde{T}^{\mu \rho}\Phi_{\frac{1}{2}}^{(s),\nu}-\widetilde{T}^{\mu \nu}\Phi_{\frac{1}{2}}^{(s),\rho}\Big)
+\varepsilon^{\mu \nu \rho \si'}\,\partial\Phi_{\frac{1}{2}}^{(s),\si'}
\nonu \\
& - & \delta^{\mu \nu}\,\Big(
  \Phi_{\frac{3}{2}}^{(s),\rho}+\frac{ i}{(1+k)}\,
  \widetilde{T}^{\rho \si'}\Phi_{\frac{1}{2}}^{(s),\si'}\,\Big)
\,\Bigg](w)
\nonu \\
& - & \delta^{\mu \rho}\, \sum_{n=2}^1 \, \frac{1}{(z-w)^n} \,
\left(\nu \;\leftrightarrow \;\rho\right)(w)+\cdots,
\nonu \\
G^{\mu}(z)\,\Phi_{\frac{1}{2}}^{(s),\nu}(w) & = & 
-\frac{1}{(z-w)^{2}}\,2\,s\:\delta^{\mu\nu}\,\Phi_{0}^{(s)}(w)
\nonu \\
& + & 
\frac{1}{(z-w)}\,\Big(-\delta^{\mu \nu}\,\partial\Phi_{0}^{(s)}+\widetilde{\Phi}_{1}^{(s),\mu \nu}\Big)(w)+\cdots,
\nonu \\
G^{\mu}(z) \, \Phi_{0}^{(s)}(w) & = & 
-\frac{1}{(z-w)} \, \Phi_{\frac{1}{2}}^{(s),\mu}(w)
+\cdots,
\nonu \\
\nonu \\ 
T^{\mu \nu}(z)\,\Phi_{2}^{(s)}(w) & = & 
\frac{1}{(z-w)^{2}}\,\Big(2\,i(s+1)\,\Phi_{1}^{(s),\mu \nu}\Big)(w)+\cdots,
\nonu \\
T^{\mu \nu}(z)\,\Phi_{\frac{3}{2}}^{(s),\rho}(w) & = &
\frac{1}{(z-w)^{2}}\Big(
- 2i(s+1)\, \varepsilon^{\mu \nu \rho \si'}\, \Phi_{\frac{1}{2}}^{(s),\si'}\Big)(w)
\nonu \\
&&-\frac{1}{(z-w)}\,i\,\Big(\delta^{\mu \rho}\,\Phi_{\frac{3}{2}}^{(s),\nu}-\delta^{\nu \rho}\,\Phi_{\frac{3}{2}}^{(s),\mu}\Big)
(w)+\cdots,
\nonu \\
T^{\mu \nu}(z)\,\Phi_{1}^{(s), \rho \si}(w)
& = & 
\frac{1}{(z-w)^{2}}\, 2\,i\,s\,\varepsilon^{\mu \nu \rho \si}\,\Phi_{0}^{(s)}(w)
\nonu \\
& - &
\frac{1}{(z-w)}\,i\,\Bigg[\,\delta^{\mu \rho}\,\Phi_{1}^{(s),\nu \si}-\delta^{\mu \si}\,\Phi_{1}^{(s),\nu \rho}-\delta^{\nu \rho}\,\Phi_{1}^{(s),\mu \si}+\delta^{\nu \si}\,\Phi_{1}^{(s),\mu \rho}\,\Bigg](w)+\cdots,
\nonu \\
T^{\mu \nu}(z)\;\Phi_{\frac{1}{2}}^{(s),\rho}(w)
& = & 
-\frac{1}{(z-w)}\,i\,\Big(\,\delta^{\mu \rho}\,\Phi_{\frac{1}{2}}^{(s),\nu}-\delta^{\nu \rho}\,\Phi_{\frac{1}{2}}^{(s),\mu}\,\Big)(w)+\cdots,
\nonu \\
T^{\mu\nu}(z)\;\Phi_{0}^{(s)}(w) & = & + \cdots.
\label{simpleope}
\eea}
For the $k \rightarrow \infty$ limit with fixed spin $s$,
the nonlinear terms
in (\ref{simpleope}) vanish.

\section{ The structure constants appearing in the OPEs
  between the lowest $16$ higher spin currents for $k=N$}

For convenience, we present
the previous coefficients appearing in Appendix $B$
for $k=N$.

$\bullet$
The coefficient in the OPE between the higher spin $1$ current and itself

The coefficient appearing in Appendix (\ref{oneone})
reduces to 
\bea
c_ {1} = \frac{k^2}{(k+1)}. \nonu
\eea

$\bullet$
The coefficients in the OPE between the higher spin $1$ current and
the higher spin $2$ currents

The coefficients appearing in Appendix (\ref{onetwo})
reduce to 
\bea
c_ {1} = 0, \qquad c_ {2} = \frac{2 i k}{(k+1)}.  \nonu
\eea

$\bullet$
The coefficients in the OPE between the higher spin $1$ current and
the higher spin $\frac{5}{2}$ currents

The coefficients appearing in Appendix (\ref{one5half})
reduce to 
\bea
c_ {1} & = & 0,\qquad c_ {2} = -\frac {1}{2}, \qquad c_ {3} = \frac{3 (13 k+10)}{4 (k+2)}, \qquad c_ {4} = -\frac{i (13 k+10)}{4 (k+1) (k+2)}, \nonu\\
c_ {5} & = & -\frac{2 i}{(k+2)},
\qquad c_ {6}  = -\frac{i (k-1) (13 k+10)}{128 (k+1)}. \nonu
\eea

$\bullet$
The coefficient in the OPE between the higher spin $1$ current and
the higher spin $3$ current

The coefficients appearing in Appendix (\ref{onethree})
reduce to 
\bea
c_{1} & = & 2, \qquad
c_{2} = -\frac{3 (13 k+10)}{2 (k+2)}, \qquad
c_{3} = \frac{(13 k+10)}{(k+1)}, \qquad
c_{4} = -\frac{(13 k+10)}{4 (k+1) (k+2)},\nonu\\
c_{5} & = & -\frac{k}{(k+1) (k+2)}. \nonu
\eea

$\bullet$
The coefficient in the OPE between the higher spin $\frac{3}{2}$ currents and
the higher spin $\frac{3}{2}$ currents

The coefficients appearing in Appendix (\ref{3half3half})
become
\bea
c_{1} & = & -\frac{2 k^2}{(k+1)}, \qquad
c_{2} = \frac{2 i k}{(k+1)}, \qquad
c_{3} = 0, \qquad
c_{4} = -2, \qquad
c_{5} = \frac{1}{(k+1)}, \nonu\\
c_{6} & = & i, \qquad
c_{7} = 0. \nonu
\eea

$\bullet$
The coefficients in the OPE between the higher spin $\frac{3}{2}$ currents and
the higher spin $2$ currents

The coefficients appearing in Appendix (\ref{3halftwo})
become
{\small
\bea
c_{1} & = & 0, \qquad 
c_{2} = \frac{(3 k+1)}{(k+1)},\qquad 
c_{3} = \frac{1}{2}, \qquad
c_{4} = -\frac{3 (13 k+10)}{4 (k+2)}, \qquad
c_{5} = \frac{(k-1) (13 k+10)}{4 (k+1) (k+2)}, \nonu\\
c_{6} & = & \frac{i (13 k+10)}{4 (k+1) (k+2)}, \qquad
c_{7} = \frac{i (3 k+2)}{(k+1) (k+2)}, \qquad
c_{8} = \frac{i}{(k+1)}, \qquad
c_{9} = \frac{(k-1)}{(k+1)}. \nonu
\eea}

$\bullet$
The coefficients in the OPE between the higher spin $\frac{3}{2}$ currents and
the higher spin $\frac{5}{2}$ currents

The coefficients appearing in Appendix (\ref{3half5half})
become
\bea
c_{1} & = & 0, \qquad  c_{2}=-\frac{8 i k}{(k+1)},  \qquad c_{3}=-2, \qquad
c_{4}  =  \frac{3 (13 k+10)}{2 (k+2)}, \qquad
c_{5}  =  -\frac{(13 k+10)}{(k+1)}, \nonu\\
c_{6} & = & \frac{(13 k+10)}{8 (k+1) (k+2)}, \qquad
c_{7}  =  \frac{(13 k+10)}{2 (k+1) (k+2)}, \qquad 
c_{8}=\frac{8 k}{(k+1) (k+2)}, \nonu\\
c_{9} & = & 0, \qquad 
c_{10}=0,  \qquad
c_{11}  =  \frac{(3 k+4) (13 k+10)}{4 (k+1) (k+2)},  \qquad
c_{12}  =  \frac{2}{(k+1)},  \nonu\\
c_{13} & = & -\frac{(3 k+4) (13 k+10)}{4 (k+1)^2 (k+2)},  \qquad
c_{14}  =  \frac{3 (13 k+10)}{4 (k+2)}, \qquad
c_{15}  =  -\frac{i (13 k+10)}{2 (k+1) (k+2)}, \nonu\\
c_{16} & = & -\frac{8 i}{(k+2)}, \qquad
c_{17}  =  -\frac{(13 k+10)}{4 (k+1) (k+2)},\qquad
c_{18}  =  -\frac{i (13 k+10) (k^2-k-3)}{4 (k+1)^2 (k+2)}, \nonu\\
c_{19} & = & \frac{4 i (k+3)}{(k+1) (k+2)}, \qquad
c_{20}  =  -\frac{4 i}{(k+1) (k+2)}, \qquad
c_{21}  =  \frac{4 i}{(k+1) (k+2)}, \nonu\\
c_{22} & = & \frac{4 i}{(k+1) (k+2)}, \qquad
c_{23}  =  \frac{(2 k+1) (13 k+10)}{4 (k+1)^2 (k+2)}, \qquad
c_{24}  =  \frac{(k+3) (13 k+10)}{4 (k+1)^2 (k+2)},\nonu\\
c_{25} & = & -\frac{4 (k-1)}{(k+1) (k+2)},\qquad
c_{26}  =  -\frac{12}{(k+1) (k+2)}, \qquad
c_{27}  =  \frac{1}{4}, \nonu\\
c_{28} & = & -\frac{3 (13 k+10)}{8 (k+2)}, \qquad
c_{29}  =  -\frac{1}{(k+1)}, \qquad
c_{30}  =  \frac{(3 k+2)}{(k+1) (k+2)}, \nonu\\
c_{31} & = & -\frac{2 i}{(k+1) (k+2)}, \qquad
c_{32}  =  \frac{i (13 k+10)}{8 (k+1)^2 (k+2)}. \nonu
\eea

$\bullet$
The coefficients in the OPE between the higher spin $\frac{3}{2}$ currents and
the higher spin $3$ current

The coefficients appearing in Appendix (\ref{3halfthree})
become
\bea
c_ {1} & = & 0, \qquad 
c_ {2} = \frac {5} {2}, \qquad
c_ {3} =  -\frac{15 (13 k+10)}{4 (k+2)}, \qquad
c_ {4}   =  \frac{5 i (13 k+10)}{4 (k+1) (k+2)},  \nonu\\
c_ {5}  & = & \frac{4 i (3 k+1)}{(k+1) (k+2)}, \qquad
c_ {6}  =  \frac{5 (k-1) (13 k+10)}{4 (k+1) (k+2)}, \qquad
c_ {7}  =  0, \qquad 
c_ {8} = 0,  \nonu\\
c_ {9} & = & 0,  \qquad 
c_ {10} = -\frac{12 i}{5 (k+1)},  \qquad 
c_ {11} = \frac{8 i}{5 (k+1)},   \qquad
c_ {12}  =  0. \nonu
\eea

$\bullet$
The coefficients in the OPE between the higher spin $2$ currents and
the higher spin $2$ currents

The coefficients appearing in Appendix (\ref{twotwo})
lead to
\bea
c_{1} & = & -\frac{2 k^2 (3 k+1)}{(k+1)^2}, \qquad
c_{2}=0,  \qquad
c_{3}=\frac{2 i k (3 k+1)}{(k+1)^2}, \qquad
c_{4}  =  0, \qquad
c_{5}  = -\frac{8 k}{(k+1)}, \nonu\\
c_{6} & = & -\frac{2}{(k+1)}, \qquad
c_{7}  =  \frac{2}{(k+1)}, \qquad 
c_{8}= \frac{k}{(k+1)^2}, \qquad
c_{9}  =  \frac{k}{(k+1)^2}, \qquad 
c_{10} = 0,  \nonu\\
c_{11} & = & 0,  \qquad
c_{12}  =  \frac{(k+2)}{(k+1)^2}, \qquad
c_{13}  =  0, \qquad
c_{14}  =  -\frac{(k+2)}{(k+1)^2}, \qquad
c_{15}  =  \frac{i (3 k^2+3 k+2)}{(k+1)^2},\nonu\\
c_{16} & = & 0, \qquad
c_{17}  =  2, \qquad
c_{18}  =  -\frac{3 (13 k+10)}{2 (k+2)}, \qquad
c_{19} =  \frac{(13 k+10)}{(k+1)}, \qquad
c_{20}  =  0,\nonu\\
c_{21} & = & -\frac{(13 k+10)}{4 (k+1) (k+2)}, \qquad
c_{22}  =  -\frac{k (3 k+2)}{2 (k+1)^2 (k+2)}, \qquad
c_{23}  =  \frac{2 k}{(k+1)^2}, \nonu\\
c_{24} & = & \frac{8 k}{(k+1) (k+2)},\qquad
c_{25}  =  -\frac{4 k}{(k+1)^2 (k+2)},\qquad
c_{26}  =  -\frac{4 k}{(k+1)^2 (k+2)},\nonu\\
c_{27} & = & -\frac{1}{2}, \qquad
c_{28}  =  \frac{3 (13 k+10)}{4 (k+2)}, \qquad
c_{29}  =  -\frac{1}{(k+1)}, \qquad
c_{30}  =  \frac{i (13 k+10)}{2 (k+1) (k+2)}, \nonu\\
c_{31} & = & \frac{8 i}{(k+2)}, \qquad
c_{32}  =  -\frac{3 (13 k+10)}{8 (k+2)}, \qquad
c_{33}  =  \frac{(13 k+10)}{8 (k+1) (k+2)}, \nonu\\
c_{34} & = & -\frac{(5 k+2)}{(k+1) (k+2)},  \qquad
c_{35}=\frac{2 i k}{(k+1)^2 (k+2)}, \qquad
c_{36}  =  -\frac{i (13 k+10)}{4 (k+1)^2 (k+2)}, \nonu\\
c_{37} & = & -\frac{i (3 k+2)}{2 (k+1)^2 (k+2)}, \qquad 
c_{38}= \frac{2 i k}{(k+1)^2 (k+2)}, \qquad
c_{39}  =  \frac{(k-2) (13 k+10)}{8 (k+1)^2 (k+2)}, \nonu\\
c_{40} & = & -\frac{(k-2) (13 k+10)}{8 (k+1)^2 (k+2)}, \qquad
c_{41}  =  \frac{(5 k^2+4 k-4)}{2 (k+1)^2 (k+2)}, \qquad 
c_{42}= -\frac{(k-2) (5 k+2)}{2 (k+1)^2 (k+2)},  \nonu\\
c_{43} & = & \frac{(5 k^2-4 k-4)}{2 (k+1)^2 (k+2)},  \qquad
c_{44}  =  -\frac{(5 k^2+4 k-4)}{2 (k+1)^2 (k+2)}, \nonu\\
c_{45} & = & -\frac{i (k^3+12 k^2+14 k+8)}{2 (k+1)^2 (k+2)},  \qquad
c_{46}  =  \frac{i (13 k+10) (2 k^2+k-2)}{8 (k+1)^2 (k+2)}, \nonu\\
c_{47} & = & -\frac{2}{(k+1)}. \nonu
\eea

$\bullet$
The coefficients in the OPE between the higher spin $2$ currents and
the higher spin $\frac{5}{2}$ currents

The coefficients appearing in Appendix (\ref{two5half})
lead to
{\small
\bea
c_{1}& = & \frac{4 (3 k+1)}{(k+1)}, \qquad
c_{2}=0,  \qquad
c_{3}= 0, \qquad
c_{4}  = \frac{4}{(k+1)}, \qquad
c_{5}  =  \frac{4 (k-1) k}{(k+1)^2}, \nonu\\
c_{6} & = & \frac{4 i k}{(k+1)^2}, \qquad
c_{7}  =  0, \qquad 
c_{8} = 0, \qquad
c_{9}  =  -\frac{5}{2}, \qquad 
c_{10} = \frac{15 (13 k+10)}{4 (k+2)},  \nonu\\
c_{11} & = & \frac{4 i (3 k+1)}{(k+1) (k+2)},  \qquad
c_{12}  =  \frac{5 i (13 k+10)}{4 (k+1) (k+2)}, \qquad
c_{13}  =  -\frac{5 (k-1) (13 k+10)}{4 (k+1) (k+2)}, \nonu\\
c_{14} & = & -\frac{1}{2}, \qquad
c_{15}  =  -\frac{3 (13 k+10)}{8 (k+2)}, \qquad
c_{16}  =  \frac{3 (13 k+10)}{4 (k+2)},\nonu\\
c_{17} & = & \frac{3 (2535 k^3+6419 k^2+5336 k+1436)}{80 (k+1) (k+2)^2}, \qquad
c_{18}  =  \frac{(7605 k^3+19337 k^2+16328 k+4628)}{80 (k+1) (k+2)^2}, \nonu\\
c_{19} & = & \frac{6 (3 k+2)}{(k+1) (k+2)}, \qquad
c_{20}  =  -\frac{2 i (7 k+6)}{(k+1)^2 (k+2)}, \nonu\\
c_{21} & = & -\frac{(2583 k^4+4444 k^3+1845 k^2+1188 k+740)}
{80 (k+1)^2 (k+2)^2},\qquad
c_{22}  =  \frac{4 (2 k+1)}{(k+1)^2 (k+2)},\nonu\\
c_{23} & = & -\frac{(13 k+10)}{4 (k+1)^2 (k+2)},\qquad
c_{24}  =  -\frac{4 (3 k+2)}{(k+1)^2 (k+2)}, \qquad
c_{25}  =  -\frac{i}{2 (k+1)}, \nonu\\
c_{26} & = & -\frac{3 (13 k+10)}{4 (k+2)}, \qquad
c_{27}  =  -\frac{1}{(k+1)}, \qquad
c_{28}  =  \frac{3 i (13 k+10)}{4 (k+1) (k+2)}, \nonu\\
c_{29} & = & -\frac{i (2183 k^3+3427 k^2-808 k-1700)}{80 (k+1)^2 (k+2)^2}, \quad
c_{30}  = -\frac{i (3063 k^3+8627 k^2+8152 k+2460)}{80 (k+1)^2 (k+2)^2}, \nonu\\
c_{31} & = & -\frac{i k (7 k+8) (13 k+10)}{4 (k+1)^2 (k+2)^2}, \qquad
c_{32}  =  -\frac{3 i (3 k+2) (13 k+10)}{4 (k+1) (k+2)^2}, \nonu\\
c_{33} & = & -\frac{(3 k+2)}{(k+1)^2 (k+2)}, \qquad
c_{34}  =  -\frac{2 (5 k+4)}{(k+1)^2 (k+2)},  \qquad
c_{35}= -\frac{(13 k+10)}{4 (k+1)^2 (k+2)}, \nonu\\
c_{36} & = & -\frac{(7 k+6)}{(k+1)^2 (k+2)}, \qquad
c_{37}  =  \frac{(13 k+10)}{4 (k+1)^2 (k+2)}, \qquad 
c_{38} = \frac{3 (3 k+2)}{(k+1)^2 (k+2)}, \nonu\\
c_{39} & = & -\frac{(5 k+2)}{(k+1)^2 (k+2)}, \qquad 
c_{40}= -\frac{i}{2 (k+1)}, \qquad
c_{41}  =  \frac{2}{(k+1)}, \nonu\\
c_{42} & = & -\frac{3 i (13 k+10)}{4 (k+1) (k+2)}, \qquad
c_{43}  =  -\frac{4}{(k+1) (k+2)}, \qquad
c_{44}  =  -\frac{(13 k+10)}{4 (k+1)^2 (k+2)},  \nonu\\
c_{45} & = & -\frac{(13 k+10)}{4 (k+1)^2 (k+2)}, \qquad
c_{46}  =  -\frac{2}{(k+1)^2}, \qquad
c_{47}  =  -\frac{2 k}{(k+1)^2 (k+2)},\nonu\\
c_{48} & = & \frac{i (13 k+10)}{4 (k+1)^2 (k+2)},  \qquad
c_{49} = \frac{i k (13 k+10)}{4 (k+1)^2 (k+2)}, \qquad
c_{50}  =  \frac{8 i (2 k+9)}{5 (k+1) (k+2)}, \nonu\\
c_{51} & = & -\frac{4 i (6 k+7)}{5 (k+1) (k+2)}, \qquad 
c_{52}= \frac{i (13 k+10)}{2 (k+1)^2 (k+2)}, \qquad
c_{53}  =  -\frac{4 i (k^2+3 k+7)}{5 (k+1)^2 (k+2)}, \nonu\\
c_{54} & = & \frac{2 i (k+6) (3 k+1)}{5 (k+1)^2 (k+2)}, \qquad
c_{55}  =  -\frac{1}{(k+1)}, \qquad 
c_{56} = \frac{3 (13 k+10)}{2 (k+1) (k+2)},  \nonu\\
c_{57} & = & \frac{3 (13 k+10)}{2 (k+1) (k+2)}, \qquad
c_{58}  = -\frac{(k-1) (13 k+10)}{4 (k+1)^2 (k+2)}.  \nonu
\eea}

$\bullet$
The coefficients in the OPE between the higher spin $2$ currents and
the higher spin $3$ current

The coefficients appearing in Appendix (\ref{twothree})
lead to
\bea
c_{1}& = & -\frac{8 i k (3 k+1)}{(k+1)^2}, \qquad
c_{2} = 0, \qquad
c_{3} = 3, \qquad
c_{4}  = -\frac{9 (13 k+10)}{2 (k+2)}, \qquad
c_{5}  =  \frac{3}{(k+1)}, \nonu\\
c_{6} & = & -\frac{16 i (2 k+1)}{(k+1) (k+2)}, \qquad
c_{7} = -\frac{3 i (13 k+10)}{(k+1) (k+2)}, \qquad 
c_{8}= \frac{3 (5 k+2)}{(k+1) (k+2)}, \nonu\\
c_{9} & = & -\frac{8 i (2 k+1)}{(k+1)^2 (k+2)}, \qquad 
c_{10}= \frac{8 i (k-1) k}{(k+1)^2 (k+2)},  \quad
c_{11}  = -\frac{3 i (13 k+10) (k^2+k-1)}{2 (k+1)^2 (k+2)},  \nonu\\
c_{12} & = & \frac{8 i (2 k+1)}{(k+1)^2 (k+2)}, \qquad
c_{13}  =  \frac{8 i (2 k+1)}{(k+1)^2 (k+2)}, \qquad
c_{14}  =  -\frac{4 (3 k+1)}{(k+1)^2}, \nonu\\
c_{15} & = & -\frac{3 k (13 k+10)}{4 (k+1)^2 (k+2)}, \qquad
c_{16}  =  \frac{9 (13 k+10)}{4 (k+2)}, \qquad
c_{17}  =  -\frac{3 (13 k+10)}{4 (k+1) (k+2)},\nonu\\
c_{18} & = & -\frac{4 i (2 k+1)}{(k+1)^2 (k+2)}, \qquad
c_{19}  =  \frac{3 i (13 k+10)}{4 (k+1)^2 (k+2)}, \qquad
c_{20}  =  0, \qquad
c_{21}  =  0,\nonu\\
c_{22} & = & \frac{32 i}{3 (k+1)}, \qquad
c_{23}  =  -\frac{16 i}{3 (k+1)}, \qquad
c_{24}  =  0, \qquad
c_{25}  =  0, \qquad
c_{26}  = -\frac{4 i (6 k+7)}{9 (k+1)^2},\nonu\\
c_{27} & = & 0, \qquad
c_{28}  =  \frac{4 i}{3 (k+1)^2}, \qquad
c_{29}  =  \frac{4 i}{3 (k+1)^2}, \qquad
c_{30}  =  -\frac{2}{3 (k+1)^2}, \nonu\\
c_{31} & = & -\frac{4 i}{3 (k+1)^2},\qquad
c_{32}  =  \frac{2 i}{3 (k+1)^2}. \nonu
\eea

$\bullet$
The coefficients in the OPE between the higher spin $\frac{5}{2}$ currents and
the higher spin $\frac{5}{2}$ currents

The coefficients appearing in Appendix (\ref{5half5half})
reduce to
{\small
\bea
c_{1}& = & \frac{8 k^2 (3 k+5)}{(k+1)^2}, \qquad
c_{2}=-\frac{8 i k (3 k+5)}{(k+1)^2}, \qquad
c_{3}= 0, \qquad
c_{4}  =  0, \qquad
c_{5}  =  -\frac{8}{(k+1)}, \nonu\\
c_{6} & = & \frac{8 (5 k^2+8 k+5)}{(k+1)^2}, \qquad
c_{7}  =  \frac{4 (3 k-1)}{(k+1)^2}, \qquad 
c_{8} = 0, \qquad
c_{9}  =  -\frac{4 (5 k+3)}{(k+1)^2}, \nonu\\
c_{10} & = & -\frac{4 i (k+3) (3 k+1)}{(k+1)^2},  \qquad
c_{11}  =  0, \qquad
c_{12}  =  -\frac{4 (8 k^2+17 k+10)}{(k+1)^2 (k+2)},  \nonu\\
c_{13} & = & \frac{4 (8 k^2+17 k+10)}{(k+1)^3 (k+2)},  \qquad
c_{14}  =  \frac{(3 k+4)}{(k+1)}, \qquad
c_{15}  =  0, \qquad
c_{16}  =  -\frac{3 (3 k+4) (13 k+10)}{2 (k+1) (k+2)},\nonu\\
c_{17} & = & \frac{4}{(k+1)}, \qquad
c_{18}  =  -\frac{2 (3 k+5)}{(k+1)^2}, \qquad
c_{19}  =  \frac{4 i}{(k+1)}, \qquad
c_{20}  =  -\frac{3 i (13 k+10)}{(k+1) (k+2)},\nonu\\
c_{21} & = & -\frac{8 i (5 k^2+10 k+4)}{(k+1)^2 (k+2)}, \qquad
c_{22}  =  -\frac{i k (13 k+10)}{(k+1)^2 (k+2)}, \qquad
c_{23}  =  \frac{2 (13 k+10)}{(k+1) (k+2)},\nonu\\
c_{24} & = & \frac{2 i (k^4+17 k^3+33 k^2+17 k-2)}{(k+1)^3 (k+2)}, \qquad
c_{25}  =  -\frac{i (13 k+10) (3 k^2+3 k-4)}{2 (k+1)^2 (k+2)},\nonu\\
c_{26} & = & -\frac{16 i}{(k+1) (k+2)}, \qquad
c_{27}  =  \frac{i (13 k+10)}{(k+1)^2 (k+2)}, \qquad
c_{28}  =  \frac{16 i}{(k+1) (k+2)}, \nonu\\
c_{29} & = & \frac{2 (9 k^2+5 k-10)}{(k+1)^2 (k+2)}, \qquad
c_{30}  =  -\frac{2 (9 k^2+12 k+5)}{(k+1)^3}, \qquad 
c_{31} = -\frac{i (13 k+10)}{(k+1)^2 (k+2)},\nonu\\
c_{32} & = & \frac{(13 k+10) (3 k^2+12 k+8)}{4 (k+1)^3 (k+2)}, \qquad
c_{33}  =  \frac{3 (3 k+4) (13 k+10)}{4 (k+1) (k+2)}, \nonu\\
c_{34} & = & -\frac{(3 k+4) (13 k+10)}{4 (k+1)^2 (k+2)},  \qquad
c_{35}= \frac{i (3 k+4) (13 k+10)}{16 (k+1)^3 (k+2)}, \nonu\\
c_{36} & = & -\frac{i (7 k^2+10 k+4)}{2 (k+1)^3 (k+2)}, \qquad
c_{37}  =  \frac{1}{2}, \qquad 
c_{38} = \frac{3 (13 k+10)}{8 (k+2)}, \qquad
c_{39}  =  -\frac{3 (13 k+10)}{4 (k+2)}, \nonu\\
c_{40} & = & -\frac{(1521 k^3+3893 k^2+3368 k+1028)}{16 (k+1) (k+2)^2}, \qquad
c_{41}  =  -\frac{(9 k+2) (169 k^2+379 k+226)}{16 (k+1) (k+2)^2}, \nonu\\
c_{42} & = & 0, \qquad
c_{43}  =  0, \qquad
c_{44}  =  \frac{1}{(k+1)}, \qquad
c_{45}  =  -\frac{3 (13 k+10)}{4 (k+2)}, \qquad
c_{46}  =  -\frac{(7 k+5)}{(k+1)^2}, \nonu\\
c_{47} & = & \frac{i}{(k+1)}, \qquad
c_{48}  =  -\frac{3 i (13 k+10)}{2 (k+1) (k+2)},  \qquad
c_{49} = -\frac{2 i}{(k+1)^2}, \nonu\\
c_{50} & = & -\frac{3 i (13 k+10)}{2 (k+1) (k+2)}, \qquad
c_{51}  =  -\frac{4 k}{(k+1)^2 (k+2)}, \qquad 
c_{52} = \frac{(13 k+10)}{(k+1)^2 (k+2)}, \nonu\\
c_{53} & = & \frac{(9 k^2+65 k+38)}{(k+1)^2 (k+2)}, \qquad 
c_{54}= \frac{10 i (3 k+2)}{(k+1)^2 (k+2)}, \qquad
c_{55}  =  -\frac{(7 k+6)}{(k+1)^3 (k+2)}, \nonu\\
c_{56} & = & -\frac{(13 k+10)}{(k+1)^3 (k+2)}, \qquad
c_{57}  =  -\frac{(443 k^4+1442 k^3+1055 k^2-860 k-724)}
{16 (k+1)^3 (k+2)^2},  \nonu\\
c_{58} & = & -\frac{(699 k^4+2498 k^3+2383 k^2-476 k-916)}
{16 (k+1)^3 (k+2)^2},  \nonu\\
c_{59}& = & -\frac{(443 k^4+1458 k^3+655 k^2-1948 k-1172)}
{16 (k+1)^3 (k+2)^2}, \nonu\\
c_{60} & = & -\frac{(699 k^4+2498 k^3+1903 k^2-1692 k-1428)}
{16 (k+1)^3 (k+2)^2},  \nonu\\
c_{61} & = & -\frac{(13 k+10) (4 k^3+7 k^2+10 k+12)}{4 (k+1)^3 (k+2)^2}, \nonu\\
c_{62} & = &  -\frac{(13 k+10) (9 k^3+18 k^2+7 k-10)}{4 (k+1)^3 (k+2)^2}, \nonu\\
c_{63} & = & \frac{(13 k+10) (2 k^3+15 k^2+15 k-2)}{4 (k+1)^3 (k+2)^2}, \qquad
c_{64}  =  \frac{12 (3 k+2)}{(k+1) (k+2)}, \nonu\\
c_{65} & = & \frac{(443 k^3+583 k^2-2024 k-1556)}{16 (k+1)^2 (k+2)}, \qquad 
c_{66}= -\frac{(7 k+6)}{(k+1)^3 (k+2)}, \nonu\\
c_{67} & = & \frac{5 (3 k+2)}{(k+1)^3 (k+2)}, \qquad
c_{68}  =  -\frac{(13 k+10)}{2 (k+1)^3 (k+2)},  \qquad
c_{69}  =  -\frac{3 (13 k+10)}{8 (k+2)},  \nonu\\
c_{70} & = & -\frac{3 i (13 k+10)}{4 (k+1) (k+2)},  \qquad
c_{71}  =  -\frac{i}{(k+1)^2}, \qquad
c_{72}  =  -\frac{5 i (3 k+2)}{(k+1)^2 (k+2)}, \nonu\\
c_{73} & = & \frac{i (2 k^2+47 k+38)}{(k+1)^3 (k+2)}, \qquad
c_{74}  =  -\frac{i (k-5) (13 k+10)}{4 (k+1)^3 (k+2)}, \qquad
c_{75}  =  -\frac{1}{2 (k+1)},\nonu\\
c_{76} & = & \frac{3 (13 k+10)}{4 (k+1) (k+2)}, \qquad
c_{77}  =  \frac{2 (2 k+1)}{(k+1)^2}, \qquad
c_{78}  =  -\frac{4}{(k+1)},\nonu\\
c_{79} & = & -\frac{4 i k}{(k+1)^2 (k+2)}, \qquad
c_{80}  =  \frac{i (13 k+10)}{2 (k+1)^2 (k+2)}, \qquad
c_{81}  =  \frac{8 i}{(k+1) (k+2)},\nonu\\
c_{82} & = & -\frac{2 (2 k^2+15 k+10)}{(k+1)^2 (k+2)}, \qquad
c_{83}  =  \frac{4 (k+4)}{(k+1) (k+2)}, \qquad
c_{84}  =  -\frac{12 i}{(k+1)^2 (k+2)},\nonu\\
c_{85} & = & \frac{2 i (3 k+2)}{(k+1)^3 (k+2)},\qquad
c_{86}  =  \frac{2 i (3 k+4)}{(k+1)^3 (k+2)}, \qquad
c_{87}  =  \frac{i (13 k+10)}{4 (k+1)^3 (k+2)},\nonu\\
c_{88} & = & \frac{i (3 k^4+26 k^3+55 k^2+34 k-4)}{3 (k+1)^3 (k+2)}, \qquad 
c_{89} = \frac{i (13 k+10) (9 k^2+7 k-6)}{24 (k+1)^3 (k+2)},\nonu\\
c_{90} & = & -\frac{i}{2 (k+1)}, \qquad
c_{91} = -\frac{1}{(k+1)}, \qquad
c_{92} = \frac{3 i (13 k+10)}{4 (k+1) (k+2)}, \qquad
c_{93} = \frac{i}{(k+1)^2}, \nonu\\
c_{94} & = & -\frac{4 (3 k+2)}{(k+1)^2 (k+2)}, \qquad
c_{95} = -\frac{i (13 k+10)}{4 (k+1)^2 (k+2)}, \qquad 
c_{96} = -\frac{i (7 k+6)}{(k+1)^2 (k+2)}, \nonu\\
c_{97} & = & \frac{(7 k+6)}{(k+1)^3 (k+2)}, \qquad 
c_{98} = \frac{(13 k+10)}{2 (k+1)^3 (k+2)},  \qquad
c_{99} = -\frac{i (k^2+30 k+24)}{(k+1)^3 (k+2)},  \nonu\\
c_{100} & = & \frac{i (k^2+18 k+16)}{(k+1)^3 (k+2)},  \qquad
c_{101} = \frac{i (k^2+28 k+24)}{(k+1)^3 (k+2)}, \qquad
c_{102} = -\frac{i (k^2+6 k+4)}{(k+1)^3 (k+2)}, \nonu\\
c_{103} & = & -\frac{i (k+3) (13 k+10)}{4 (k+1)^3 (k+2)}, \qquad
c_{104} = \frac{i (13 k+10)}{4 (k+1)^2 (k+2)}, \qquad
c_{105} = \frac{i (13 k+10)}{4 (k+1)^3}, \nonu\\
c_{106} & = & -\frac{12 i}{(k+1)^2 (k+2)}, \qquad 
c_{107} = -\frac{i (k^2+8 k+8)}{(k+1)^3 (k+2)}, \qquad
c_{108} = -\frac{i (k^2+12 k+8)}{(k+1)^3 (k+2)}, \nonu\\
c_{109} & = & -\frac{i (k+4) (13 k+10)}{4 (k+1)^3 (k+2)}, \qquad
c_{110} = -\frac{(k^3-5 k^2-3 k+6)}{(k+1)^3 (k+2)}, \nonu\\
c_{111} & = & -\frac{(k^3-2 k^2-18 k-12)}{(k+1)^3 (k+2)}, \qquad
c_{112} = -\frac{(13 k+10)}{4 (k+1) (k+2)},  \nonu\\
c_{113} & = & -\frac{(13 k+10) (k^2+3 k+1)}{4 (k+1)^3 (k+2)}, \qquad
c_{114} = \frac{(13 k^3+48 k^2+98 k+52)}{(k+1)^3 (k+2)}, \nonu\\
c_{115} & = & \frac{(7 k+6)}{(k+1)^3 (k+2)},  \qquad
c_{116} = -\frac{3 (13 k+10)}{4 (k+1) (k+2)}, \qquad
c_{117} = \frac{(13 k+10)}{4 (k+1)^2 (k+2)}, \nonu\\
c_{118} & = & \frac{i (13 k+10)}{4 (k+1)^3 (k+2)}. \nonu
\eea}

$\bullet$
The coefficients in the OPE between the higher spin $\frac{5}{2}$ currents and
the higher spin $3$ current

The coefficients appearing in Appendix (\ref{5halfthree})
reduce to
\bea
c_{1} & = & -\frac{4 (15 k+1)}{(k+1)}, \qquad
c_{2} = 0, \qquad
c_{3} = \frac{8}{(k+1)}, \qquad
c_{4} = -\frac{8 i (k+2)}{(k+1)^2}, \qquad
c_{5} = 0, \nonu\\ 
c_{6} & = & -\frac{4 (k-1) (5 k+7)}{(k+1)^2}, \qquad
c_{7} = \frac{7}{2}, \qquad 
c_{8} = \frac{21 (13 k+10)}{8 (k+2)}, \qquad
c_{9} = -\frac{21 (13 k+10)}{4 (k+2)}, \nonu\\
c_{10} & = & -\frac{3 (3549 k^3+9041 k^2+7688 k+2228)}{16 (k+1) (k+2)^2}, \nonu\\
c_{11} & = & -\frac{(10647 k^3+26755 k^2+21592 k+5212)}
{16 (k+1) (k+2)^2},  \nonu\\
c_{12} & = & \frac{21 (13 k+10)}{4 (k+2)}, \qquad
c_{13} = \frac{(7 k+6)}{(k+1)^2 (k+2)},  \nonu\\
c_{14} & = & \frac{i (3949 k^3+10497 k^2+8968 k+2164)}{16 (k+1)^2 (k+2)^2}, \nonu\\
c_{15} & = & \frac{i (3197 k^3+5809 k^2+1416 k-204)}{16 (k+1)^2 (k+2)^2},\nonu\\
c_{16} & = & \frac{(49 k+22)}{(k+1)^2 (k+2)}, \qquad
c_{17} = \frac{1}{2 (k+1)}, \qquad
c_{18} = \frac{i}{2 (k+1)}, \nonu\\
c_{19} & = & -\frac{3 i (13 k+10)}{4 (k+1) (k+2)}, \qquad
c_{20} = \frac{(13 k+10)}{4 (k+1)^2 (k+2)},\nonu\\
c_{21} & = & \frac{i (13 k+10) (63 k^2+103 k+38)}{8 (k+1)^2 (k+2)^2}, \qquad
c_{22} = \frac{i (11 k+1) (13 k+10)}{2 (k+1) (k+2)^2}, \nonu\\
c_{23} & = & -\frac{2 (55 k+26)}{(k+1) (k+2)}, \qquad
c_{24} = \frac{(3469 k^4+5492 k^3+1335 k^2+716 k+652)}
{16 (k+1)^2 (k+2)^2},\nonu\\
c_{26} & = & 0, \qquad
c_{27} = 0, \qquad
c_{28} = 0,  \qquad
c_{29} = -\frac{40}{7 (k+1)}, \qquad
c_{30} = \frac{12}{7 (k+1)}, \nonu\\
c_{31} & = & \frac{4}{7 (k+1)}, \qquad
c_{32} = 0, \qquad
c_{33} = 0, \qquad
c_{34} = \frac{64}{7 (k+1)},  \qquad
c_{35} = -\frac{48}{7 (k+1)}, \nonu\\
c_{36} & = & \frac{4 (6 k^2-11 k-11)}{21 (k+1)^2}, \qquad
c_{37} = 0, \qquad 
c_{38} = -\frac{4 i (k^2+16 k+16)}{7 (k+1)^2 (k+2)}, \nonu\\
c_{39} & = & \frac{8 i (5 k+14)}{7 (k+1)^2}, \qquad 
c_{40} = -\frac{12 i (k^2+2 k+4)}{7 (k+1)^2 (k+2)}, \qquad
c_{41} = \frac{48}{7 (k+1)^2 (k+2)}, \nonu\\
c_{42} & = & -\frac{32}{7 (k+1)^2 (k+2)}, \qquad
c_{43} = -\frac{32}{7 (k+1)^2 (k+2)}, \qquad
c_{44} = \frac{32}{7 (k+1)^2 (k+2)},  \nonu\\
c_{45} & = & -\frac{8 (7 k+10)}{7 (k+1)^2 (k+2)}, \qquad
c_{46} = \frac{8 (7 k+4)}{7 (k+1)^2 (k+2)}, \qquad
c_{47} = \frac{16}{7 (k+1)},\nonu\\
c_{48} & = & -\frac{12}{7 (k+1)}, \qquad
c_{49} = -\frac{5 i}{7 (k+1)}, \qquad
c_{50} = \frac{2 i}{7 (k+1)}, \qquad
c_{51} = -\frac{3 i (13 k+10)}{7 (k+1) (k+2)}, \nonu\\
c_{52} & = & 0, \qquad
c_{53} = \frac{(13 k+10)}{7 (k+1)^2 (k+2)}, \qquad 
c_{54} = \frac{i (2 k-5) (13 k+10)}{14 (k+1)^2 (k+2)}, \nonu\\
c_{55} & = & -\frac{i (5 k-7) (13 k+10)}{14 (k+1)^2 (k+2)}, \qquad 
c_{56} = -\frac{(13 k+10)}{(k+1)^2 (k+2)}, \qquad
c_{57} = -\frac{i (13 k+10)}{7 (k+1)^2 (k+2)}. \nonu
\eea

$\bullet$
The coefficients in the OPE between the higher spin $3$ current and
the higher spin $3$ current

The coefficients appearing in Appendix (\ref{threethree})
reduce to
{\small
\bea
c_{1} & = & \frac{8 k^2 (15 k+1)}{(k+1)^2}, \qquad
c_{2} = 0,  \qquad
c_{3} = -\frac{24}{(k+1)}, \qquad
c_{4} = \frac{16 (15 k^2+8 k-4)}{(k+1)^2}, \nonu\\
c_{5} & = & -\frac{4 (3 k-2)}{(k+1)^2}, \qquad
c_{6} = 0, \qquad
c_{7} = 4, \qquad 
c_{8} = \frac{3 (13 k+10)}{k+2}, \nonu\\
c_{9} & = & -\frac{6 (13 k+10)}{(k+2)}, \qquad
c_{10} = -\frac{6 (13 k+10)}{(k+2)}, \qquad
c_{11} = -\frac{3 (13 k+10)}{4 (k+2)}, \qquad
c_{12} = \frac{24}{(k+1)}, \nonu\\
c_{13} & = & -\frac{(1521 k^3+3901 k^2+3400 k+1060)}{2 (k+1) (k+2)^2}, \qquad
c_{14} = -\frac{3 (507 k^3+1279 k^2+1048 k+268)}{2 (k+1) (k+2)^2}, \nonu\\
c_{15} & = & 0, \qquad
c_{16} = 0, \qquad
c_{17} = -\frac{3 i (13 k+10)}{(k+1) (k+2)}, \qquad
c_{18} = \frac{128 (2 k+1)}{(k+1) (k+2)},\nonu\\
c_{19} & = & -\frac{32 (2 k+1)}{(k+1)^2 (k+2)}, \qquad
c_{20} = 0, \qquad
c_{21} = \frac{3 (193 k^3+413 k^2-56 k-92)}{2 (k+1)^2 (k+2)},\nonu\\
c_{22} & = & \frac{32 i (2 k+1)}{(k+1)^2 (k+2)}, \qquad
c_{23} = \frac{8 (7 k^2+33 k+14)}{(k+1)^2 (k+2)}, \qquad
c_{24} = \frac{2 (2 k+1)}{(k+1)^3 (k+2)},\nonu\\
c_{25} & = & -\frac{(483 k^4+1562 k^3+1103 k^2-796 k-468)}
{4 (k+1)^3 (k+2)^2},\nonu\\
c_{26} & = & -\frac{(579 k^4+1946 k^3+1839 k^2+36 k-212)}
{4 (k+1)^3 (k+2)^2}, \qquad
c_{27} = -\frac{3 k^2 (13 k+10)}{(k+1)^2 (k+2)^2}, \nonu\\
c_{28} & = & -\frac{3 k (3 k+2) (13 k+10)}{2 (k+1)^2 (k+2)^2}, \qquad
c_{29} = 0, \qquad
c_{30} = \frac{1}{(k+1)}, \nonu\\
c_{31} & = & -\frac{2 (15 k^2+8 k-4)}{3 (k+1)^2}, \qquad
c_{32} = \frac{(3 k-2)}{6 (k+1)^2}, \qquad
c_{33} = 0. \nonu
\eea}
One can analyze these coefficients under the $k \rightarrow \infty$ limit.


\end{document}